國立臺灣大學理學院物理學研究所

博士論文

Department of Physics

College of Science

National Taiwan University

Doctoral Dissertation

# 一維 Sine-Gordon 方程式和量子自旋傳輸

# Sine-Gordon Equation and Quantum spin transport in one-dimension

研究生：郭南宏

Student: Nan-Hong Kuo

指導教授：胡崇德 博士

Advisor: Dr. Chong-Der Hu

# 中文摘要

我研究一維自旋 1/2 反鐵磁系統的量子自旋傳輸機制. 我先將此自旋鏈轉換成費米子，再做玻色化近似。最後變成(雙) Sine-Gordon 方程式，在第二章我證明解的等價性。考慮外場的變化，我加入絕熱相位並考慮在不同的邊界條件下此方程式的解。還原此解到原來的自旋鏈物理系統. 我觀察到在一般的固定邊界條件下有自旋=1 通過整個系統。而此結論不同於一般的結論——自旋是由邊界態來傳輸。這寫在第三章。而在其他等價性不同的邊界條件下，自旋是累積在邊界。其機制不同於前，我也提出一個拓樸圖像。這是第四章的內容。最後在第五章，我利用 Mobius 轉換找到雙 Sine-Gordon 方程式含絕熱相位的數值精確解。在有限與無限系統，相對應解也有不同。另外我也討論微擾解的方法。

**Figure captions: (圖目錄)**

**Table captions: (表目錄)**

**Sine-Gordon Equation and Quantum spin transport in one-dimension**

Nan-Hong Kuo

## Abstract

We studied the spin transport mechanism in a S=1/2 antiferromagnetic chain. The spin chain is mapped into a fermion system, where equation of motion is transformed into a (**Double**) **Sine-Gordon Equation** ((**D**)**SGE**) with the approach of bosonization. We studied first, the non-interacting case. By varying adiabatically a phase angle $\varphi$ which comes from external fields, the spin states change between the Néel state and dimer state and a quantized spin $S = 1$ is transported by the bulk state from one end of the spin chain to the other. We have also considered the interacting case. I found that it is equivalent to the situation of twisted boundary condition. The spin states possess topological meaning. I also transform the solutions of **SGE** in Wazwaz [**20**] into another form we are familiar with. Finally, I use **Möbius transformation** to numerically solve asymmetric **DSGE**, which was not solved before.

# Table of contents:

# Chapter 1. Introduction:

One-dimensional systems have been studied for a long time. They have uniques physical properties which are very different from those of two or three dimensional systems. Many special techniques, such as Bethe ansatz, [**1**], bosonization, [**2**], and matrix product, [**3**], had been developed to study one-dimensional systems.

There are also physical systems which are one-dimensional, such as spin chains, [**4**], optical fibers, [**5**], and polymers, [**6**]. Hence, this is still a relevant field with very rich physics to be studied.

One of the important equations of one -dimensional systems is the Sine-Gordon Equation (**SGE**). It will be introducted in this chapter. I will transform the one-dimensional spin chain problem into a **SGE** and study the spin transport. Double Sine-Gordon Equation (**DSGE**) is also a hot topic in one-dimension if our system has more complicate interactions or when we want to know the excited states in optics systems. I study **DSGE** with an asymmetric phase. It is different from **SGE** because the asymmetric phase is not changed by change variable. But I was able to solve this "non-integrable" equation by mathematical techinque and the solutions of asymmetric **DSGE** finally become roots of several polynomial equations. I think this is a break through and also a pioneer work because a non-linear equation, such as asymmetric **DSGE**, become a " linear" one although several parameters must be found numerically.

The plan of each chapter is the following: In Chapter **1.1**, one-dimensional quantum physics mathematical tool-bosonization is introduced. In Chapter **1.2**, I introduce the theory of **SGE**. Chapter **1.1** and **1.2** are background knowledge. In Chapter **2**, I transform **Wazwaz's solutions** of **SGE** [**20**] into familiar forms if we specify the condition. In Chapter **3** , I solve **SGE** with adiabatic parameter: $\varphi$ in certain boundary condition, I study and discuss the static soliton, to show spin pump may also happen through bulk state. This not a usual phenomena in 1+1 ( space + $\varphi$) dimensional physics because in quantum Hall effect charge are transported through edge states. In Chapter **4**, I treat the problem with twisted boundary condition **SGE** which have many aspects different from before. In Chapter **5**, in the first part, I numericaliy solve **asymmetric DSGE** for infinite system using **Möbius transformation**. In the second part of Chapter **5**, I study asymmetric **DSGE** in finite system. This one have some different aspects from the infinite case, including **1**, no phase transition and **2**, the perturbation method which is different from usual perturbation in field theory seems not bad if the system is finite.

Each chapter can be thinked a independent chapter. Readers who are interested in one subject of my Ph. D. thesis can begin with each chapter.

## 1.1 Quantum physics in one dimension and bosonization

### 1.1.1 Quantum physics in one dimension

The difference between one and higher dimension is that perturbation theory is wrong in one dimension. Because in high dimension nearly free quasiparticle excitations exist, but in one dimension, particle that tries to move has to push its neighbors because of interactions. So in one dimension, any excitation is a collective one.

We can use formal way to illustrate this thing. In one dimension, we can use Luttinger fields. That means the spectrum is linear. Because of particle-hole spectrum in one dimension in low energy, this approximation is good. I will describe this point in detail in the next section. We can suppose the free Hamiltion as: $H_{free} = \sum_k \varepsilon_k c^+_{r,k} c_{r,k}$. We can see what happen if we add usual perturbation term in the Hamiltion like:

$$H_{dens} = \int d^d x V(x,t)\rho(x) \qquad 1.1.1.1$$

and the susceptibility measures the response due to perturbation:

$$\chi(q,w) = \frac{1}{\Omega}\sum_k \frac{f_F(\xi_k) - f_F(\xi_{k+q})}{w + \xi(k) - \xi(k+q) + i\delta} \qquad 1.1.1.2$$

where $\xi(k) = \varepsilon_k - \varepsilon_F$ and $\varepsilon_F$ is the Fermi energy, $\Omega$ is the volume of the system and $f_F$ is the Fermi-Dirac distribution function. If one can find a wavevector $Q$ such that both $\xi(k)$ and $\xi(k+Q)$ are zero, this leads to singularity. In high dimensions, this occurs only for a very limited set of points. Because of the integration over $k$ in **Eq**.(**1**.**1**.**2**), the singularity of the denominator is smoothed out. But in one dimension, the Fermi surfaces consists of only two points and this leads to serious singularity at $Q = 2k_F$.

Another difference between one and higher dimension is particle-hole spectrum. When one destroys a particle with momentum $k$ and creates a particle with momentum $k+q$. The energy of the particle-hole excitations is a function of their momentum $q$. One has in general a continuum of energies. In higher dimensions, for $q < 2k_F$, the particle-hole excitations lead to a continuum extending to zero energy for all $q$ vectors smaller than $2k_F$. In one dimension, the Fermi surface is reduced to two Fermi points. There is no angular degree of freedom. The only way to get a low-energy excitation is to destroy and create pairs close to the Fermi points, the only places where the particle-hole energy can vanish for $q = 0$ and for $q = 2k_F$. We look at excitations at small $q$, The energy of a particle-hole excitation is

$$E_k(q) = \xi(k+q) - \xi(k) \qquad 1.1.1.3$$

where $\xi(k)$ should be occupied and $\xi(k+q)$ empty and:

$$\xi(k) = \frac{k^2 - k_F^2}{2m} \qquad 1.1.1.4$$

For $k \in [k_F - q, k_F]$. The dispersion is

$$\delta E(q) = \max(E_k(q)) - \min(E_k(q)) = [\xi(k_F+q) - \xi(k_F)] - [\xi(k_F) - \xi(k_F - q)] = \frac{q^2}{m} \qquad 1.1.1.5$$

Let **1**. $E(q) = \frac{k_F q}{m} = v_F q$ be the average value $E(q)$ of $E_k(q)$. **2**. $E = \frac{p^2}{2m} = \frac{(mv_F)^2}{2m}$

$$\delta E(q) = \frac{q^2}{m} = \frac{E(q)^2}{mv_F^2} = \frac{E(q)^2}{2E} \qquad 1.1.1.6$$

What I show above can explain in one dimension, we can calculate the energy perturbation ( i.e. not good for state perturbation because the collective mode in one dimension), such as Form-Factor perturbation theory, see [**5**.**5**], [**5**.**6**], [**5**.**7**] and [**5**.**8**] for detail. The particle-hole excitations represent one of the collective mode of "bosonic-excitation", It destroy and create a fermion each once one Reader can see in the next section, **eq**. (**1**.**1**.**2**.**1**), the particle-hole excitation operator: $b_q$ or $b_q^+$ are sum of all $k$ of $c^+_{k+q}c_k$. That means collective states, The particle-hole excitation in **eq**. (**1**.**1**.**2**.**1**) is actually a coherence state. and it has the energy dispersion: $\delta E(q)$. So in one dimension, we can perturb and count correct by the dispersion. This "particle-hole" excitation as a "bonsonic-excitation" is the origin idea to make one-dimension fermions by pair to become a boson. And it is also the origin idea of techinque of "bosonization" and I will show the "particle-hole excitation" in one dimension is actually a bosonic excitation in the next section.

## 1.1.2 Luttinger liquids and Bosonization

In one dimension, we can bosonize the fermion by the following procedure:

$$b_q^+ := \frac{i}{\sqrt{n_q}} \sum_{k=-\infty}^{\infty} c_{k+q}^+ c_k \qquad 1.1.2.1a$$

$$b_q := \frac{-i}{\sqrt{n_q}} \sum_{k=-\infty}^{\infty} c_k^+ c_{k+q} \qquad 1.1.2.1b$$

where $q = \frac{2\pi}{L} n_q$ with $n_q$ being an integer. We can check the commutation relations to confirm $b_q$ is a boson operator:

$$[b_g, b_{q'}] = [b_q^+, b_{q'}^+] = 0 \qquad 1.1.2.2a$$

$$[b_q, b_{q'}^+] = \frac{\delta_{q,q'}}{n_q} \sum_{k=-\infty}^{\infty} (c_k^+ c_k - c_{k+q}^+ c_{k+q}) = \frac{\delta_{q,q'}}{n_q} \sum_{k=-\infty}^{\infty} (\Theta(-k) - \Theta(-k-q)) = \frac{\delta_{q,q'}}{n_q} n_q = \delta_{q,q'} \qquad 1.1.2.2b$$

The equality: $\sum_{k=-\infty}^{\infty} (\Theta(-k) - \Theta(-k-q)) = n_q$ is only correct for infinite Luttinger system. Define $\rho_R =: R^+(x)R(x)$ : which is the right-moving fermion densities and $\rho_L(x) =: L^+(x)L(x)$ : which is the left-moving fermion densities, then

$$\rho(x) = \rho_R(x) + \rho_L(x) \qquad 1.1.2.3a$$

$$j(x) = \rho_R(x) - \rho_L(x) \qquad 1.1.2.3b$$

One can further define $\phi_R(x)$ and $\phi_L(x)$ as:

$$\rho_R(x) = \frac{\partial_x \phi_R(x)}{2\pi} \qquad 1.1.2.4a$$

$$\rho_L(x) = \frac{\partial_x \phi_L(x)}{2\pi}. \qquad 1.1.2.4b$$

Hence

$$\phi_R(x) = \frac{2\pi}{\sqrt{L}} \sum_{p>0} e^{-\frac{\alpha p}{2}} \frac{1}{ip} \{e^{ipx} \rho_R(p) - e^{-ipx} \rho_R(-p)\} \equiv \phi_R^-(x) + \phi_R^+(x) \qquad 1.1.2.5a$$

$$\phi_L(x) = \frac{2\pi}{\sqrt{L}} \sum_{p>0} e^{-\frac{\alpha p}{2}} \frac{1}{ip} \{e^{ipx} \rho_L(p) - e^{-ipx} \rho_L(-p)\} \equiv \phi_L^-(x) + \phi_L^+(x) \qquad 1.1.2.5b$$

where $\alpha$ is a cut off with $\alpha \sim 1$. We express $\rho_R$ and $\rho_L$ as bosonic creation and annihilation operators $b^+$ and $b$. With $p > 0$, we obtain:

$$\rho_R(p) = \sqrt{\frac{p}{2\pi}} b_{1,p}; \rho_R(-p) = \sqrt{\frac{p}{2\pi}} b_{1,p}^+ \qquad 1.1.2.6a$$

$$\rho_L(p) = \sqrt{\frac{p}{2\pi}} b_{2,-p}^+; \rho_L(-p) = \sqrt{\frac{p}{2\pi}} b_{2,-p}^+ \qquad 1.1.2.6b$$

where $[b_{i,p}, b_{j,p'}^+] = \delta_{i,j} \delta_{p,p'}$. This corresponds to the anomalous commentator equality mensioned above. We further define

$$\theta_+(x) = \phi_R(x) + \phi_L(x) \qquad 1.1.2.7a$$

$$\theta_-(x) = \phi_R(x) - \phi_L(x) \qquad 1.1.2.7b$$

So **Eq.** (**1.1.2.4**) can expressed as:

$$\rho(x) = \frac{\partial_x \theta_+(x)}{2\pi} \quad 1.1.2.8a$$

$$j(x) = \frac{\partial_x \theta_-(x)}{2\pi} \quad 1.1.2.8b$$

and the commutations is

$$[\theta_\pm(x), \theta_\pm(x')] = 0; [\theta_+(x), \theta_-(x')] = 2\pi i sgn(x-x') \quad 1.1.2.9$$

The momentum is

$$\Pi(x) = \frac{-1}{4\pi}\partial_x \theta_-(x) \quad 1.1.2.10$$

And right and left going fermions can be written as

$$R(x) = \frac{1}{\sqrt{2\pi\alpha}}\eta_1 e^{i\phi_R(x)} \quad 1.1.2.11a$$

$$L(x) = \frac{1}{\sqrt{2\pi\alpha}}\eta_2 e^{-i\phi_L(x)} \quad 1.1.2.11b$$

where $\{\eta_i, \eta_j\} = 2\delta_{ij}$, $\eta_i$ is called Majorana-fermions or Kelin factor in order to keep the statistic property of fermions.

Another direct way to see the bosonization of fermion is propositional to the exponential of boson operator is to see their commutator.

## 1.1.3 Construct soliton operator for the quantized Sine-Gordon Equation

When we do problem in one dimensional quantum antiferromagnets, we can construct Kink operator when we do Jordon-Wigner transformation, which is:

$$K(n) = \exp(i\pi \sum_{j=1}^{n-1} S^+(j)S^-(j)) \quad 1.1.3.1$$

It has the property like:

$$K(n)|\frac{1}{2}\cdots\frac{1}{2}\rangle = i^{n-1}|-\frac{1}{2}\cdots-\frac{1}{2},\frac{1}{2}\cdots\frac{1}{2}\rangle \quad 1.1.3.2$$

This is just the soliton operator. And fermions $c(n)$ and $c^+(n)$ are related to spin operators $S^+(n)$ and $S^-(n)$ by:

$$c(n) \equiv K(n)S^-(n) = e^{i\pi \sum_{j=1}^{n-1} S^+(j)S^-(j)} S^-(n) \quad 1.1.3.3$$

$$c^+(n) \equiv S^+(n)K^+(n) = S^+(n)e^{-i\pi \sum_{j=1}^{n-1} S^+(j)S^-(j)} \quad 1.1.3.4$$

We can bosonization the fermion operator sfter applying Jordon-Wigner transformations. The soliton operator can also be constructed directly from [**18**]. An operator $\psi(x)$ annihilating a soliton at $x$ must increase $\theta_+$ by $2\pi\beta$ in the region to the left of $x$. and $\beta$ is an arbitrary real number.

$$[\theta_+(y), \psi(x)] = 2\pi\beta\psi(x), \text{ when } (y < x) \quad 1.1.3.5$$

$$[\theta_+(y), \psi(x)] = 0, \text{ when } (y > x) \quad 1.1.3.6$$

How do I construct $\psi(x)$? The result is shown in **eq**. (**1.1.3.13**). To see it more directly,

$$O_\alpha = e^{i\alpha\theta_+(x)} \quad 1.1.3.7$$

$$Q_\beta = e^{i\beta\int_{-\infty}^{x_1} \Pi(x_0, x_1')dx_1'} \quad 1.1.3.8$$

Where $\Pi(x_0, x_1) = \partial_{x_0}\theta_+(x_0, x_1)$, $Q_\beta$ will shift the value of $\theta_+(x_0, x_1')$ to $\theta_+(x_0, x_1') + \beta$ for all $x_1' < x_1$. Thus, $Q_\beta(x)$ creates a coherent state which we can call a soliton.

$$Q_\beta(x)|\{\theta_+(x_0, x_1')\}\rangle = |\{\theta_+(x_0, x_1') + \beta\theta(x_1 - x_1')\}\rangle \quad 1.1.3.9$$

where $\theta(x_1 - x_1')$ is the step function by usual sense. Because it is like the "time-ordering" of two operator: $Q_\beta(x)$ and $\theta_+(x_0, x_1')$, so by usual field theory construction, **Eq**.(**2.3.9**) can be transformed into a communator, means:

$$[Q_\beta(x_0, x_1), \theta_+(x_0, x_1')] = i\beta Q_\beta(x_0, x_1)\theta(x_1 - x_1') \quad 1.1.3.10$$

which is basically the form and meaning of **Eq**. (**1.1.3.5**) and (**1.1.3.6**). But I can go further, by **Baker-Hausdorff** formula:

$$e^{\hat{A}}e^{\hat{B}} = e^{\hat{A}+\hat{B}+\frac{1}{2}[\hat{A},\hat{B}]} = e^{\hat{B}}e^{\hat{A}}e^{[\hat{A},\hat{B}]} \quad 1.1.3.11$$

and from bosonization, we know:

$$[\theta_+(x), \Pi(x')] = i\delta(x - x') \quad 1.1.3.12$$

One can construct an operator like those in **eq**. (**1.1.3.3**) and (**1.1.3.4**).

$$\psi_{\alpha,\beta}(x) =: O_\alpha(x)Q_\beta(x) := e^{\frac{-i}{2}\alpha\beta}e^{i\alpha\theta_+(x)+i\beta\int_{-\infty}^{x_1} \Pi(x_0, x_1')dx_1'} \quad 1.1.3.13$$

and

$$\psi_{\alpha,\beta}(x)\psi_{\alpha,\beta}(x') = \psi_{\alpha,\beta}(x')\psi_{\alpha,\beta}(x)e^{-i\Phi(x,x')} \quad 1.1.3.14$$

Again by **Baker-Hausdorff** formula, we can obtain:

$$i\Phi(x, x') = -i\alpha\beta \quad 1.1.3.15$$

By requiring $\psi_{\alpha,\beta}(x)$ to be a fermion operator, the relation must hold:

$$\alpha\beta = \pm\pi \quad 1.1.3.16$$

It is useful to write left and right components of the Fermi field in the form:

$$R(x) = \psi_1(x) = -i\sqrt{\frac{c\mu}{2\pi}}\, e^{\frac{\mu}{8\epsilon}} : e^{-i2\pi\beta\int_{-\infty}^{x_1} \Pi(x_0, x_1')dx_1' + \frac{i}{2\beta}\theta_+(x)} : \quad 1.1.3.17$$

$$L(x) = \psi_2(x) = \sqrt{\frac{c\mu}{2\pi}}\, e^{\frac{\mu}{8\epsilon}} : e^{-i2\pi\beta\int_{-\infty}^{x_1} \Pi(x_0, x_1')dx_1' - \frac{i}{2\beta}\theta_+(x)} : \quad 1.1.3.18$$

where $\psi_1$ and $\psi_2$ in **eq**. (**1.1.3.17**), (**1.1.3.18**) is the two kinds of $\psi_{\alpha,\beta}$ in **eq**. (**1.1.3.13**) by

choosing different sign of **eq**. (**1**.**1**.**3**.**16**). And $\beta$ in **eq**. (**1**.**1**.**3**.**17**) and (**1**.**1**.**3**.**18**) is an arbitrary constant. The phase factor and constants in front of the operators $R$ and $L$ are chosen so that $\gamma$-matrices have the canonical form:

$$\gamma^0 = \sigma^1, \gamma^1 = i\sigma_2, \gamma^5 = \gamma^0\gamma^1 = -\sigma_3 \quad 1.1.3.19$$

I also show below the same content but with different representation which are often used in many books. There are abnormal commutators used at infinite one-dimension system. Let $R(i), L(i)$ represent right-going and left-going fermions, then:

$$\ll R(1)R^+(2) \gg = G_R(\bar{z}_{12}) = \frac{1}{2\pi\bar{z}_{12}} \quad 1.1.3.20a$$

$$\ll L(1)L^+(2) \gg = G_L(z_{12}) = \frac{1}{2\pi z_{12}} \quad 1.1.3.20b$$

$z_{12} = |z_1 - z_2|$ and $z_j = x_j + it_j, j = 1,2$. If we let $\bar{J}(q) = \sum_q R^+(p+q)R(p)$, and simply change the variables without considering the infinity, we will get $[\bar{J}(q), \bar{J}(p)] = 0$. The reason is the same as **eq**. (**1**.**1**.**2**.**2b**). We should not simply change $k+q$ into $k$ and let $\sum_{k=-\infty}^{\infty}(\Theta(-k) - \Theta(-k-q))$ become 0! Instead we consider Feymann diagram and using **eq**. (**1**.**1**.**3**.**20**) and obtain the result:

$$\ll J(z_1)J(z_2) \gg = \frac{1}{4\pi^2|z_{12}|^2} \quad 1.1.3.21$$

I can find the connection between abnormal commutator, **eq**. (**1**.**1**.**2**.**4**) and bosonized relation, **eq**. (**1**.**1**.**2**.**2b**). Because we know:

$$\ll [A(x), B(y)] \gg = \lim_{\tau \to 0^+} [\ll A(\tau, x)B(0, y) \gg - \ll A(-\tau, x)B(0, y) \gg] \quad 1.1.3.22$$

so

$$\ll [J(x), J(y)] \gg = \frac{1}{4\pi^2}\left\{\frac{1}{[0 + i(x-y)]^2} - \frac{1}{[-0 + i(x-y)]^2}\right\} \quad 1.1.3.23$$
$$= \frac{1}{4\pi^2}\partial_x\left[\frac{1}{-i0 + x - y} - \frac{1}{i0 + x - y}\right] = \frac{-i}{2\pi}\partial_x\delta(x-y).$$

If we do Fourier transform, then $-i\partial_x \to q$, and remember **eq**. (**1**.**1**.**2**.**1**), $b_q \propto \frac{1}{\sqrt{n_q}} \propto \frac{1}{\sqrt{q}}$, the only difference between **Eq**.(**1**.**1**.**2**.**2b**) and (**1**.**1**.**3**.**23**) is our definition: $[b_q, b^+_{q'}]$ having additional multiplication factor $\propto \frac{1}{q}$. So both **Eq**. (**1**.**1**.**2**.**2b**) and (**1**.**1**.**3**.**23**) are the same.

## 1.2 Introduction of Sine-Gordon Equation ( SGE ):

### 1.2.1 Traveling solution of Sine-Gordon Equation

It can be found in chapter **19** of K. Huang [**8**], or chapter **10** of L. H. Ryder [**9**], that the **SGE** has the form:

$$\partial_t^2\theta - v^2\partial_x^2\theta = g\cos(\theta_+ + \varphi) \qquad 1.2.1.1$$

We can simplify this two-variable equation by defining $\xi = x - v't$ . Then it becomes a one-variable equation which is easier to solve by direct integration. It is called the traveling wave solution. I do that by the following steps. **eq**. (**1.2.1.1**) becomes:

$$\frac{d^2\theta}{d\xi^2} = \frac{g}{v'^2 - v^2}\cos(\theta(\xi) + \varphi) = -\frac{\partial V}{\partial\theta} \qquad 1.2.1.2$$

where we introduce the potential $V$

$$V = V(\theta) = \frac{-g}{v'^2 - v^2}\sin(\theta + \varphi) \qquad 1.2.1.3$$

and the ”conserved energy”

$$C = \frac{1}{2}(\frac{d\theta}{d\xi})^2 + V(\theta) \qquad 1.2.1.4$$

So from **eq**. (**1.2.1.4**), we obtain:

$$\int \frac{d\theta}{\sqrt{2(C - V(\theta))}} = \xi + \xi_0 \qquad 1.2.1.5$$

If we choose $C = \frac{-g}{v'^2 - v^2}$, which is the maximum of the potential $V$, then

$$-\sqrt{\frac{v^2 - v'^2}{4g}}\int \frac{d\Theta}{\sqrt{\sin^2(\frac{\Theta}{2})}} = \xi + \xi_0 \qquad 1.2.1.6$$

where $\Theta = \frac{\pi}{2} - (\theta + \varphi),$ and the solution is obtained

$$\Theta = 4\tan^{-1}(e^{-\sqrt{\frac{g}{v^2 - v'^2}}(\xi + \xi_0)}) \qquad 1.2.1.7$$

### 1.2.2 The separability of Sine-Gordon Equation

The separability of **SGE** is discussed in **ref**. [**10**]. It is important for finding solutions of a finite system with the boundary conditions in the following chapters. The solutions must be separable in time and space and are usually called standing wave solution. Since Sine-Gordon Equation is a nonlinear equation, the time and space parts are separable inside some function. In the following lemma, we will find this function.

**Lemma** (**3.2.1**) *Assume Sine-Gordon equation has ”separability” inside some function. Let* $\psi(x,t) = X(x)\cdot T(t)$ *and* $\theta = g(\psi)$ *be some transform of* $\psi$*. If* $\theta = g(\psi)$ *is the solution of Sine-Gordon Equation, then* $\theta = 2\cos^{-1}\{sn[\frac{\ln(\alpha\psi)}{k\beta}, k]\}$*, where sn is one of the Jacobi Elliptic function* (**JEF**) *and k is its parameter.*

**Proof** Solving for $g$ by directly inserting it into the equation: $\theta_{xx} + \epsilon\theta_{tt} = \sin\theta$ $(\epsilon = \pm 1)$, we get:

$$(\psi_{xx} + \epsilon\psi_{tt}) \cdot g'(\psi) + (\psi_x^2 + \epsilon\psi_t^2) \cdot g''(\psi) = \sin[g(\psi)] \qquad 1.2.2.1$$

Because $\psi(x,t) = X(x) \cdot T(t)$, we let $(X')^2 = \sum_{r=0}^{\infty} 2a_r \cdot X^r$ and $(T')^2 = \sum_{r=0}^{\infty} 2b_r \cdot T^r$ and above equation becomes

$$\begin{aligned} \sin[g(X \cdot T)] &= \{\sum_{r=1}^{\infty} r(a_r X^{r-1} \cdot T + \epsilon b_r T^{r-1} \cdot X)\} g'(X \cdot T) \\ &+ \{\sum_{r=0}^{\infty} 2(a_r X^r \cdot T^2 + \epsilon b_r T^r \cdot X^2)\} g''(X \cdot T) \end{aligned} \qquad 1.2.2.2$$

$g'(X \cdot T)$ , $g''(X \cdot T)$ and $\sin[g(X \cdot T)]$ are also expanded in power of $(X \cdot T)^r$. If one consider only those terms with $r \leq 2$, then

$$\beta^2 \psi \{g'(\psi) + \psi g''(\psi)\} = \sin[g(\psi)] \qquad 1.2.2.3$$

where $\beta^2 = 2(a_2 + \epsilon b_2)$. Multiplying both sides by $2g'$ and integrating the equation, we have:

$$(\psi g')^2 = C - \frac{2}{\beta^2} \cos(g) \qquad 1.2.2.4$$

We define $h = \cos\frac{g}{2}$ and $z = \frac{\ln(\alpha\psi)}{k\beta}$ , where $\alpha$ is introduced to take care of the constant of integration. By requiring constant $C$ to satisfy: $k^2(2 + C\beta^2) = 4$, above equation becomes

$$(\frac{dh}{dz})^2 = (1 - h^2) \cdot (1 - k^2 h^2) \qquad 1.2.2.5$$

This can be satisfied by one of the **JEF** if $|k| \leq 1$

$$h = sn(z;k) \qquad 1.2.2.6$$

So we have proved:

$$\theta = 2\cos^{-1}\{sn[\frac{\ln(\alpha\psi)}{k\beta}, k]\} \qquad 1.2.2.7$$

**End of Proof**.

From above we can obtain our familiar solution if we let $k = 1$ and $\beta = -n$ , because $sn(u;1) = \tanh(u)$

$$\psi = \frac{1}{\alpha} \tan^n(\frac{\theta}{4}) \qquad 1.2.2.8$$

Actually, we can generalize the **Lemma** to a $n$ –dimensional equation like the following form (usually called Kelin-Gordon Equation):

$$\sum_{i=1}^{n} \epsilon_i \theta_{x_i x_i} = f(\theta) \qquad 1.2.2.9$$

Let $\theta = g(\psi)$ be the transform of a separable function $\psi$. By inserting it into above equation, we get

$$\{\sum_{i=1}^{n} \epsilon_i \psi_{x_i x_i}\} \cdot g'(\psi) + \{\sum_{i=1}^{n} \epsilon_i \psi_{x_i}^2\} \cdot g''(\psi) = f(g) \qquad 1.2.2.10$$

Let

$$\psi(x_1, \ldots, x_n) = \prod_{i=1}^{n} X_i(x_i) \qquad 1.2.2.11$$

and

$$(X_i')^2 = \sum_{r=0}^{n} 2a_{ir} \cdot X_i^r \qquad 1.2.2.12$$

we get similar formula as before:

$$\beta^2\psi\{g'(\psi) + \psi g''(\psi)\} = f[g(\psi)]. \qquad 1.2.2.13$$

Following the same procedure, we get

$$(\psi g')^2 = A + \frac{2}{\beta^2}\int f(g)dg. \qquad 1.2.2.14$$

The integration can only be done case by case. We can define

$$\pm\sqrt{A + \frac{2}{\beta^2}\int f(g)dg} = h(g) \qquad 1.2.2.15$$

so we obtain $\psi g' = h(g) \Rightarrow \int \frac{dg}{h(g)} = \int \frac{d\psi}{\psi} = \ln(\alpha\psi)$. Possibily we can integrate out $\int \frac{dg}{h(g)}$ and obtain the required transformation, $\phi = g(\psi)$ for a particular $f(\phi)$.

## 1.2.3 Separable solution of Sine-Gordon Equation and its complex extensions

The details of this subsection can be found in [**11**], [**12**], [**13**], [**14**] and also the standard solutions of [**28**]. In this subsection, I will extend the domain of $(x,t)$ from $R\times R$ to $C\times R$ and $R\times C$, even $C\times C$, which means I complexify the domain. It is not only interesting in mathematics by itself, in fact, it is relate to the edge state and even spin pump phenomena as shown in later chapters. If $\theta$ is the solution of the standard **SGE** and if $\psi(x,t) = \tan(\frac{\theta}{4}) = X(x)\cdot T(t)$ is its separable part, then:

$$(X')^2 = pX^4 + mX^2 + q \qquad 1.2.3.1$$

$$(T')^2 = -qT^4 + (m-1)T^2 - p \qquad 1.2.3.2$$

If $p,q \neq 0$, we can do a scale transformation for $X$ and $T$

$$X \to \sqrt[4]{\frac{q}{p}}\,X \qquad 1.2.3.3$$

$$T \to i\sqrt[4]{\frac{p}{q}}\,T \qquad 1.2.3.4$$

so

$$X\cdot T \to iX\cdot T \qquad 1.2.3.5$$

and the differential equations of $X$ and $T$ become

$$(X')^2 = \sqrt{pq}\,X^4 + mX^2 + \sqrt{pq} \qquad 1.2.3.6$$

and

$$(T')^2 = \sqrt{pq}\,T^4 + (m-1)T^2 + \sqrt{pq} \qquad 1.2.3.7$$

So we have the solutions

$$X = \sqrt{k_1}\, sn(\alpha_1 x, k_1) \qquad 1.2.3.8$$

and

$$T = \sqrt{k_2}\, sn(\alpha_2 t, k_2). \qquad 1.2.3.9$$

Since I complexify the domain, $sn(\alpha_i z_i, k_i)$ and even $k_i$ can be complex numbers. Inserting (**1.2.3.8**) and (**1.2.3.9**) into (**1.2.3.6**) and (**1.2.3.7**), we obtain following relations:

$$\sqrt{pq}\, k_1^2 + mk_1 + \sqrt{pq} = 0, \qquad 1.2.3.10$$

$$\sqrt{pq}\, k_2^2 + (m-1)k_2 + \sqrt{pq} = 0, \qquad 1.2.3.11$$

$$\alpha_1^2 k_1 = \sqrt{pq} \qquad 1.2.3.12$$

and

$$\alpha_2^2 k_2 = \sqrt{pq}\,. \qquad 1.2.3.13$$

Actually, the solutions of $k_1$ and $k_2$ are:

$$k_1 = -\frac{2\sqrt{pq}}{m + \sqrt{m^2 - 4pq}} \qquad 1.2.3.14$$

and

$$k_2 = -\frac{2\sqrt{pq}}{(m-1) + \sqrt{(m-1)^2 - 4pq}}. \qquad 1.2.3.15$$

Since the coefficient of $X^4$ and constant term in **eq**. (**1.2.3.6**) are the same, $\pm X, \pm\frac{1}{X}$ are also solutions and so are $\pm T, \pm\frac{1}{T}$. We have four types of solutions:

$$\psi_1 = X \cdot T; \psi_2 = \frac{1}{\psi_1}; \psi_3 = \frac{X}{T}; \psi_4 = \frac{1}{\psi_3} \qquad 1.2.3.16$$

and

$$\psi_1 = \sqrt{k_1 k_2}\, sn(\alpha_1 x, k_1) sn(\alpha_2 t, k_2) = 1/\psi_2 \qquad 1.2.3.17a$$

$$\psi_3 = \sqrt{\frac{k_1}{k_2}}\, \frac{sn(\alpha_1 x, k_1)}{sn(\alpha_2 t, k_2)} = 1/\psi_4. \qquad 1.2.3.17b$$

Special cases occur when $pq = 0$, (i.e., when the scaling transformation of (**1.2.3.3**) or (**1.2.3.4**) are not defined.) This includes two structural groups: (1) $p = q = 0$, this contains the single soliton solutions, and (2) $pq = 0$ with either $p$ or $q \neq 0$, this contains the two-soliton solutions. ( Note: the $N$-soliton solutions with $N \geq 3$ are not contained in the solution set of (**1.2.3.6**) and (**1.2.3.7**). The reasons will be explained in later subsection with topology of the differential equation).

Note that the four expressions are not distinct at each point of the $(pq, m)$. As we have known $\psi_1$ is related to $\psi_2$ by change $\varphi \to \varphi + 2\pi$ (up to sign), so is $\psi_3$ related to $\psi_4$. For (1) if $pq > 0$ and $(m-1) > 2\sqrt{pq}$, then four types are related by real $x$ or $t$ translations, leaving only one basic form. Whereas (2) if $pq > 0$ and $m > 2\sqrt{pq} > (m-1)$, then ($\psi_4$ and $\psi_1$) and ($\psi_3$ and $\psi_2$) are related by real $x$ translations leading to two basic forms in that region because $sn(\theta + iK', k) = \frac{1}{ksn(\theta,k)}$, where $2K'$ is the imaginary period of $sn(\theta, k)$. Using the relations

(**1.2.3.14**) and (**1.2.3.15**), the solutions (**1.2.3.17**) may be written as:

$$\psi_1 = \sqrt{-pq}\,\frac{sn(\alpha_1 x, k_1)}{\alpha_1}\,\frac{sn(\alpha_2 t, k_2)}{\alpha_2} \qquad 1.2.3.18a$$

$$\psi_3 = i\frac{\frac{sn(\alpha_1 x,k_1)}{\alpha_1}}{\frac{sn(\alpha_2 t,k_2)}{\alpha_2}} \qquad 1.2.3.18b$$

because $\frac{sn(\alpha_1 x,k_1)}{\alpha_1}$ and $\frac{sn(\alpha_2 t,k_2)}{\alpha_2}$ are real for all (real) $pq$ and $m$. Thus, $\psi_1$ and $\psi_2$ are real when $pq < 0$ and purely imginary when $pq > 0$, whilst $\psi_3$ and $\psi_4$ are pure imaginary for all $pq \neq 0$. We consider the limit forms $\psi_j^L (j = 1, 2, 3, 4)$ of these basic $\psi$ solutions as $|pq| \to 0$. These are (module signs):

$$\psi_1^L = 0, \psi_2^L = \infty \qquad 1.2.3.19a$$

$$\psi_3^L = i\sqrt{1-\frac{1}{m}}\,\frac{\sinh(\sqrt{m}\,x)}{\sinh(\sqrt{m-1}\,t)}, \psi_4^L = \frac{1}{\psi_3^L} \qquad 1.2.3.19b$$

There are three points about (**1.2.3.19**) to be noted. Firstly, $4\tan^{-1}[\frac{\psi_i^L}{4}]$ is an odd function, but they are surely solutions of **SGE**. Secondly, they are solutions of $pq = 0$, but they are disconnected and different from two-soliton solutions. But below I will show (**1.2.3.19**) will become our familiar two-soliton solutions by "further complexifying", which means they go back to the real forms. The readers can check it directly:

**Case** *1: If* $\sqrt{m-1}\,t \to \sqrt{m-1}\,t + i\frac{\pi}{2}$, *then* $4\tan^{-1}\psi_3^L$ *becomes* $\varphi_{SS} = 4\tan^{-1}[\frac{\sqrt{1-\frac{1}{m}}\sinh\sqrt{m}\,x}{\cosh\sqrt{m-1}\,t}]$

**Case** *2: If* $\sqrt{m}\,x \to \sqrt{m}\,x - i\frac{\pi}{2}$, *then* $4\tan^{-1}\psi_3^L$ *becomes* $\varphi_{SA} = 4\tan^{-1}[\frac{\sqrt{1-\frac{1}{m}}\cosh\sqrt{m}\,x}{\sinh\sqrt{m-1}\,t}]$

**Case** *3: If* $\sqrt{m}\,x \to \sqrt{m}\,x + i\frac{\pi}{2}$, *then* $4\tan^{-1}\psi_3^L$ *becomes* $\varphi_B = 4\tan^{-1}[\sqrt{\frac{1}{m}-1}\,\cosh\sqrt{m}\,x \csc\sqrt{1-m}\,t]$

So if we extend the domain from $R \times R$ to $R \times C$ or $C \times R$, we can find familiar solutions: $\varphi_{ss}, \varphi_{sa}, \varphi_b$ and three of them are actually the special solutions after we do special complexifization of the general solutions and surely take $|pq| \to 0$. Also, we can do the inverse process, which means we can go from our familiar real solutions and complexify them:

**Case** *1:* $p = q = 0$

*The only real solutions here occur for* $m \geq 1$ *and are the one-parameter set of single-soliton solutions:*

$$\varphi_S = 4\tan^{-1}\{\exp[\pm\sqrt{m}\,(x \pm \sqrt{1-\frac{1}{m}}\,t)]\} \qquad 1.2.3.20$$

**Case** *2: Either p or* $q = 0$

*Real solutions occur only for* $m \geq 0$*:*

***(A)*** *When* $m \geq 1$, *these are the soliton-soliton and soliton-antisoliton scattering states:*

$$\varphi_{SS} = 4\tan^{-1}[\frac{\sqrt{1-\frac{1}{m}}\,\sinh\sqrt{m}\,x}{\cosh\sqrt{m-1}\,t}] \qquad 1.2.3.21$$

$$\varphi_{SA} = 4\tan^{-1}[\frac{\sqrt{1-\frac{1}{m}}\cosh\sqrt{m}x}{\sinh\sqrt{m-1}t}] \quad 1.2.3.22$$

***(B)*** *When* $1 \geq m \geq 0$, *these are the soliton-antisoliton bound states, or breathers:*

$$\varphi_{B} = 4\tan^{-1}[\sqrt{\frac{1}{m}-1}\cosh\sqrt{m}x\csc\sqrt{1-m}t] \quad 1.2.3.23$$

**Case** *3:* $pq \neq 0$ *and* $X(0) = T(0) = 0$

*In this case real solutions occur only for* $pq < 0$ *and return to the preivious form:*

$$\varphi = 4\tan^{-1}\{\sqrt{-pq}[\frac{sn(\alpha_1 x;k_1)}{\alpha_1}\frac{sn(\alpha_2 x;k_2)}{\alpha_2}]\} \quad 1.2.3.18a$$

*But for* $pq < 0$, *the conditions on the parameters pq and m mean that* $k_j \in i(0,\infty)$, *i.e., pure imaginary.*

**1.2.3.2 Complex Extensions**:

**Case** ***1***: $p = q = 0$

*I complexify (**1.2.3.20**) by following*

$$\varphi_S = 4\tan^{-1}\{\exp[\pm\sqrt{m}(x \pm \sqrt{1-\frac{1}{m}}t + i\theta)]\} \quad 1.2.3.24$$

*with* $\theta$ *real.*

***(A)*** $m > 1$

$\varphi_s^C = \varphi_R + i\varphi_I$

$$\varphi_R = (1-\mu)\pi - 2\mu\tan^{-1}[|\cos\sqrt{m}\theta|\csc h(\sqrt{m}x \pm \sqrt{m-1}t) \quad 1.2.3.25$$

$$\varphi_I = 2\tanh^{-1}[\sin\sqrt{m}\theta\sec h(\sqrt{m}x \pm \sqrt{m-1}t)] \quad 1.2.3.26$$

*where* $\mu = sgn\cos\sqrt{m}y$. *This function is periodic in* $\theta$ *and has the following structure:*

*(1): real solitons occur when* $\sqrt{m}\theta = 2n\pi$, $n \in Z$.

*(2): real antisolitons occur when* $\sqrt{m}\theta = (2n+1)\pi$

*(3): when* $\sqrt{m}\theta = (n+\frac{1}{2})\pi$, $\varphi_R$ *is discontinuous, while* $\varphi_I$ *is singular.*

*The intermediate values of* $\theta$ *determine transition zones in which both* $\varphi_R$ *and* $\varphi_I$ *are smooth solitary waves.* $\varphi_R$ *looking like the real soliton, while* $\varphi_I$ *is bell-shaped.*

***(B)*** $0 \leq m \leq 1$

*In this case, the argument of (**1.2.3.20**) is complex and so the only influence of general phase* $\theta$ *is to change the origin of x and t.*

$\varphi_s^C = \varphi_R + i\varphi_I$

$$\varphi_R = (1-\mu)\pi - 2\mu\tan^{-1}[|\cos\sqrt{1-m}\,t|\operatorname{csc}h(\sqrt{m}\,x) \qquad 1.2.3.27$$

$$\varphi_I = 2\tanh^{-1}[\sin\sqrt{1-m}\,t\operatorname{sec}h(\sqrt{m}\,x)] \qquad 1.2.3.28$$

*where* $\mu = sgn(\cos\sqrt{\frac{1}{m}-1}\,t)$. *This is a standing-wave solution which is solitary in x and periodic in t. Compare **(1.2.3.27)** and **(1.2.3.28)** with **(1.2.3.25)** and **(1.2.3.28)**, we can see the t- behavior here corresponding to the θbehavior of the previous case. Thus, as time develops,* $\varphi_R$ *here oscillates between the soliton and antisoliton states.*

***(C)*** $m \le 0$

*Once again the phase θ has no influence on the structure of the solutions in this case, its effect being to change the origin of x and t.*

$\varphi_s^L = \varphi_R + i\varphi_I$

$$\varphi_R = \pi \qquad 1.2.3.29$$

$$\varphi_I = 2\tanh^{-1}[\sin\sqrt{-m}\,(x \pm \sqrt{1-\frac{1}{m}}\,t)] \qquad 1.2.3.30$$

*We can see in this case* $\varphi_I$ *is a singular periodic wave travelling at a speed faster than that of light (i.e.* $\sqrt{1-\frac{1}{m}} > 1$*).* $\varphi_R$ *on the other hand, remains constant at* $\pi$.

**Case** *2:* $pq = 0$ *with* $p$ *or* $q \ne 0$

***(A)*** $m > 1$

*In this case, the real solutions are **(1.2.3.21)**,* $\varphi_{SS} = 4\tan^{-1}[\frac{\sqrt{1-\frac{1}{m}}\sinh\sqrt{m}\,x}{\cosh\sqrt{m-1}\,t}]$ *and **(1.2.3.22)**,* $\varphi_{SA} = 4\tan^{-1}[\frac{\sqrt{1-\frac{1}{m}}\cosh\sqrt{m}\,x}{\sinh\sqrt{m-1}\,t}]$.

*We complexify them by the substitution:* $x \to x + i\theta$ *with θ real. Of particular interest are the singular limits at* $\sqrt{m}\,\theta = (n+\frac{1}{2})\pi$ *which take the forms:*

$$\varphi^C_{SS/SA}|_{\sqrt{m}\theta=(n+\frac{1}{2})\pi} = (-1)^n\begin{cases} (-1)^\mu 2\pi + i4\coth^{-1}\psi\,, & (|\psi|>1, x<0) \\ i4\tanh^{-1}\psi\,, & (|\psi|<1) \\ 2\pi + i4\coth^{-1}\psi\,, & (|\psi|>1, x>0) \end{cases} \qquad 1.2.3.31$$

*Where* $\mu = 1$ *and* $\psi = \sqrt{1-\frac{1}{m}}\cosh\sqrt{m}\,x\operatorname{sec}h\sqrt{m-1}\,t$ *for* $\varphi^c_{SS}$.

$\mu = 2$ *and* $\psi = \sqrt{1-\frac{1}{m}}\sinh\sqrt{m}\,x\operatorname{csc}h\sqrt{m-1}\,t$ *for* $\varphi^c_{SA}$.

***(B)*** $0 \le m \le 1$

*The only real solutions in this case are the soliton-antisoliton bound state or breather, **(1.2.3.23)*** $\varphi_B = 4\tan^{-1}[\sqrt{\frac{1}{m}-1}\cosh\sqrt{m}\,x\csc\sqrt{1-m}\,t]$ *for* $0 < m < 1$. *Its limit form as* $m \to 1$ *is* $\varphi_B|_{m=1} = 4\tan^{-1}[\frac{\cosh x}{t}]$.

*The complex breather can be obtained from (**1.2.3.23**) by the substitution:* $x \to x + i\theta$ *with* $\theta$ *real. When* $\sqrt{m}\,\theta = (n + \frac{1}{2})\pi$, *where* $\varphi_b^c$ *has the singular form:*

$$\varphi_B^c|_{\sqrt{m}\theta=(n+\frac{1}{2})\pi} = 4\tan^{-1}[\frac{i\sqrt{\frac{1}{m}-1}\,\sinh\sqrt{m}\,x}{\sin\sqrt{1-m}\,t}] \qquad 1.2.3.32$$

***(c)*** $m < 0$

*There is only one type of solution in this class and this is a complex solution which takes the form:*

$$\varphi = 4\tan^{-1}[\frac{i\sqrt{1-\frac{1}{m}}\,\sinh\sqrt{-m}\,x}{\sin\sqrt{1-m}\,t}] \qquad 1.2.3.33$$

## 1.2.4 Exact N-Soliton Solitons of Sine-Gordon Equation

According to chapter **2** of [**7**], we can have multisoliton solutions, for **N**=**2**, where **N** is the soliton numbers, its form is

$$\theta_{SA} = 4\tan^{-1}[\frac{\sinh(\frac{vt}{\sqrt{1-v^2}})}{v\cosh(\frac{x}{\sqrt{1-v^2}})}] \qquad 1.2.4.1$$

where **S** means "soliton", and **A** means "antisoliton". The solution above has the following properties.

$$\lim_{t->+\infty}\theta_{SA} \simeq \theta_S(x+vt+\Delta) + \theta_A(x-vt-\Delta) \qquad 1.2.4.2$$

$$\lim_{t->-\infty}\theta_{SA} \simeq \theta_S(x+vt-\Delta) + \theta_A(x-vt+\Delta) \qquad 1.2.4.3$$

where the phase shift

$$\Delta = \sqrt{1-v^2}\,\ln(v) \qquad 1.2.4.4$$

So from this form, (**1**.**2**.**4**.**1**) is actually the **N**=**2** multisoliton solution. The next step is to use Hirota's formula: [**15**], [**16**] to construct **N** = **2** multisoliton solution first, then try to extend the form to the case $N > 2$. We consider the solitons in pairs via their corresponding center-of-velocity frames. Hirota's formula for the two-soliton solution is:

$$\tan(\frac{\theta}{4}) = \frac{\sum_{i=1}^{2}\exp(\gamma_i\cdot\xi_i+\alpha_i)}{1+\exp[B_{12}+\sum_{i=1}^{2}(\gamma_i\cdot\xi_i+\alpha_i)} \qquad 1.2.4.5$$

where $\xi_i = x - u_i t$, $\gamma_i = \frac{1}{\sqrt{1-u_i^2}}$, $\alpha_i$ is a phase factor, and $\exp(B_{12}) = \frac{(\gamma_1-\gamma_2)^2-(\gamma_1u_1-\gamma_2u_2)^2}{(\gamma_1+\gamma_2)^2-(\gamma_1u_1+\gamma_2u_2)^2}$. How do we understand this formula? By the approximated forms of **eq**.(**1**.**2**.**4**.**2**) and (**1**.**2**.**4**.**3**), we may naively gauss two-soliton solution has the form $\theta = \theta_1 + \theta_2$. So we get:

$$\tan\frac{\theta}{4} = \tan\frac{\theta_1+\theta_2}{4} = \frac{\tan\frac{\theta_1}{4}+\tan\frac{\theta_2}{4}}{1-\tan\frac{\theta_1}{4}\cdot\tan\frac{\theta_2}{4}} = \frac{\exp(\gamma_1\cdot\xi_1+\alpha_1)+\exp(\gamma_2\cdot\xi_2+\alpha_2)}{1-\exp(\gamma_1\cdot\xi_1+\alpha_1)\cdot\exp(\gamma_2\cdot\xi_2+\alpha_2)} \quad 1.2.4.6$$

where $\theta_1$ and $\theta_2$ are single soliton travelling solutions. **Eq**. (**1.2.4.6**) is actually similar to the Hiota's formula. **Eq**. (**1.2.4.5**) has the additional interacting phase ($i.e.\exp(B_{12})$) between two solitons. Usually we identify $u^2 = \exp(B_{12})$, so above formula change to:

$$\tan\frac{\theta}{4} = \frac{\tan\frac{\theta_1}{4}+\tan\frac{\theta_2}{4}}{1-u^2\cdot\tan\frac{\theta_1}{4}\cdot\tan\frac{\theta_2}{4}} \quad 1.2.4.7$$

or

$$\tan\frac{\theta}{4} = \frac{\exp[\gamma(x-vt)+\alpha_1]+\exp[\gamma(x+vt+\alpha_2)}{1-v^2\cdot\exp(2\gamma x)\cdot\exp(\alpha_1+\alpha_2)} \quad 1.2.4.8$$

By setting $e^{\alpha_2} = -e^{\alpha_1} = \frac{1}{v}$, Hirota's formula, **eq**. (**1.2.4.8**) is reduced to the **N**=**2** soliton solution. Next we construct the **N**-soliton solution from a linear superposition of **N** solitons by considering them in pairs. Let $\tan(\frac{\theta}{4}) = \frac{g}{f}$, among them:

$$f = \sum_{\mu=0,1}^{(e)} \exp[\sum_{i<j}^{N} B_{ij}\mu_i\mu_j + \sum_{i=1}^{N}\mu_i(r_i\xi_i+\alpha_i)] \quad 1.2.4.9$$

$$g = \sum_{\mu=0,1}^{(o)} \exp[\sum_{i<j}^{N} B_{ij}\mu_i\mu_j + \sum_{i=1}^{N}\mu_i(r_i\xi_i+\alpha_i)] \quad 1.2.4.10$$

where $\sum_{\mu=0,1}^{(e)}$ and $\sum_{\mu=0,1}^{(o)}$ represent summing over all possible combinations of $\mu_i$ such that $\mu_1 = 0,1; \mu_2 = 0,1; \ldots; \mu_N = 0,1$ , so that $\sum_{i=1}^{N}\mu_i$ is even or odd. However, for the **N**-soliton solution, with **N**>**2**, there is no unique choice of frame. But we can consider the solitons in pairs via their corresponding center-of-velocity frames. Why are we sure the construction above is actually the multisoliton solution of the equation? Because **SGE** is an integrable equation. Inverse Scattering Method, see **chapter 10** of **Ref**. [**39**], says integrable equation can divide many-body scattering into pairs of two-body scattering. Take **N**=**3** for example, the solution of three solitons' solution is:

$$\tan\frac{\theta}{4} = \frac{\sum_{i=1}^{3}\tan\frac{\theta_i}{4} - (u_{12}u_{23}u_{31})^2\prod_{i=1}^{3}\tan\frac{\theta_i}{4}}{1-\sum_{i\neq j}u_{ij}^2\cdot\tan\frac{\theta_i}{4}\cdot\tan\frac{\theta_j}{4}} \quad 1.2.4.11$$

where $u_{ij}$ is the common speed of the $i_{th}$ and $j_{th}$ in their center-of -velocity frame.

### 1.2.5 Algebraic Geometry ( finite-zone) solutions of Sine-Gordon Equation

Take [**19**] as one of the references, the more general solutions of **SGE** is described by Riemann theta functions. The solution has the form

$$\theta = 2i\ln\left(\frac{\Theta(\vec{l} + \frac{\vec{1}}{2}|\overleftrightarrow{B})}{\Theta(\vec{l}|\overleftrightarrow{B})}\right) \qquad 1.2.5.1$$

where Riemmian theta function $\Theta$ is defined as:

$$\Theta(\vec{l}, \overleftrightarrow{B}) = \sum_{\vec{m}\in Z\times Z} e^{2\pi i(\vec{l}\cdot\vec{m})+\pi i(\vec{m}^{+}\cdot\overleftrightarrow{B}\cdot\vec{m})} \qquad 1.2.5.2$$

This function is similar to the partition function in field theory. ( i.e. $\overleftrightarrow{B}^{-1}$ the propagator and $\vec{l}$ is the external source). We also have following properties:

$$\Theta(\vec{l} + \frac{\vec{1}}{2}|\overleftrightarrow{B}) = \Theta(\vec{l}|\overleftrightarrow{B})^{*} \qquad 1.2.5.3$$

We give two examples, one phase and two phase **SGE** solutions which are x-periodic and t-periodic and **even** in x:.

**Case** ***1: Even, periodic one-phase solutions: the pendulum***

*The classical pendulum solutions are given by:*

$$\theta(t;H;t_0) = 2\sin^{-1}(\sqrt{H/2}\, sn(t-t_0;\sqrt{H/2}\,))$$

*where sn is the Jacobi elliptic function with modulus* $\sqrt{H/2}$, $t_0$ *is the phase shift and H is the conserved energy,*

$$H = \frac{1}{2}\theta_t^2 + 1 - \cos\theta$$

*This means we take* $\overleftrightarrow{B}$ *and* $\vec{l}$ *in* ***eq. (1.2.5.2)*** *to be* $1 \times 1$, *or scalar numbers.( i.e.,* $N = 1$ *variable). Here we only consider one variable, say t ( or* $\xi = x - vt$*). In this case,*

$$\Theta(l,B) = \sum_{m\in Z} e^{2i\pi ml+i\pi Bm^2} \qquad 1.2.5.4$$

*Then we take*

$$B = \oint_b \frac{dE}{R(E)} \cdot C = \frac{1}{2} + i\,\mathrm{Im}(B) \qquad 1.2.5.5$$

*and*

$$l(t;t_0) = 8\pi iC(t-t_0),\ t_0 \in [0, T = \frac{2\pi}{w}) \qquad 1.2.5.6$$

*where* $l(t;t_0), B$ *( and therefore the theta function) are specified by the simple periodic spectrum* $\sum_{N=1}^{(s)}$ *:*

$$\sum_{N=1}^{(s)} = \{E_1, E_2 | E_1 = \frac{1}{16}e^{i\phi}, E_2 = E_1^{*}, 0 < \phi < \pi\} \qquad 1.2.5.7$$

$$H(\sum_{N=1}^{(s)}) = 1 - 8(E_1 + E_2) = 1 - \cos\phi \in (0,2) \qquad 1.2.5.8$$

*is the pendulum (this solution) potential. The coefficients are:*

$$C = (\oint_a \frac{dE}{R(E)})^{-1} \qquad 1.2.5.9$$

$$R^2 = E(E - E_1)(E - E_2) \qquad 1.2.5.10$$

*$a, b$ cycles are defined on the Riemann surface **(E, R(E))** and **Eq. (1.2.5.5), (1.2.5.9), (1.2.5.10)** show the topology property. Here genus=1, and the line connect $E_1$ and $E_2$ is the branch cut and $\oint_a, \oint_b$ means integral around the circle of a and b, the two independent integral parts of **genus=1**.*

**Case** *2: **Even, two phase**, **x- and t-periodic breather trains**:*

*This case means we choose* $\mathbf{N} = \mathbf{2}$ *variables, say x and t. We have two cases in this* $\sum_{N=2}^{(s)}$:

$$\sum_{N=2}^{(s)} = \{E_1, E_2, E_3, E_4 | E_1 = E_2^* = re^{i\phi}, \qquad 1.2.5.11a$$
$$E_1 = \frac{1}{16}e^{i\phi}, E_3 = E_4^* = \frac{1}{16^2 r}e^{i\phi}, 0 < r < \frac{1}{16}, 0 < \phi < \pi\}$$

$$\sum_{N=2}^{(s)} = \{E_1, E_2, E_3, E_4 | E_j = \frac{1}{16}e^{i\phi_j}, \phi_{2j} = -\phi_{2j-1}, 0 < \phi_1 < \phi_2 < \pi\} \qquad 1.2.5.11b$$

*Now $\overleftrightarrow{B}$ is a $2 \times 2$ normalized period matrix of **genus=2** Riemann surface defined by the simple periodic spectrum $\sum_{N=2}^{(s)}$,*

$$R^2(E) = E\prod_{j=1}^{4}(E - E_j)$$

*In **case (a)**, we choose* $\overrightarrow{l} = \begin{pmatrix} iax+ibt \\ iax-ibt \end{pmatrix}; \overleftrightarrow{B} = \begin{pmatrix} \frac{1}{2} + i\beta & i\alpha \\ i\alpha & \frac{1}{2} + i\beta \end{pmatrix}$, *then*

$$\Theta(\overrightarrow{l}, \overleftrightarrow{B}) = \theta_4(2iax;\tau_1)\theta_4(2ibt;\tau_2) + i\theta_2(2iax;\tau_1)\theta_2(2ibt;\tau_2) \qquad 1.2.5.12$$

*where $\tau_1 = i\frac{K'(k_1)}{K(k_1)}, \tau_2 = i\frac{K'(k_2)}{K(k_2)}$. So the solution **(1.2.5.2)** is:*

$$\theta = 2i\ln\left(\Theta(\overrightarrow{l}, \overleftrightarrow{B})\right) = 4\tan^{-1}[\frac{\theta_2(2iax;\tau_1)\theta_2(2ibt;\tau_2)}{\theta_4(2iax;\tau_1)\theta_4(2ibt;\tau_2)}] \qquad 1.2.5.13$$

*In **case (b)**, we choose* $\overrightarrow{l} = \begin{pmatrix} iax \\ iax-ibt \end{pmatrix}; \overleftrightarrow{B} = \begin{pmatrix} \frac{1}{2} + i\alpha & i\alpha \\ i\alpha & \frac{1}{2} + i\beta \end{pmatrix}$, *then*

$$\Theta(\overrightarrow{l}, \overleftrightarrow{B}) = \theta_3(2iax;\tau_1)\theta_4(2ibt;\tau_2) + i\theta_2(2iax;\tau_1)\theta_3(2ibt;\tau_2) \qquad 1.2.5.14$$

*where $\tau_1, \tau_2$ the same as before. So the solution **(1.2.5.2)** is:*

$$\theta = 2i\ln\Big(\Theta(\overrightarrow{l}, \overleftrightarrow{B})\Big) = 4\tan^{-1}\left[\frac{\theta_2(2iax;\tau_1)\theta_3(2ibt;\tau_2)}{\theta_3(2iax;\tau_1)\theta_4(2ibt;\tau_2)}\right] \qquad 1.2.5.15$$

*Both **Eq.(1.2.5.13)** and (**1.2.5.15**) are separable solutions we expected in finite-zone. Here I defined theta functions $\theta_j$ as:*

$$i\theta_1(z,b) \equiv \Theta[\frac{1}{2},\frac{1}{2}](z,b) = \sum_{n\in z} e^{\pi i(n+\frac{1}{2})^2 b + 2\pi i(n+\frac{1}{2})(z+\frac{1}{2})} \qquad 1.2.5.16a$$

$$\theta_2(z,b) \equiv \Theta[\frac{1}{2},0](z,b) = \sum_{n\in z} e^{\pi i(n+\frac{1}{2})^2 b + 2\pi i(n+\frac{1}{2})z} \qquad 1.2.5.16b$$

$$\theta_3(z,b) \equiv \Theta[0,0](z,b) = \sum_{n\in z} e^{\pi i n^2 b + 2\pi i n z} \qquad 1.2.5.16c$$

$$\theta_4(z,b) \equiv \Theta[0,\frac{1}{2}](z,b) = \sum_{n\in z} e^{\pi i n^2 b + 2\pi i n(z+\frac{1}{2})} \qquad 1.2.5.16d$$

*From $\sum_{N=2}^{(s)}$ in **Eq.(1.2.5.11)**, we can see the interesting topology structure of these solutions. They are genus=2 solutions in topology.*

# 2. Sine-Gordon equation with asymmetric phase

## 2.1 Introduction:

A Combined Sine-Cosine-Gordon Equation had been analyszed by A. M. Wazwaz using variable separated ODE method. Initially, I find his results can help me find different kinds of solutions of **SGE** with adiabatic phase. By analysis of his solutions, I find that they can be transformed into **SGE** with only adiabatical phase change. A paper of A.M.Wazwaz, [**20**], give the travelling wave solutions for combined Sine-Cosine-Gordon equations. They are the solution that combine time and space variables in an infinite system. But maybe we can do some limit approximation, e.g.soliton velocity $\to 0$ or system size $\to \infty$ to use them in a finite system. The advantage for his solutions is not simply putting $\varphi$ outside $4\tan^{-1}$, which would be only a simple translation. He use the method of the variable separated ODE method,which I will state below.

### 2.1.1 List Wazwaz's solutions

In this chapter, I will repeat some process and solutions of Wazwaz [**20**], we will generally use his notation for convenience of discussion. Physical problem, like quantum spin transport problem [**34**] require us to solve Combined Sine-Cosine-Gordon equation:

$$u_{tt} - ku_{xx} + \alpha\sin(u) + \beta\sin(u) = 0 \qquad 2.1.1.1$$

Here $\alpha$ and $\beta$ are arbitrary number, this is a generalization of Sine-Gordon Equation with adiabatical phase [**34**], which further require $\alpha^2 + \beta^2 = R^2$. Let $\xi = x - ct$, and $u' = \frac{du}{d\xi}$, we can obtain ODE to obtain traveling wave:

$$(c^2 - k)u'' + \alpha\sin(u) + \beta\cos(u) = 0 \qquad 2.1.1.2$$

and we also require $k \neq c^2$. The wonderful step of Wazwaz [**20**] is the variable separated **ODE** method. Assume $u(\xi)$ satisfies the ODE given by:

$$u'(\xi) = a\sin(\frac{u}{2}) + b\cos(\frac{u}{2}) \qquad 2.1.1.3$$

Differential again, we get:

$$u''(\xi) - \frac{a^2 - b^2}{4}\sin(u) - \frac{ab}{2}\cos(u) = 0 \qquad 2.1.1.4$$

Compare **Eq**.(**2.1.1.2**) and (**2.1.1.4**),we obtain the relation:

$$ab = \frac{2\beta}{k - c^2} \qquad 2.1.1.5a$$

$$a^2 - b^2 = \frac{4\alpha}{k - c^2} \qquad 2.1.1.5b$$

Also define $\gamma$ for latter used:

$$\gamma = \sqrt{\alpha^2 + \beta^2} - \alpha \qquad 2.1.1.6$$

So from Wazwaz [**20**], we can write four kinds of solutions directly by doing indefinite integrate, **Eq**.(**2.1.1.3**):

$$\frac{1}{a\sin(\frac{u}{2}) + b\cos(\frac{u}{2})}du = d\xi \quad 2.1.1.7$$

$$\frac{4}{\sqrt{b^2+a^2}}\tanh^{-1}(\frac{b\tan\frac{u}{4} - a}{\sqrt{b^2+a^2}}) = \xi + \xi_0 \quad 2.1.1.8a$$

$$\frac{4}{\sqrt{b^2+a^2}}\tanh^{-1}(\frac{b\cot\frac{u}{4} + a}{\sqrt{b^2+a^2}}) = \xi + \xi_0 \quad 2.1.1.8b$$

$$\frac{4}{\sqrt{b^2+a^2}}\coth^{-1}(\frac{b\tan\frac{u}{4} - a}{\sqrt{b^2+a^2}}) = \xi + \xi_0 \quad 2.1.1.8c$$

$$\frac{4}{\sqrt{b^2+a^2}}\coth^{-1}(\frac{b\cot\frac{u}{4} + a}{\sqrt{b^2+a^2}}) = \xi + \xi_0 \quad 2.1.1.8d$$

Take solution (**2.1.1.8a**) for example, we obtain the solution directly:

$$u = 4\tan^{-1}(\frac{a}{b} + \frac{\sqrt{a^2+b^2}}{b}\tanh(\frac{\sqrt{a^2+b^2}}{4}\xi)) \quad 2.1.1.9$$

At first, these four kinds of the four solutions looks independent. But there are actually relations between them. **Solution** (**2.1.1.8b**) is related to (**2.1.1.8a**) by $u \to u + 2\pi$,and $x \to -x, t \to -t$. This transformation don't change the original equation, so does **Solution** (**2.1.1.8d**) is related to (**2.1.1.8c**). And (**2.1.1.8a**) and (**2.1.1.8c**) are actually two kinds of results by integrate **Eq**.(**2.1.1.7**), although look a little different, but we can check by: $\frac{d\coth^{-1}(x)}{dx} = \frac{d\tanh^{-1}(x)}{dx} = \frac{1}{1-x^2}$, they are actually related by: $\coth^{-1}(x) - \tanh^{-1}(x) = \frac{i\pi}{2}$. We can transform **Solution** (**2.1.1.8a~d**) from $a, b$ to $\alpha, \beta$ by calculate;

$$\frac{\sqrt{a^2+b^2}}{4} = \frac{1}{2\sqrt{k-c^2}}\frac{\sqrt{\beta^2+\gamma^2}}{\sqrt{2\gamma}} \quad 2.1.1.10$$

$$= \frac{1}{2\sqrt{k-c^2}}\frac{\sqrt{\alpha^2+\beta^2-\alpha\sqrt{\alpha^2+\beta^2}}}{\sqrt{\sqrt{\alpha^2+\beta^2}-\alpha}} = \frac{1}{2\sqrt{k-c^2}}\sqrt[4]{\alpha^2+\beta^2}$$

inside tanh or coth ,and other coefficients are easier to calculate, so we get:

$$u(x,t) = 4\tan^{-1}(\frac{2\sqrt[4]{\beta^2+\alpha^2}}{\sqrt{2\gamma}}\tanh[\frac{\sqrt[4]{\alpha^2+\beta^2}}{2\sqrt{k-c^2}}(x-ct)+\xi_0)] + \frac{\beta}{\gamma}) \quad 2.1.1.11a$$

$$u(x,t) = 4\cot^{-1}(\frac{2\sqrt[4]{\beta^2+\alpha^2}}{\sqrt{2\gamma}}\tanh[\frac{\sqrt[4]{\alpha^2+\beta^2}}{2\sqrt{k-c^2}}(x-ct)+\xi_0)] - \frac{\beta}{\gamma}) \quad 2.1.1.11b$$

$$u(x,t) = 4\tan^{-1}(\frac{2\sqrt[4]{\beta^2+\alpha^2}}{\sqrt{2\gamma}}\coth[\frac{\sqrt[4]{\alpha^2+\beta^2}}{2\sqrt{k-c^2}}(x-ct)+\xi_0)] + \frac{\beta}{\gamma}) \quad 2.1.1.11c$$

$$u(x,t) = 4\cot^{-1}(\frac{2\sqrt[4]{\beta^2+\alpha^2}}{\sqrt{2\gamma}}\coth[\frac{\sqrt[4]{\alpha^2+\beta^2}}{2\sqrt{k-c^2}}(x-ct)+\xi_0)] - \frac{\beta}{\gamma}) \qquad 2.1.1.11d$$

If we use these solutions apply to static case ( i.e.; c=0 ), and also set k=1, also let $\alpha^2+\beta^2 = R^2$ in order to study nontrivial solution involved $\varphi$. This means:

$$\alpha = R\cos\varphi$$
$$\beta = R\sin\varphi$$

From **Eq**.(**2**.**1**.**1**.**5**), (**2**.**1**.**1**.**6**), we get:

$$a = 2\sqrt{\frac{R}{k-c^2}}\cos\frac{\varphi}{2}$$
$$b = 2\sqrt{\frac{R}{k-c^2}}\sin\frac{\varphi}{2}$$

So $\frac{\beta}{\gamma} = \frac{\sin\varphi}{1-\cos\varphi} = \frac{\cos\frac{\varphi}{2}}{\sin\frac{\varphi}{2}}$,and $\frac{2\sqrt[4]{\alpha^2+\beta^2}}{\sqrt{2r}} = \frac{1}{|\sin\frac{\varphi}{2}|}$,so the four solutions become:

$$u(x) = 4\tan^{-1}\{\frac{1}{|\sin\frac{\varphi}{2}|}\tanh[\frac{\sqrt{R}}{2}(x-x_0)] + \frac{\cos\frac{\varphi}{2}}{\sin\frac{\varphi}{2}}\} \qquad 2.1.1.12a$$

$$u(x) = 4\cot^{-1}\{\frac{1}{|\sin\frac{\varphi}{2}|}\tanh[\frac{\sqrt{R}}{2}(x-x_0)] - \frac{\cos\frac{\varphi}{2}}{\sin\frac{\varphi}{2}}\} \qquad 2.1.1.12b$$

$$u(x) = 4\tan^{-1}\{\frac{1}{|\sin\frac{\varphi}{2}|}\coth[\frac{\sqrt{R}}{2}(x-x_0)] + \frac{\cos\frac{\varphi}{2}}{\sin\frac{\varphi}{2}}\} \qquad 2.1.1.12c$$

$$u(x) = 4\cot^{-1}\{\frac{1}{|\sin\frac{\varphi}{2}|}\coth[\frac{\sqrt{R}}{2}(x-x_0)] - \frac{\cos\frac{\varphi}{2}}{\sin\frac{\varphi}{2}}\} \qquad 2.1.1.12d$$

## 2.2 Further study of the states of solitons

### 2.2.1 Mathematics calculation and lemmas:

Take **eq**.(**2**.**1**.**1**.**12a**) for example: using the inequalitys, $1 \geq \tanh[\frac{\sqrt{R}}{2}(x-x_0)] \geq -1 \Rightarrow |\frac{\sin\frac{\varphi}{4}}{\cos\frac{\varphi}{4}}| \geq \frac{\tanh[\frac{\sqrt{R}}{2}(x-x_0)]}{|\sin\frac{\varphi}{2}|} - \frac{\cos\frac{\varphi}{2}}{|\sin\frac{\varphi}{2}|} \geq -|\frac{\cos\frac{\varphi}{4}}{\sin\frac{\varphi}{4}}|$, we get $4\tan^{-1}(|\cot\frac{\varphi}{4}|) = 4\tan^{-1}(|\tan(\frac{\pi}{2}-\frac{\varphi}{4})| \geq u(x,t) \geq 4\tan^{-1}(-|\tan\frac{\varphi}{4}|) \Rightarrow 2\pi-\varphi \geq u(x,t) \geq -\varphi$, so the difference between the maximal and minimal values are equal to $2\pi$ for arbitrary $\varphi$. Other three solutions have the same conclusion. If we let R=1 for convenient, so take **Eq**.(**2**.**1**.**1**.**12a**) for example again, we can calculate its energy, by: $-\frac{\partial^2 u}{\partial x^2} + \sin(u+\varphi) = 0$, we get

$$\frac{-1}{2}(\frac{\partial u}{\partial x})^2 - \cos(u+\varphi) + \frac{1}{2}(\frac{\partial u}{\partial x})^2|_{x=x_0} + \cos(u+\varphi)|_{x=x_0} = 0 \qquad 2.2.1.1$$

I use **eq**.(**2**.**1**.**1**.**12a**) to get

$$u|_{x=x_0} = 4\tan^{-1}(\cot\frac{\varphi}{2}) = 4\tan^{-1}(\tan(\frac{\pi}{2} - \frac{\varphi}{2})) = 2\pi - 2\varphi \Rightarrow \cos(u+\varphi)|_{x=x_0} = \cos(2\pi - \varphi). \quad 2.2.1.2$$

The derivative of both sides of

$$\tan\frac{u}{4} = \frac{1}{|\sin\frac{\varphi}{2}|}\tanh(\frac{x-x_0}{2}) + \frac{\cos\frac{\varphi}{2}}{\sin\frac{\varphi}{2}} \quad 2.2.1.3$$

We get

$$\frac{\sec^2(\frac{u}{4})}{4}\frac{\partial u}{\partial x} = \frac{1}{|\sin\frac{\varphi}{2}|}\frac{2e^{x-x_0}}{(e^{x-x_0}+1)^2} \quad 2.2.1.4$$

Because

$$\cos\frac{u}{4}|_{x=x_0} = \cos(\frac{\pi}{2} - \frac{\varphi}{2})|_{x=x_0} = \sin\frac{\varphi}{2} \quad 2.2.1.5$$

Combine these, so $\frac{1}{2}(\frac{\partial u}{\partial x})^2|_{x=x_0} = 2\sin^2(\frac{\varphi}{2})$. So we get:

$$\begin{aligned}\frac{1}{2}(\frac{\partial u}{\partial x})^2 &= \cos\varphi + 2\sin^2(\frac{\varphi}{2}) - \cos(u+\varphi) \quad 2.2.1.6\\ &= \frac{1}{2}(\frac{\partial u}{\partial x})^2|_{x=x_0} + V(\varphi) = 1 - \cos(u+\varphi)\end{aligned}$$

$V(\varphi) = \cos\varphi - \cos(u+\varphi)$and because

$$\frac{\partial u}{\partial x} = \sqrt{2 - 2\cos(u+\varphi)} \quad 2.2.1.7$$

So we can get $V_s$

$$V_s = \int_{-\infty}^{\infty}\frac{1}{2}(\frac{\partial u}{\partial x})^2 dx = \sqrt{2}\int_{-\varphi}^{2\pi-\varphi}\sqrt{1-\cos(u+\varphi)}\,du = \sqrt{2}\int_{0}^{2\pi}\sqrt{1-\cos(u')}\,du' \quad 2.2.1.8$$

Where $u' = u + \varphi$. Because Hamiltion $H = \int_{-\infty}^{\infty}[\frac{1}{2}(\frac{\partial u}{\partial x})^2 + V(\varphi)]dx$, so total energy $H = V_s + V_p$, where

$$V_s = \int_{-\infty}^{\infty} 2\sin^2(\frac{\varphi}{2})dx + V_p \quad 2.2.1.9$$

$$V_p = \int_{-\infty}^{\infty}(\cos\varphi - \cos(u+\varphi))dx \quad 2.2.1.10$$

**Case** *1*. ***If*** $a = 0, b = 1$, ***because*** $\int\frac{du}{\cos\frac{u}{2}} = 4\tanh^{-1}(\tan\frac{u}{4})$, ***which meet the solution (2.1.1.8a), so in limit, it is right.***

**Case** *2*. ***If*** $a = 1, b = 0$, ***actually this go back to our familiar Sine-Gordon Equation, because this case imply*** $\beta = 0$, ***and*** $\int\frac{du}{\sin\frac{u}{2}} = 2\ln(\tan\frac{u}{4}) = \xi + \xi_0 \Rightarrow u = 4\tan^{-1}(e^{\frac{\xi+\xi_0}{2}})$. $\xi = x - t$. ***But this not the limit of any solution (2.1.1.8a~d).***

Because we require $\alpha^2 + \beta^2 = R$, we will have the solution directly just by translating $\varphi$. These form will be described in the next subsection, is like

$$u = -\varphi + 4\tan^{-1}\frac{F(z)}{G(\tau)} = 4\tan^{-1}\frac{F(z)\cos\frac{\varphi}{4} - G(\tau)\sin\frac{\varphi}{4}}{F(z)\sin\frac{\varphi}{4} + G(\tau)\cos\frac{\varphi}{4}} \quad 2.2.1.11$$

if we let $\alpha = R\cos\varphi$ and $\beta = R\sin\varphi$ If we begin with Sine-Gordon solution

$u = 4\tan^{-1}(e^{\frac{\xi}{2}}) = 4\tan^{-1}\frac{e^{\frac{1}{2}x}}{e^{\frac{1}{2}t}}$, and want to write down the solution involving $\varphi$, we get $u = -\varphi + 4\tan^{-1}\frac{e^{\frac{1}{2}x}}{e^{\frac{1}{2}t}} = 4\tan^{-1}\frac{e^{\frac{\xi}{2}}\cos\frac{\varphi}{4} - \sin\frac{\varphi}{4}}{e^{\frac{\xi}{2}}\sin\frac{\varphi}{4} + \cos\frac{\varphi}{4}}$. But this is different from any solution of **Eq**.(**2**.**1**.**1**.**11a~d**), which involved $\tanh\frac{\xi}{2} = \frac{e^{\xi}-1}{e^{\xi}+1}$ or $\coth\xi$ inside $4\tan^{-1}$. So we can see they are two different kinds of solutions. But I can go further by take the limit:

**Lemma** *1*

$$\lim_{x\to 0} \mathbf{tanh}^{-1}(x-1) \simeq \lim_{x\to 0} \frac{1}{2}\ln(x) \qquad 2.2.1.12$$

**Proof** Let $y = \tanh^{-1}(x-1)$, so $x - 1 = \tanh(y) = \frac{e^{y}-e^{-y}}{e^{y}+e^{-y}}$.
But at $x \to 0$, which imply $y \to -\infty$.
So $\lim_{y\to-\infty} \frac{e^{y}-e^{-y}}{e^{y}+e^{-y}} \simeq \lim_{y\to-\infty} \frac{e^{y}-e^{-y}}{e^{-y}} \simeq \lim_{y\to-\infty} e^{2y} - 1$,
and $\lim_{y\to-\infty} x \simeq \lim_{y\to-\infty} e^{2y} \Rightarrow y = \frac{1}{2}\ln(x)$ at $x \to 0$ or $y \to -\infty$.
**End of Proof**

We can apply the lemma to **solution** (**2**.**1**.**1**.**8a**), and see if **case 2**’**s** solution can be reduced to the original **SGE**:

$$\lim_{b\to 0} \frac{4}{\sqrt{b^2+a^2}}\mathbf{tanh}^{-1}\left(\frac{b\tan\frac{u}{4} - a}{\sqrt{b^2+a^2}}\right) \simeq \frac{4}{a}\lim_{b\to 0}\mathbf{tanh}^{-1}\left(\frac{b}{a}\tan\frac{u}{4} - 1\right) \simeq \lim_{b\to 0}\frac{2}{a}\ln\left(\frac{b}{a}\tan\frac{u}{4}\right) \qquad 2.2.1.13$$

The last step come from the **Lemma 1**. In **case 2**, we also let $a = 1$, so we have the solution:

$$2\ln\left(b\tan\frac{u}{4}\right) = 2\ln\left(\tan\frac{u}{4}\right) = \xi + \xi_0 \qquad 2.2.1.14$$

$$\Rightarrow u = 4\tan^{-1}\left(e^{\frac{\xi+\xi_0}{2}}\right)$$

We let $\xi_0 = 2\ln(b)$, so it is just the standard traveling solution of **SGE**! In Next **Lemma**, I will prove the solutions of Wazwaz are actually the solutions of ordinary **SGE** minusing $\varphi$.

**Lemma** *2*

$$u(\xi) = 4\tan^{-1}\left\{\frac{1}{|\sin\frac{\varphi}{2}|}\tanh\left[\frac{1}{4}(\xi+\xi_0)\right] + \frac{\cos\frac{\varphi}{2}}{\sin\frac{\varphi}{2}}\right\} = -\varphi + 4\tan^{-1}\left(e^{\frac{\xi}{2}}\right), \qquad 2.2.1.15$$

*if* $\sin\frac{\varphi}{2} \neq 0$, *which means* $b \neq 0$ *and* $\varphi \neq 2n\pi$

**Proof** Original $u(\xi) = 4\tan^{-1}\left\{\frac{1}{|\sin\frac{\varphi}{2}|}\tanh\left[\frac{\sqrt{R}}{2}(\xi+\xi_0)\right] + \frac{\cos\frac{\varphi}{2}}{\sin\frac{\varphi}{2}}\right\}$, so we need to decide what the value of $\sqrt{R}$?

From $\int\frac{du}{\sin\frac{u}{2}} = 2\ln(\tan\frac{u}{4}) = \xi + \xi_0$, which imply $a = 1, b = 0$.

Also by $\frac{\sqrt{a^2+b^2}}{4} = \frac{1}{2\sqrt{k-c^2}}\sqrt[4]{\alpha^2+\beta^2}$, if we let $\alpha^2+\beta^2 = R^2, k = 1, c = 0$, so $\sqrt{R} = \frac{1}{2}$ (i.e. If $\sqrt{R} \neq \frac{1}{2}$ in original equation, we can scale $\xi$).

Because $\tanh(\frac{\xi+\xi_0}{4}) = \frac{e^{\frac{\xi+\xi_0}{2}}-1}{e^{\frac{\xi+\xi_0}{2}}+1}$, so inside $4\tan^{-1}$ we obtain:

$$\frac{1}{\sin\frac{\varphi}{2}}\frac{e^{\frac{\xi}{2}}e^{\frac{\xi_0}{2}}-1}{e^{\frac{\xi}{2}}e^{\frac{\xi_0}{2}}+1} + \frac{\cos\frac{\varphi}{2}}{\sin\frac{\varphi}{2}} = \frac{e^{\frac{\xi}{2}}e^{\frac{\xi_0}{2}}(1+\cos\frac{\varphi}{2})+\cos\frac{\varphi}{2}-1}{\sin\frac{\varphi}{2}(e^{\frac{\xi}{2}}e^{\frac{\xi_0}{2}}+1)} = \frac{e^{\frac{\xi}{2}}e^{\frac{\xi_0}{2}}\cos^2(\frac{\varphi}{4})-\sin^2(\frac{\varphi}{4})}{\sin\frac{\varphi}{4}\cos\frac{\varphi}{4}e^{\frac{\xi}{2}}e^{\frac{\xi_0}{2}}+\sin\frac{\varphi}{4}\cos\frac{\varphi}{4}}$$

If we choose $e^{\frac{\xi_0}{2}} = \frac{\sin\frac{\varphi}{4}}{\cos\frac{\varphi}{4}}$, and also $\varphi \neq 2n\pi \Rightarrow \sin\frac{\varphi}{4} \neq 0$,

So inside $4\tan^{-1}$ we get $\frac{e^{\frac{\xi}{2}}\cos\frac{\varphi}{4}-\sin\frac{\varphi}{4}}{e^{\frac{\xi}{2}}\sin\frac{\varphi}{4}+\cos\frac{\varphi}{4}} = -\varphi + 4\tan^{-1}(e^{\frac{\xi}{2}})$.

And $4\tan^{-1}\frac{e^{\frac{\xi}{2}}\cos\frac{\varphi}{4}-\sin\frac{\varphi}{4}}{e^{\frac{\xi}{2}}\sin\frac{\varphi}{4}+\cos\frac{\varphi}{4}} = -\varphi + 4\tan^{-1}(e^{\frac{\xi}{2}})$.

**End of Proof**.

At $\varphi = 2n\pi$, which means $b = 0$, this is **case 2**, So nontrivial solutions of Wazwaz have proved to be simply translation of Sine-Gordon solutions.

### 2.2.2 Conclusions and Results of this section:

We have discussed the solutions of Wazwaz [**20**] We also have shown that they can be transformed into a more physically transparent form in **eqs**. (**2.1.1.12**) where a phase $\varphi$ is introduced. The reference point of the soliton $\xi_0$ is affected by $\varphi$. The form, when applied to a spin chain system, exhibits interesting property. The phase $\varphi$ is related to external fields. By varying external fields adiabatically, and hence $\varphi$, the $\xi_0$ is also changed. This implies that the solitons are moved by external fields. Since the solitons, in turn, are related the spins. We are able to reach the conclusion that the spins can be transported by external fields.

# Chapter 3: Evident of spin transport through bulk state by solving asymmetric Sine-Gordon Equation

## 3.0 Introduction

An adiabatic quantum pump is a device that generates a dc current by a cyclic variation of some system parameters, the variation being slow enough so that the system remains close to the ground state throughout the pumping cycle. After the pioneering work of Thouless [**21**] and Niu and Thoulesss [**22**], the quantum adiabatic pumping physics gets more attention. It is applied to the systems like open quantum dots [**23-25**], superconducting quantum wires [**26**, **27**], the Luttinger quantum wire [**28**], the interacting quantum wire [**29**] and of course the spin systems.

In recent years, spintronics become an exciting new field of research. Various proposals of generating spin current have been studied. Among them, an adiabatic spin transport process is most interesting. Quantum spin transport physics probably has been inspired by the phenomenal work of Thouless [**21**], which is clearly related to the topological explanation of quantum Hall effect by Thouless *et. al.* [**30**]. However Halperin [**31**] pointed out before that the quasi-one dimensional edge states played an important role in quantum Hall effect. Hatsugai [**32**, **33**] showed that the edge states indeed have topological meaning and thus confirm their importance.

Shindou [**34**] has shown that the origin of spin transport is due to the edge state of the system. Fu, Kane and Mele [**35**] and Fu and Kane [**36**] studied a similar problem, also see [**46**]. Among their contribution, they found that the edge state crossing (Kramers degeneracy) is essential for spin transport. So the possibility of spin transport through the bulk states of the system is not revealed from these studies and leave this bulk state spin transport as a open problem.

Here we mention very briefly the basic theme of spin transport in adiabatic process. Suppose we consider a spin chain and constructed a parameter space with $(h_{st}, \Delta) = R(\cos\varphi, \sin\varphi)$ where $h_{st}$ is the applied magnetic field $\Delta$ the dimer states bond strength. Fixing $R$ and varying $\varphi$ adiabatically in time, one can argue that a line integral of $\mathbf{A}_n(\mathbf{K}) = (i/2\pi) < n(\mathbf{K})|\nabla_{\mathbf{K}}|n(\mathbf{K}) >$, where $n(\mathbf{K})$ is the Bloch function for the n-th band, and $\mathbf{K} = (k, \Delta, h_{st})$, on a closed loop yields exactly $\pm 1$ due to the singularity at the origin. In other words, $\mathbf{A}$ is related to a fictitious magnetic field $\mathbf{B}_n(\mathbf{K}) = \nabla_{\mathbf{K}} \times \mathbf{A}_n(\mathbf{K})$. One with the Stokes' theorem, can express the line integral in terms of surface integral ($\int \mathbf{B} \cdot d\mathbf{S}$) where the integration is on two dimensional closed surface enclosing the origin. This is exactly the quantization of particle transport proposed by Thouless [**21**]. It is well known to us that one can express spin chain problem into a spin less fermion problem with Jordan-Wigner transformations. One can use this kind of adiabatic variation of parameters as a tool of quantized spin transport. Shindou considered the spin polarization $P_{S^z} = \frac{1}{N} \sum_{j=1}^{N} j S_j{}^z$ and divided it into two parts, bulk state part and edge state part and he concluded that edge state part of spin polarization contribute to spin transport. Fu and Kane [**36**] considered a similar system with an additional interaction of spin-orbit coupling. They calculated the energy bands of the bulk states and end (edge) states and were able to show clearly that whenever there is Kramers degeneracy of end states, there is spin transport and it has a $Z_2$ symmetry.

We have already mentioned that in all previous studies of spin transport, the contribution is coming from the edge states. Here, most probably first in the literature, we raise the question, whether the edge states are indispensable in spin transport?. We do the rigorous analytical exercises to complete the search of this question. One can see during our analytical derivation that spin transport is nothing but the transport of soliton in the system.

## 3.1 Hamiltonian and Continuum field theoretical studies

We consider a spin 1/2 chain of finite length, described by the Heisenberg Hamiltonian similar to that of Shindou [**34**]. A controlled dimerization amplitude and applied magnetic field are also present. The total Hamiltonian has three parts:

$$H(t) = H_0 + H_{\dim} + H_{st} \qquad 3.1.1$$

where

$$H_0 = J\sum_{i=1}^{N} \mathbf{S}_i \cdot \mathbf{S}_{i+1} \qquad 3.1.2$$

$$H_{\dim} = \frac{\Delta(t)}{2}\sum_{i=1}^{N}(-1)^i(S_i^+S_{i+1}^- + S_i^-S_{i+1}^+) \qquad 3.1.3$$

and

$$H_{st} = h_{st}(t)\sum_i(-1)^i S_i^z \qquad 3.1.4$$

$H_{\dim}$ is the bond alternation term which can be induced by applying an electric field to the spin chain to alter the exchange interaction. It introduces into the system the strength of dimerization $\Delta(t)$. $H_{st}$ is the coupling of the system to a staggered external field $h_{st}(t)$. The time-dependent bond strength $\Delta(t)$ and staggered field $h_{st}(t)$ can be varied adiabatically so as to create a parameter space for Berry phase. We write $\Delta$ and $h_{st}$ as $(h_{st}, \Delta) = R(\cos\varphi, \sin\varphi)$, with $R$ fixed. Varying $\varphi$ adiabatically, we expect spins to be transported. We shall also argue that going through one cycle along the loop, there will be quantized spin component transported from one end to the other. The method of bosonization [**37**~**40**] has been used successfully to treat various one-dimensional systems, including the spin chains. It is suitable for the system we are considering. To this end, we first make the Jordon-Winger transformation to represent spins by fermion field $f_i$ and $f_i^\dagger$. Then, the bosonizations of $f_i$ and $f_i^\dagger$ will performed.

$$f_j = \exp(i\pi\sum^{j-1} S_i^+S_i^-)S_j^- \qquad 3.1.5$$

$$f_j^\dagger = S_j^+\exp(-i\pi\sum^{j-1} S_i^+S_i^-) \qquad 3.1.6$$

and

$$f_j \backsimeq R(x_j)e^{ik_Fx_j} + L(x_j)e^{-ik_Fx_j} \qquad 3.1.7$$

$$f_j^\dagger \backsimeq R^\dagger(x_j)e^{-ik_Fx_j} + L^\dagger(x_j)e^{ik_Fx_j} \qquad 3.1.8$$

where

$$R(x) = \frac{1}{\sqrt{2\pi\alpha}}\eta_1 e^{i[\theta_+(x)+\theta_-(x)]/2} \qquad 3.1.9$$

and

$$L(x) = \frac{1}{\sqrt{2\pi\alpha}}\eta_2 e^{i[-\theta_+(x)+\theta_-(x)]/2} \qquad 3.1.10$$

are the slowly varying fields, and $\eta_1$ and $\eta_2$ are the Klein factors. Here, $k_F$ is the Fermi wave vector and and $\alpha$ is the lattice constant. For a half filled system we have $k_F = \pi/2\alpha$. In the following derivations, we left out the details because they can be found in many textbooks [**37~40**].

$$\begin{aligned} S_j^z &= f_j^+ f_j - \frac{1}{2} \\ &= \frac{\partial_x\theta_+(x_j)}{2\pi} - (-1)^j\frac{1}{\pi\alpha}\sin\theta_+(x_j) \end{aligned} \qquad 3.1.11$$

and

$$\begin{aligned} S_i^+S_{i+1}^- + S_i^-S_{i+1}^+ &= f_j^+ f_{j+1} + f_{j+1}^+ f_j \\ &= -\alpha[4\pi\Pi^2 + \frac{1}{4\pi}(\partial_x\theta_+)^2] - (-1)^j\frac{1}{\pi\alpha}\cos\theta_+(x_j) \end{aligned} \qquad 3.1.12$$

Here $\theta_+ =$ is bosonization phase and $\Pi(x) = -(1/4\pi)\partial_x\widehat{\theta}_-(x)$ is the conjugate momentum of $\theta_+(x)$. Substituting **eqs**. (**5.1.11**) and (**5.1.12**) into **eqs**. (**5.1.1~4**), and dropping the rapidly varying components such as $\sum_j(-1)^j\cos\theta_+(x_j)$, we obtained

$$H = \int dx\{v[\pi\eta\Pi^2 + \frac{1}{4\pi\eta}(\partial_x\theta_+)^2] - \frac{R}{\pi\alpha^2}\sin(\theta_+ + \varphi) + \frac{J}{2\pi^2\alpha^3}\cos 2\theta_+\} \qquad 3.1.13$$

where the velocity

$$v = J\sqrt{1 + \frac{2}{\pi}} \qquad 3.1.14$$

and the quantum parameter

$$\eta = \frac{2J}{v} \qquad 3.1.15$$

were discussed in ref. [**37~40**]. Thus, we have the equation of motion

$$\partial_t^2\theta_+ = v^2\partial_x^2\theta_+ + \frac{4JR}{\alpha^2}\cos(\theta_+ + \varphi) + \frac{4|\gamma|J^2}{\pi\alpha^3}\sin 2\theta_+ \qquad 3.1.16$$

The term of $\sin 2\theta_+$ is irrelevant in the sense of renormalization group analysis, so we consider only the part

$$\partial_\tau^2\theta_+ = \partial_z^2\theta_+ + \cos(\theta_+ + \varphi) \qquad 3.1.17$$

where we have change variables: $z = 2\sqrt{JR}\,x/v\alpha$ and $\tau = 2\sqrt{JR}\,t/\alpha$. It is similar to the standard Sine-Gordon Equation

$$\partial_\tau^2\theta_+ - \partial_z^2\theta_+ + \sin\theta_+ = 0 \qquad 3.1.18$$

which has been well-studied. However, for our purpose which is to study the spin transport, we will solve it on a chain of finite length on which the phase $\varphi$ is no longer a trivial constant but introduces new meaning to the solution. This way, one can recognize the motion of spins from one end to the other.

## 3.2 Analysis of sine-Gordon equation on a finite chain

We shall analyze **eq**. (**3.1.18**) first. The result can be applied to **eq**. (**3.1.17**). **Eq**. (**3.1.18**) has many kinds of solutions. The traveling-wave solutions, such as $\tan^{-1}[\exp(\gamma(z-v\tau))]$ is not suitable for our purpose because they cannot meet fixed boundary conditions. For the finite-length systems, we consider the so called ”separable solutions” [**41~43**].

$$\theta_{+}(z,\tau) = 4\tan^{-1}(A\frac{f(\beta z)}{g(\Omega\tau)}) \qquad 3.2.1$$

and $f(\beta z)$ and $g(\Omega\tau)$ must satisfy the following equations:.

$$(\partial_z f)^2 = (\frac{1}{\beta^2})[-\kappa A^2 f^4 + \mu f^2 + (\frac{\lambda}{A^2})] \qquad 3.2.2$$

and

$$(\partial_\tau g)^2 = (\frac{1}{\Omega^2})[-\lambda g^4 + (\mu-1)g^2 + \kappa] \qquad 3.2.3$$

with the requirements $\mu^2 + 4\kappa\lambda \geq 0$ and $(\mu-1)^2 + 4\kappa\lambda \geq 0$. $A$, $\mu$, $\kappa$, $\lambda$ $\beta$ and $\Omega$ are mutually related constants. We will show how they are determined in a while. First, we would like to put forward the observation that $f(\beta z)$ and $g(\Omega\tau)$ satisfying **eqs**. (**3.2.2**) and (**3.2.3**) are **Jacobi elliptic functions** (**JEF**) [**44**]. Jacobi elliptic functions are defined as the following:

$$u = \int_0^{sn(u)} \frac{dt}{\sqrt{(1-t^2)(1-k^2t^2)}} \qquad 3.2.4$$

where $sn(u)$ is one of the **JEF**s and $k$ is a constant in the range $\in [0.1]$. The second **JEF** is $cn(u)$ where $sn^2(u) + cn^2(u) = 1$. There are more **JEF**s. They can be found in Appendix A. The ones we are going to encounter are $sc(u) = sn(u)/cn(u)$ and $dn^2(u) = 1 - k^2 sn^2(u)$. Both $f$ and $g$ are **JEF**s and their constants are denoted by $k_f$ and $k_g$. $\mu$, $\kappa$ and $\lambda$ are constants determined by $k_f$ and $k_g$. The relations are different for different **JEF**.

Here we give an example of $f(\beta z) = cn(\beta(z-z_0))$ and $g(\Omega\tau) = cn(\Omega\tau)$ where $z_0$ is a constant. With the equation for $cn(u)$ ( see **Table A.1** of **Appendix A**)

$$(\partial_u cn(u))^2 = [1 - cn(u)^2][1 - k^2 + k^2 cn(u)^2] \qquad 3.2.5$$

we found from comparison with **eq**. (**3.2.2**) that $\kappa A^2 = k_f^2$, $\mu = 2k_f^2 - 1$ and $\lambda/A^2 = 1 - k_f^2$. As a result, we get

$$k_f = \frac{A^2}{1+A^2} + \frac{A^2}{\beta^2(1+A^2)^2} \qquad 3.2.6$$

and,

$$k_g = \frac{A^2}{1+A^2} - \frac{A^2}{\Omega^2(1+A^2)^2} \qquad 3.2.7$$

where

$$\Omega^2 = \beta^2 + \frac{1-A^2}{1+A^2} \qquad 3.2.8$$

If we choose the fixed boundary condition $\theta_{+}(z=0) = \theta_{+}(z=L) = 0$ with $L$ being the length of the system, then we will have

$$\beta L = 4lK(k_f) \qquad 3.2.9$$

with

$$K = \int_0^1 \frac{dt}{\sqrt{(1-t^2)(1-k^2t^2)}} \qquad 3.2.10$$

being the complete elliptic integral of the first kind, $z_0 = L/4l$ and $l$ is an integer. Not all the combinations of **JEF**s can satisfy the **SGE**. A table of the differential equations for all the **JEF**s is given in **Appendix A**. We will discuss the solutions of the **SGE** in **eq**. (**3**.**1**.**17**) on a finite system under various boundary conditions. Although finite-length solutions are well-known, different boundary conditions and the presence of $\varphi$ will impose restrictions on the solutions and infuse physical meaning to the wave forms.

**Case** *1: **Periodic boundary condition*** $\theta_+(z,\tau) = \theta_+(z+L,\tau)$

The first boundary condition coming to mind is the periodic boundary condition. There are many combinations of **JEF**s that can satisfy the periodic boundary condition. Here are two examples.

**1a**:

$$\theta_+(z,\tau) = \frac{\pi}{2} - \varphi + 4\tan^{-1}\{Acn[\beta(z-z_0);k_f]cn[\Omega\tau;k_g]\} \qquad 3.2.11$$

where $\beta L = 4lK(k_f)$ and

**1b**:

$$\theta_+(z,\tau) = \frac{\pi}{2} - \varphi + 4\tan^{-1}\{Asc[\beta(z-z_0);k_f]dn[\Omega\tau;k_g]\} \qquad 3.2.12$$

where $\beta L = 2lK(k_f)$ and $z_0$ is a arbitrary constant. This idea come from Laughlin [**5.25**]. For givengiven boundary conditions, and $\varphi$ increase by $2\pi$, $z_0$ will change to fit the boundary conditions. For this boundary condition, there is no spin transport if one varies the parameter $\varphi$ adiabatically. The reason is quite simple. Increasing $\varphi$ only gives a constant change to $\theta_+(z)$ everywhere. Thus the fermion field operators on every site from Jordan-Wigner transformation acquire a constant phase and the spins remain the same.

**Case** *2: **Fixed boundary condition*** $\theta_+(z=0,\tau) = \theta_+(z=L,\tau) = 0$

It seems that we can have the solutions like

$$\theta_+(z,\tau) = \frac{\pi}{2} - \varphi + 4\tan^{-1}\{Acn[\beta(z-z_0);k_f]cn[\Omega\tau;k_g]\} \qquad 3.2.13$$

where $\beta L = 2lK(k_f)$. However, the presence of the adiabatic change term $\pi/2 - \varphi$ in front requires that $cn[\beta(z-z_0);k_f]$ to be finite. Then the function $cn[\Omega\tau;k_g]$ makes the inverse tangent function varying with time and hence, the forms in solution , **eq**.(**3**.**2**.**13**) can not satisfy the fixed boundary condition.

**Case** *3: **Free end boundary condition*** $(\partial\theta_+(z,\tau)/\partial z)|_{z=0} = (\partial\theta_+(z,\tau)/\partial z)|_{z=L} = 0$

The solution is

$$\theta_+(z,\tau) = \frac{\pi}{2} - \varphi + 4\tan^{-1}\{Adn[\beta(z-z_0);k_f]sn[\Omega\tau;k_g]\} \qquad 3.2.14$$

where $\beta L = 2K_f$ and $\beta z_0 = K_f$. The energy is equal to $16\beta E(K)$ where

$$E(K) = \int_0^{\pi/2} \frac{\sqrt{1-k^2t^2}}{\sqrt{1-t^2}} dt \qquad 3.2.15$$

is the complete elliptic integral of the second kind. This solution cannot provide the system with spin transport for the same reason as that in **case 1**.

## 3.3 Detailed analysis of the static soliton case

### 3.3.1 The solutions fit the fixed boundary conditions and the energy

It is most interesting to study the solution of the static soliton of **eq**. (**3.1.17**). Let us first consider the boundary condition

$$\theta_+(z=0) = 0 \qquad 3.3.1.1$$
$$\theta_+(z=L) = 2\pi$$

The phase difference $2\pi$ implies that the fermion field or the spins have same boundary conditions at both ends and hence, a common case for a finite spin chain. On the other hand, it is a fixed boundary condition of $\theta_+$. Therefore, different values of $\varphi$ will induce distinct solutions. The soliton has the form

$$\theta_+(z,\tau) = \frac{\pi}{2} - \varphi + 4\tan^{-1}\{Asc[\beta(z-z_0);k_f]dn[\Omega\tau;k_g]\} \qquad 3.3.1.2$$

where

$$k_f^2 = 1 - A^2 + \frac{A^2}{\beta^2(1-A^2)} \qquad 3.3.1.3$$

$$k_g^2 = 1 - \frac{1}{A^2} + \frac{1}{\Omega^2(1-A^2)} \qquad 3.3.1.4$$

$$\Omega = A\beta \qquad 3.3.1.5$$

and

$$\beta L = K(k_f) \qquad 3.3.1.6$$

**Eqs**. (**3.3.1.3~3.3.1.5**) can be derived by comparing coefficients with standard **JEF**s, *sc* and *dn*. and **eq**. (**3.3.1.6**) comes from the boundary conditions in eq. (**3.3.1.1**). We seek the static solution because it can always satisfy above boundary conditions. In this case, we require $k_g = 0$, $dn(\Omega\tau, k_g = 0) = 1$ and $A$ takes a special value $A_{th}$. In view of **eqs**. (**3.3.1.3**) and (**3.3.1.4**), the static soliton is

$$\theta_+(z,\tau) = \frac{\pi}{2} - \varphi + 4\tan^{-1}\{A_{th}sc[\beta(z-z_0);k_f]\} \qquad 3.3.1.7$$

with

$$\beta = 1/(1-A_{th}^2) \qquad 3.3.1.8$$

and

$$k_f = \sqrt{1-A_{th}^4} \qquad 3.3.1.9$$

The boundary condition at $z = 0$ requires that

$$\tan(\frac{\varphi}{4} - \frac{\pi}{8}) = A_{th}sc(-\beta z_0) \qquad 3.3.1.10$$

which determines $z_0$. For the boundary condition at $z = L$, we have derived the following lemma: For the solution in **eq**. (**3.3.1.7**) with

$$\beta L = K(k_f = \sqrt{1 - A_{th}^4}) = \int_0^1 \frac{dt}{\sqrt{(1-t^2)[1-(1-A_{th}^4)t^2]}} \quad 3.3.1.11$$

the difference of $\theta_+(z = 0)$ and $\theta_+(z = L)$ is always equal to $2\pi$. The derivation is given in **Appendix B**. **Eqs**. (**3.3.1.7**) and (**3.3.1.11**) are the main result of this paper. **Eq**. (**3.3.1.11**) can be generalized as

$$\beta L = lK \quad 3.3.1.12$$

where $l$ is any nonzero integer. As The larger the magnitude of $l$, the higher the energy. In **Fig**. **1** ,$A_{th}$ evaluated with **eq**. (**3.3.1.8**) and (**3.3.1.11**) is plotted. It shows that $A_{th}$ decreases rapidly with increasing $L$. The magnitude of $A_{th}$ is closely related to the wave form of $\theta_+$. A small $A_{th}$ results in a steep change in $\theta_+$, or a sharp domain wall. As it will be shown later, it related to the quantum spin transport.

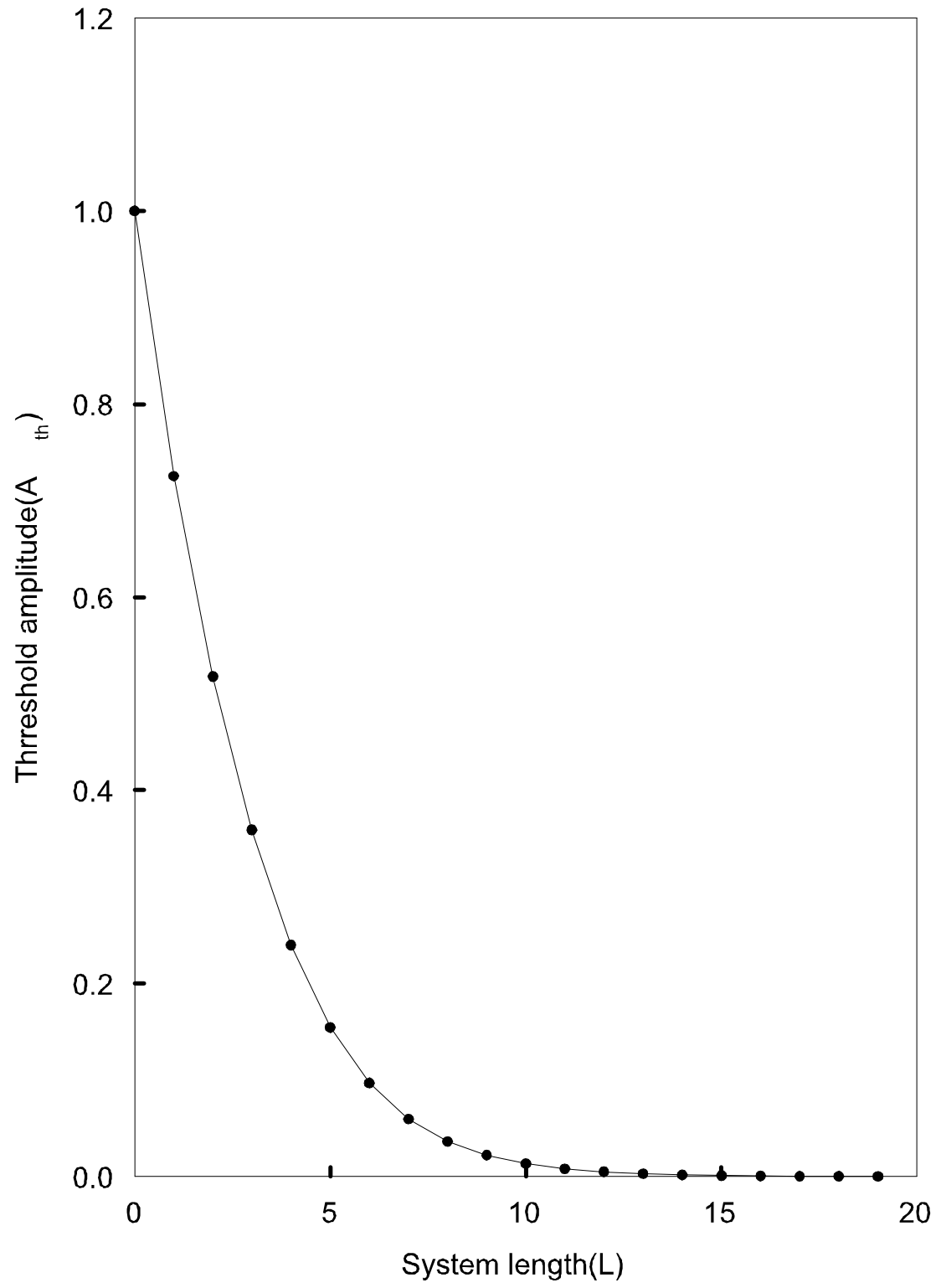

**Fig**. **1** : Threshold amplitude ($A_{th}$) versus system length $L$ for the static soliton solution.

The energy of the static soliton can be calculated with

$$\mathcal{E} = \int_0^L dz[\frac{1}{2}(\frac{\partial\theta_+}{\partial\tau})^2 + \frac{1}{2}(\frac{\partial\theta_+}{\partial z})^2 - \sin(\theta_+ + \varphi)] \quad 3.3.1.13$$

It can be shown that

$$\frac{1}{2}(\frac{\partial\theta_+}{\partial z})^2 + \sin(\theta_+ + \varphi) - \frac{1}{2}(\frac{\partial\theta_+}{\partial\tau})^2|_{z=z_0} - \sin(\theta_+ + \varphi)|_{z=z_0} \quad 3.3.1.14$$

$$= \frac{1}{2}(\frac{\partial\theta_+}{\partial z})^2 + \sin(\theta_+ + \varphi) - \frac{8A_{th}^2}{(1-A_{th}^2)^2} - 1 = 0$$

Hence, **eq**. (**3.3.1.12**) becomes

$$\mathcal{E} = \int_0^L dz[1 + \frac{8A_{th}^2}{(1 - A_{th}^2)^2} - 2\sin(\theta_+ + \varphi)] \qquad 3.3.1.15$$

where the term of the time derivative is dropped for we are considering the static case. We can change the variable of integration and get

$$\mathcal{E} = \sqrt{2}\int_0^{2\pi} d\theta_+[1 + \frac{8A_{th}^2}{(1 - A_{th}^2)^2} - \sin(\theta_+ + \varphi)]^{1/2} - L[1 + \frac{8A_{th}^2}{(1 - A_{th}^2)^2}] \qquad 3.3.1.16$$

Therefore the total energy $\mathcal{E}$ is independent of $z_0$ and $\varphi$ because the integration is over an entire period. $\mathcal{E}$ depends on only one parameter, $\beta$, for static soliton because $A = A_{th}$ and $A_{th}$ is also determined by $\beta$. The spectrum is plotted in **Fig**. **2** with **eq**. (**3.3.1.12**). It is very similar to that of standing wave with $\beta$ being the wave vector.

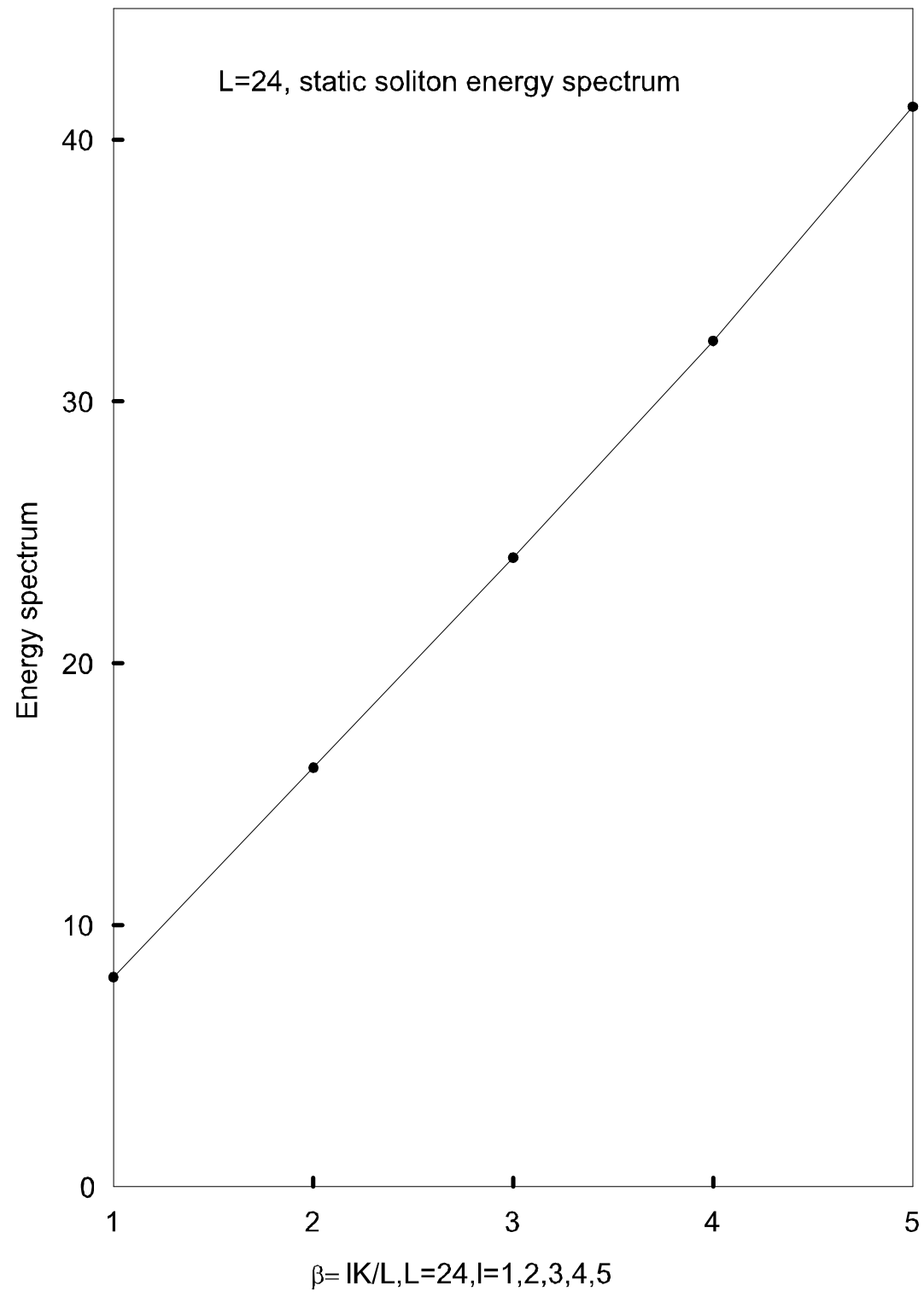

**Fig**. **2** :Energy spectrum of the static soliton versus $\beta L = lK$ with the boundary condition $\theta_+(z = 0) = 0$ and $\theta_+(z = L) = 2l\pi$ when $L = 24$.

In the limit $L \to \infty$, $A_{th}$ becomes vanishingly small as it can be seen from **eqs**. (**3.3.3.11**) and (**3.3.3.8**). We thus have $\beta$, $k_f \to 1$ and $K(k_f) \simeq \ln(4/A_{th}^2)$. With **eq**. (**3.3.3.8**) we get

$$A_{th} \simeq 2\exp(-L/2) \qquad 3.3.1.17$$

The magnitude of $A_{th}$ is small even for a modest length of $L$. For example,when $L = 24$, $A_{th} \simeq 2/e^{12} \simeq 1.23 \times 10^{-5}$. Since $A_{th}$ can be viewed as the amplitude of the nonlinear wave, the wave form on a long spin chain becomes flat everywhere except for narrow regions $cn(z - z_0) \sim 0$. Therefore, one can expect a sudden change of $\theta_+$ or a sharp domain wall.

### 3.3.2 Other solutions with the same $A_{th}, x_0, energy$

The solution of

$$\partial_\tau^2\theta_+ = \partial_z^2\theta_+ + \cos(\theta_+ + \varphi) \quad 3.1.17$$

is

$$\theta_+(z,\tau) = \frac{\pi}{2} - \varphi + 4\tan^{-1}\{A_{th}sc[\beta(z-z_0);k_f]\} \quad 3.3.1.7$$

Although the period of $\varphi$ is $4\pi$. But the period of the equation in **Eq**. (**3.1.17**) is $2\pi$. So there must be another solution which is degenerate to **Eq**. (**3.3.1.7**), which is:

$$\begin{aligned}\theta'_+(z,\tau) &= \frac{\pi}{2} - \varphi - \mathbf{2\pi} + 4\tan^{-1}\{A'_{th}sc[\beta'(z'-z'_0);k_f]\} \\ &= \frac{\pi}{2} - \varphi + 4\tan^{-1}\{-\frac{1}{A'_{th}}\mathbf{cs}[\beta'(z'-z'_0);k_f]\} \\ &= \frac{\pi}{2} - \varphi + 4\tan^{-1}\{A'_{th}sc[\beta'(z'-z'_0)+\mathbf{K};k_f]\}\end{aligned} \quad 3.3.2.1$$

where $\beta' = \frac{1}{1-A'^2_{th}}$ and $k_f = \sqrt{1-A'^4_{th}}$, so $k'_f = \sqrt{1-k_f^2} = A'^2_{th}$. I also used the property: $-2\pi + 4\tan^{-1}(\frac{f}{g}) = 4\tan^{-1}(\frac{-g}{f})$. Finally the properties of the Jacobi elliptic functions as **Appendix B** is used:

$$sn(u+K) = \frac{cn(u)}{dn(u)}, cn(u+K) = -k'\frac{sn(u)}{dn(u)} \quad 3.3.2.2$$

$$\Rightarrow sc(u+K) = \frac{sn(u+K)}{cn(u+K)} = \frac{-cn(u)}{A^2_{th}sn(u)} = \frac{-1}{A^2_{th}}cs(u)$$

Since the period of $sc$ is $2K$, so $sc(u+K) = sc(u-K)$. Similiarly, the period of $4\tan^{-1}$ is $4\pi$, so $2\pi + 4\tan^{-1} = -2\pi + 4\tan^{-1}$. Because $4\tan^{-1}\{A_{th}sc[\beta(z-z_0);k_f]\}$ is an monotonically increasing function. We can also consider a monotonically decreasing function $\widetilde{\theta}_+(z,\tau)$

$$\widetilde{\theta}_+(z,\tau) = \frac{\pi}{2} - \varphi - 4\tan^{-1}\{A_{th}sc[\beta(z-z_0);k_f]\} \quad 3.3.2.3$$

Its boundary condition is"opposite" to **eq**.(**3.3.1.1**):

This is interesting because in **Eqs**. (**3.3.2.3**), $A_{th}$ and $z_0$ are the same as those in **eq**. (**3.3.1.1**). So $\widetilde{\theta}_+$ and $\theta_+$ are actually degenerate states with the same energy, but with opposite boundary condition. if we put this back into the original systems with **eqs**. (**3.1.11**) and (**3.1.12**), but not to consider the alternative term and momentum term ( i.e. because this case is static):

$$S_j^z \simeq \frac{\partial_x\theta_+(x_j)}{2\pi} \quad 3.3.2.4a$$

and

$$S_i^+S_{i+1}^- + S_i^-S_{i+1}^+ \simeq \frac{-\alpha}{4\pi}(\partial_x\theta_+)^2 \quad 3.3.2.4b$$

From **Eq**. (**3.3.2.4a**) and the boundary conditions of $\theta_+$, **eq**. (**3.3.1.1**) gives spin up configuration because the derivative is positive from the left end to the right end. While the boundary conditions of $\widetilde{\theta}_+$, **Eq**. (**3.3.2.2**) give spin down configuration because the derivative is negative from the left end to the right end. So we can describe spin up and spin down in our system.

## 3.4 Spin transport

In this section, we present the results of our calculation above and show how the spin is transported. Furthermore, we show that the spin transported is quantized and the transport is via bulk states. First of all, we plotted $z_0$ versus $\varphi$ in **Fig. 3** with the length of the system $L = 24$. It is worth mentioning that $z_0$ changes sign at $\varphi = \pi/2$ and the slope is steep. $|z_0|$ approaches $L/2$ as $\varphi$ is a little distant from $\pi/2$. Since $sc(u) \approx \sinh(u)$ and $\varphi - \pi/2 = 4\tan^{-1}\{A_{th}sc[\beta(-z_0);k_f]\}$ (the boundary condition at $z = 0$), the argument of the $\tan^{-1}$ function changes exponentially with position. Recall also that $A_{th} \simeq 2\exp(-L/2)$. For the magnitude of $\delta = \varphi - \pi/2 = 4\tan^{-1}\{A_{th}sc[\beta(-z_0);k_f]\}$ to increase from 0 to a finite value, $|z_0|$ has to increase from 0 to approximately $L/2$, a rapid change. This is the reason why the slope of $z_0$ is so steep. In view of **eq. (3.3.1.7)**, $z_0$ can be viewed as a reference point of the solution. Its movement is a clear indication that the solitons is set in motion by $\varphi$. Its motion is not smooth as one can easily see that there is an abrupt change near $\varphi = \pi/2$, a manifestation of the nonlinearity of the solution. Finally, it decreases by a distance $L$ when $\varphi$ increases by $2\pi$.

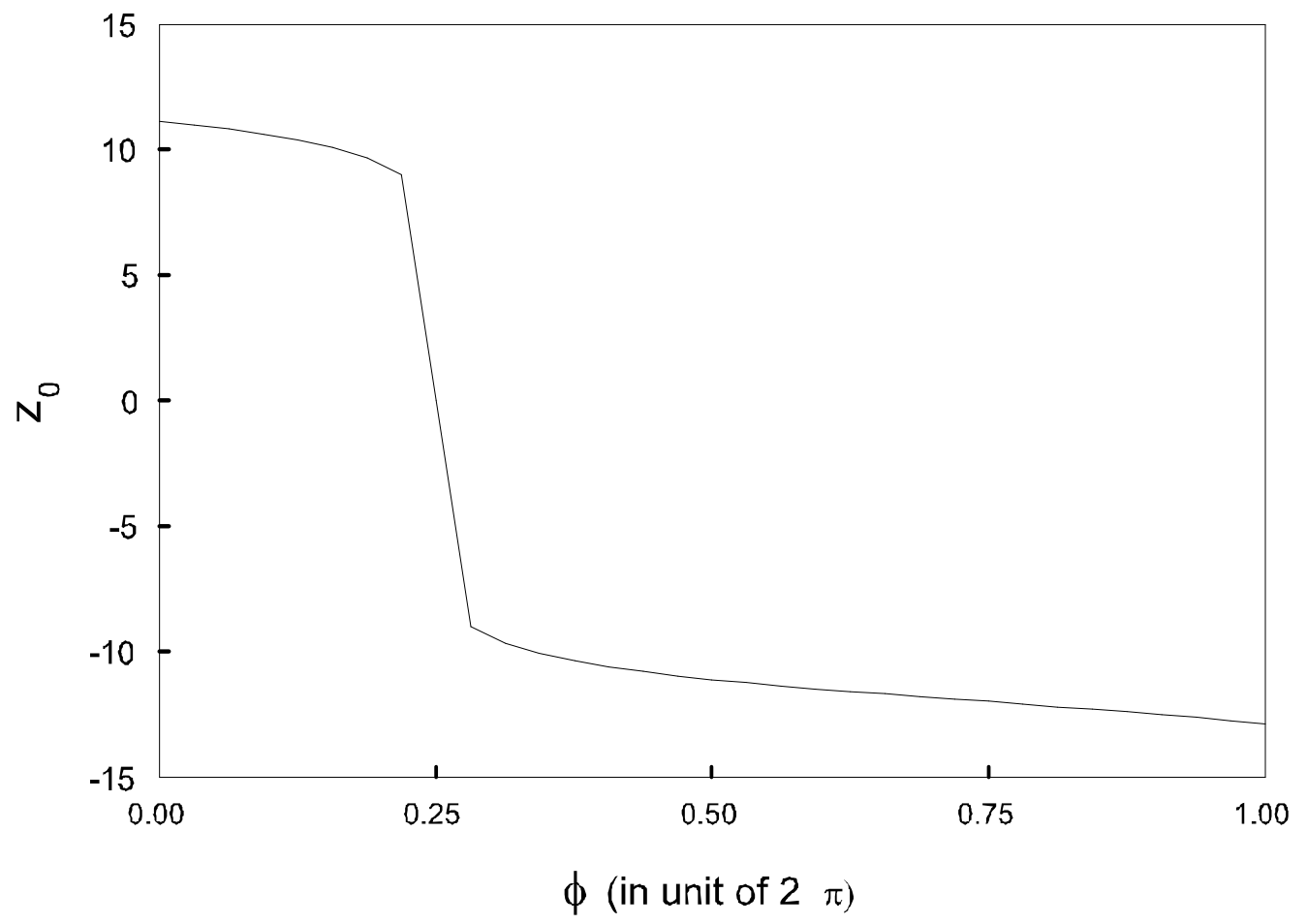

**Fig. 3** : $z_0$ versus adiabatical paramter $\varphi$ with $L = 24$ for static soliton and boundary conditions $\theta_+(z = 0) = 0$ and $\theta_+(z = L) = 2\pi$.

To see more clearly how the soliton moves when $\varphi$ varies, we plotted $\theta_+$, $S_j^z$ and $S_{j+1}^+S_j^- + S_{j+1}^-S_j^+$ versus $z$ in Figures 4 for various values of $\varphi$ and $L = 24$. We can go back to the original spin system to see how spins are transported by utilizing **eqs. (3.1.11)** and **(3.1.12)**. $S_j^z$ contains a term proportional to the gradient of $\theta_+$ and this is the dominant term. Hence, the region where $\theta_+$ changes abruptly can be viewed as a domain wall and it is the place where the peak of $S_j^z$ situates. The behavior of $\theta_+$ is interesting. At $\varphi = 0$ in **Fig. 4(a)**, it is almost flat except at the right end where there is a domain wall. As $\varphi$ increases to $\pi/4$ in **Fig. 4(b)**, the domain wall tends to move away from the right end. When $\varphi$ increases to very near $\pi/2$ in **Fig. 4(c)**, the domain wall moves deep into the bulk. At $\varphi = \pi/2$ in **Fig. 4(d)** the domain wall situated at $z = L/2$. (For comparison, that of $L = 14$ is shown in **Fig. 4(k)**. The edge effect is more important here.) Fig. 4(e) shows that the domain wall moves further to left as $\varphi$ increases by a small amount. When $\varphi = 3\pi/4$ in **Fig. 4(f)**, the domain wall moves to a place near the left end. In the range $3\pi/4 < \varphi < 2\pi$ in **Fig. 4(g-j)**, there is no apparent movement of domain wall but the plateau of $\theta_+$ is lowered. *Thus, the peak of $S_j^z$ moves together with domain wall and we presented clear evidence that spins moves from right to left with varying $\varphi$.* The reason why the domain wall or the peak of $S_j^z$ moves and moves so swift is the following. Physically, Figures 4 show that the static soliton solution really is a Néel state in the spin chain, except in the neighborhood of $\varphi = \pi/2$. In this region, the Néel state becomes unstable due to the dimer coupling $\Delta(t)$ we added. This is manifest in **Figs**. **4b-4f** where $S_{j+1}^{+}S_j^{-} + S_{j+1}^{1}S_j^{+}$ which is proportional to the dimer state amplitude is large. Recall that $\varphi$ is defined in $(h_{st}, \Delta) = R(\cos\varphi, \sin\varphi)$. The dimer strength $\Delta$ is the the largest when $\varphi \simeq \pi/2$. It is when the Néel state becomes unstable and the dimer state amplitude becomes significant that the transport of spin becomes possible. To see the spin transport quantitatively, we consider the case of large $L$ which can be simulated very well by the case $L = 24$. The following equation will be very useful for our purpose

$$\frac{\partial\theta_+}{\partial z} = \frac{4A_{th}\beta dn(\beta(z-z_0))}{cn^2(\beta(z-z_0)) + A_{th}^2 sn^2(\beta(z-z_0))} \qquad 3.4.1$$

since it is the dominant contribution to $S^z$. From **eqs**. (**3.3.1.8**), (**3.3.1.17**), (**A-11**), (**A-17**) **and** (**A-18**), we find that

$$\frac{\partial\theta_+}{\partial z} \simeq \frac{8e^{-L/2}\cosh(z-z_0)}{1+4e^{-L}\sinh^2(z-z_0)} \qquad 3.4.2$$

The peak of $\partial\theta_+/\partial z$ or $S^z$ is at

$$z - z_0 \simeq L/2 \qquad 3.4.3$$

The larger $L$, the narrower the peak. $S_j^z$ and $\partial\theta_+/\partial z$ increase and decrease exponentially near the peak. When $\delta = \varphi - \pi/2$ where $\delta$ is a small in magnitude but finite, we have

$$\tan(\frac{\delta}{4}) = A_{th}sc(-\beta z_0) \simeq A_{th}\sinh(-\beta z_0) \qquad 3.4.4$$

by the boundary condition at $z = 0$.If $\delta < 0$ then $z_0 \simeq L/2$. The resulting $S_j^z$, due to **eqs**. (**3.1.11**) and (**3.4.1**), has a peak at the right end and vanishes everywhere else. When $\varphi = \pi/2$, **eq**. (**3.4.4**) gives $z_0 = 0$ and the peak of $S_j^z$ moves to the center of the spin chain. If $\delta > 0$, then $z_0 \simeq -L/2$, and the peak moves to the left end. As $\varphi$ increases onward from $\pi/2 + \delta$, there is no spin transport in the bulk. Nevertheless, the plateau of $\theta_+$ is lowered (see **Figs**. **4g-4j**) and as a result, the peak of $S^z$ at the left end decreases and a peak of $S^z$ at the right end start to grow. This kind of change continues until $\varphi$ reaches $5\pi/2 - \delta$. At this stage the state of soliton returns to that of $\varphi = \pi/2 - \delta$. We will elaborate more on the quantity of spins being transported. This can be done with **eq**. (**3.1.11**). The spin polarization is

$$P_{S^z} = \frac{1}{L}\int_0^L zS^z(z)dz \qquad 3.4.5$$

By integration by parts, we found

$$P_{S^z} = \int_0^L S^z(z')dz' - \frac{1}{L}\int_0^L\int_0^z S^z(z')dz'dz \qquad 3.4.6$$

To find the variation of $P_{S^z}$ due to $\varphi$ we note that the first term remain constant as $\varphi$ varies. This can be seen by substituting **eq**. (**3.1.11**) into the integration. The contribution of the oscillatory term vanishes as $L \to \infty$ and the term of the derivative gives unity due to our boundary condition, no matter what the value of $\varphi$ is. Thus, denoting the variation of $P_{S^z}$ due to the adiabatic change of $\varphi$ by $\delta P_{S^z}$, we have

$$\delta P_{S^z} = -\frac{1}{L}\{[\int_0^L\int_0^z S^z(z')dz'dz]|_{\varphi=\varphi_2} - [\int_0^L\int_0^z S^z(z')dz'dz]|_{\varphi=\varphi_1}\} \qquad 3.4.7$$

$$\simeq -\frac{1}{2\pi L}\{[\int_0^L \theta_+(z)dz]|_{\varphi_2} - [\int_0^L \theta_+(z)dz]|_{\varphi_1}\}$$

where the second step can be reached by using the approximation

$$S^z \simeq (\partial\theta_+/\partial z)/2\pi \qquad 3.4.8$$

in integration for large $L$. In view of **Figs**. **4(b-f)** where $\varphi$ increases exceeding $\pi/2$, we found that $\theta_+$ increases in the entire length of the system by amount approaching $2\pi$ if $L \longrightarrow \infty$ and hence, $\delta P_{S^z} \simeq -1$ around $\varphi = \pi/2$ and a spin *unity* is moved from right to left around $\varphi = \pi/2$. In **Figs**. **4g-4j**, where $\theta_+$ is almost constant away from ends, we did not see any spin movement in the bulk but rather, there are changes of spins at both ends. **Thus a quantized spin is transported**. From above analysis we conclude that there is a swift spin transport in the bulk during the short interval between $\varphi = \pi/2 - \delta$ and $\varphi = \pi/2 + \delta$ where $\delta$ can be made arbitrarily small if $L \rightarrow \infty$. The net spin transported is unity and as shown by **Figs**. **4b-4f**, **it is transported through bulk**.

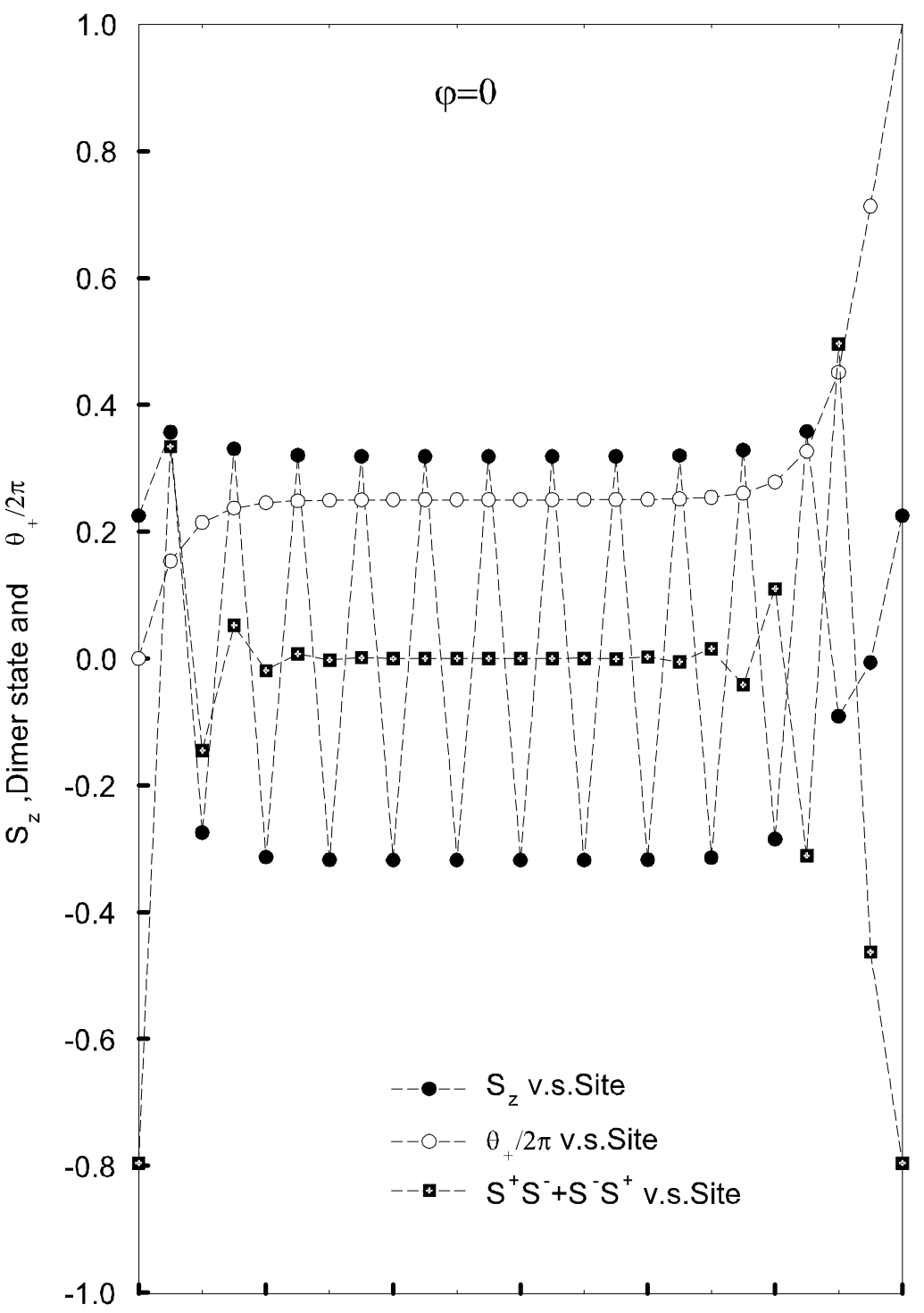

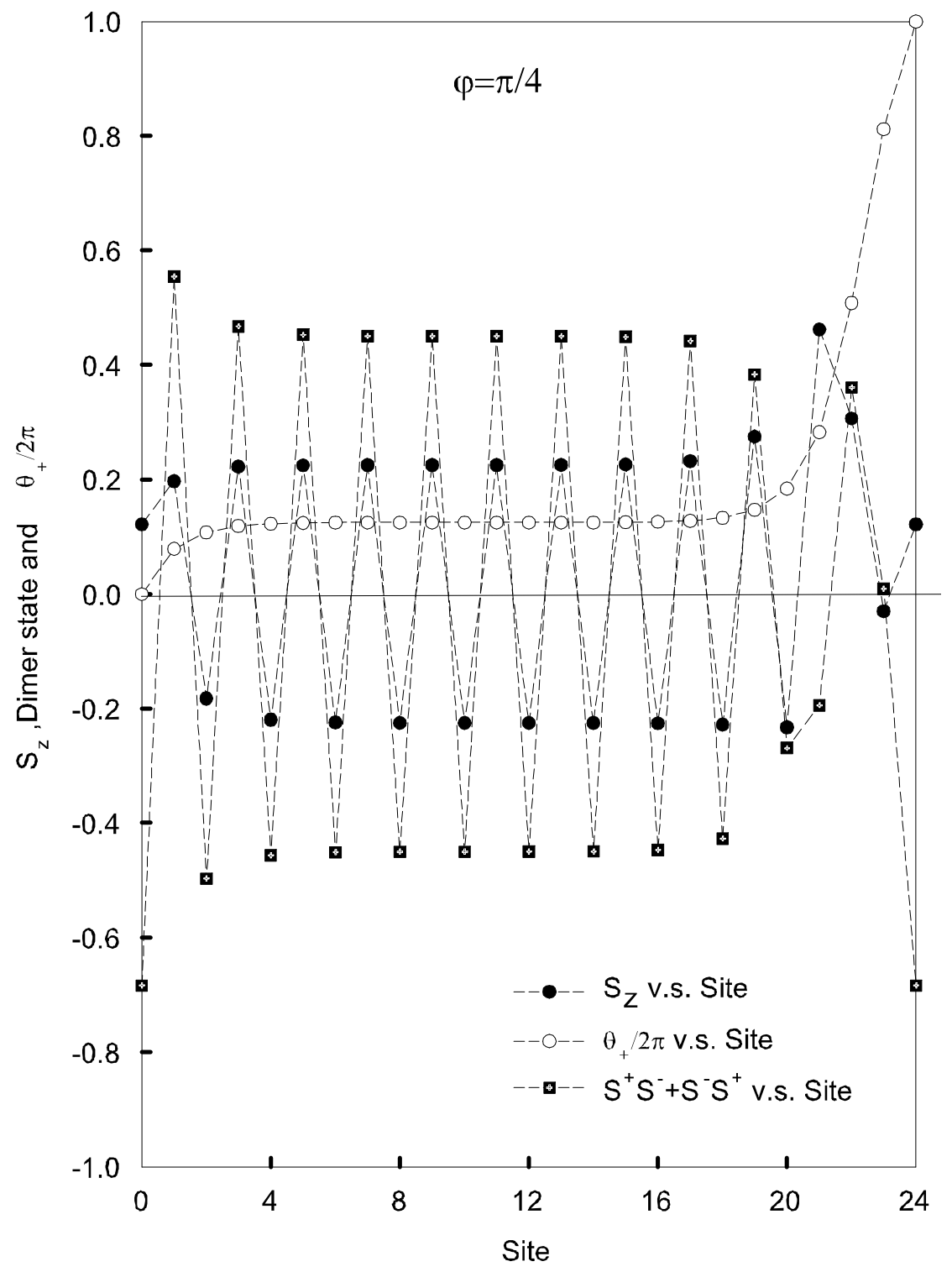

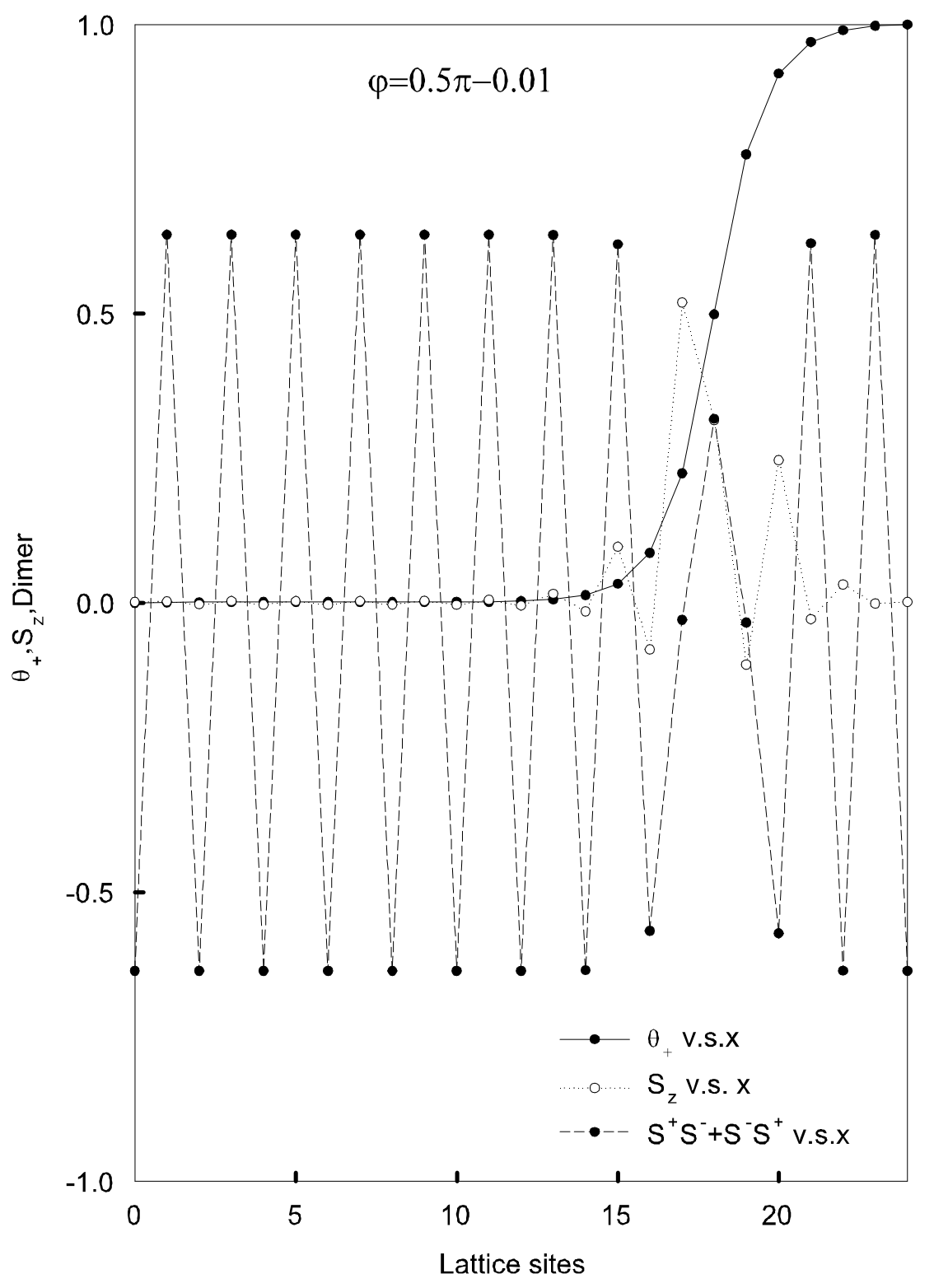

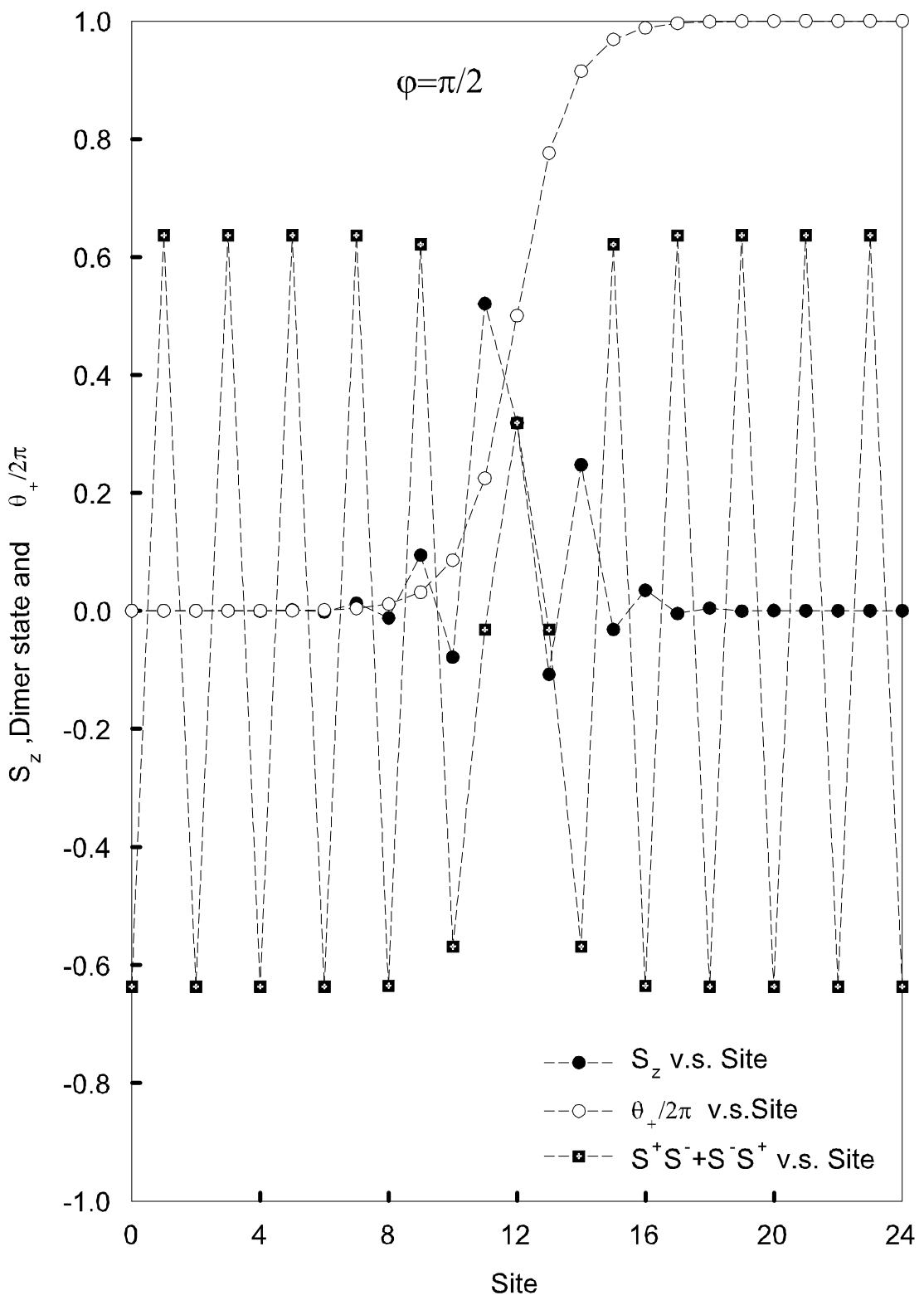

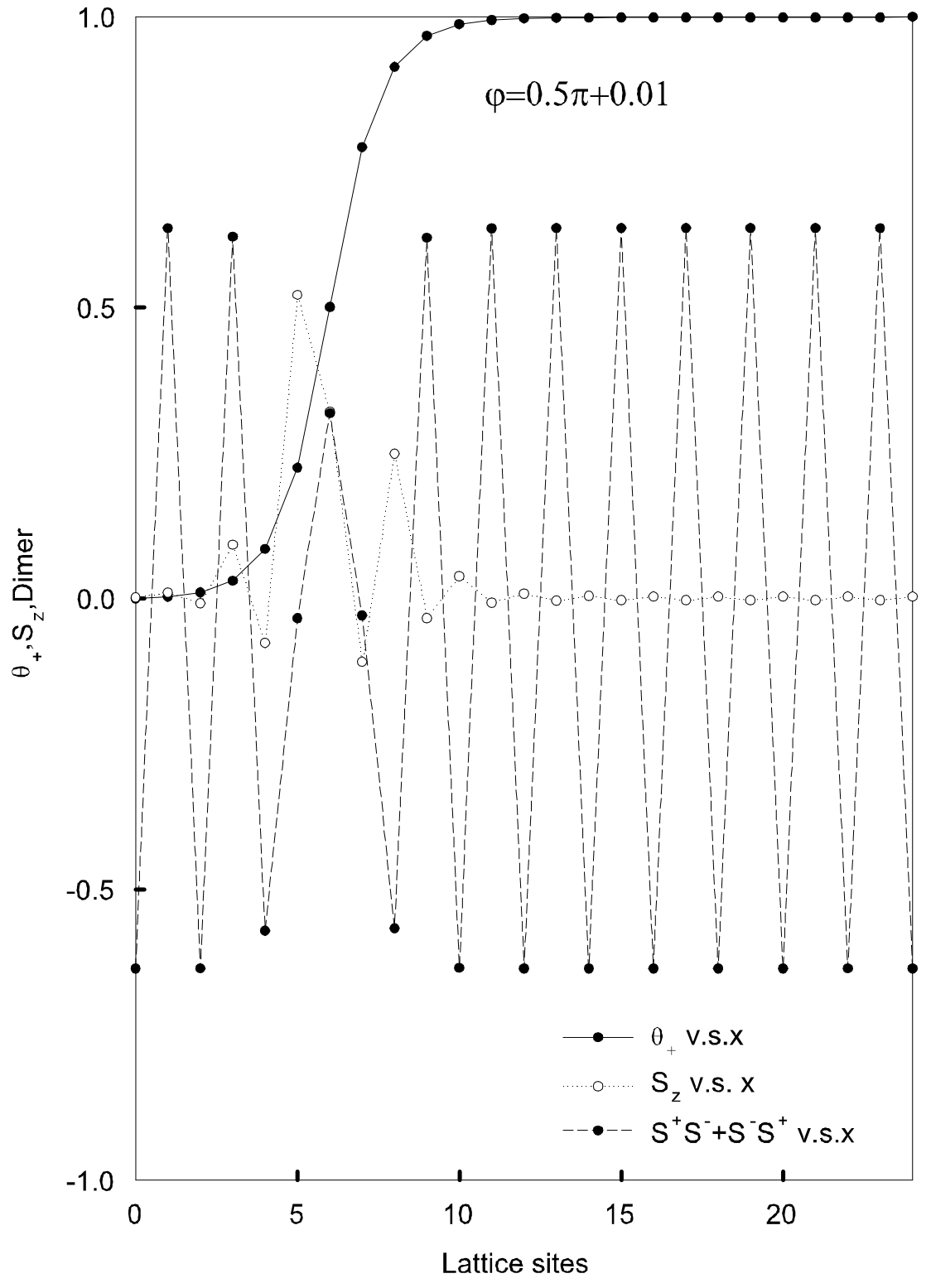

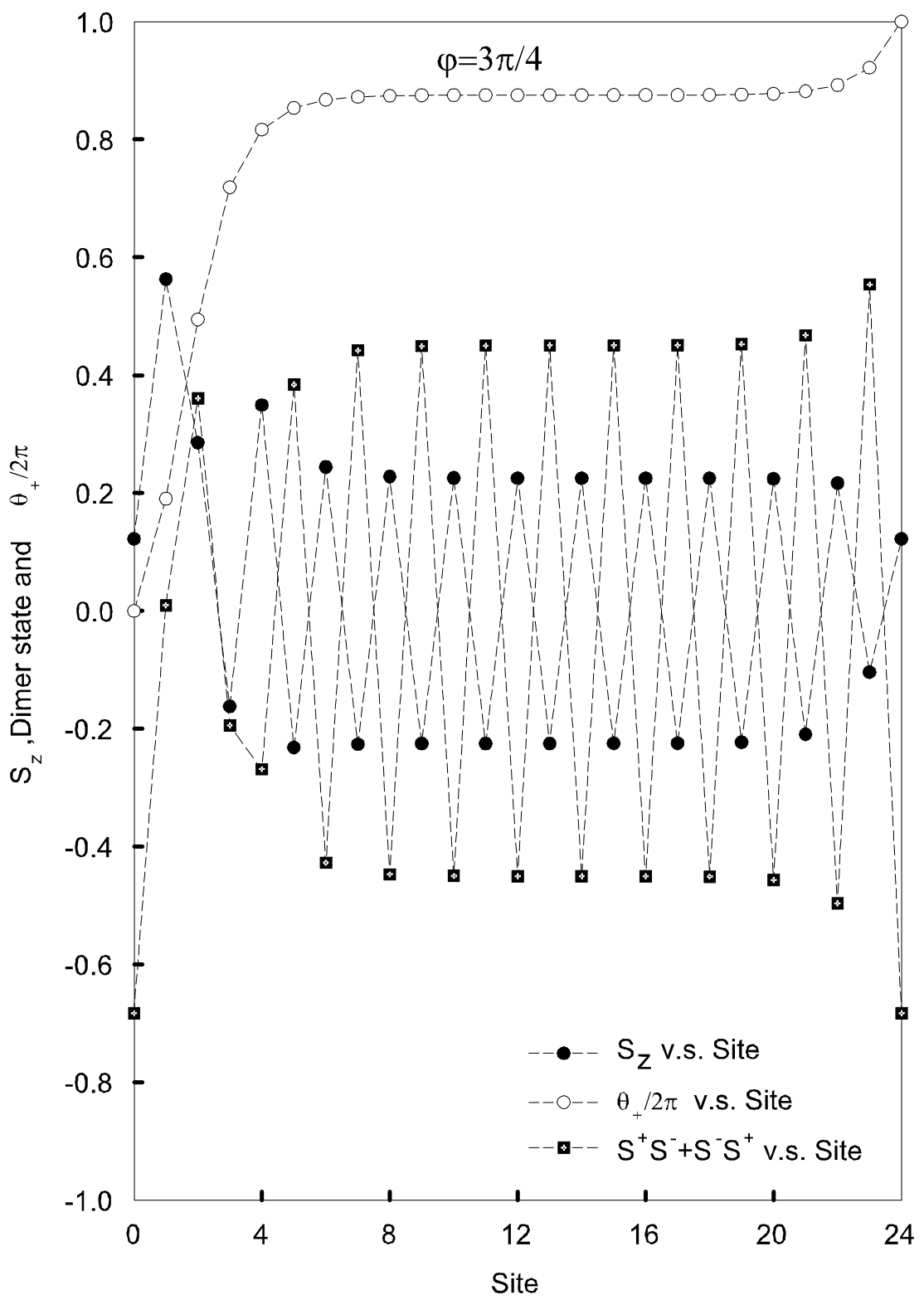

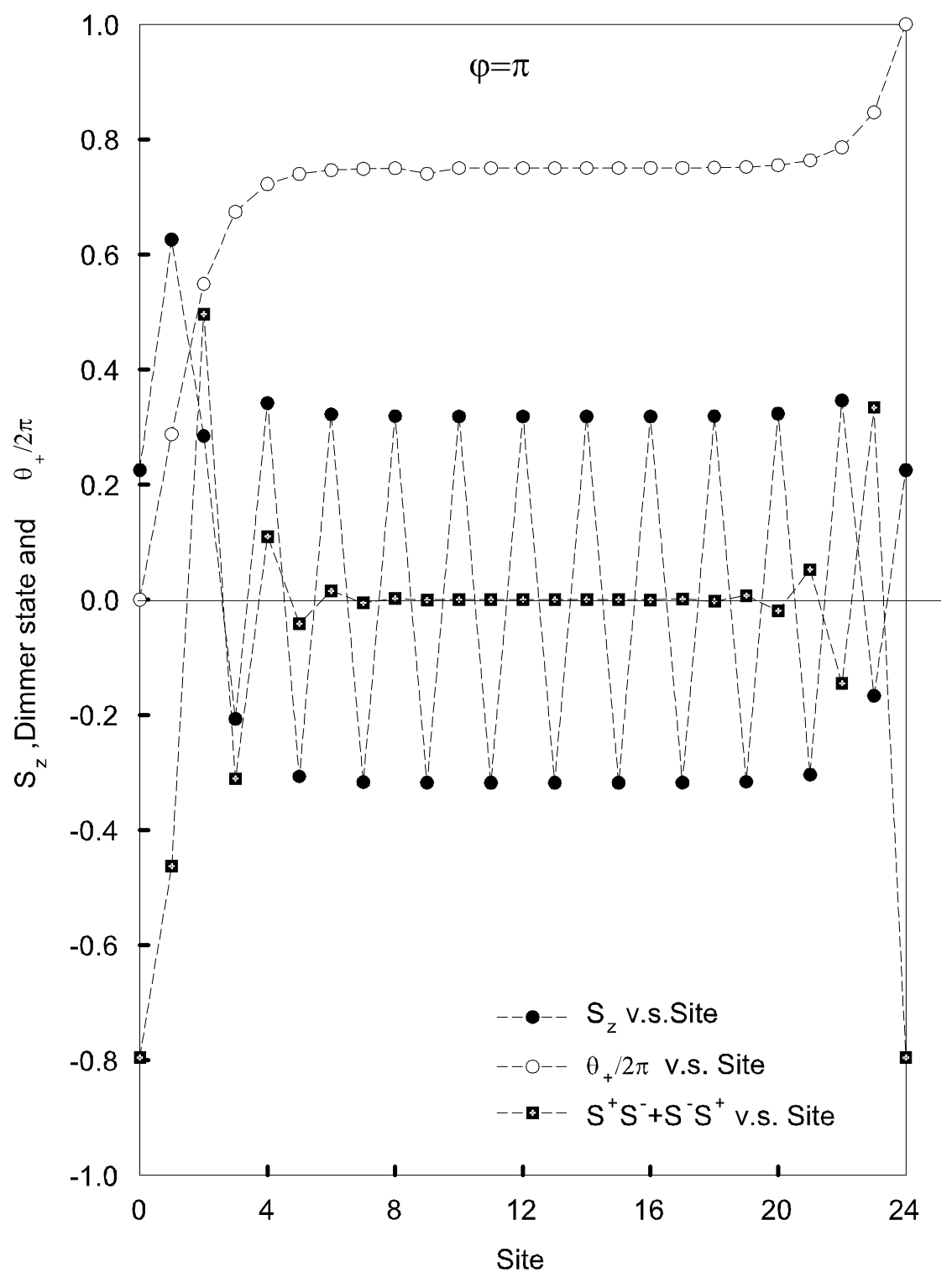

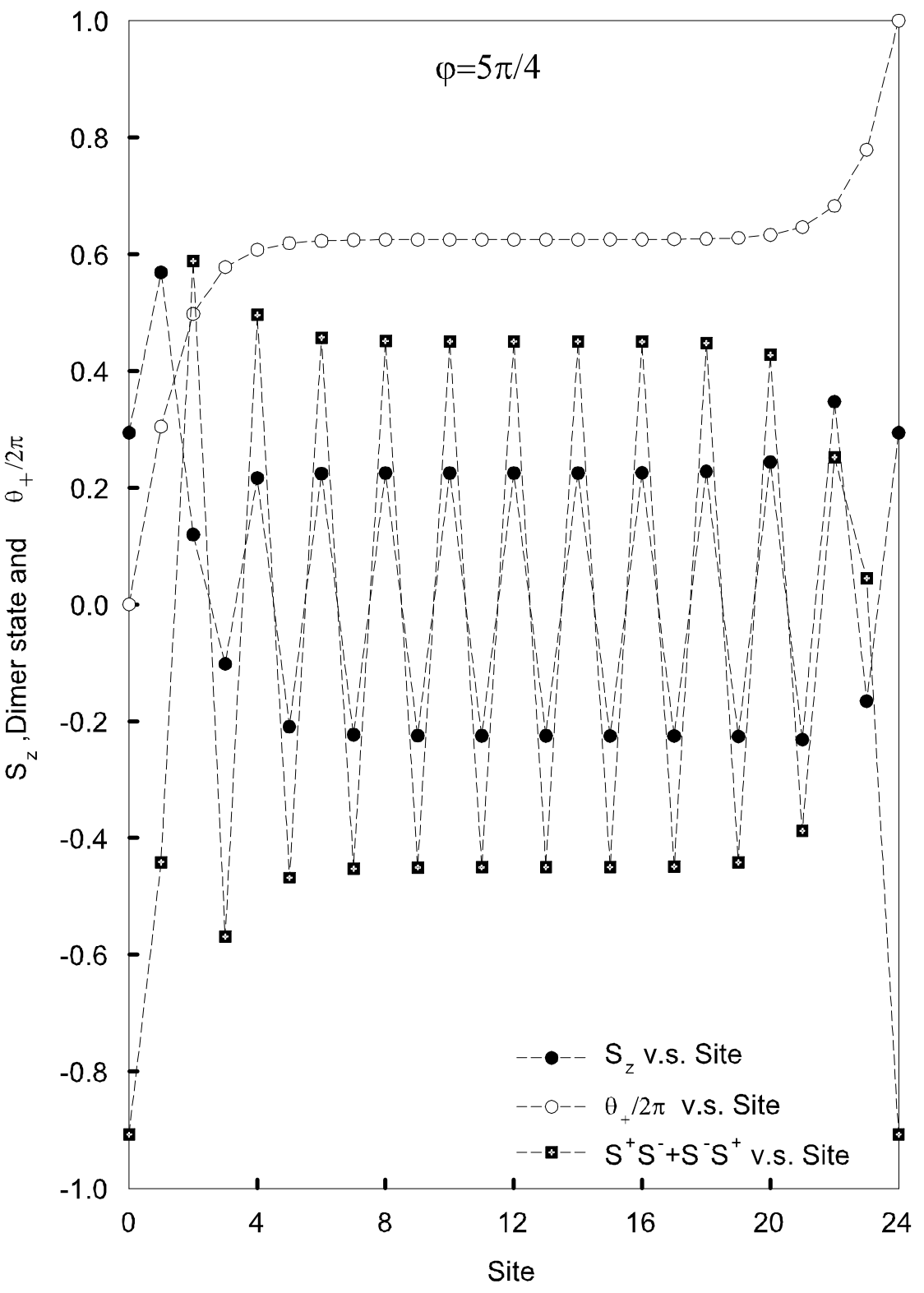

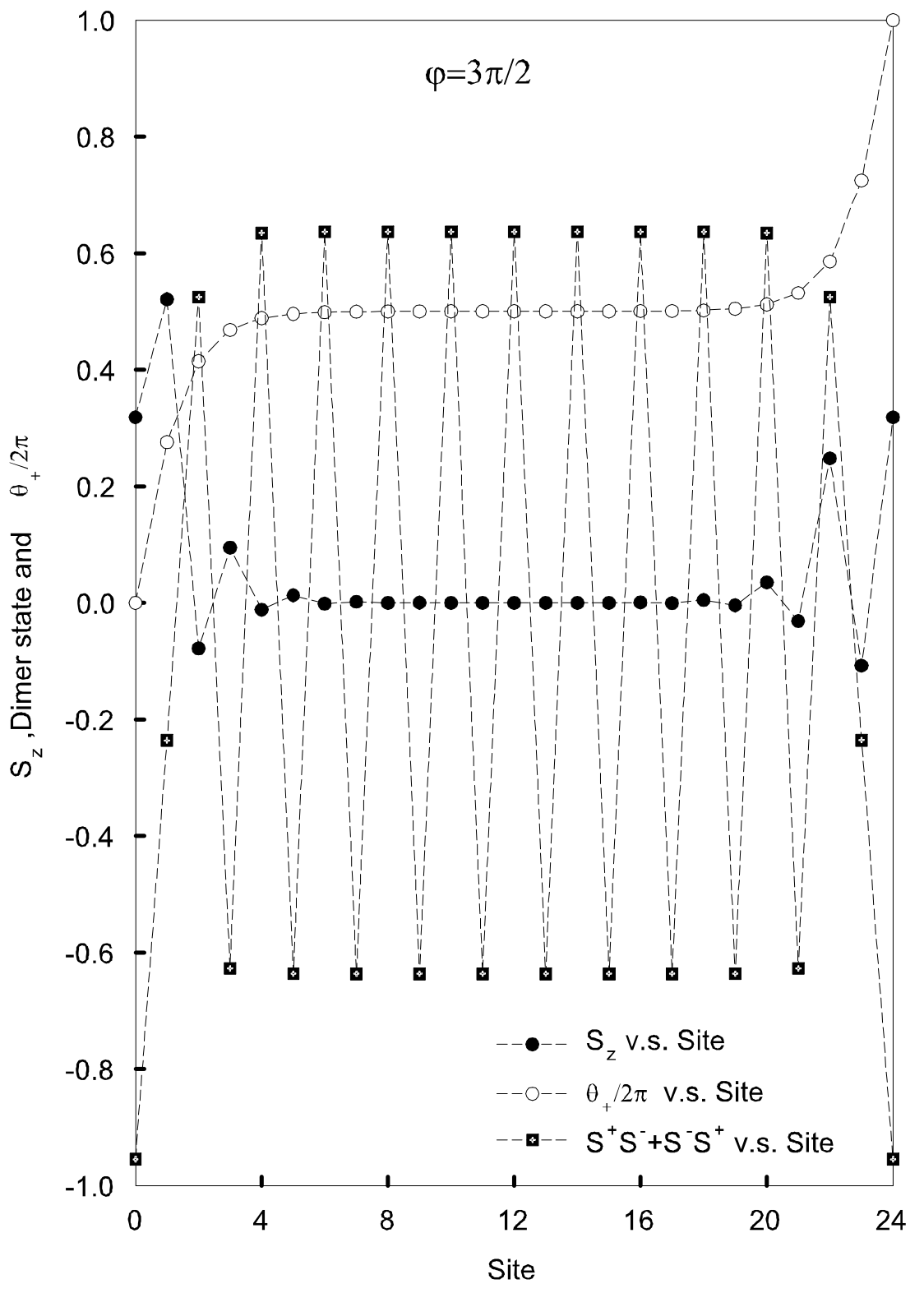

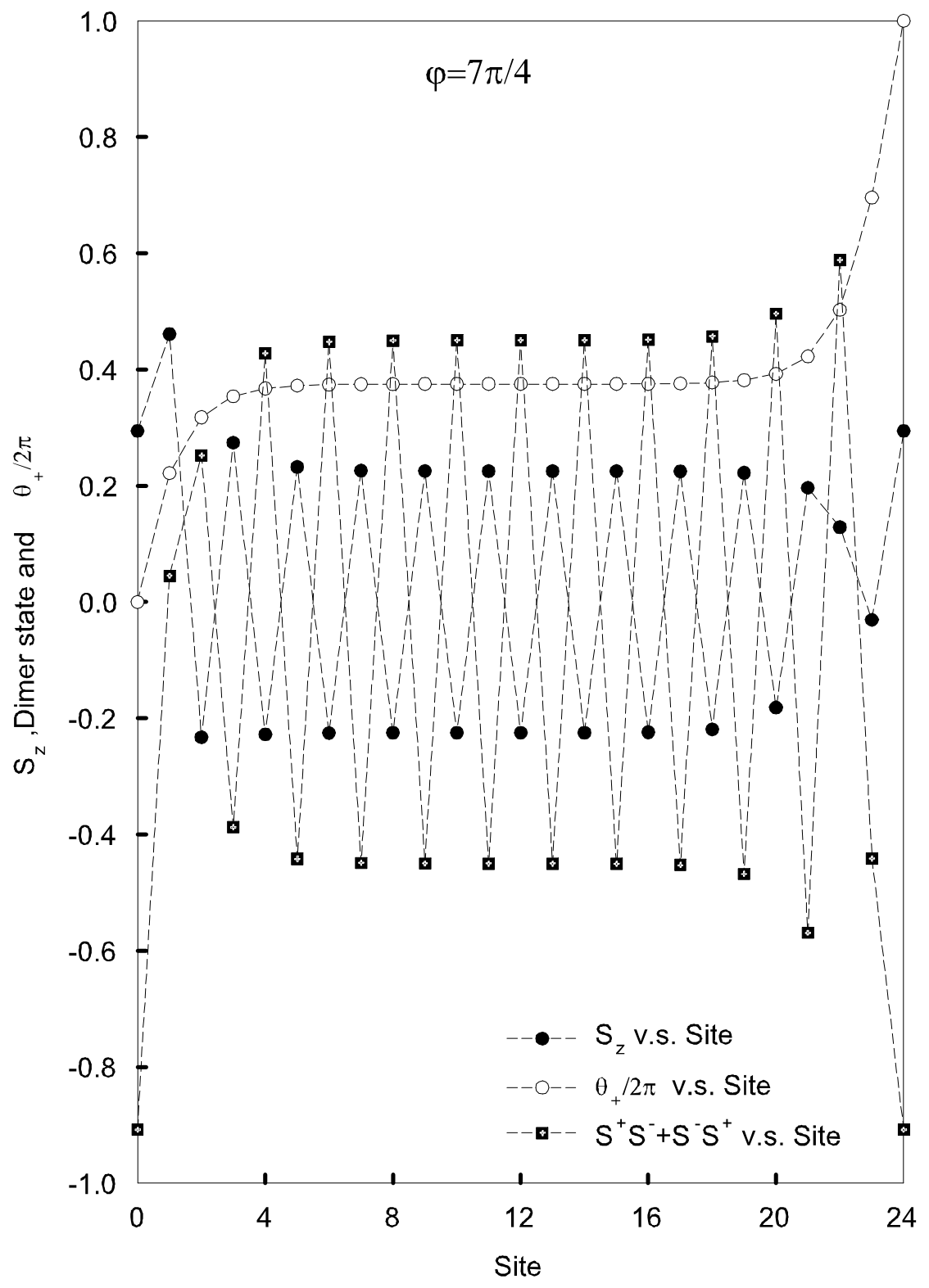

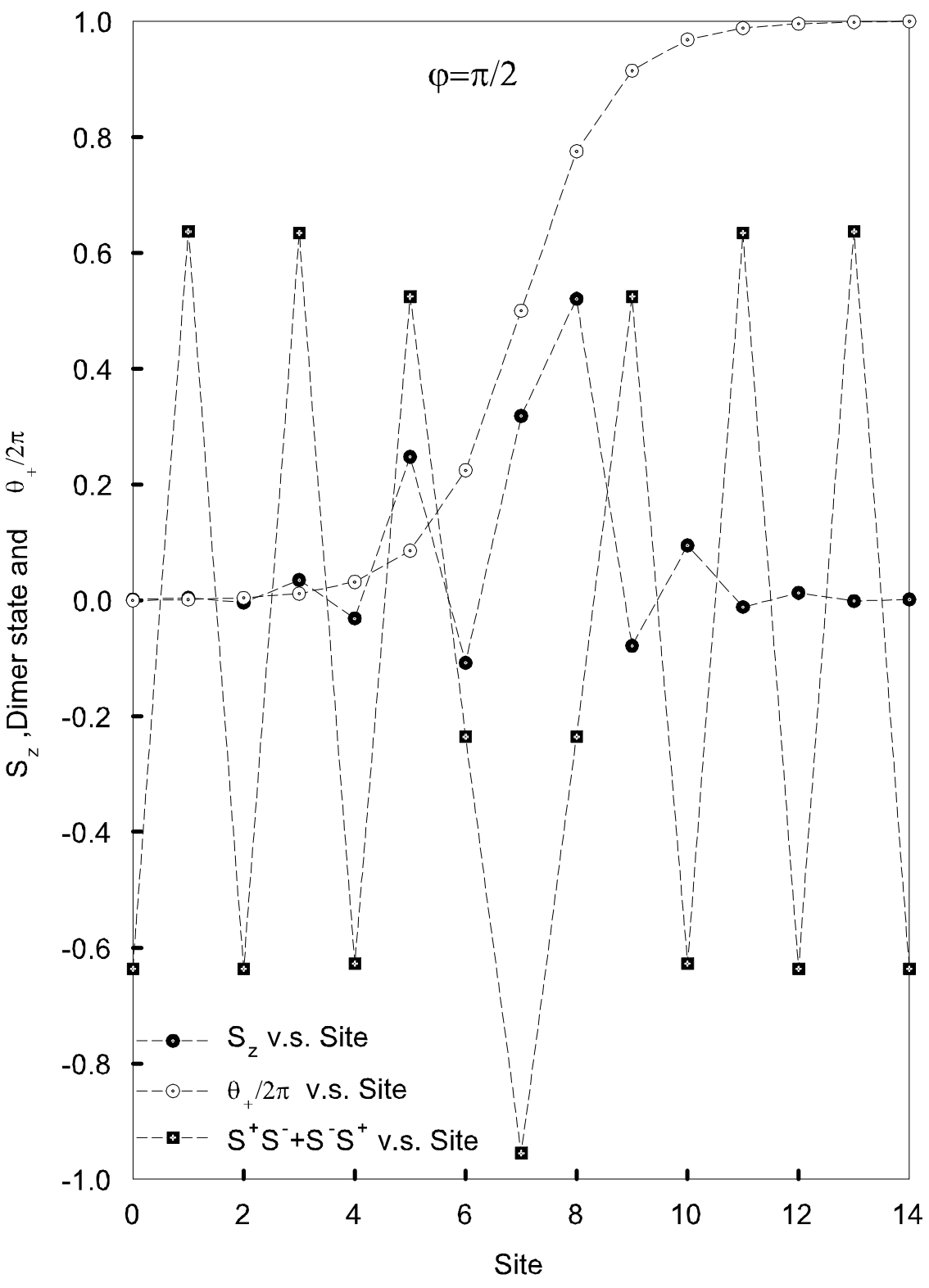

**Fig. 4(a)**: $\varphi = 0$; **4(b)**: $\varphi = \pi/4$; **4(c)**: $\varphi = \pi/2 - 0.01$; **4(d)**: $\varphi = \pi/2$; **4(e)**: $\varphi = \pi/2 + 0.01$; **4(g)**: $\varphi = 3\pi/4$; **4(f)**: $\varphi = \pi$; **4(h)**: $\varphi = 5\pi/4$;**4(i)**: $\varphi = 3\pi/2$; **4(j)**: $\varphi = 7\pi/4$; **4(k)**: $\varphi = \pi/2$ and $L = 14$

## 3.5 Spin transport phenomena connecting to source and drain

Either from **Figs**. **4**(**a**-**j**) or from **eq**. (**3**.**3**.**1**.**7**) one can see that the soliton returns to the starting state if $\varphi$ increases by $2\pi$. Thus the question will inevitably be raised: How can spin be transported in practice? In a realistic system, two ends of the spin chain must be connected to leads. The leads ought to serve as a spin source and a spin drain. Thus it is reasonable to envisage the following picture: Starting at $\varphi = \pi/2 + \delta$ the left end can start to dump spin into a spin drain and the right end can start to extract spin from the source. When $\varphi = 5\pi/2 - \delta$ which is equivalent to $\varphi = \pi/2 - \delta$, the dumping of spin at the left end is complete and the peak of spin at the right end has grown into saturation. Then an unity of spin is transported from the right end to the left end when $\varphi$ increases from $\varphi = \pi/2 - \delta$ to $\varphi = \pi/2 + \delta$. This is in all intent and purpose, same as a physical system of spin transport.

We provide the following picture of source and drain. The spin source and drain are consist of many spin chains. Each chain is identical to our system which is of length $L$. Thus, in the ideal case the source extended from $L$ to $+\infty$ and drain from 0 to $-\infty$. In the process of spin transport, we assume that the source and drain have many solitons as that in our system. These solitons are connected smoothly. Our boundary conditions are $\theta_+(z = 0) = 0$ and $\theta_+(z = L) = 2\pi$ in the system. In the source and drain $\theta_+(z = 2mL) = 2m\pi$ where integer $m > 1$ for the source and $m < 0$ for the drain. This can be seen easily from **eq**. (**3**.**3**.**1**.**11**) or **Appendix B**. Since $\beta L = K$ and the period of **JEF** $sc(u)$ is $2K$, the angle given by $\tan^{-1}$ in **eq**. (**3**.**3**.**1**.**7**) increases by $\pi/2$ if $z$ increases by $L$. Hence, $\theta_+(z + mL) = \theta_+(z) + 2m\pi$.

Above result is shown in **Figs**. **5**. As $\varphi$ increases, $\theta_+$ in the system varies and so do the $\theta_+$ in the source and drain accordingly. In **Fig**. **5**(**a**), $\varphi$ increases from $\pi/4$ to $3\pi/4$, one domain wall moves from right to left inside the system. Recall that the peaks of $S^z_j$ situate near the domain walls. Thus, both of them move continuously from source into the system and from system into the drain. For $\varphi$ increases from $3\pi/4$ to $9\pi/4 = \pi/4 + 2\pi$, as shown in **Fig**. **5**(**b**), the variation of $\theta_+$ inside the system is such that its plateau is descending. There is no spin transport in bulk. However, at $z = 0$ and $z = L$, the heights of domain walls change and thus, the heights of the peaks of $S^z_{j=0,24}$. This means that the system extracts spin from source or dump spin into the drain. In view of **eq**. (**3**.**3**.**1**.**7**), we found that $\theta_+$ keeps pace with $\varphi$ since $z_0$ changes very little in this range. Hence, the plateau descends uniformly with the variation of $\varphi$. Once the plateau comes close to 0, another plateau moves into the system from right and spin is transport. The spin movement in bulk is also evidenced by the variation of $z_0$. It starts near $L/2$ at $\varphi = \pi/4$ and decreases continuously to $-L/2$ at $\varphi = 3\pi/4$. As $\varphi$ increases further, $z_0$ decreases such that $z_0 < -L/2$. Since the peak position is at $z_0 + L/2$, it moves from the source into the system and then into the drain. **Hence**, **in the limit** $L \to \infty$, **a quantized spin is transported**. Now I will give a more accurate mathematical description on this phenomena, by $\theta_+$ function form:

$$\theta_+(z,\tau) = \frac{\pi}{2} - \varphi + 4\tan^{-1}\{A_{th}sc[\beta(z - z_0(\varphi)); k_f]\} \qquad 3.5.1$$

and by the picture of **Fig**.**5**, I give a "step by step" picture with this $\theta_+$ v.s. $z$ figure. As parameter $\varphi$ increase,we know $4\tan^{-1}\{A_{th}sc[\beta(z - z_0(\varphi)); k_f]\}$ is translated along z-axis and $\frac{\pi}{2} - \varphi$ is translated parallel $\theta_+$-axis. To satisfy the static soliton boundary condition as $\varphi$ increase,we can imagine that $\theta_+ - curve$ is fixed. But by "laboratory coordinate", the $\theta_+$ v.s. $z$ coordinate frame is move along the curve. Because $\frac{\partial\theta_+}{\partial\varphi} = -1,$ this picture scrols uniformly along $\theta_+ - axis$. Further polarization decreases uniformly with $\varphi$. So by this accuate picture, we can along explain why when $\varphi$ is not near $\frac{\pi}{2}$, the $\theta_+$ plateau descends nearly uniformly. It is because the steep part of $\theta_+$ is nearly parallel to the movie scroll ( i.e. $\theta_+ - axis$) is at $\theta_+ = 0, 2\pi$. Scroll on the other hand,when $\varphi$ is near $\frac{\pi}{2}$, the nearly zero-slop part is at $\theta_+ = 0, 2\pi$, but still movie score goes down uniformly, so we can see "spin transport" movie soon from right to left because it will compense the nearly no-slop part of $\theta_+$.

By this short mathematical describe, we can see not only near $\varphi = \frac{\pi}{2}$ can spin is ready to pump, actually the uniformly decrease polarization tell us the system pumps spin at every $\varphi$.

The spin transport in our picture is different from those of Shindou's [**34**] and Fu and Kane's [**35**; **36**] in one point. In these authors' picture, the spin transport is carried out by end states and the level crossing of the end (edge) states is essential while in ours it through a bulk state. It is important because the quantized spin transport is protected by the fact that it is through a bulk state. Connecting to spin reservoir cannot destroy the quantization as it will do the transport due to end states.

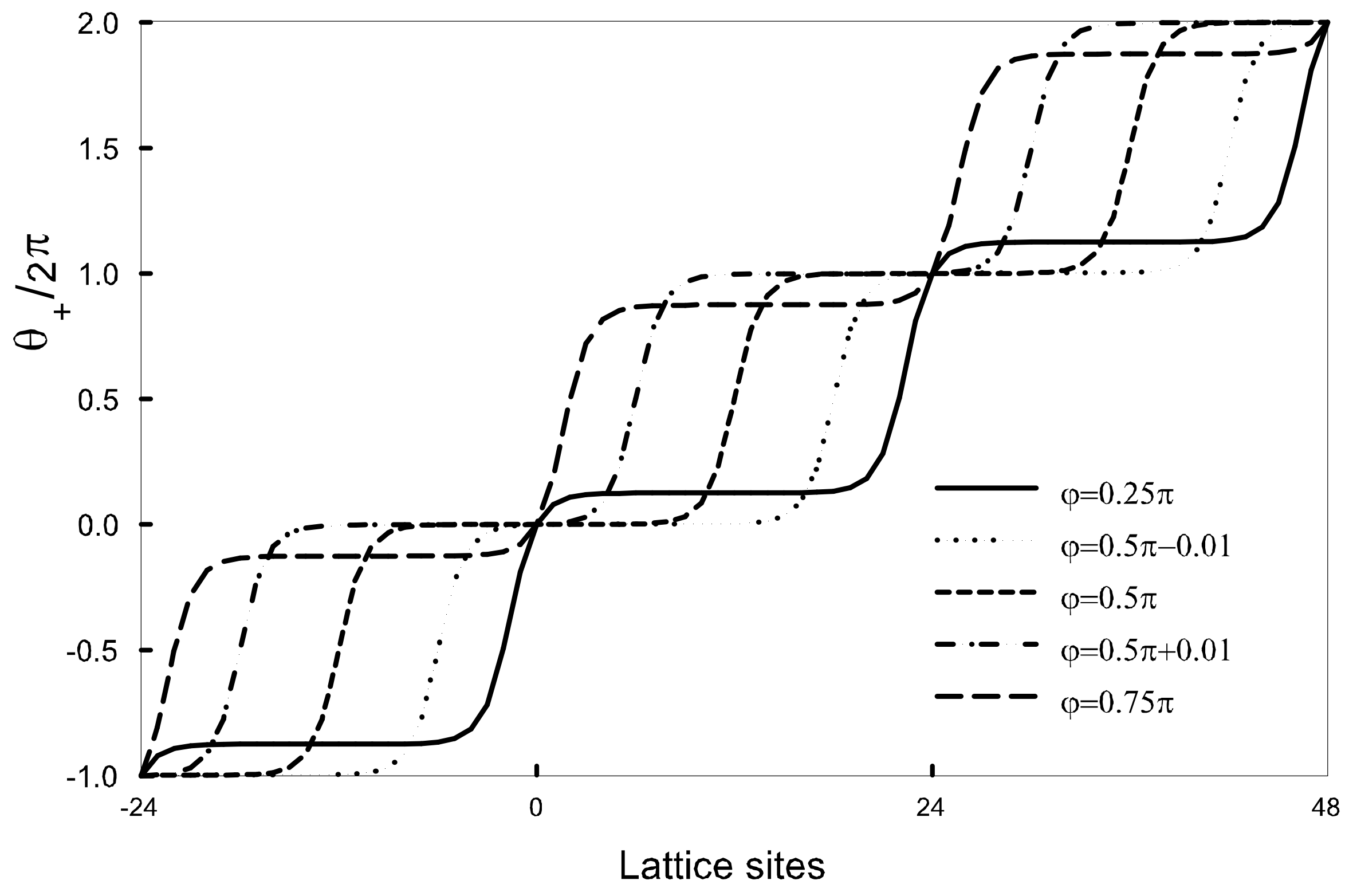

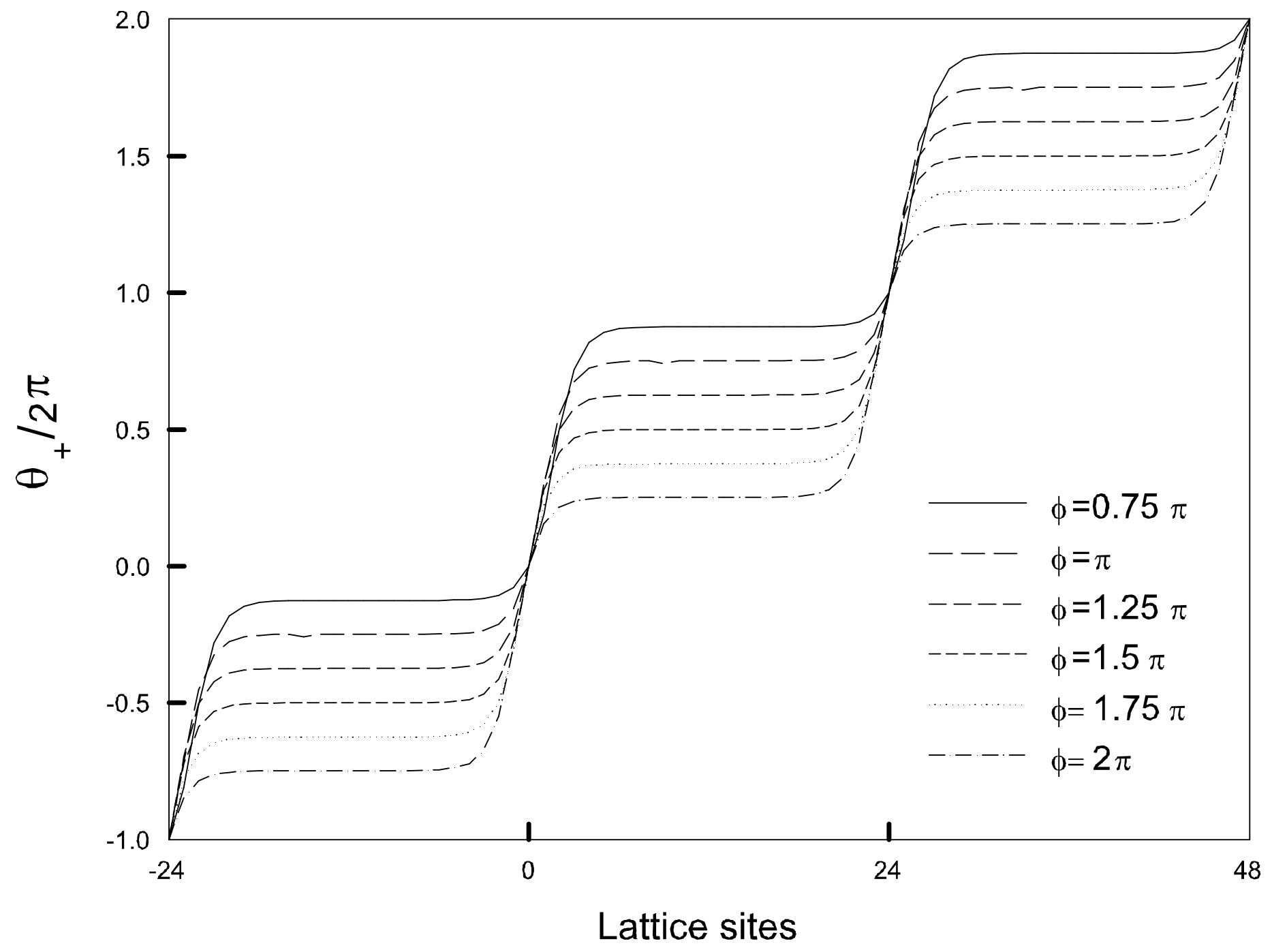

**Fig**. **5**: $\theta_+$ vs position for different $\varphi$. The system occupies $0 \le z \le 24$, the spin source occupies $z > 24$ and drain $z < 0$.

(**a**): $\pi/4 \le \varphi \le 3\pi/4$ (**b**):$3\pi/4 \le \varphi \le 2\pi$

# Chapter 4. Sine-Gordon Equation with twisted boundary condition

## 4.0 Introduction

In this chapter, I solve **SGE** on a finite chain with twisted boundary conditions and we consider the static soliton because it is compatible with the physical system. By twisted boundary conditions, I mean that the spin at one end is different from the spin at the other end. I define the adiabatic parameter on the applied electric and magnetic field plane. As the parameter changes,spin states change between the Néel state and dimer state and a quantized spin $S = 1$ is transported by the bulk state from one end of the system to the other. The motivation and potential physical application is that the states and the conditions in source and drain ends are different in general. Hence, in electronic and spintronic circuit, my theoretical method is very realistic. It is helpful to device design and circuit analysis.

## 4.1 Hamiltonian

In [**47**], we scale $x, t$ so that it become the standard **SGE**. We consider a different equation. It is not in the standard form of **SGE** alothough it can be transformed to the standard **SGE** by scale $\theta_+$. In fact, it is equivalent to standard **SGE** with twisted boundary conditions. So its physics application is that the states at two ends can be different. In [**47**], we consider the states at two ends are the same and discuss the transport properties of spin transport in that boundary conditions. Here due to different boundary conditions, we have much different physics than those in [**47**].

We start from the Hamiltion:

$$H = \int dx\{v[\pi\eta\Pi^2 + \frac{1}{4\pi\eta}(\partial_x\theta_+)^2] - \frac{R}{\pi\alpha^2}\sin(\theta_+ + \varphi) + \frac{J}{2\pi^2\alpha^3}\cos 2\theta_+\} \qquad 3.1.1.13$$

and obtain the equation of motion:

$$\partial_t^2\theta_+ = v^2\partial_x^2\theta_+ + \frac{4JR}{\alpha^2}\cos(\theta_+ + \varphi) + \frac{4|\gamma|J^2}{\pi\alpha^3}\sin 2\theta_+ \qquad 3.1.1.16$$

with the same symbol definitions in §**3.1**. We also neglect the second potential in this chapter because it is a irrevalent operator. But here I give different scaling, that means: $\theta_+ = \sqrt{\eta}\,\theta$, so that we will get the equation of motion

$$\partial_\tau^2\theta = \partial_z^2\theta + \sqrt{\eta}\cos(\sqrt{\eta}\,\theta_+ + \varphi) \qquad 4.1.1$$

where we have change variables:

$$z = \sqrt{2vR}\,\sqrt[4]{\eta}\,x/v\alpha\ ,\ \tau = \sqrt{2vR}\,\sqrt[4]{\eta}\,t/\alpha. \qquad 4.1.2$$

It is similar to the standard sine-Gordon equation

$$\partial_\tau^2\theta_+ - \partial_z^2\theta_+ + \sin\theta_+ = 0 \qquad 4.1.3$$

which has been well-studied. However, for our purpose which is to study the spin transport, we will solve it on a chain of finite length on which the phase $\varphi$ is no longer a trivial constant but introduces new meaning to the solution. This way, one can recognize the motion of spins from one end to the other.

## 4.2 Different views of relation between $\beta'$ and $A_{th}$

In this chapter, we want to solve the following equation because the motivation mention above:

$$\frac{\partial^2\theta}{\partial\tau^2} - \frac{\partial^2\theta}{\partial z^2} - \sqrt{\eta}\cos[\sqrt{\eta}\theta + \varphi] = 0 \qquad 4.2.1a$$

We can obtain the standard form by let $\theta' = \sqrt{\eta}\theta,\ z' = \sqrt{\eta}z,\ t' = \sqrt{\eta}t$

$$\frac{\partial^2\theta'}{\partial\tau'^2} - \frac{\partial^2\theta'}{\partial z'^2} - \cos(\theta' + \varphi) = 0 \qquad 4.2.1b$$

But we focus on the static solution which is the main topic in the previous chapter:

$$\theta = \frac{1}{\sqrt{\eta}}(\frac{\pi}{2} - \varphi + 4\tan^{-1}\{A * sc[\beta\sqrt{\eta}(z - z_0)] * dn[\Omega\sqrt{\eta}\tau]\}) \qquad 4.2.2$$

Here we have two points of views of relations between $\beta$ and $A_{th}$ :

**1**. In view of the form of **eq**. (**4.2.2**), let $\beta' = \beta\sqrt{\eta}$ and $\Omega' = \Omega\sqrt{\eta}$, but z not scaled:

For $f = sc$,

$$(\frac{df}{dz})^2 = (f^2 + 1) * (1 + k_f'^2 f^2) \qquad 4.2.3$$

with $k_f'^2 = 1 - k_f^2$, and for $f = dn$,

$$(\frac{df}{d\tau})^2 = (f^2 - 1) * (1 - k_g^2 - f^2) \qquad 4.2.4$$

we have the relations like before:

$$\Omega' = A'\beta' \qquad 4.2.5$$

$$k_g^2 = 1 - \frac{1}{A'^2} + \frac{1}{\Omega'^2(1 - A'^2)} \qquad 4.2.6$$

Here the prime of $A'$ is just to indicate that is the first point of view.

For static soliton, we let $k_g = 0$ and from **eq**. (**4.2.5**), (**4.2.6**). and we obtain the solution. We denote $A'$ by $A'_{th}$ for static soliton, and $\beta' = \frac{1}{1-A'^2}$

$$\theta(z) = \frac{1}{\sqrt{\eta}}(\frac{\pi}{2} - \varphi + 4\tan^{-1}\{A'_{th} * sc[\frac{1}{1 - A'^2_{th}}(z - z_0)]\}) \qquad 4.2.7$$

**2**. We can also $z$ and $\tau$ instead of $\beta$ and $\Omega$, which means $z' = \sqrt{\eta}z$ and $\tau' = \sqrt{\eta}\tau$

For $f = sc$,

$$(\frac{df}{dz'})^2 = (f^2 + 1) * (1 + k_f'^2 f^2) \qquad 4.2.8$$

comparing with **eq**. (**3.2.2**), (**3.2.3**). For $f = dn$,

$$(\frac{df}{d\tau'})^2 = (f^2 - 1) * (1 - k_g^2 - f^2) \qquad 4.2.9$$

we have relation as before:

$$\Omega = A\beta \qquad 4.2.10$$

$$k_g^2 = 1 - \frac{1}{A^2} + \frac{1}{\Omega^2(1-A^2)} \qquad 4.2.11$$

Also, for static solitons, $k_g = 0$ and hence $\beta = \frac{1}{1-A^2}$, where $A$ is to be denoted by $A_{th.}$
And we get the state a solution different from before:

$$\theta(z') = \frac{1}{\sqrt{\eta}}(\frac{\pi}{2} - \varphi + 4\tan^{-1}\{A_{th} * sc[\frac{1}{1-A_{th}^2}(z' - z_0')]\}) \qquad 4.2.12$$

**Eq**. (**4.2.7**) and (**4.2.12**) have the same form, but the variables are different. The reason is that in the procedure, $z$, $\tau$, $\beta$ and $\Omega$ are dimensionless. They are just only numbers. Especially, $z$ and $z'$ should be regard as "site". So the boundary condition: $\theta(z = L) = 2\pi$, which means $\theta($ final site$) = 2\pi$. The concept of site is independent of scaling of $z$. Therefore, there is no contradiction in these two points of view.

## 4.3 Twisted boundary condition of Sine-Gordon Equation:

The boundary conditions give the following equations:

$$\theta(z = 0) = \frac{1}{\sqrt{\eta}}(\frac{\pi}{2} - \varphi + 4\tan^{-1}\{A_{th} * sc[\frac{1}{1-A_{th}^2}(-z_0)]\}) = 0 \qquad 4.3.1$$

$$\theta(z = L) = \frac{1}{\sqrt{\eta}}(\frac{\pi}{2} - \varphi + 4\tan^{-1}\{A_{th} * sc[\frac{1}{1-A_{th}^2}(L - z_0)]\}) = 2\pi \qquad 4.3.2$$

We solve **eq**. (**4.3.1**), (**4.3.2**) for large enough L ( i.e. $L \geq 10$). This requires that $k_f \to 1$ and the approximation $sc(u) \simeq \sinh(u)$ can be used. **Eq**. (**4.3.1**), (**4.3.2**) become

$$\tan(\frac{\varphi - \frac{\pi}{2}}{4}) = A_{th} * \frac{e^{-\frac{x_0}{1-A_{th}^2}} - e^{\frac{x_0}{1-A_{th}^2}}}{2} \qquad 4.3.3$$

$$\tan(\frac{\varphi - \frac{\pi}{2}}{4} + \frac{\pi\sqrt{\eta}}{2}) = A_{th} * \frac{e^{\frac{L-x_0}{1-A_{th}^2}} - e^{\frac{x_0-L}{1-A_{th}^2}}}{2} \qquad 4.3.4$$

We can solve (**4.3.3**), (**4.3.4**) analytically to get:

$$z_0 = \frac{1-A_{th}^2}{2}\ln[\frac{e^{\frac{L}{1-A_{th}^2}}\tan(\frac{\varphi-\frac{\pi}{2}}{4}) - \tan(\frac{\varphi-\frac{\pi}{2}}{4} + \frac{\pi\sqrt{\eta}}{2})}{e^{\frac{-L}{1-A_{th}^2}}\tan(\frac{\varphi-\frac{\pi}{2}}{4}) - \tan(\frac{\varphi-\frac{\pi}{2}}{4} + \frac{\pi\sqrt{\eta}}{2})}] \qquad 4.3.5$$

$$\approx \frac{1}{2}\{L + \ln[\frac{-\tan(\frac{\varphi-\frac{\pi}{2}}{4})}{\tan(\frac{\varphi-\frac{\pi}{2}}{4} + \frac{\pi\sqrt{\eta}}{2})}]\}$$

$$A_{th} = 2\tan(\frac{\varphi-\frac{\pi}{2}}{4})\{\sqrt{\frac{e^{-\frac{L}{1-A_{th}^2}}\tan(\frac{\varphi-\frac{\pi}{2}}{4})-\tan(\frac{\varphi-\frac{\pi}{2}}{4}+\frac{\pi\sqrt{\eta}}{2})}{e^{\frac{L}{1-A_{th}^2}}\tan(\frac{\varphi-\frac{\pi}{2}}{4})-\tan(\frac{\varphi-\frac{\pi}{2}}{4}+\frac{\pi\sqrt{\eta}}{2})}} \qquad 4.3.6$$

$$-\sqrt{\frac{e^{\frac{L}{1-A_{th}^2}}\tan(\frac{\varphi-\frac{\pi}{2}}{4})-\tan(\frac{\varphi-\frac{\pi}{2}}{4}+\frac{\pi\sqrt{\eta}}{2})}{e^{-\frac{L}{1-A_{th}^2}}\tan(\frac{\varphi-\frac{\pi}{2}}{4})-\tan(\frac{\varphi-\frac{\pi}{2}}{4}+\frac{\pi\sqrt{\eta}}{2})}}\}^{-1}$$

$$\approx 2e^{-\frac{L}{2}}\sqrt{-\tan(\frac{\varphi-\frac{\pi}{2}}{4})\tan(\frac{\varphi-\frac{\pi}{2}}{4}+\frac{\pi\sqrt{\eta}}{2})}$$

From above very good approximation, we can see there are two kinds of solutions essentially:

**1**. $\varphi \in [\frac{\pi}{2}-2\pi\sqrt{\eta}+2n\pi, \frac{\pi}{2}+2n\pi]$: This will be called the "allowed region". In this region, $\tan(\frac{\varphi-\frac{\pi}{2}}{4})\tan(\frac{\varphi-\frac{\pi}{2}}{4}+\frac{\pi\sqrt{\eta}}{2}) < 0$. We can have real values of $A_{th}$ and $z_0$ and the solutions represent "bulk states".

**2**. $\varphi \in [\frac{\pi}{2}+2(n-1)\pi, \frac{\pi}{2}+2n\pi-2\pi\sqrt{\eta}]$: This is called the "forbidden region". In this region, $\tan(\frac{\varphi-\frac{\pi}{2}}{4})\tan(\frac{\varphi-\frac{\pi}{2}}{4}+\frac{\pi\sqrt{\eta}}{2}) > 0$

We can't have real values of $A_{th}$ and $z_0$, and the solutions represent "edge states", which will be clear later.

In both regions, $z_0 \simeq L/2$, $|A_{th}| \simeq e^{-L/2}$. Hence $k_f = \sqrt{1-A_{th}^2} \simeq 1$, $K(k_f) \simeq \ln(\frac{4}{A_{th}^2})$ and $K \simeq L$.

Because the period of $\sqrt{\eta}\theta$ is $4\pi$, so after $\varphi$ increase by $2\pi$, we have another solution besides **eq**. (**4**.**2**.**7**):

$$\theta = \frac{1}{\sqrt{\eta}}(\frac{\pi}{2}-\varphi+4\tan^{-1}\{A_{th} * sc[\frac{1}{1-A_{th}^2}(z-z_0)-\mathbf{K}]\}) \qquad 4.3.7$$
$$= \frac{1}{\sqrt{\eta}}(\frac{\pi}{2}-\varphi+4\tan^{-1}\{-\frac{1}{A_{th}} * cs[\frac{1}{1-A_{th}^2}(z-z_0)\}\})$$
$$= \frac{1}{\sqrt{\eta}}(\frac{\pi}{2}-\varphi+\mathbf{2\pi}+4\tan^{-1}\{A_{th} * sc[\frac{1}{1-A_{th}^2}(z-z_0)]\})$$

This is not equivalent to the previous one.

Finding two solutions in the forbidden region is important. We will have two edge states coming from the two solutions. In each allowed region, one of the two solutions gives lower energy ( i.e. ground state), another one gives excited state. This picture is good because it appear that adiabatic phase works in twised boundary condition of **SGE**, while the adiabatic phase seem no function in the ordinary fixed boundary condition in **Chapter 3**.

For the second solution, we have the following twised boundary conditions

$$\theta(z=0) = \frac{1}{\sqrt{\eta}}(\frac{\pi}{2}-\varphi+\mathbf{2\pi}+4\tan^{-1}\{A_{th} * sc[\frac{1}{1-A_{th}^2}(-z_0)]\}) = 0 \qquad 4.3.8a$$

$$\theta(z=L) = \frac{1}{\sqrt{\eta}}(\frac{\pi}{2}-\varphi+\mathbf{2\pi}+4\tan^{-1}\{A_{th} * sc[\frac{1}{1-A_{th}^2}(L-z_0)]\}) = 2\pi \qquad 4.3.8b$$

In **Table 4**.**1** and **4**.**2**, I listed the values of $z_0$,$A_{th}$, and energy of solution in **eq**.(**4**.**2**.**7**) and **eq**.(**4**.**3**.**7**) resplictively for different values of $\varphi$ at L=10. The entries in bold characters are those in the "allowed region" and others are those in the "forbidden region".

**Table 4.1 :**

| $\varphi\backslash$ | $A_{th}$ | $z_0$ | *Total energy* |
|---|---|---|---|
| **$-0.5\pi$** | **$9.07999 \times 10^{-5}$** | **10** | **8** |
| **$-0.25\pi$** | **0.00491267** | **5.60576** | **5.55462** |
| **0** | **0.00558127** | **5** | **4.69549** |
| **$0.25\pi$** | **0.00491267** | **4.39424** | **5.55462** |
| **$0.5\pi$** | **$9.07999 \times 10^{-5}$** | **0** | **8** |
| $0.75\pi$ | $0.00735344i$ | $3.99072 + 1.57088i$ | 11.6609 |
| $\pi$ | $0.0134882i$ | $4.11821 + 1.57108i$ | 15.968 |
| $1.25\pi$ | $0.0248198i$ | $3.98824 + 1.57576i$ | 20.2213 |
| $1.5\pi$ | $9.07999 \times 10^{-5}$ | $-10$ | × |

**Table 4.2 :**

| $\varphi\backslash$ | $A_{th}$ | $z_0$ | *Total energy* |
|---|---|---|---|
| $0.5\pi$ | $9.07999 \times 10^{-5}$ | 20 | × |
| $0.75\pi$ | $-0.0248198i$ | $6.01176 + 1.57176i$ | 20.2213 |
| $\pi$ | $-0.0134882i$ | $5.88179 + 1.57108i$ | 15.968 |
| $1.25\pi$ | $-0.00735344i$ | $6.00928 + 1.57088i$ | 11.6609 |
| **$1.5\pi$** | **$9.07999 \times 10^{-5}$** | **10** | **8** |
| **$1.75\pi$** | **0.00491267** | **5.60576** | **5.55462** |
| **$2\pi$** | **0.00558127** | **5** | **4.69549** |
| **$2.25\pi$** | **0.00491267** | **4.39424** | **5.55462** |
| **$2.5\pi$** | **$9.07999 \times 10^{-5}$** | **0** | **8** |

# 4.4 Real solutions in the forbidden region

In this section, we analyze the solutions in the "forbidden region" and show they are real. Further analysis gives their physical implication.

## 4.4.1 Arguments that the state in the forbidden region is real

I give here a simple proof that $\theta(x)$ is real for large enough L. For Large L,

$$z_0 \approx \frac{1}{2}\{L + \ln[\frac{-\tan(\frac{\varphi-\frac{\pi}{2}}{4})}{\tan(\frac{\varphi-\frac{\pi}{2}}{4} + \frac{\pi\sqrt{\eta}}{2})}]\} \quad 4.4.1.1$$

and

$$A_{th} \approx 2e^{-\frac{L}{2}}\sqrt{-\tan(\frac{\varphi-\frac{\pi}{2}}{4})\tan(\frac{\varphi-\frac{\pi}{2}}{4} + \frac{\pi\sqrt{\eta}}{2})} \quad 4.4.1.2$$

**Case** *1: In the forbidden region: Let* $\alpha_1 = \tan(\frac{\varphi-\frac{\pi}{2}}{4})$ *and* $\alpha_2 = \tan(\frac{\varphi-\frac{\pi}{2}}{4} + \frac{\pi\sqrt{\eta}}{2})$, $\alpha_1\alpha_2 > 0$

***Eq.(4.3.5)** and **(4.3.6)** give*

$$z_0 \approx \frac{1}{2}\{L + i\pi + \ln[\frac{\alpha_1}{\alpha_2}]\} \qquad \textit{4.4.1.3a}$$

$$A_{th} \approx i2e^{-\frac{L}{2}}\sqrt{\alpha_1\alpha_2} \qquad \textit{4.4.1.3b}$$

If L large enough, then $k_f \to 1$, so $sc(u,k_f) \to \sinh(u)$

$$\begin{aligned}\theta(z) &= \frac{1}{\sqrt{\eta}}[\frac{\pi}{2} - \varphi + 4\tan^{-1}\{i2e^{-\frac{L}{2}}\sqrt{\alpha_1\alpha_2} \cdot sc[z - \frac{L}{2} - i\frac{\pi}{2} - \ln\sqrt{\frac{\alpha_1}{\alpha_2}}]\}] \\ &\simeq \frac{1}{\sqrt{\eta}}[\frac{\pi}{2} - \varphi + 4\tanh - 1\{i2e^{-\frac{L}{2}}\sqrt{\alpha_1\alpha_2} \cdot \frac{e^{z-\frac{L}{2}-i\frac{\pi}{2}-\ln\sqrt{\frac{\alpha_1}{\alpha_2}}} - e^{-z+\frac{L}{2}+i\frac{\pi}{2}+\ln\sqrt{\frac{\alpha_1}{\alpha_2}}}}{2}\}] \\ &= \frac{1}{\sqrt{\eta}}[\frac{\pi}{2} - \varphi + 4\tan^{-1}(\alpha_2 e^{z-L} + \alpha_1 e^{-z}) \qquad 4.4.1.4\end{aligned}$$

Therefore, $\theta$ remains to be real even if $A_{th}$ is imaginary and $z_0$ is complex. For completeness, I give the approximated form in the "allowed region":

**Case** *2: In the allowed region:* $\alpha_1\alpha_2 < 0$

***Eq**. **(4.3.5)** and **(4.3.6)** give*

$$z_0 \approx \frac{1}{2}\{L + \ln[\frac{-\alpha_1}{\alpha_2}]\} \qquad \textit{4.4..1.5a}$$

$$A_{th} \approx 2e^{-\frac{L}{2}}\sqrt{-\alpha_1\alpha_2} \qquad \textit{4.4.1.5b}$$

In view of **eq**. (**4**.**2**.**7**)

$$\begin{aligned}\theta(z) &= \frac{1}{\sqrt{\eta}}[\frac{\pi}{2} - \varphi + 4\tan^{-1}\{2e^{-\frac{L}{2}}\sqrt{-\alpha_1\alpha_2} \cdot sc[z - \frac{L}{2} - \ln\sqrt{\frac{-\alpha_1}{\alpha_2}}]\}] \\ &\simeq \frac{1}{\sqrt{\eta}}[\frac{\pi}{2} - \varphi + 4\tan^{-1}\{2e^{-\frac{L}{2}}\sqrt{-\alpha_1\alpha_2} \cdot \frac{e^{z-\frac{L}{2}-\ln\sqrt{\frac{-\alpha_1}{\alpha_2}}} - e^{-z+\frac{L}{2}+\ln\sqrt{\frac{-\alpha_1}{\alpha_2}}}}{2}\}] \\ &= \frac{1}{\sqrt{\eta}}[\frac{\pi}{2} - \varphi + 4\tan^{-1}(\alpha_2 e^{z-L} + \alpha_1 e^{-z}) \qquad 4.4.1.6\end{aligned}$$

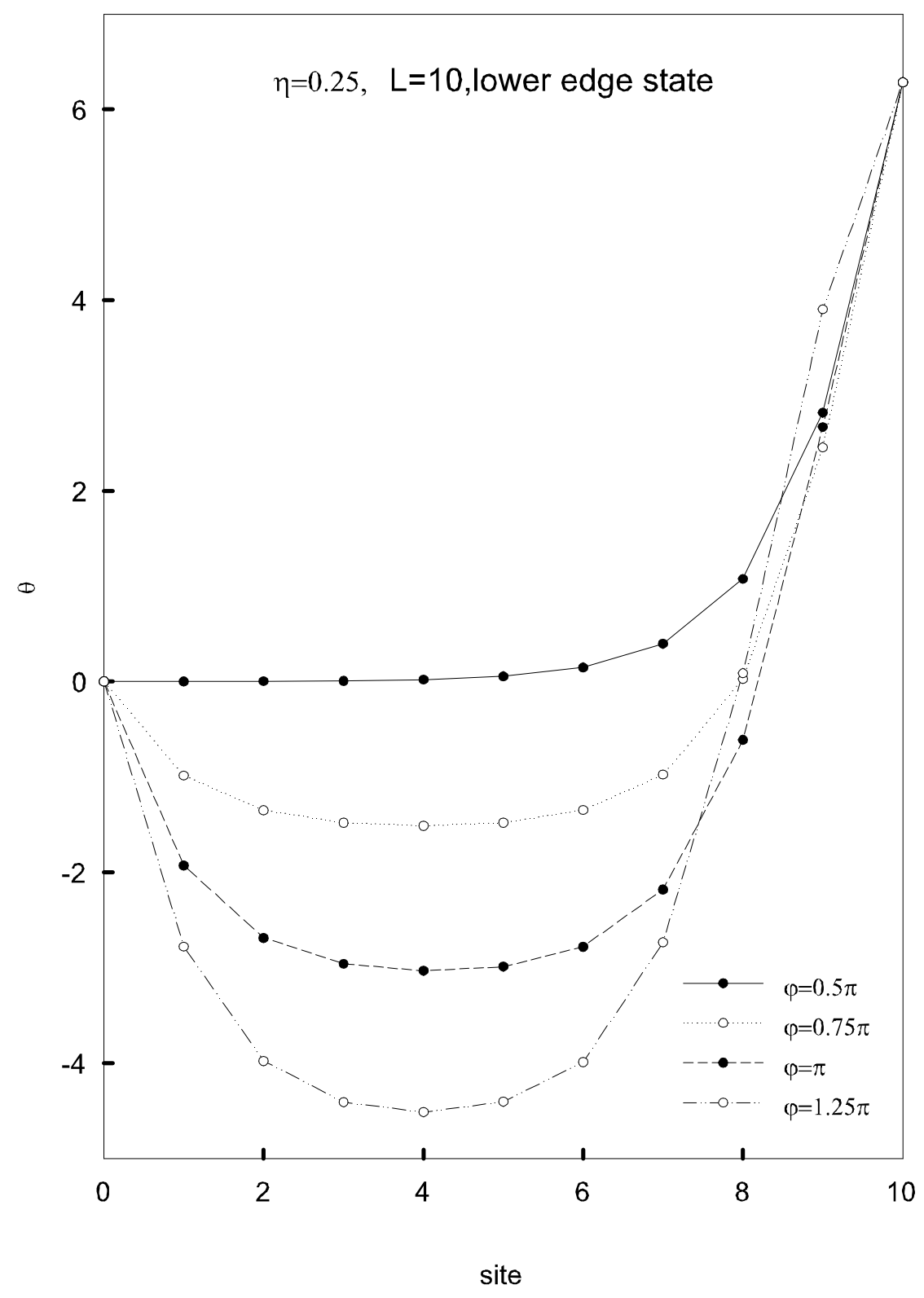

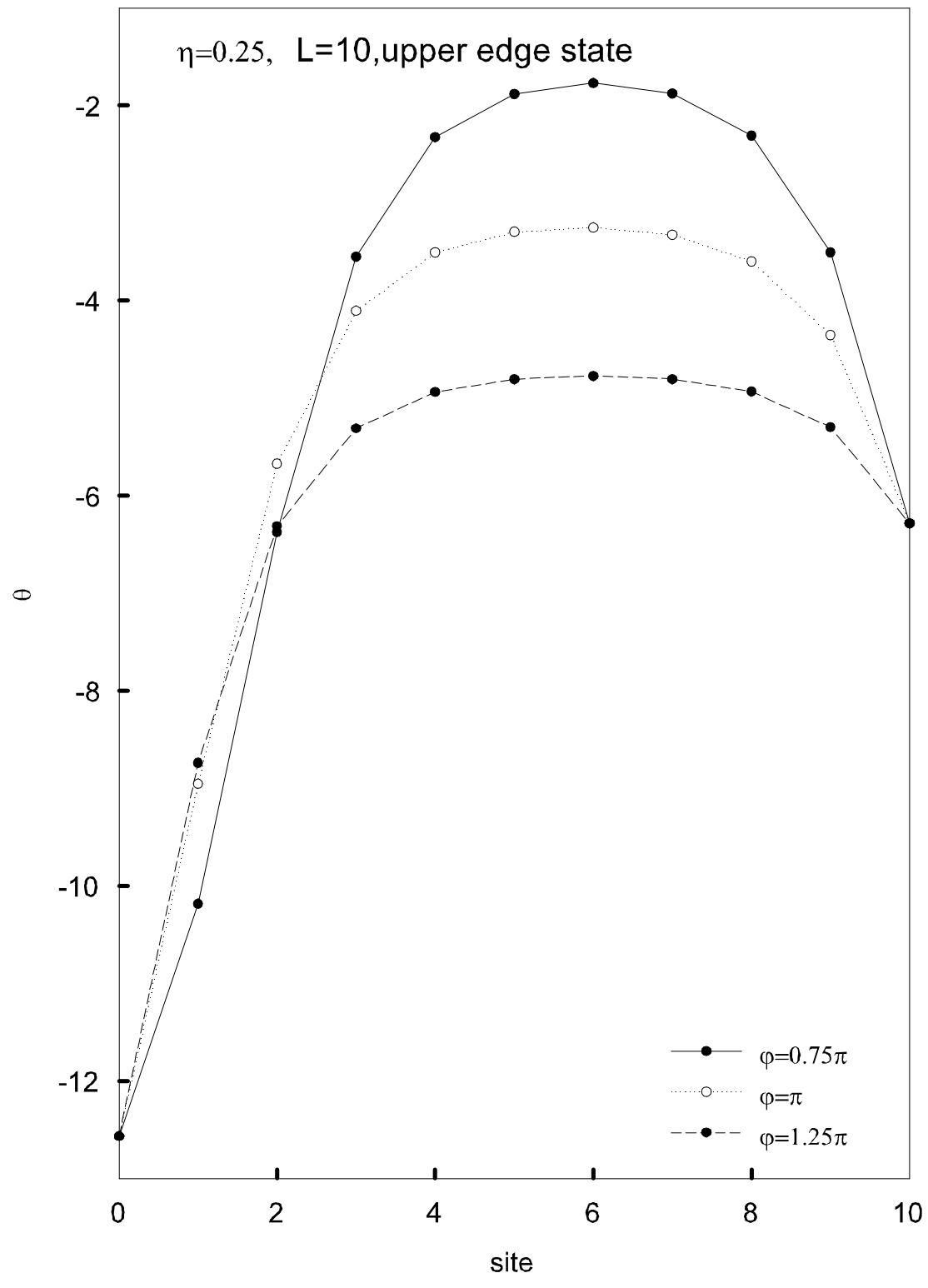

**Fig**. **6**(**a**): $\eta = \frac{1}{4}$, L=10, lower edge state ; **Fig**. **6**(**b**): $\eta = \frac{1}{4}$, L=10, upper edge state

**Eq**. (**4.4.1.4**) and (**4.4.1.6**) are the same, both are real and satisfy our boundary condition. I plot in **Fig**. **6**(**a**), (**b**) to show two different solutions in the forbidden regions. In **Fig**. **6** (**a**), I use **eqs**. (**4**.**3**.**1**; **4**.**3**.**2**) as our boundary conditions, but in **Fig**. **6**(**b**), I use **eqs**.(**4**.**3**.**8a**) as our boundary condition. In fact, the term $2\pi$ in (**4**.**3**.**8a**) can be absorbed into the arctan function as an additional term of argument. Hence, **Fig**. **6a** and **6b** are distinct solutions.

## 4.4.2 The energy formula in twised boundary condition SGE:

If I start from the Hamiltion:

$$H = \int_0^L dz[\frac{\Pi^2}{2} + \frac{(\partial\theta)^2}{2} - \sin(\sqrt{\eta}\theta + \varphi)] \qquad 4.4.2.1$$

then the E.O.M. should be:

$$\frac{\partial^2\theta}{\partial t^2} - \frac{\partial^2\theta}{\partial z^2} - \sqrt{\eta}\cos(\sqrt{\eta}\theta + \varphi) = 0 \qquad 4.4.2.2$$

The solution is:

$$\theta(z) = \frac{1}{\sqrt{\eta}}\{\frac{\pi}{2} - \varphi + 4\tan^{-1}[A_{th}sc[\beta(z - z_0)]]\} \qquad 4.4.2.3$$

where $\beta = \frac{1}{1-A_{th}^2}$

**Case** *1: $\varphi \in$ allowed region, which means $A_{th}$ and $z_0$ are both real:*

*Using the method in (**3.2.1.1**), we get*

$$\frac{1}{2}(\frac{\partial\theta}{\partial x})^2|_{z=z_0} = \frac{1}{\sqrt{\eta}}\frac{8A_{th}^2}{(1 - A_{th}^2)^2}, \qquad 4.4.2.4a$$

$$\sin(\sqrt{\eta}\theta + \varphi)|_{z=z_0} = 1, \qquad 4.4.2.4b$$

*and*

$$V_s = \int_0^L [1 + \frac{1}{\sqrt{\eta}}\frac{8A_{th}^2}{(1 - A_{th}^2)^2} - \sin(\sqrt{\eta}\theta + \varphi)]dz \qquad 4.4.2.5$$

*Since*

$$\frac{1}{2}(\frac{\partial\theta}{\partial z})^2 = 1 + \frac{1}{\sqrt{\eta}}\frac{8A_{th}^2}{(1 - A_{th}^2)^2} - \sin(\sqrt{\eta}\theta + \varphi) \qquad 4.4.2.6$$

*I find*

$$V_s = \frac{1}{\sqrt{2}}\int_0^{2\pi}[1 + \frac{1}{\sqrt{\eta}}\frac{8A_{th}^2}{(1 - A_{th}^2)^2} - \sin(\sqrt{\eta}\theta + \varphi)]d\theta \qquad 4.4.2.7$$

*and*

$$V_s = \frac{L}{\sqrt{\eta}}\frac{8A_{th}^2}{(1 - A_{th}^2)^2} + V_p \qquad 4.4.2.8$$

*where $V_p = \int_0^L [1 - \sin(\sqrt{\eta}\theta + \varphi)]dz$, so total energy*

$$E = V_s + V_p = 2V_s - \frac{L}{\sqrt{\eta}}\frac{8A_{th}^2}{(1 - A_{th}^2)^2} \qquad 4.4.2.9$$

**Case** *2: $\varphi \in$ forbidden region, which means $A_{th}$ is pure imaginary and* $\mathrm{Im}(z_0) \simeq \pm\frac{\pi}{2}$ :

*In this case, we should not take $z_0$ as our reference point, because it is not real. As in the **Fig.6 (a)**, **(b)**; the $\theta(z)$ is still real but has an extremum ( i.e. $\theta(z)$ is no longer a monotonic function), we take the point of extremum as our reference point because $\frac{\partial\theta}{\partial z} = 0$. This point can be shown to be at $z = \mathrm{Re}(z_0)$.*

$$\frac{\partial\theta}{\partial z} = \frac{4A_{th}}{1-A_{th}^2}\frac{dn[\beta(z-z_0)]}{cn^2[\beta(z-z_0)] + A_{th}^2 sn^2[\beta(z-z_0)]} \quad 4.4.2.10$$

*For $z = \mathrm{Re}(z_0)$, the numerator $dn[\beta(z-z_0)]$ can be written as $dn(iv)$. Since $dn(iv,k) = dn(v,k')/cn(v,k')$, where $k' = \sqrt{1-k^2}$, I find that $dn[\beta(z-z_0)]|_{z=\mathrm{Re}(z_0)} \sim 0$ since $v = \mathrm{Im}(z_0) \simeq \pi/2$ and $dn(v,k') \sim k'$ which is vanishingly small. A little proof can also show that the minimum of $-\sin(\sqrt{\eta}\theta+\varphi)$ is also at $z = \mathrm{Re}(z_0)$ because*

$$\sin(\sqrt{\eta}\theta+\varphi) = \cos[4\tan^{-1}\{A_{th}sc[\beta(z-z_0)]\}] \quad 4.4.2.11$$

*Since $sc[\beta(z-z_0)]|_{z=z_0}$ is an extremum, so is $\tan^{-1}\{A_{th}sc[\beta(z-z_0)]\}$ and hence, the r. h. s. is a maximum. As a result, the two edge states have different energies:*

**Case** *(1): $\theta[\mathrm{Re}(z_0)]$ is the minimum*

$$E = T+V = 2T = \sqrt{2}\left(\int_{\theta[\mathrm{Re}(z_0)]}^{0} + \int_{\theta[\mathrm{Re}(z_0)]}^{2\pi}\right)\sqrt{S-\sin(\sqrt{\eta}\theta+\varphi)}\,d\theta \quad 4.4.2.12a$$

**Case** *(2): $\theta[\mathrm{Re}(z_0)]$ is the maximum*

$$E = \sqrt{2}\left(\int_{0}^{\theta[\mathrm{Re}(z_0)]} + \int_{2\pi}^{\theta[\mathrm{Re}(z_0)]}\right)\sqrt{S-\sin(\sqrt{\eta}\theta+\varphi)}\,d\theta \quad 4.4.2.12b$$

See **Fig.7(a)**, (**b**) are examples for $\eta = \frac{1}{4}$ and $\frac{1}{3}$ at L=10. I show the interesting energy crossing phenomena between both edge states and excited state.

So in the twisted boundary condition, the spin accumulate phenomena on the edge is similiar to what Shindou [**34**] and Fu & Kane [**35**] got in their paper.

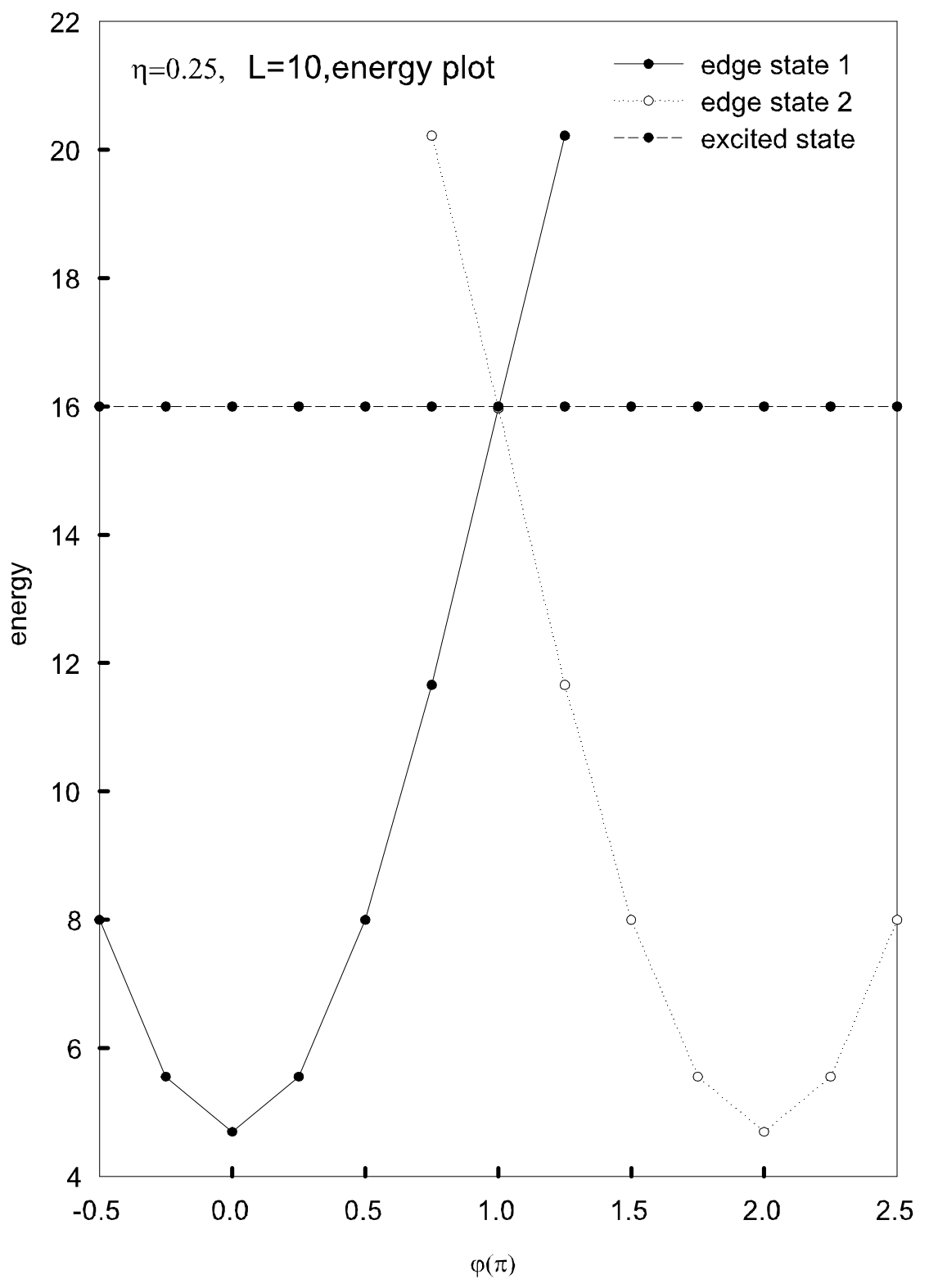

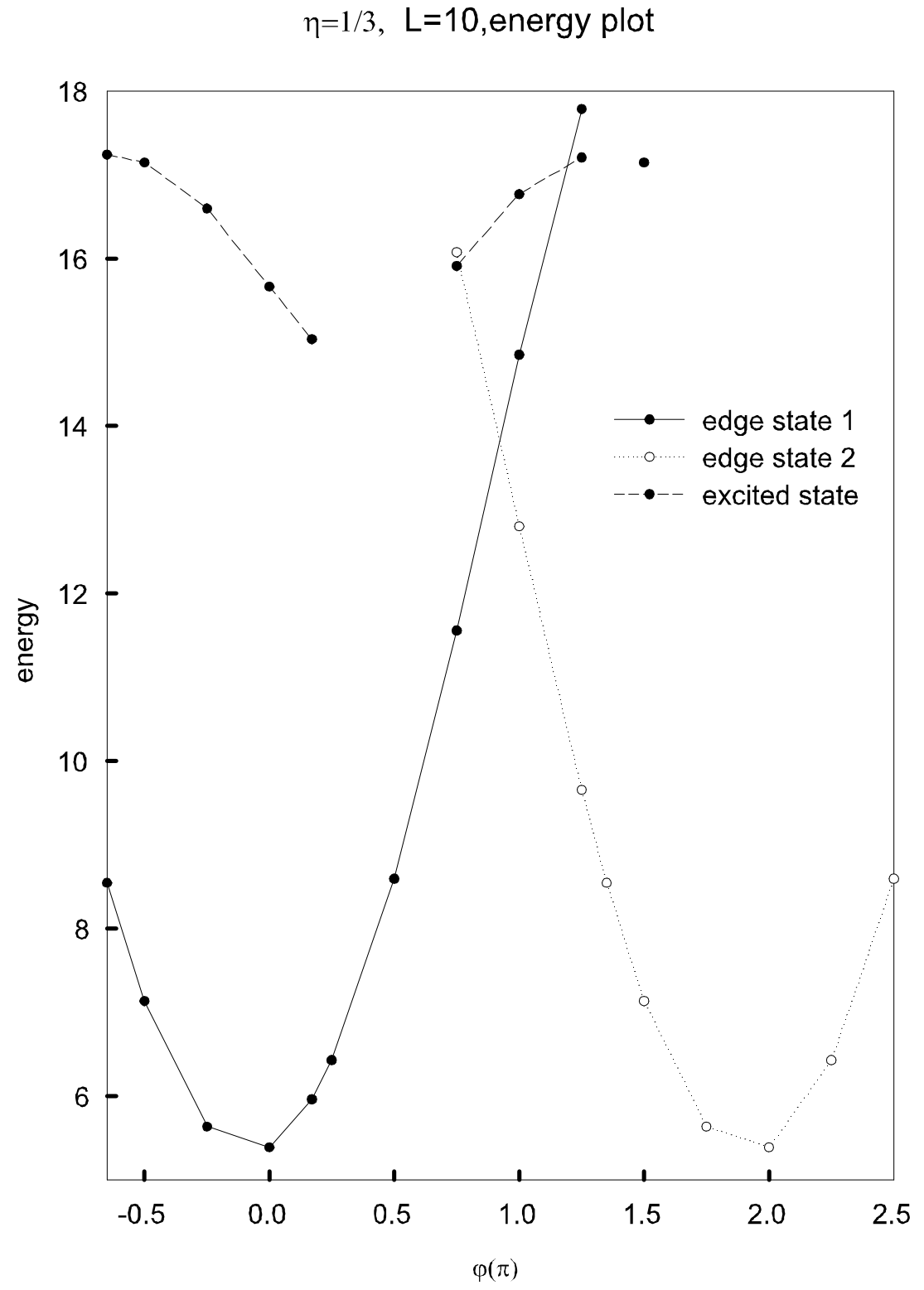

**Fig**. **7**(**a**): $\eta = \frac{1}{4}$, L=10: edge and excited state energy crossing phenomena; **Fig**. **7**(**b**): $\eta = \frac{1}{3}$, L=10: edge and excited state energy crossing phenomena

# 4.5 Another observation in the forbidden region

For the equation

$$\partial_\tau^2\theta - \partial_z^2\theta - \sqrt{\eta}\cos(\sqrt{\eta}\theta + \varphi) = 0 \qquad 4.5.1$$

we have the static solution:

$$\theta(z) = \frac{1}{\sqrt{\eta}}\left(\frac{\pi}{2} - \varphi + 4\tan^{-1}\{A_{th}sc[\beta(z - z_0); k_f]\}\right) \qquad 4.5.2$$

and the potential is:

$$V = 1 - \cos(\sqrt{\eta}\theta + \varphi - \frac{\pi}{2}) \qquad 4.5.3$$

Classically or quantum mechanically, the system will be at the point of lowest potential. If we impose the boundary condition: $\theta(z = 0) = 0$, $\theta(z = L) = 2\pi$, and $\theta(z = L) - \theta(z = 0) = 2\pi$ one can see easily in the example of $\eta = \frac{1}{4}$. **1**. When $\varphi = \frac{-\pi}{2}$, the lowest potential is at $z = L$. **2**. When $\varphi = 0$, the lowest potential is at $z = \frac{L}{2}$. **3**. When $\varphi = \frac{\pi}{2}$, the lowest potential is at $z = 0$. The lowest potential is whitin the system ( i.e. $z \in [0, L]$ ). Hence, $\varphi \in [\frac{-\pi}{2}, \frac{\pi}{2}]$ is the "allowed" region and the solutions give the bulk states. On the other hand, in the forbidden region, the lowest point of the potential is beyond the system. Since the system still want to have energy as low as possible, the state will concerntrate on one of the edge. So from potential analysis, we conclude there will be a "edge state" in forbidden region if we just consider the boundary conditions. But if we use the loosen boundary condition, which means we can add one period in one of the boundary condition as pervious subchapter mention:

**1**. If we use $\theta(z = L) = 10\pi$ instead of $2\pi$ for $\varphi \in (\frac{\pi}{2}, \frac{3\pi}{2})$.

**2**. Or we use $\theta(z = 0) = -8\pi$ instead of 0 for $\varphi \in (\frac{-3\pi}{2}, \frac{-\pi}{2})$.

Then the lowest point of potential is within the system and we get bulk state. We can generalize above argument to $0 < \eta \leq 1$. The lowest point of the potential is $\sqrt{\eta}\theta + \varphi - \frac{\pi}{2} = 2n\pi, n \in Z$. If we keep the original boundary condition, $\theta \in [0, 2\pi]$, The allowed region is $\varphi \in [\frac{\pi}{2} + 2n\pi - 2\pi\sqrt{\eta}, \frac{\pi}{2} + 2n\pi]$. We get bulk states..

Another observation is that our present boundary condition is equivalent to the non-interacting case with twisted boundary condition, namely, if the new variable is $\widetilde{\theta}$, then $\widetilde{\theta}(z = 0) = 0$ and $\widetilde{\theta}(z = L) = 2\pi\sqrt{\eta}$. So one can see the interaction in one-dimension is equivalent to twisted boundary condition.

## 4.6 Arguments that the spin state in the forbidden region is edge state and the topology view of this case:

In "forbidden region", we can see the $\theta(x)$ curve in these regions, see **Fig**. **6** (**a**), (**b**) for $\eta = \frac{1}{4}$ and L=10 for example and we find particular steep curve in the edge. If transformed into the spin state this gives spin accumulation at the edge. So we have "edge states" in forbidden region!

If spin state, the physical state, not $\theta(x)$ become edge state in the forbidden region, then we have an elegent topology picture as **Ref**. [**32**; **33**]. Our "edge states" here is different from those in **Ref**. [**32**, **33**]. We have the branch cut by complexifying $\varphi$. The edge state will bring us from the ground state to a excited state. This concept is the same as the branch cut in the phase space, once we go to the branch cut, we will go to another Rieman surface (i.e. another excited state).

Take (**4**.**4**.**1**.**1**) for example, we have $\varphi = \frac{\pi}{2}, \frac{-\pi}{2}$ as the "zeros" of two tan function product,also we have $\varphi = \frac{3\pi}{2}, \frac{-3\pi}{2}$ as the "poles" inside one fundamental bi-period region. So in one period of $\varphi$ ( i.e. $4\pi$ ), we have two zeros and poles. ( But in (**4**.**4**.**1**.**1**) we will have infinitely many pairs of zeros and poles if we do not restrict $\varphi$ in finite period ). So we have **genus**=**1** ( one kind of Riemann Surface ) in complexified $\varphi$, the same beautiful topology as **Ref**. [**32**; **33**].

In **chapter 3**, the $\theta_+(x)$ also have steep slop near the edge in nearly all $\varphi$, that means we have spin at the ends because there is a derivative term in spin and dimer state. Why do we say that is a bulk state in physical spin state sense? Let us repeat the static soliton in the repeated zone of the system, like the periodic property mensioned in **chapter 3**, then they satisfy the property of bulk state. If this is the right argument, then $\theta(x)$ in allowed region in twised boundary condition is also the same as those in the previous chapter. Then in allowed region, we can say the ground state is still bulk state in spin physical state. In the forbidden region, the state become edge state, not only because it has steeper slop than bulk near edge, but this state has no periodic property. So I say this one is the edge state. This is not strange for spin state to change from bulk state to edge state because we can see **Fig**.**1** of Fu and Kane's paper [**35**]. Intially, the whole system is in bulk state. But as $\varphi$ increase, some spin accumulates on the edge, so the state become the edge state and the energy increases from the ground state. So our solution in different region show the same phenomena as that in their paper. In **Fig**.**8**, I show the Rieman surface to explain the idea. Here, I choose top surface represent one edge state, while below surface represent another edge state and branch cut is bulk states. But we can also exchange their roles because edge and bulk states here are conjugate from topological view. That means branch cut is edge states, while one of the surface is the ground state and another is the excited state of bulk states.

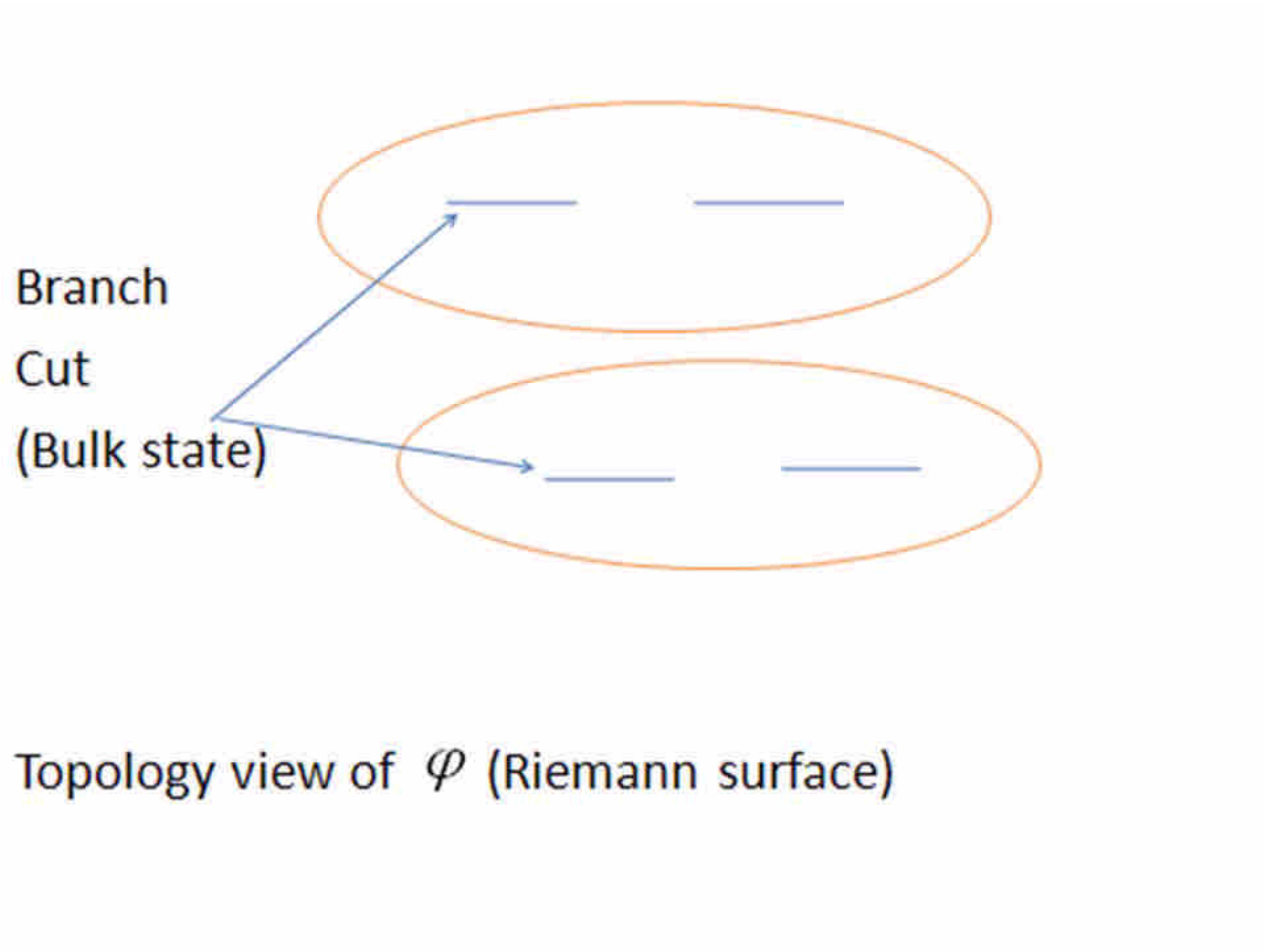

Fig.8. Topological view of $\varphi$

## 4.7 Conclusions:

The existence of forbidden regions due to the twisted boundary conditions gives very different aspects from our previous work, **ref**. [**47**]. This boundary condition is not like the ordinary boundary condition in **Chapter 3**. The twisted boundary condition give us two edge state and their energies cross each other and cross with excited state, see **Fig**: **7** (**a**), (**b**). This phenonema also appear in [**34**], [**36**], and it is realized not by numerically calculate as they do, but by solving **SGE** with special boundary condition in my thesis.

# Chapter 5. Asymmetrical double-sine-Gordon equation

In this Chapter, we present solutions of asymmetric **DSGE** of an infinite system based on **Möbius transformation** and numerical exercise. This method is able to give the forms of the solutions for all the region on the $\varphi - \eta$ parameter plane where $\varphi$ is an additional phase and $\eta$ is the ration of the magnitudes of two sine terms. We are also able to show how the deconfinement occurs near $\varphi = (1/2 + n)\pi$ and $\varphi = n\pi$. We predict different kind of solutions and transitions among them in different parts of the parameter space of this equation.

## 5.1 Introduction:

The **SGE** is a rare example of integrable system. However, the lagrangian of a realistic physical system often gives a more comlicated equation of motion than **SGE**. For example, a quantum spin chain is mapped into a lagrangian with several potential terms. Systems with nonlinear optical properties also give rise to more complicated wave equations. Thus a more complete model is desirable. This leads to **DSGE**. It loses the integrability of **SGE**. The main difficulty of **DSGE** is that it is non-integrable. Analytical form of the solutions of asymmetric **DSGE** is important because it can provide deeper insight than numerical results. I sought an analytical form of the solution which can serve as an anchor to numerical analysis or approximations. It can also be springboard to study quantum fluctustions. Physically, this equation is closely related to the problem of a quantum spin chain, ( see **eq**. (**3**.**1**.**16**)). The additional phase is an adiabatic parameter given by applied external field [**47**], [**34**], [**36**]. Hence asymmetric **DSGE** is important in both physica and mathematics. The mathematical solution of **DSGE** can be found in **Ref**. [**48**], [**49**], [**50**].

In this section, I briefly summerize the work of [**52**], [**53**], [**54**], [**55**]. Because **DSGE** is not an integrable equation, so the works of them are using perturbation method -Form factor perturbation theory (**FFPT**) to solve asymmetric **DSGE** by using intergrable equation as basis. And here integrable model means **SGE**. Its spectrum.are well-known. Although this idea is good, but we must remember, for example, see [**40**], [**39**], that perturbation is false in one dimension because their excitation is not quasi-particle like in higher dimension, but the collective excitation. This phenomena can be describe by susceptibility, which measure the response due to external parameter. Another reason is the well known: Renormalization group in **SGE**.

Due to the reason above, if we want to obtain the state of asymmetric **DSGE**, we don't use perturbation theory, especially for infinite length in one dimension. We must "exactly" solve this nonintegrable equation. This seem to be out of question, but we take "the middle way". We find the theoretical form of the solutions first, then we use **Möbius transformation** to simplify it. Finally, numerical method is applied to calculate the parameters.

In perivious work, our main interest is finite systems and fixed boundary conditions ( i.e. $\theta(x = 0) = 0, \theta(x = L) = 2\pi$ ). It is more complex in **DSGE**. We solve the equation for an infinite system. The solutions contain hyperbolic functions, which are related to **JEF**. For example, $sc(u,k) \to \sinh(u)$ if $k \to 1$.

In §**5.2**, I list classical solutions and energy of **DSGE** depend on the second potential coefficient $\eta$. It is helpful for us to study further asymmetric **DSGE**. In §**5**.**3**, we give mathematical solution by **Möbius transform** and see how **Möbius** parameter change with adiabatical parameter in **Fig**. **10 and Fig**.**11** (**a**), (**b**) at $|\eta| < \frac{1}{4}$ and $|\eta| > \frac{1}{4}$ corresponding. I also plot potential at $\eta = 0.15.0.35, -0.35$ in **Fig**. **9** (**a**),(**b**),(**c**) for compare.

## 5.2 Classical solutions of Double Sine-Gordon Equation

### 5.2.1 List of classical solutions of DSGE

In [**48**], [**49**] for example, also see **Chapter 3** of [**51**], we can see several solutions of Double Sine-Gordon which does not contain the asymmetric phase.

We begin with the equation:

$$\theta_{tt} - c^2\theta_{xx} + (\frac{c}{d^2})^2(\sin\theta + 2\eta\sin 2\theta) = 0 \qquad 5.2.1.1$$

If we want to find the traveling wave solution with speed $v$, it is convenient to define $s = \frac{\gamma}{d}(x - vt)$ and $\gamma = \frac{1}{\sqrt{1-(\frac{v}{c})^2}}$. Then the equation become:

$$\frac{d^2\theta}{d^2s} = \sin(\theta) + 2\eta\sin(2\theta) \qquad 5.2.1.2$$

The kink solutions $\theta(s)$ satisfy the boundary condition: $\theta(s = \pm\infty) = \theta_a, \theta_b$ and $\frac{d\theta}{ds}(s = \pm\infty) = 0$. Here $\theta_a, \theta_b$ are the minima of the potential $V(\theta)$

$$V(\theta) = -\cos\theta - \eta\cos 2\theta \qquad 5.2.1.3$$

The kind of traveling solution depend on $\eta$ because their potential are different at $\eta < -\frac{1}{4}$, $|\eta| < \frac{1}{4}$, and $\eta > \frac{1}{4}$ :

**Case** ***1***: $\eta < -\frac{1}{4}$

*We have two kinds of travelling kinks:*

$$\theta^{>} = 2\tan^{-1}[\pm\sqrt{\frac{4|\eta|-1}{4|\eta|+1}}\coth(\sqrt{\frac{16|\eta|^2-1}{16|\eta|}}s)] \qquad \mathit{5.2.1.4}$$

$$\theta^{<} = 2\tan^{-1}[\pm\sqrt{\frac{4|\eta|-1}{4|\eta|+1}}\tanh(\sqrt{\frac{16|\eta|^2-1}{16|\eta|}}s)] \qquad \mathit{5.2.1.5}$$

*The minima of* $V(\theta)$ *are located at* $\theta_{\min} = \cos^{-1}(\frac{-1}{4\eta}) + 2n\pi$ *, where n is an integer.* $V(\theta_{\min}) = \frac{1}{8\eta} + \eta$;

*The absolute maxima of* $V(\theta)$ *are located at* $\theta_{abs.\max} = (2n+1)\pi$ . $V(\theta_{abs.\max}) = 1 - \eta$;

*The relative maxima of* $V(\theta)$ *are located at* $\theta_{rel.\max} = 2n\pi$ . $V(\theta_{abs.\max}) = -1 - \eta$.

**Case** ***2***: $|\eta| < \frac{1}{4}$

*We have one travelling kink:*

$$\theta^{>} = 2\tan^{-1}[\pm\sqrt{1+4\eta}\,csch(\sqrt{1+4\eta}\,s] \qquad \mathit{5.2.1.6}$$

*Because in this region,* $V(\theta)$ *is similar to the case when* $\eta = 0$, *there is no relative maxima or minima.*

*The minima of* $V(\theta)$ *are located at* $\theta_{\min} = 2n\pi$. $V(\theta_{\min}) = -1 - \eta$;

*The maxima of* $V(\theta)$ *are located at* $\theta_{\max} = (2n+1)\pi$. $V(\theta_{\max}) = 1 - \eta$.

**Case** *3:* $\eta > \frac{1}{4}$

*We have two kinds of travelling kinks:*

$$\theta^{>} = 2\tan^{-1}[\pm\sqrt{1+4\eta}\,csch(\sqrt{1+4\eta}\,s] \qquad 5.2.1.7$$

$$\theta^{B} = 2\tan^{-1}[\pm\frac{1}{\sqrt{4\eta-1}}\cosh(\sqrt{4\eta-1}\,s)] \qquad 5.2.1.8$$

*The absolute minima of* $V(\theta)$ *are located at* $\theta_{abs.\min} = 2n\pi$ . $V(\theta_{abs.\min}) = -1 - \eta$;

*The relative minima of* $V(\theta)$ *are located at* $\theta_{rel.\min} = (2n+1)\pi$ . $V(\theta_{rel.\min}) = 1 - \eta$;

*The maxima of* $V(\theta)$ *are located at* $\theta_{\max} = \cos^{-1}(\frac{-1}{4\eta}) + 2n\pi$. $V(\theta_{abs.\max}) = \frac{1}{8\eta} + \eta$.

I can show some limiting behavior of these solutions:

**Case** *1:* $\eta = 0$ :

***DSGE*** *become* ***SGE****. so* ***Eq****. (****5.2.1.6****) should equal to one of the travelling wave solution of* ***SGE****. If* $\theta_{ss} = \sin\theta$*, ( i.e.* ***SGE****), then the solution is* $\theta = 4\tan^{-1}e^{\pm s} \Rightarrow \theta = 2\tan^{-1}[\pm csch(s)]$. *This is just the solution, (****5.2.1.6****) at* $\eta = 0$.

**Case** *2:* $\eta = \frac{-1}{4}$ :

*Does (****5.2.1.4****) equal to (****5.2.1.6****) and (****5.2.1.5****) vanish? Expanding (****5.2.1.5****),* $\tan(\frac{\theta^{<}}{2}) = \pm\sqrt{\frac{-4\eta-1}{-4\eta+1}} \bullet \frac{e^{\sqrt{\frac{16\eta^2-1}{-16\eta}}s} - e^{-\sqrt{\frac{16\eta^2-1}{-16\eta}}s}}{e^{\sqrt{\frac{16\eta^2-1}{-16\eta}}s} + e^{-\sqrt{\frac{16\eta^2-1}{-16\eta}}s}}$*, we can easily find* $\tan(\frac{\theta^{<}}{2}) = 0$ *when* $\eta = \frac{-1}{4}$*. Expand (****5.2.1.4****):* $\tan(\frac{\theta^{>}}{2}) = \pm\sqrt{\frac{-4\eta-1}{-4\eta+1}} \bullet \frac{e^{\sqrt{\frac{16\eta^2-1}{-16\eta}}s} + e^{-\sqrt{\frac{16\eta^2-1}{-16\eta}}s}}{e^{\sqrt{\frac{16\eta^2-1}{-16\eta}}s} - e^{-\sqrt{\frac{16\eta^2-1}{-16\eta}}s}}$*, and also (****5.2.1.6****):* $\tan(\frac{\theta^{>}}{2}) = \pm\sqrt{1+4\eta}\,\frac{2}{e^{\sqrt{1+4\eta}s} - e^{-\sqrt{1+4\eta}s}}$*. I find they are the same if* $\eta \to \frac{-1}{4}$.

## 5.2.2 List of energies of classical solutions in DSGE

From **Eq**.(**5.2.1.2**), we multiply $\theta_s$ on both sides and integrate respect to $s$:

$$\int \theta_s \bullet \theta_{ss} ds = \int \theta_s(\sin\theta + 2\eta\sin 2\theta)ds + S \qquad 5.2.2.1$$

we get:

$$\frac{1}{2}(\frac{d\theta}{ds})^2 + \cos\theta + \eta\cos 2\theta = S \qquad 5.2.2.2$$

where $S$ is called "**action**" because $S = T - V$. In $s \to \pm\infty$, then we choose the constant action $S = -V_{\min}$, Whether it is $V_{abs,\min}$ or $V_{rel.\min}$ depends on what solution we want. The energy is;

$$H = \int_{-\infty}^{\infty}(\frac{\theta_x^2}{2} - \cos\theta - \eta\cos 2\theta)dx \qquad 5.2.2.3$$

From **Eq**.(**5.2.2.2**), if I consider only the static case, then:

$$\frac{\theta_x^2}{2} = S - \cos\theta - \eta\cos 2\theta \qquad 5.2.2.4$$

imply

$$\frac{d\theta}{dx} = \sqrt{2S - 2\cos\theta - 2\eta\cos 2\theta} \qquad 5.2.2.5$$

So

$$V_s = \int_{-\infty}^{\infty} \frac{\theta_x^2}{2} dx = \frac{1}{\sqrt{2}} \int_0^{2\pi} \sqrt{S - \cos\theta - \eta\cos 2\theta}\, d\theta \qquad 5.2.2.6$$

and

$$V_p = \int_{-\infty}^{\infty} (-\cos\theta - \eta\cos 2\theta) dx \qquad 5.2.2.7$$

imply

$$V_s = S \bullet x|_{-\infty}^{\infty} + V_p \qquad 5.2.2.8$$

But the first term of **Eq.**(**5.2.2.8**) is diverges if we consider infinite system. So if we don't think this infinity because we can do energy shift to cancel this term, then.

$$V_s = V_p \qquad 5.2.2.9$$

and energy is equal to twice of $V_s$

$$H = V_s + V_p = 2V_s \qquad 5.2.2.10$$

The energy also depends on $\eta$, described as following:

**Case 1**: $\eta < -\frac{1}{4}$

*We choose* $S = -V(\theta_{\min}) = -\eta - \frac{1}{8\eta}$ *and define* $\phi_0$ *is one of* $\theta_{\min}$*:*

$$\phi_0 = \cos^{-1}\left(\frac{-1}{4\eta}\right) \qquad 5.2.2.11$$

*Here we have two kinds of kinks.*

*(**A**): Large kink:*

$\theta \in [\phi_0, 2\pi - \phi_0]$*, and it must varies cross over one of absolute maximum points*

$$: V_s^{>} := \frac{1}{\sqrt{2}} \int_{\phi_0}^{2\pi-\phi_0} \sqrt{-\eta - \frac{1}{8\eta} - \cos\theta - \eta\cos 2\theta}\, d\theta = \frac{1}{\sqrt{2}} \int_{\phi_0}^{2\pi-\phi_0} \sqrt{-2\eta(\cos\theta + \frac{1}{4\eta})^2}\, d\theta \qquad 5.2.2.12$$

$$= \sqrt{-\eta} \int_{\phi_0}^{2\pi-\phi_0} (\cos\theta + \frac{1}{4\eta}) d\theta = \sqrt{-\eta}\,(-\sin\theta + \frac{\theta}{4\eta})\Big|_{\theta=\phi_0}^{\theta=2\pi-\phi_0} = \frac{1}{2\sqrt{-\eta}}\left(\sqrt{16\eta^2 - 1} + \pi - \phi_0\right)$$

*(**B**): Small kink:*

$\theta \in [-\phi_0, \phi_0]$*, and it must varies cross one of relative maximum points*

$$: V_s^{<} := \frac{1}{\sqrt{2}} \int_{\phi_0}^{-\phi_0} \sqrt{-\eta - \frac{1}{8\eta} - \cos\theta - \eta\cos 2\theta}\, d\theta \qquad 5.2.2.13$$

$$= \sqrt{-\eta}\,(-\sin\theta + \frac{\theta}{4\eta})|_{\theta=\phi_0}^{\theta=-\phi_0} = \frac{1}{2\sqrt{-\eta}}(\sqrt{16\eta^2 - 1} - \phi_0)$$

**Case 2**: $|\eta| < \frac{1}{4}$

*We choose* $S = -V(\theta_{\min}) = 1 + \eta$

$$: V_s^{>} := \frac{1}{\sqrt{2}} \int_{0}^{2\pi} \sqrt{1 + \eta - \cos\theta - \eta\cos 2\theta}\, d\theta \qquad 5.2.2.14$$

*The integral also depend on wether* $\eta$ *is lager or smaller than* $0$

***(A)**: **If*** $\frac{1}{4} > \eta > 0$ :

$$: V_s^{>} := 2\sqrt{4\eta + 1} + \frac{\ln(2\sqrt{\eta} + \sqrt{4\eta + 1})}{\sqrt{\eta}} \qquad 5.2.2.15$$

***(B)**: **If*** $0 > \eta > -\frac{1}{4}$ :

$$: V_s^{>} := 2\sqrt{4\eta + 1} + \frac{\sin^{-1}(2\sqrt{-\eta})}{\sqrt{-\eta}} \qquad 5.2.2.16$$

**Case 3**: $\eta > \frac{1}{4}$

*Because one kind of solution and its absolute minimum is identitcal to those of (A) of case 2, so:*

$$: V_s^{>} := 2\sqrt{4\eta + 1} + \frac{\ln(2\sqrt{\eta} + \sqrt{4\eta + 1})}{\sqrt{\eta}} \qquad 5.2.2.17$$

*Another is the bubble solution* $\theta(s) \to (2n + 1)\pi$ *(the relative minimum) as* $s \to \pm\infty$. *In this case,* $S = -V(\theta_{rel,\min}) = \eta - 1$

$$: V_s := 2\frac{1}{\sqrt{2}} \int_{2\tan^{-1}(\frac{1}{\sqrt{4\eta-1}})}^{\pi} \sqrt{\eta - 1 - \cos\theta - \eta\cos 2\theta}\, d\theta = 2\sqrt{4\eta - 1} - \frac{\ln(2\sqrt{\eta} + \sqrt{4\eta - 1})}{\sqrt{\eta}} \qquad 5.2.2.18$$

*One must notice is the upper and lower bound of the integral. We have divided the integral into two part. The upper bound* $\pi$ *is the relative minimum while lower bound* $2\tan^{-1}(\frac{1}{\sqrt{4\eta-1}})$ *is directly from* $\theta^B(s = 0) = 2\tan^{-1}[\pm\frac{1}{\sqrt{4\eta-1}}\cosh(\sqrt{4\eta - 1}\, s)]|_{s=0} = \pm 2\tan^{-1}(\frac{1}{\sqrt{4\eta-1}})$ *This is the minimum. From the potential point of view, this state does not touch the maximum point of potential ( i.e.* $\theta = \cos^{-1}(\frac{-1}{4\eta})$*).*

## 5.2.3 A method to construct solutions and action $E$ of Double Sine-Gordon Equation

Authors of [**50**] list two types of solutions, hyperbolic functions for infinite system and Jacobi elliptic function for finite system. Combining theirs and my method together, I developed a systematic method to construct the solutions listed in §**5.2.1** and the actions, which is used to calculate the energy, list in §**5.2.2**.

We start from **Eq**. (**5.2.2.2**):

$$\frac{1}{2}(\frac{d\theta}{ds})^2 + \cos(\theta) + \eta\cos(2\theta) = S \qquad 5.2.2.2$$

and assume the solutions have to form:

$$\theta = 2\tan^{-1}[f(s)] \qquad 5.2.3.1$$

Substituting **Eq**. (**5.2.3.1**) into (**5.2.2.2**), we obtain:

$$(\frac{df}{ds})^2 = (\frac{S+1-\eta}{2})f^4 + (S+3\eta)f^2 + (\frac{S-1-\eta}{2}) \qquad 5.2.3.2$$

**Eq**. (**5.2.3.2**) can be satisfied by various kinds of functions. I listed those which will be used later:

**Case** *1:* $y = csch(s)$ *which satisfies:* $y'^2 = y^4 + y^2$

*Let*

$$f(s) = \alpha \bullet csch(\beta s) \qquad 5.2.3.3$$

*so that*

$$(\frac{df}{ds})^2 = \alpha^2\beta^2[csch^4(\beta s) + csch^2(\beta s)] = \frac{\beta^2}{\alpha^2}f^4 + \beta^2 f^2 \qquad 5.2.3.4$$

*comparing **Eq**. (**5.2.3.2**) and (**5.2.3.4**), we obtain:*

$$\frac{S+1-\eta}{2} = \frac{\beta^2}{\alpha^2}; S+3\eta = \beta^2; \frac{S-1-\eta}{2} = 0 \qquad 5.2.3.5$$

***Eq**. (**5.2.3.5**) is solved and the solutions list in §**5.2.1**, §**5.2.2** can be found:*

$$S = \eta + 1; \beta = \pm\sqrt{1+4\eta}; \alpha = \pm\sqrt{1+4\eta} \qquad 5.2.3.6$$

**Case** *2:* $y = \cosh(s)$ *which satisfies:* $y'^2 = y^2 - 1$

*Let*

$$f(s) = \alpha \bullet \cosh(\beta s) \qquad 5.2.3.7$$

*so that*

$$(\frac{df}{ds})^2 = \alpha^2\beta^2[\cosh^2(\beta s) - 1] = \beta^2 f^2 - \alpha^2\beta^2 \qquad 5.2.3.8$$

*comparing **Eq**. (**5.2.3.2**) and (**5.2.3.8**), we obtain:*

$$\frac{S+1-\eta}{2} = 0; S+3\eta = \beta^2; \frac{S-1-\eta}{2} = -\alpha^2\beta^2 \qquad 5.2.3.9$$

***Eq**. (**5.2.3.9**) is solved and the solutions list in §**5.2.1**, §**5.2.2** can be found:*

$$S = \eta - 1; \beta = \pm\sqrt{4\eta - 1}; \alpha = \frac{\pm 1}{\sqrt{4\eta - 1}} \quad 5.2.3.10$$

**Case** *3:* $y = \tanh(s); \coth(s)$ *both satisfies:* $y'^2 = y^4 - 2y^2 + 1$

*Two types of solutions will share same number of* $\alpha, \beta, E$. *Let*

$$f(s) = \alpha \cdot \tanh(\beta s) \quad 5.2.3.11$$

*so that*

$$(\frac{df}{ds})^2 = \alpha^2\beta^2[\tanh^4(\beta s) - 2\tanh^2(\beta s) + 1] = \frac{\beta^2}{\alpha^2}f^4 - 2\beta^2 f^2 + \alpha^2\beta^2 \quad 5.2.3.12$$

*comparing* ***Eq****.* ***(5.2.3.2)*** *and* ***(5.2.3.12)****, we obtain:*

$$\frac{S + 1 - \eta}{2} = \frac{\beta^2}{\alpha^2}; S + 3\eta = -2\beta^2; \frac{S - 1 - \eta}{2} = \alpha^2\beta^2 \quad 5.2.3.13$$

***Eq****.* ***(5.2.3.13)*** *is solved and the solutions list in* *§****5.2.1****, §****5.2.2*** *can be found:*

$$S = -\eta - \frac{1}{8\eta}; \beta = \pm\sqrt{\frac{1 - 16\eta^2}{16\eta}}; \alpha = \pm\sqrt{\frac{4\eta + 1}{4\eta - 1}} \quad 5.2.3.14$$

**Case** *4:* $y = \mathrm{sec}\,h(s)$ *which satisfies:* $y'^2 = -y^4 + y^2$

*Let*

$$f(s) = \alpha \cdot sech(\beta s) \quad 5.2.3.15$$

*so that*

$$(\frac{df}{ds})^2 = \alpha^2\beta^2[-sech^4(\beta s) + sech^2(\beta s)] = \frac{-\beta^2}{\alpha^2}f^4 + \beta^2 f^2 \quad 5.2.3.16$$

*comparing* ***Eq****.* ***(5.2.3.2)*** *and* ***(5.2.3.16)****, we obtain:*

$$\frac{S + 1 - \eta}{2} = \frac{-\beta^2}{\alpha^2}; S + 3\eta = \beta^2; \frac{S - 1 - \eta}{2} = 0 \quad 5.2.3.17$$

***Eq****.* ***(5.2.3.17)*** *is solved and the solutions list in* *§****5.2.1****, §****5.2.2*** *can be found:*

$$S = \eta + 1; \beta = \pm\sqrt{1 + 4\eta}; \alpha = \pm\sqrt{-1 - 4\eta} \quad 5.2.3.18$$

**Case** *5:* $y = \sinh(s)$ *which satisfies:* $y'^2 = y^2 + 1$

*Let*

$$f(s) = \alpha \cdot \sinh(\beta s) \quad 5.2.3.19$$

*so that*

$$(\frac{df}{ds})^2 = \alpha^2\beta^2[\sinh^2(\beta s) + 1] = \beta^2 f^2 + \alpha^2\beta^2 \quad 5.2.3.20$$

*comparing* ***Eq****.* ***(5.2.3.2)*** *and* ***(5.2.3.20)****, we obtain:*

$$\frac{S+1-\eta}{2} = 0; S+3\eta = \beta^2; \frac{S-1-\eta}{2} = \alpha^2\beta^2 \quad 5.2.3.21$$

***Eq**. (**5.2.3.21**) is solved and the solutions list in §**5.2.1**, §**5.2.2** can be found:*

$$S = \eta - 1; \beta = \pm\sqrt{4\eta - 1}; \alpha = \pm\frac{1}{\sqrt{1-4\eta}} \quad 5.2.3.22$$

I will also list finite type solutions which will be used in § **5.6**. We can compare the following results with those in **case 1~5**. Using the following behavior: when $L \to \infty$, then $k_f \to 1$.

$$\begin{aligned} sn(x, k_f) &\to \tanh(x) \quad 5.2.3.23 \\ sc(x, k_f) &\to \sinh(x) \\ cn(x, k_f), dn(x, k_f) &\to sech(x) \end{aligned}$$

**Case** *6:* $y = cn(s)$ *which satisfy:* $y'^2 = -k_f^2 y^4 + (2k_f^2 - 1)y^2 + 1 - k_f^2$

*Let*

$$f(s) = \alpha \cdot cn(\beta s) \quad 5.2.3.24$$

*so that*

$$\begin{aligned} \left(\frac{df}{ds}\right)^2 &= \alpha^2\beta^2[-k_f^2 cn^4(\beta s) + (2k_f^2 - 1)cn^2(\beta s) + 1 - k_f^2] \quad 5.2.3.25 \\ &= \frac{-\beta^2}{\alpha^2}k_f^2 f^4 + \beta^2(2k_f^2 - 1)f^2 + \alpha^2\beta^2(1 - k_f^2) \end{aligned}$$

*comparing **Eq**. (**5.2.3.2**) and (**5.2.3.25**), we obtain:*

$$\frac{S+1-\eta}{2} = \frac{-\beta^2}{\alpha^2}k_f^2; S+3\eta = \beta^2(2k_f^2-1); \frac{S-1-\eta}{2} = \alpha^2\beta^2(1-k_f^2) \quad 5.2.3.26$$

***Eq**. (**5.2.3.26**) can be simplified by defining* $A = \frac{-2k_f^2}{2k_f^2-1}$ *and* $B = \frac{2k_f^2-1}{2(1-k_f^2)}$*:*

$$\alpha^2 = A \cdot \frac{S+3\eta}{S+1-\eta} = B \cdot \frac{S-1-\eta}{S+3\eta} \quad 5.2.3.27$$

*From **Eq**. (**5.2.3.27**), we obtain the equation of S:*

$$(A-B)\mathbf{S}^2 + 2\eta(3A+B)\mathbf{S} + (9A-B)\eta^2 + B = 0 \quad 5.2.3.28$$

$$S = \frac{-(3A+B)\eta \pm \sqrt{16AB\eta^2 - AB + B^2}}{A-B} \quad 5.2.3.29$$

*Put A and B into **Eq**. (**5.2.3.29**), we get:*

$$S = \eta \mp \sqrt{(1-2k_f)^2 + (256k_f^8 - 512k_f^6 + 320k_f^4 - 64k_f^2)\eta^2} \quad 5.2.3.30$$

*I choose the positive squirt because* $S|_{k_f \to 1} = 1 + \eta$*, which is just the sech case as we expected.*

*Put **Eq**. (**5.2.3.25**) into **Eq**. (**5.2.3.27**) and (**5.2.3.26**), we also obtain the explicitly formula of α and β, their limit ( $k_f \to 1$) also meet our expected.*

**Case** *7: $y = sc(s)$ which satisfies: $y'^2 = (1-k_f^2)y^4 + (2-k_f^2)y^2 + 1$*

*Let*

$$f(s) = \alpha \cdot sc(\beta s) \qquad 5.2.3.31$$

*so that*

$$(\frac{df}{ds})^2 = \alpha^2\beta^2[(1-k_f^2)sc^4(\beta s) + (2-k_f^2)sc^2(\beta s) + 1] = \frac{\beta^2}{\alpha^2}(1-k_f^2)f^4 + \beta^2(2-k_f^2)f^2 + \alpha^2\beta^2 \qquad 5.2.3.32$$

*comparing **Eq**.(**5.2.3.2**) and (**5.2.3.32**), we obtain:*

$$\frac{S+1-\eta}{2} = \frac{\beta^2}{\alpha^2}(1-k_f^2); S+3\eta = \beta^2(2-k_f^2); \frac{S-1-\eta}{2} = \alpha^2\beta^2 \qquad 5.2.3.33$$

*So we obtain the solutions list in §**5.2.1**, §**5.2.2** by solve **Eq**. (**5.2.3.33**).*

*in this case, we get:*

$$S = \frac{-(k_f^4 - 16k_f^2 + 16)\eta \pm \sqrt{(-64k_f^6 + 320k_f^4 - 512k_f^2 + 256)\eta^2 + (k_f^8 - 4k_f^6 + 4k_f^4)}}{-k_f^4} \qquad 5.2.3.34$$

*I choose the positive squirt because $S|_{k_f \to 1} = \eta - 1$, which is just the* sinh *case as we expected. Put **Eq**. (**5.2.3.34**) into **Eq**. (**5.2.3.27**) and (**5.2.3.33**), we also obtain the explicitly formula of α and β, their limit( $k_f \to 1$) also meet our expected.*

**Case** *8: $y = sn(s)$ which satisfies: $y'^2 = k_f^2 y^4 + (-k_f^2 - 1)y^2 + 1$*

*Let*

$$f(s) = \alpha \cdot sn(\beta s) \qquad 5.2.3.35$$

*so*

$$(\frac{df}{ds})^2 = \alpha^2\beta^2[k_f^2 sn^4(\beta s) + (-k_f^2 - 1)sn^2(\beta s) + 1] = \frac{\beta^2}{\alpha^2}k_f^2 f^4 + \beta^2(-k_f^2 - 1)f^2 + \alpha^2\beta^2 \qquad 5.2.3.36$$

*compare **Eq**. (**5.2.3.2**) and (**5.2.3.36**), we obtain:*

$$\frac{S+1-\eta}{2} = \frac{\beta^2}{\alpha^2}k_f^2; S+3\eta = -\beta^2(-k_f^2 - 1); \frac{S-1-\eta}{2} = \alpha^2\beta^2 \qquad 5.2.3.37$$

*so we obtain the solutions list in §**5.2.1**, §**5.2.2** by solve **Eq**. (**5.2.3.37**):*

*From **Eq**. (**5.2.3.37**), and I let $A = \frac{-2k_f^2}{2k_f^2-1}, B = \frac{2k_f^2-1}{2(1-k_f^2)}$ for convenient. So the same equation as **Eq**. (**5.2.3.27**), (**5.2.3.28**) and (**5.2.3.29**):*

*Again following the same procedure, we get in this case:*

$$S = \frac{(k_f^4 + 14k_f^2 + 1)\eta \pm \sqrt{(64k_f^6 + 128k_f^4 + 64k_f^2)\eta^2 + (k_f^8 - 2k_f^4 + 1)}}{(k_f^2 - 1)^2} \qquad 5.2.3.38$$

*This formula is correct for $0 \leq k_f < 1$, but not correct for its limit ( $k_f = 1$). The reason is explained by the following example. If we want to find solution of $ax^2 + bx + c = 0$, its solution $x = \frac{-b\pm\sqrt{b^2-4ac}}{2a}$, but it is right for $a \neq 0$. The solution is $x = \frac{-c}{b}$ if $a = 0$. This is what* ***Eq.(5.2.3.28)*** *show, because $(A - B)|_{k_f=1} = 0$! So when $k_f = 1$, from* ***Eq.(5.2.3.28)***, *we obtain:*

$$S|_{k_f=1} = \frac{-(9A - B)\eta^2 - B}{2(3A + B)\eta} = -\eta - \frac{1}{8\eta} \qquad 5.2.3.39$$

*This just the value of* tanh *we expect! Inserting back into* ***Eq. (5.2.3.38)***, *we also can obtain the limit values of $\alpha$ and $\beta$ as we expect.*

# 5.3 General Mathematical Analysis include easier case:$\eta \leq \frac{1}{4}$

## 5.3.1 Potential analysis and related topics

Because the potential of double sine-gordon equation is:

$$V = -\cos(\theta + \varphi) - \eta\cos(2\theta) \qquad 5.3.1.1$$

I list some potential figures v.s. $\theta$ for $\varphi = 0, \pi/4, \pi/2$ at **Fig. 9.(a)** $\eta = -0.35$; **(b)**. $\eta = 0.15$; **(c)**. $\eta = 0.35$.

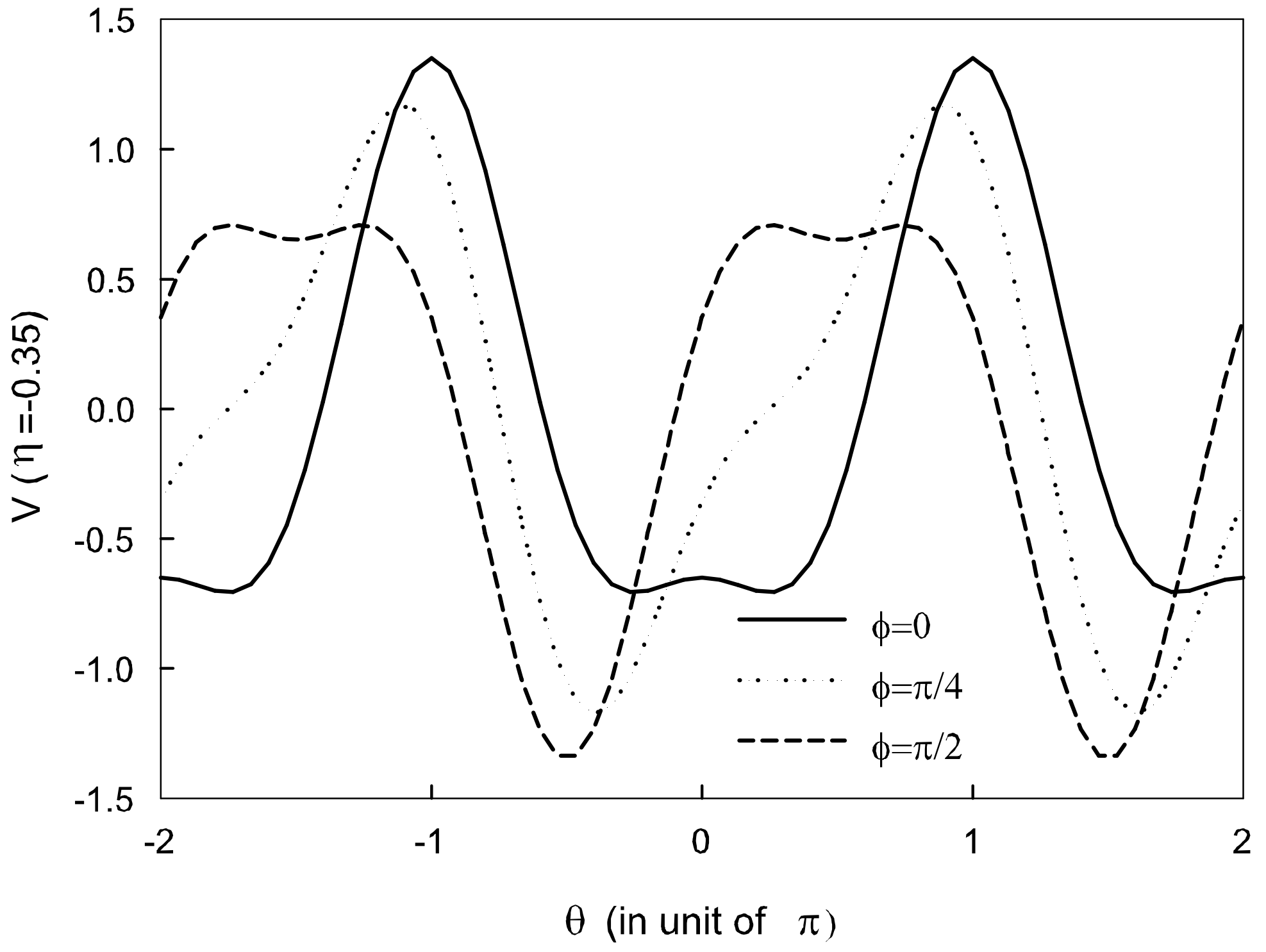

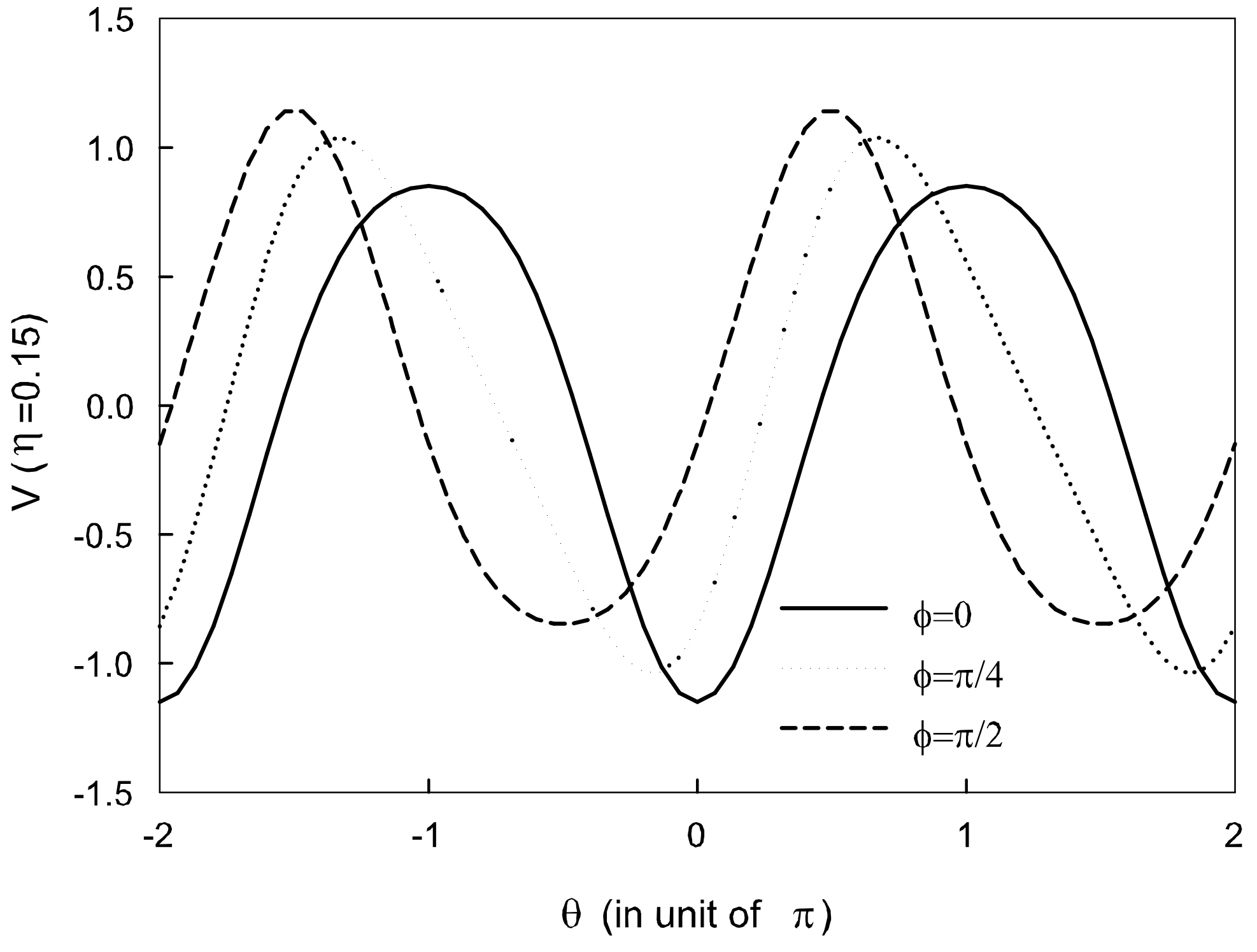

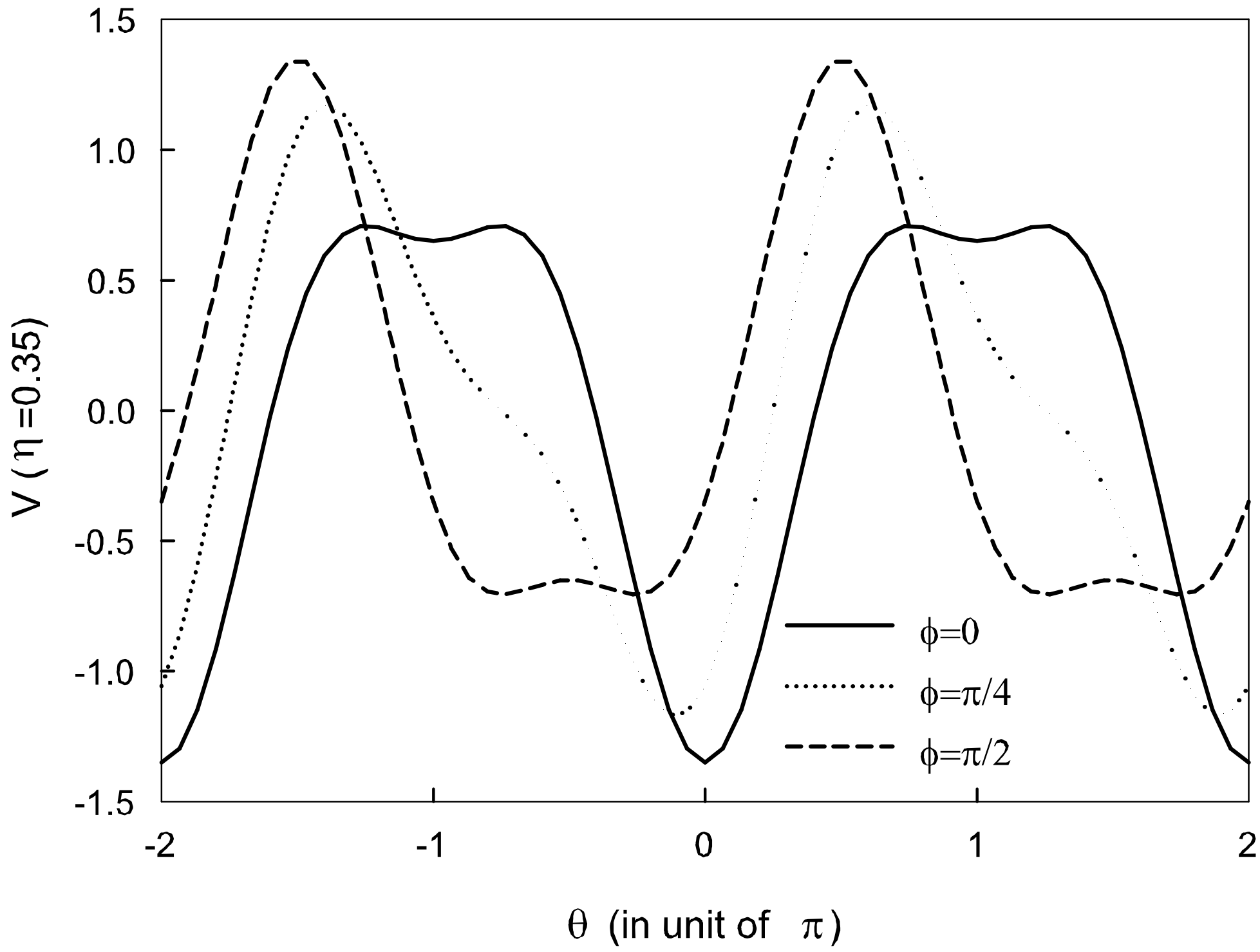

Fig 9. Potential of **DSGE**: (a) $\eta = -0.35$; (b). $\eta = 0.15$; (c). $\eta = 0.35$

So when $\varphi = 0$, $V = -\cos(\theta) - \eta\cos(2\theta)$;
but when $\varphi = \frac{\pi}{2}$, $V = -\cos(\theta + \frac{\pi}{2}) - \eta\cos(2\theta) = -\cos(\widetilde{\theta}) + \eta\cos(2\widetilde{\theta})$; where $\widetilde{\theta} = \theta + \frac{\pi}{2}$.
So compare the potential at $\varphi = 0$ and $\frac{\pi}{2}$, we find $V$ seems change sign of $\eta$ between them. Especially when $|\eta| > \frac{1}{4}$, different sign of $\eta$ may even give us different solution.

For example, if we start from $\varphi = 0$ and $\eta > \frac{1}{4}$, then from previous subsection, we know the solutions are:

$$\theta^{>} = 2\tan^{-1}[\pm\sqrt{1+4\eta}\,csch(\sqrt{1+4\eta}\,s] \qquad 5.2.1.7$$

$$\theta^{B} = 2\tan^{-1}[\pm\frac{1}{\sqrt{4\eta-1}}\cosh(\sqrt{4\eta-1}\,s)] \qquad 5.2.1.8$$

But by previous observation, we know when $\varphi$ adiabatical change to $\frac{\pi}{2}$, then the solution will become:

$$\theta^{>} = 2\tan^{-1}[\pm\sqrt{\frac{4|\eta|-1}{4|\eta|+1}}\coth(\sqrt{\frac{16|\eta|^2-1}{16|\eta|}}\,s)] - \frac{\pi}{2} \qquad 5.3.1.2a$$

$$\theta^{<} = 2\tan^{-1}[\pm\sqrt{\frac{4|\eta|-1}{4|\eta|+1}}\tanh(\sqrt{\frac{16|\eta|^2-1}{16|\eta|}}\,s)] - \frac{\pi}{2} \qquad 5.3.1.2b$$

Notice we minus $\frac{\pi}{2}$ due to observation above, so a little different from **Eq**. (**5.2.1.7**) **and** (**5.2.1.8**).

So this is just a example when $\varphi = \frac{\pi}{2}$, the allowed solution will only in change-sign $\eta$ region, but fourtually in $|\eta| < \frac{1}{4}$, there is one kind of solution(include its Möbius form).

Also from **eq**. (**5**.**3**.**1**.**1**), I can prove the period of physical quantities and phenomena is $\pi$ in asymmetric **DSGE**, not $2\pi$ as in **DSGE**. We can let $\widetilde{\theta} = \theta + \varphi$, so **eq**. (**5**.**3**.**1**.**1**) becomes:

$$V = -\cos\widetilde{\theta} - \eta\cos(2\widetilde{\theta} - 2\varphi) \qquad 5.3.1.3$$

This give the explain why the period of energy and spin transport below subsection is $\pi$ because these two ones are not related to $\theta$ translation. So from **eq**. (**5**.**1**.**4**.**3**), the period of potential $V$ is equal to $2\varphi$, which is $2\pi$. So when $\varphi$ go around $\pi$, the system goback to its origin.

Here, I will show how to calculate the total energy in general $\varphi$. The same doing with §**5**.**2**.**2**, in there we only calculate total energy of $\varphi = 0$. The first thing is to calculate $V_s$:

$$V_s = \int_{-\infty}^{\infty} \frac{\theta_x^2}{2} dx = \frac{1}{\sqrt{2}} \int_0^{2\pi} \sqrt{S - \cos(\theta + \varphi) - \eta\cos 2\theta}\, d\theta \qquad 5.3.1.4$$

Then the same delete the infinity constant, so total energy $H = 2V_s$. Notice because $E = -V_{\min}$, so action $E$ is different at each $\varphi$. Especially thing is when $\eta > 1/4$ and $\varphi = \pi/2$, because now the solution become large and small kinks. But the total range of both kinks is still $2\pi$. So if we consider total energy, then **Eq**. (**5**.**3**.**1**.**4**) still right and continuous at $\varphi = \pi/2$. But if we only consider large kink part, then at $\varphi = \pi/2$, the energy of large kink part is discontinuous become small kink carry some part of energy. It can show in the following equation:

$$\begin{aligned} V_s &= \frac{1}{\sqrt{2}}\left(\int_{-\phi_0}^{\phi_0} \sqrt{S - \cos(\theta + \varphi) - \eta\cos 2\theta}\, d\theta + \int_{\phi_0}^{2\pi-\phi_0} \sqrt{S - \cos(\theta + \varphi) - \eta\cos 2\theta}\, d\theta\right) \\ &= \frac{1}{\sqrt{2}} \int_0^{2\pi} \sqrt{S - \cos(\theta + \varphi) - \eta\cos 2\theta}\, d\theta \end{aligned} \qquad 5.3.1.5$$

Also from the figures of each $\varphi$, I have find some thing: In $|\eta| > \frac{1}{4}$, we have known there is a bubble solution: $\theta_B$. Because from the figure of the potential, $\theta_B$ exist in two neighbor $\theta_{\max}$ and among them pass through $\theta_{rel,\min}$. That is because $\theta_{rel,\min}$ is only semi-stable for state to exist, not like $\theta^{>}$ is stable because its range go from one $\theta_{abs,\min}$ to the next $\theta_{abs,\min}$ from $s \to -\infty$ to $s \to \infty$. And the existence of the bubble also can be seen in general $\varphi$, if we can see some "valley" in the potential figure in each $\varphi$. But the occur of the "valley" seems no physical rule to guess. Because in each $\eta$ and $\varphi$, sometimes we can find "valley", but sometimes not find.

## 5.3.2 The equation to be solved

I want to solve :

$$\frac{d^2\theta}{d^2 s} = \sin(\theta + \varphi) + 2\eta\sin(2\theta) \qquad 5.3.2.1$$

Integrate and expand **Eq**. (**5**.**3**.**2**.**1**), we get:

$$\frac{1}{2}\left(\frac{d\theta}{ds}\right)^2 + \cos(\varphi)\cos(\theta) - \sin(\varphi)\sin(\theta) + \eta\cos(2\theta) = S \qquad 5.3.2.2$$

Where E is a constant, called "action", $S = -V_{\min}$. And $V$ is the potential:
$V = -\cos(\theta + \varphi) - \eta\cos(2\theta)$

Suppose

$$\theta = 2\tan^{-1}[f(s)] \qquad 5.3.2.3$$

Then we have following equation:

$$2\left(\frac{df}{ds}\right)^2 = (S + \cos(\varphi) - \eta)f^4 + 2\sin(\varphi)f^3 + (2S + 6\eta)f^2 + 2\sin(\varphi)f + (S - \cos(\varphi) - \eta) \qquad 5.3.2.4$$

The above Eq. is similar the partial equation of Jacobi elliptic function( **JEF** ) or hypertriangle function except the terms the odd power of $(f^3, f)$. By elliptic function theory ( **Appendix C**), we can do " **Möbius transformation**" to transform **Eq**. (**5.3.2.4**) to the stand form for **JEF**.

I means choose suitable coefficients: a, b, c, d; and let $g(s) = \frac{a*f(s)+b}{c*f(s)+d}$ , the by the theory, it can transform to:

$$(\frac{dg}{ds})^2 = a_4 f^4 + a_2 f^2 + a_0 \qquad 5.3.2.5$$

## 5.3.3 Function form and equations for infinite systems

In any $\varphi$ and in infinite system with no boundary conditions. It is reasonable to use hypertriangle function. So I let

$$f(s) = \frac{a\sinh(rs) + b}{c\sinh(rs) + d} \qquad 5.3.3.1$$

Put **eq**. (**5.3.3.1**) into **eq**. (**5.3.2.4**) and compare the powers of sinh, also restrict the scaling:

$$(ab - cd)^2 = 1 \qquad 5.3.3.2a$$

We have following equations:

$$(a^2 + c^2)^2 S + (a^4 - c^4)\cos\varphi + 2ac(a^2 + c^2)\sin\varphi + (-a^4 + 6a^2c^2 - c^4)\eta = 0 \qquad 5.3.3.2b$$

$$\begin{aligned} &4(ab + cd)(a^2 + c^2)S + 4(a^3b - c^3d)\cos\varphi \\ &+ 2(a^3d + 3a^2bc + 3ac^2d + bc^3)\sin\varphi \\ &+ [-4a^3b + 6(2a^2cd + 2abc^2) - 4c^3d]\eta \\ &= 0 \end{aligned} \qquad 5.3.3.2c$$

$$\begin{aligned} &[6a^2b^2 + 2(a^2d^2 + 4abcd + b^2c^2) + 6c^2d^2]S \\ &+ (6a^2b^2 - 6c^2d^2)\cos\varphi + 6(a^2bd + ab^2c + acd^2 + bc^2d)\sin\varphi + \\ &6(-a^2b^2 + a^2d^2 + 4abcd + b^2c^2 - c^2d^2)\eta \\ &= 2r^2 \end{aligned} \qquad 5.3.3.2d$$

$$\begin{aligned} &4(ab + cd)(b^2 + d^2)S + 4(b^3a - d^3c)\cos\varphi + 2(b^3c + \\ &3b^2ad + 3bd^2c + ad^3)\sin\varphi + \\ &[-4b^3a + 6(2d^2ab + 2cdb^2) - 4d^3c]\eta \\ &= 0 \end{aligned} \qquad 5.3.3.2e$$

$$(b^2 + d^2)^2 S + (b^4 - d^4)\cos\varphi + 2bd(b^2 + d^2)\sin\varphi + (-b^4 + 6b^2d^2 - d^4)\eta = 2r^2 \qquad 5.3.3.2f$$

As we see, it is hard to find analytical solutions from **eq**. (**5.3.3.2**) because the equations are complex. So I solve these equations by numerically and I show the datas in **Table 5.1**.

And because the Equations above are four power, so by the mathematical theory, we should have four roots. And my observation is that if $(a, b, c, d)$ is a solution, then $(-a, b, -c, d), (a, -b, c, -d)$ and $(-a, -b, -c, -d)$ are also solutions. This show some aspect of high-symmetry of these equations.

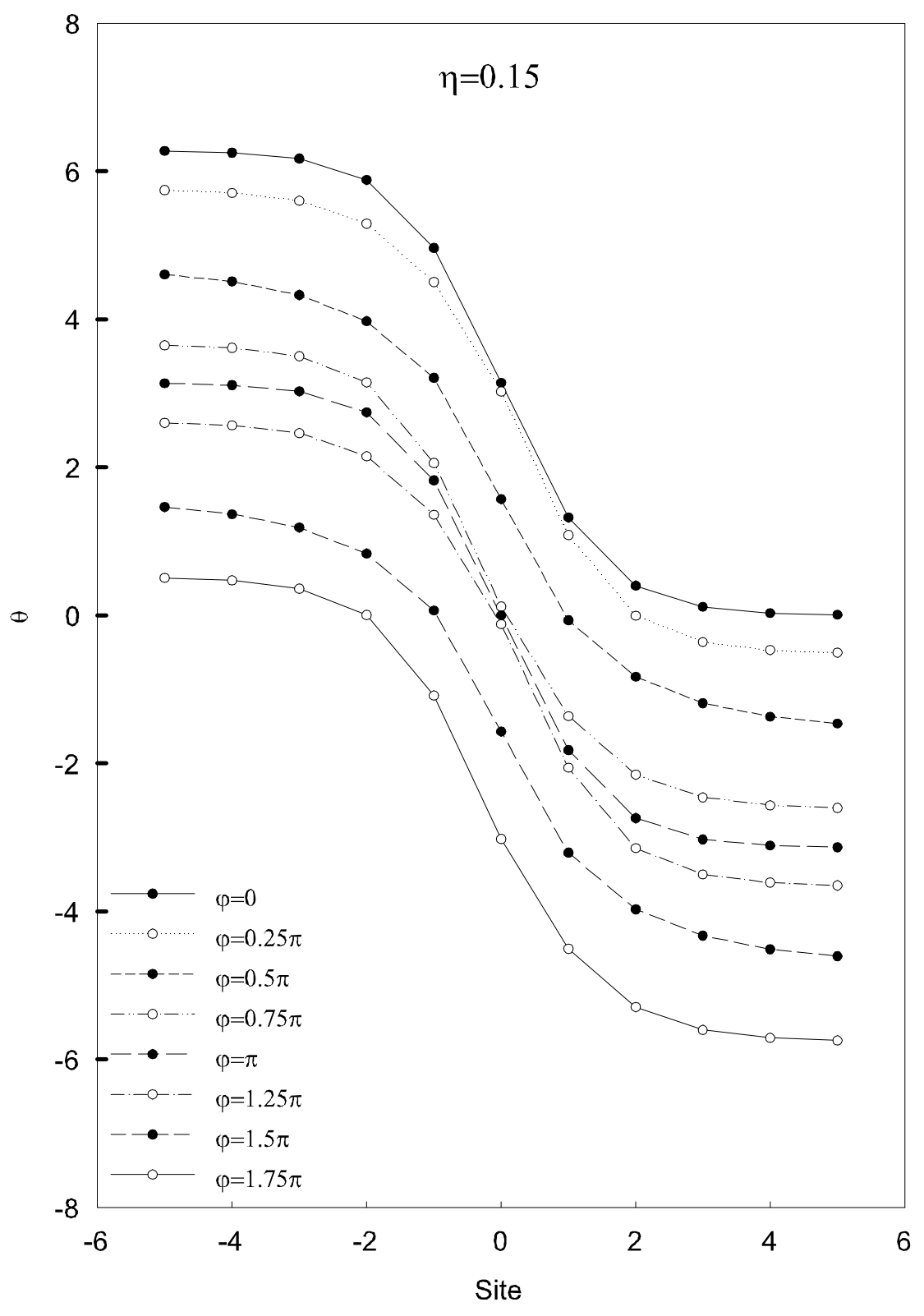

**Fig. 10**: Solutions (**eq.** (**5.3.3.1**)) for $\eta = 0.15$ and $\varphi = 0 \sim 7\pi/8$.

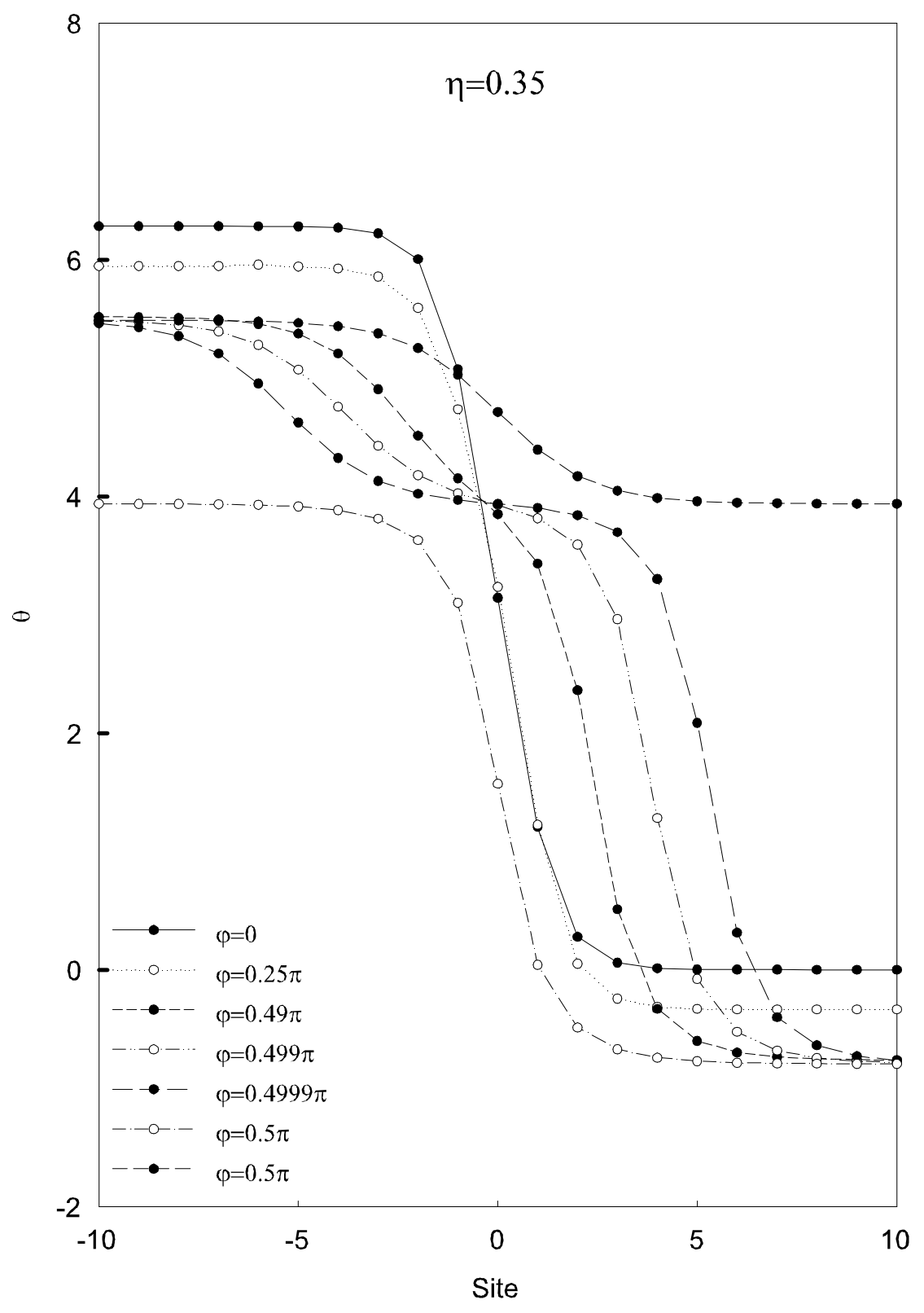

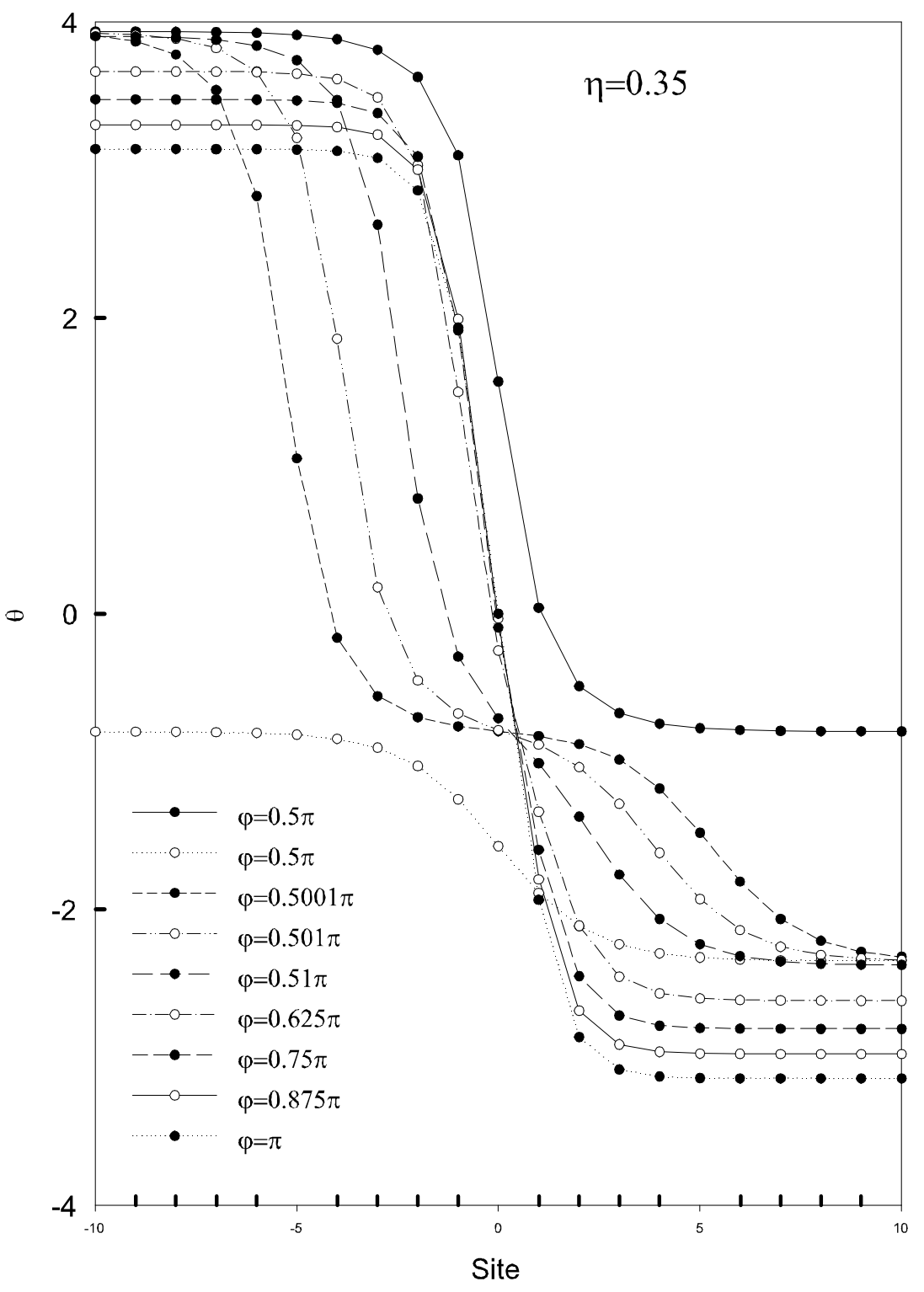

**Fig**. **11** (**a**), (**b**): Solutions (**eq**. (**5.3.3.1**)) for $\eta = 0.35$ and $\varphi = 0\sim\pi/2$. and $\varphi = \pi/2\sim\pi$ At $\varphi = \pi/2$ and the solutions have the form of **eq**. (**5.3.4.4**) which gives large and small kinks.

**Table 5.1**: $L = \infty$, no boundary conditions
List datas of $a, b, c, d, r, E$ for **Eq. (5.3.3.2)**: $f(s) = \frac{a\sinh(rs)+b}{c\sinh(rs)+d}$ at each $\varphi$

**1**. $\varphi = 0$

| $\eta\diagdown$ | $a$ | $b$ | $c$ | $d$ | $r$ | $S$ | energy of large kink | energy of bubble |
|---|---|---|---|---|---|---|---|---|
| −0.25 | × | × | × | × | × | 0.75 | $2\pi$ | × |
| −0.15 | 0 | 0.7953 | 1.2574 | 0 | ±0.5325 | 0.85 | 7.1055 | × |
| 0 | 0 | 1 | 1 | 0 | ±1 | 1 | 8 | × |
| 0.15 | 0 | 1.1247 | 0.8891 | 0 | ±1.2649 | 1.15 | 8.74 | × |
| 0.25 | 0 | 1.1892 | 0.8409 | 0 | ±1.4142 | 1.25 | 9.1824 | × |
| 0.35 | 0 | 1.2447 | 0.8034 | 0 | ±1.5492 | 1.35 | 9.5949 | 0.5134 |
| 0.45 | 0 | 1.2936 | 0.7731 | 0 | ±1.6733 | 1.45 | 9.9836 | 1.1785 |
| 0.55 | 0 | 1.3375 | 0.7477 | 0 | ±1.7889 | 1.55 | 10.3523 | 1.8272 |

**2**. $\varphi = \frac{\pi}{8}$

| $\eta\diagdown$ | $a$ | $b$ | $c$ | $d$ | $r$ | $S$ | E(large-kink) | E(bubble) |
|---|---|---|---|---|---|---|---|---|
| −0.25 | −0.3828 | 0.8053 | 0.7914 | 0.9472 | ±1.0501 | 0.9309 | 7.6508 | × |
| −0.15 | −0.34 | 0.8202 | 0.9431 | 0.666 | ±0.9187 | 0.9277 | 7.6103 | × |
| 0 | −0.1951 | 0.9808 | 0.9808 | 0.1951 | ±1 | 1 | 8 | × |
| 0.15 | −0.1112 | 1.121 | 0.8887 | 0.034 | ±1.2316 | 1.1215 | 8.609 | × |
| 0.25 | −0.0835 | 1.1923 | 0.8389 | −0.0023 | ±1.3797 | 1.2118 | 9.019 | × |
| 0.35 | −0.0659 | 1.2525 | 0.7994 | −0.0198 | ±1.5162 | 1.3053 | 9.4148 | × |
| 0.45 | −0.054 | 1.305 | 0.7675 | −0.0287 | ±1.6425 | 1.4007 | 9.7946 | 3.5892 |
| 0.55 | −0.0455 | 1.3517 | 0.7409 | −0.0334 | ±1.7601 | 1.4973 | 10.1586 | 3.86 |

**3**. $\varphi = \frac{\pi}{4}$

| $\eta\diagdown$ | $a$ | $b$ | $c$ | $d$ | $r$ | $S$ | E(large kink) | E(bubble) |
|---|---|---|---|---|---|---|---|---|
| −0.25 | −0.4667 | 0.8361 | 0.7122 | 0.8667 | ±1.2696 | 1.1009 | 8.52 | × |
| −0.15 | −0.461 | 0.8285 | 0.7977 | 0.7357 | ±1.1254 | 1.0409 | 8.2206 | × |
| 0 | −0.3827 | 0.9239 | 0.9239 | 0.3827 | ±1 | 1 | 8 | × |
| 0.15 | −0.2381 | 1.106 | 0.89 | 0.0656 | ±1.1254 | 1.0409 | 8.2206 | × |
| 0.25 | −0.1736 | 1.2041 | 0.8336 | −0.0216 | ±1.2696 | 1.1009 | 8.52 | × |
| 0.35 | −0.133 | 1.2828 | 0.7858 | −0.0601 | ±1.4132 | 1.1746 | 8.8607 | × |
| 0.45 | −0.1063 | 1.3485 | 0.7476 | −0.0772 | ±1.548 | 1.2563 | 9.2124 | × |
| 0.55 | −0.0878 | 1.4053 | 0.7169 | −0.0846 | ±1.6734 | 1.3428 | 9.563 | 6.254 |

**4**. $\varphi = \frac{3\pi}{8}$

| $\eta$＼ | $a$ | $b$ | $c$ | $d$ | $r$ | $S$ | E(large kink) | E(bubble) |
|---|---|---|---|---|---|---|---|---|
| −0.25 | −0.5342 | 0.8414 | 0.6522 | 0.8447 | ±1.3797 | 1.2118 | 9.019 | × |
| −0.15 | −0.5498 | 0.8167 | 0.7071 | 0.7686 | ±1.2316 | 1.1215 | 8.609 | × |
| 0 | −0.5556 | 0.8315 | 0.8315 | 0.5556 | ±1 | 1 | 8 | × |
| 0.15 | −0.4265 | 1.0509 | 0.9073 | 0.1091 | ±0.9187 | 0.9269 | 7.6103 | × |
| 0.25 | −0.289 | 1.2393 | 0.8303 | −0.1003 | ±1.0501 | 0.9309 | 7.6508 | × |
| 0.35 | −0.2021 | 1.3768 | 0.7516 | −0.1721 | ±1.2208 | 0.9684 | 7.8697 | × |
| 0.45 | −0.1515 | 1.4798 | 0.6953 | −0.1905 | ±1.3818 | 1.0266 | 8.1692 | 8.6583 |
| 0.55 | −0.1198 | 1.5626 | 0.6546 | −0.1911 | ±1.5275 | 1.0971 | 8.5004 | 8.4736 |

**5**. $\varphi = \frac{\pi}{2}$

| $\eta$＼ | $a$ | $b$ | $c$ | $d$ | $r$ | $S$ | E(large kink) | E(bubble) |
|---|---|---|---|---|---|---|---|---|
| −0.25 | −0.5946 | 0.8409 | 0.5946 | 0.8409 | ±1.4142 | 1.25 | 9.1824 | × |
| −0.15 | −0.6287 | 0.7953 | 0.6287 | 0.7953 | ±1.2649 | 1.15 | 8.74 | × |
| 0 | −0.7071 | 0.7071 | 0.7071 | 0.7071 | ±1 | 1 | 8 | × |
| 0.15 | −0.8891 | 0.5623 | 0.8891 | 0.5623 | ±0.6325 | 0.85 | 7.1055 | × |
| 0.25 | × | × | × | × | × | 0.75 | $2\pi$ | × |
| 0.35 | × | × | × | × | × | 0.7071 | 5.6561 | × |
| 0.45 | × | × | × | × | × | 0.7278 | 5.4508 | × |
| 0.55 | × | × | × | × | × | 0.7773 | 5.3966 | × |

From **Table 5.1**, I can plot two figures.

In **Fig**. **12**, I show how large kink energy varing with different $\eta$, especially, when $\eta < 0$, the energy of large kink will increase with $\varphi$ within $\varphi \in [0, \frac{\pi}{2}]$, but when $\eta > 0$, the energy of large kink will decrease with $\varphi$ within $\varphi \in [0, \frac{\pi}{2}]$. And noted in $\varphi = \frac{\pi}{2}$, we do not include the small kink energy.

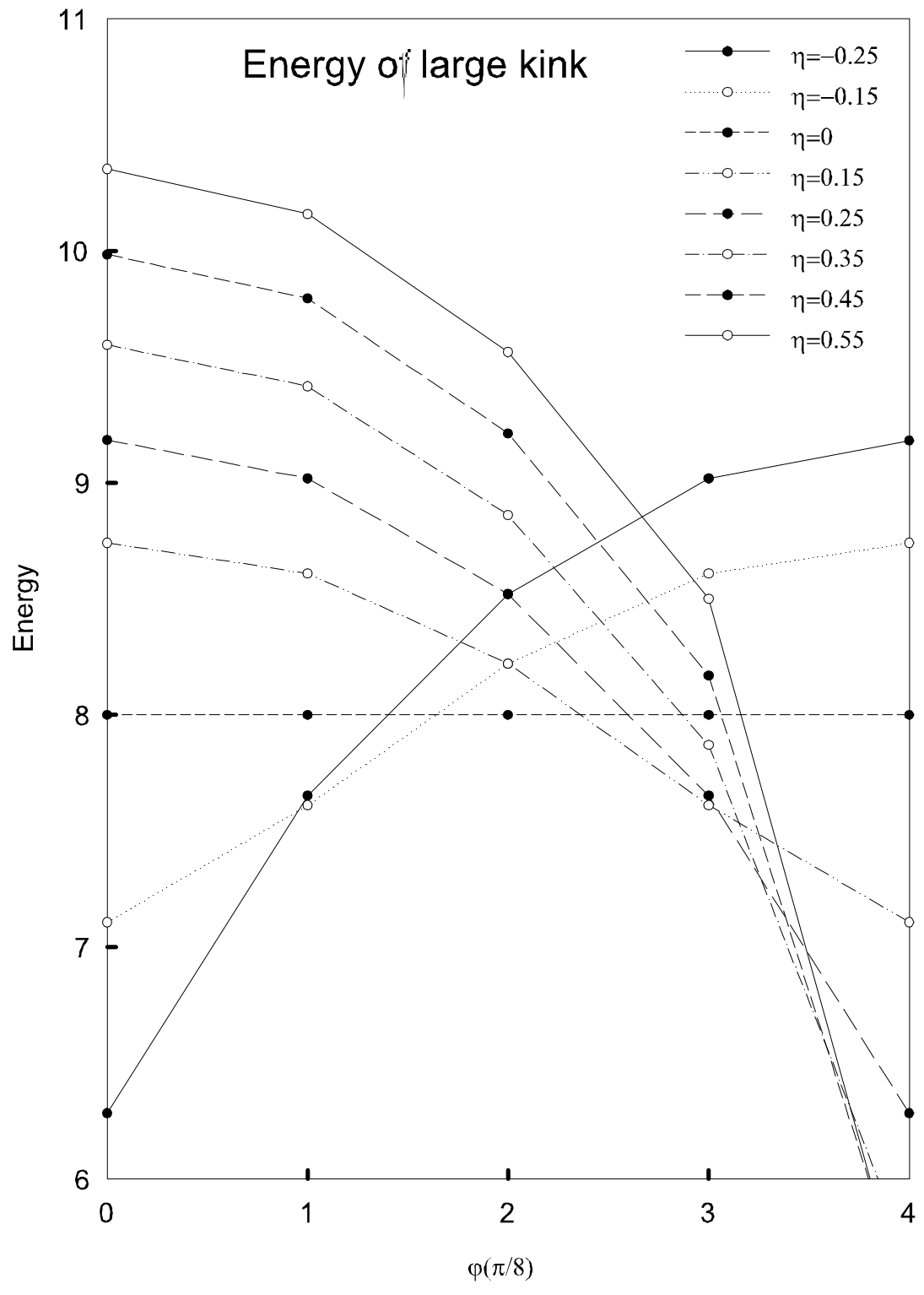

Fig. 12: Energy of large kink for different $\eta$ and $\varphi$.

In **Fig**. **13**, I show the energy crossing between large kink and bubble in some $\eta$, because $\eta$ has lower bound to appear bubble energy, See the phase diagram, **Fig**. **15**. In **Fig**.**13**, we can see energy crossing between two states with the same physical boundary condition. So I conclude that in some $\eta$, we have the ground state degenerate phenomena, like the phenomena of edge state case in **Chapter 4**.

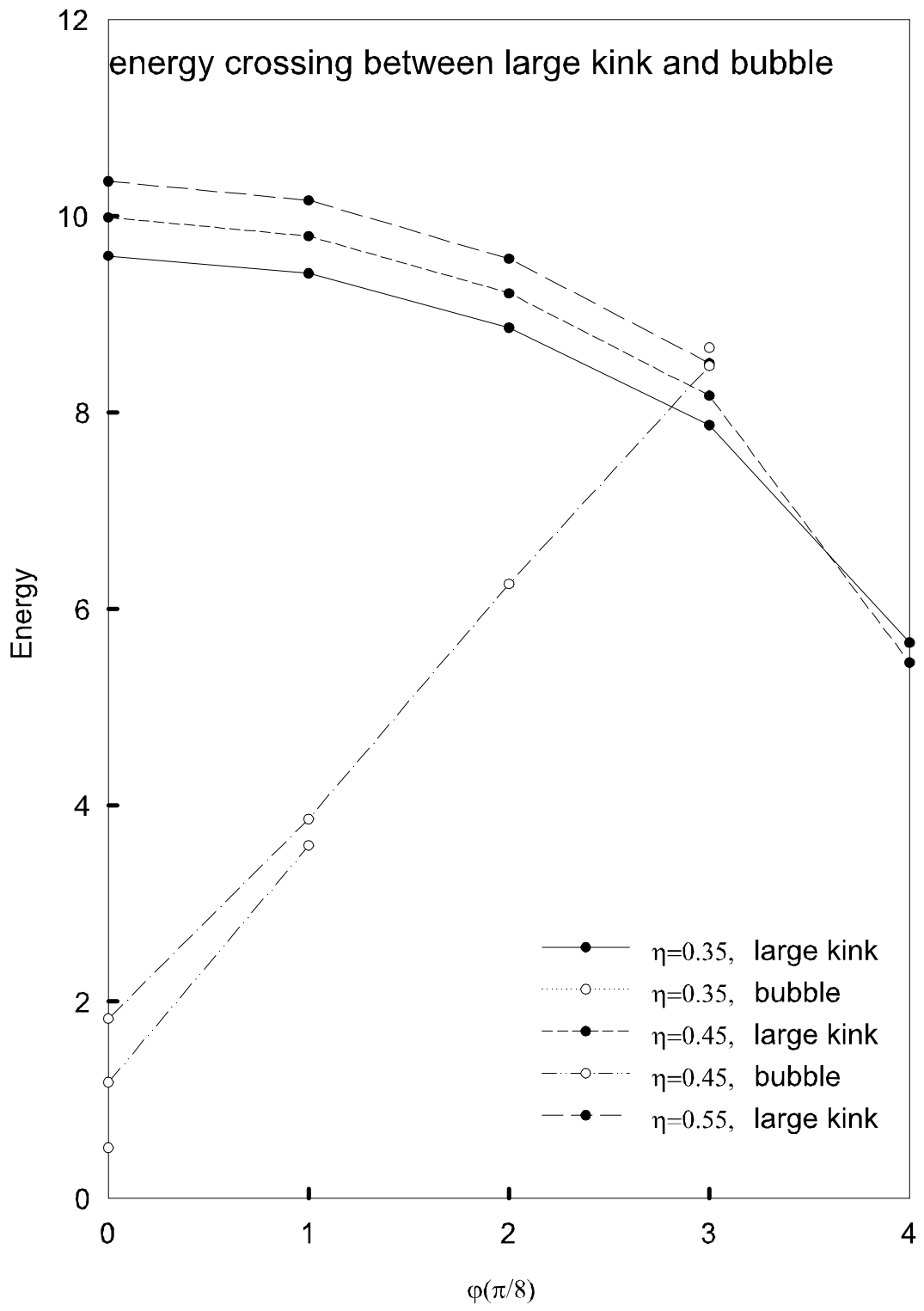

Fig.13. energy crossing between large kink and bubble.

### 5.3.4 Particular problem occur in $\eta > \frac{1}{4}$:

In this I want to explain why I can't find real solutions of **Eq**. (**5**.**3**.**3**.**2**) at $\varphi = \frac{\pi}{2}$ and $\eta > \frac{1}{4}$. If $\pi > \varphi > \pi/2$, then from **Eq**.(**5**.**3**.**3**.**2**), we have following symmetry:

| $\varphi$ | $\to$ | $\pi - \varphi$ |
|---|---|---|
| $a$ | $\to$ | $-c$ |
| $b$ | $\to$ | $d$ |
| $c$ | $\to$ | $-a$ |
| $d$ | $\to$ | $b$ |
| $r$ | $\to$ | $r$ |
| $S$ | $\to$ | $S$ |

5.3.4.1

But $r$ and $E$ are the same at $\varphi$ and $\pi - \varphi$. This symmetry also happen at other hypertriangle and Jacobi elliptic functions (used for finite system). We suppose we have only one sets of datas in each real solutions. This assuming is right in infinite system, but wrong in finite system. Suppose we have real solution of **Eq**. (**5**.**3**.**3**.**2**) at $\varphi = \frac{\pi}{2}$ and $\eta > \frac{1}{4}$, then the real solutions must satisfy $a = -c$ ; $b = d$ at $\varphi = \pi/2$ by continuous property of parameter $a, b, c, d$. ( i.e. if $a \neq -c$ or $b \neq d$ at $\varphi = \frac{\pi}{2}$, then they violate continuous!) Put these into **Eq**. (**5**.**3**.**3**.**2**), we get:

$$a^4(S + 1 - \eta) = 0 \tag{5.3.4.2a}$$

$$a^2 b^2 = \frac{r^2}{4S - 12\eta} \tag{5.3.4.2b}$$

$$b^4(S+1+\eta) = \frac{r^2}{2} \qquad 5.3.4.2c$$

Also the scaling restriction Eq.:

$$4a^2b^2 = 1 \qquad 5.3.4.2d$$

**Eq**. (**5**.**3**.**4**.**2d**) means $a \neq 0$, so **Eq**. (**5**.**3**.**4**.**2a**) imply $S+\eta = 1$. Put these into **Eq**. (**5**.**3**.**4**.**2b**,**c**), we get the equation:

$$\frac{r^2}{1-4\eta} = 1 \qquad 5.3.4.3$$

So we get contradiction from **Eq**. (**5**.**3**.**4**.**3**) , if $\eta > \frac{1}{4}$, which means we don't have such real solutions satisfy $a = -c, b = d$ in $\varphi = \pi/2$ and $\eta > \frac{1}{4}$.

For complete, we also list equations suit for $\eta > \frac{1}{4}$ and $\varphi = \frac{\pi}{2}$. But we will not use $f(s) = \frac{a\sinh(rs)+b}{c\sinh(rs)+d}$, instead we use

$$f(s) = \frac{a\sinh(rs)+b\cosh(rs)}{c\sinh(rs)+d\cosh(rs)} \qquad 5.3.4.4$$

due to the easy proof above. The equations are:

$$(ab-cd)^2 = 1 \qquad 5.3.4.5a$$

$$(a^2+c^2)^2S + (a^4-c^4)\cos\varphi + 2ac(a^2+c^2)\sin\varphi + (-a^4+6a^2c^2-c^4)\eta = 2r^2 \qquad 5.3.4.5b$$

$$\begin{aligned}&4(ab+cd)(a^2+c^2)S + 4(a^3b-c^3d)\cos\varphi \\ &+ 2(a^3d+3a^2bc+3ac^2d+bc^3)\sin\varphi \\ &+ [-4a^3b+6(2a^2cd+2abc^2)-4c^3d]\eta = 0\end{aligned} \qquad 5.3.4.5c$$

$$\begin{aligned}&[6a^2b^2+2(a^2d^2+4abcd+b^2c^2)+6c^2d^2]S \\ &+ (6a^2b^2-6c^2d^2)\cos\varphi \\ &+ 6(a^2bd+ab^2c+acd^2+bc^2d)\sin\varphi \\ &+ 6(-a^2b^2+a^2d^2+4abcd+b^2c^2-c^2d^2)\eta = -4r^2\end{aligned} \qquad 5.3.4.5d$$

$$\begin{aligned}&4(ab+cd)(b^2+d^2)S + 4(b^3a-d^3c)\cos\varphi + \\ &2(b^3c+3b^2ad+3bd^2c+ad^3)\sin\varphi + \\ &[-4b^3a+6(2d^2ab+2cdb^2)-4d^3c]\eta = 0\end{aligned} \qquad 5.3.4.5e$$

$$(b^2+d^2)^2S + (b^4-d^4)\cos\varphi + 2bd(b^2+d^2)\sin\varphi + (-b^4+6b^2d^2-d^4)\eta = 2r^2 \qquad 5.3.4.5f$$

Notice the only different are right hand side of **Eq**. (**5**.**3**.**4**.**5b**), (**5**.**3**.**4**.**5d**) are different with those in **Eq**.(**5**.**3**.**3**.**2b**), (**5**.**3**.**3**.**2d**).

We plot the kink solutions in **Fig**. **10** for $\eta = 0.15$ and $\varphi = 0\sim 1.75\pi$. The solutions at $\eta = 0.35$ and $\varphi = 0\sim 0.5\pi$ and $\varphi = 0.5\pi\sim\pi$ are shown in **Fig**. **11** (**a**) and **Fig**. **11** (**b**) respectively. The kinks for $\eta < 1/4$ ( **Fig**. **10**) change smoothly and retain their shapes as $\varphi$ moves across $\pi/2$. The total change of $\theta$, $\Delta\theta = \theta(s=\infty) - \theta(s=-\infty)$, is equal to $2\pi$. This can also be deduced from **eq**. (**5**.**3**.**3**.**1**) by tracing the variation of $\theta$ with respect to $s$. On the other hand, the solutions for $\eta > 1/4$ (**Fig**. **3a and 3b**) develop a second kink as $\varphi$ approaches $\pi/2$. At $\varphi = \pi/2$, the form in **Eq**. (**5**.**3**.**3**.**1**) is no longer applicable. **Eq**. (**5**.**3**.**4**.**4**) has to be used and it gives a large kink and a small kinks which are the decedents of the connected kinks at $\varphi = \pi/2 - \varepsilon$ where $\varepsilon$ is an infinitesimal positive number. Note also that the total change of $\theta$ of neither the large kink nor the small kink is equal to $2\pi$, but rather the sum of them is. This can also be seen from **Eq**. (**5**.**3**.**4**.**4**).

We here propose a classical explanation. It is related to the minima of the potential. As we argued in section 2, $d\theta/ds|_{s=\pm\infty} = 0$ and $V(\theta(\pm\infty)) = V_{\min}$. The solutions in **Eqs**. (**5**.**3**.**3**.**1**) **and** (**5**.**3**.**4**.**4**) extend from one minimum to another. The large kink coming from **eq**. (**5**.**3**.**3**.**1**) start from one absolute minimum of the potential, passing through a major peak and ends at another absolute minimum. In the special case with the solutions coming from **Eq**. (**5**.**3**.**4**.**4**), there are a large kink and a small kink. Both connect two absolute minima but the former passes a major peak of the potential and the latter passes a minor peak. The large kink extends from $\theta = 2\tan^{-1}(b/d)$ to $\theta = 2\pi - 2\tan^{-1}(b/d)$ and the small kink extend from $\theta = -2\tan^{-1}(a/c)$ to $\theta = 2\tan^{-1}(a/c)$. However, the form we proposed in **Eq**. (**5**.**3**.**3**.**1**) has set $\Delta\theta = 2\pi$ *in prior*. Hence, the form in **eq**. (**5**.**3**.**4**.**4**) has to be used. We will elaborate more on this in next section.

Finally, we give the form of the bubble solution. Instead of **eq**. (**5**.**3**.**3**.**1**), we use

$$f(s) = \frac{a\cosh(rs) + b}{c\cosh(rs) + d} \qquad 5.3.4.6$$

and follow the same procedure, we are able to obtain the bubble solution, similar to that in **eq**. (**5**.**2**.**1**.**8**). Its shape is shown in **Fig**. **14**. by the dashed line which connects two relative minima of potential. The bubble solution can be found only when the relative minima exist.

In the proof that we have no solution of **eq**. (**5**.**3**.**3**.**1**) in certain region, we can see the pure mathematical reason in **eq**. (**5**.**3**.**4**.**2**) **and** (**5**.**3**.**4**.**3**). But why is that in physical reason? I have one classical viewpoint. See the potential ( i.e. see the potential plot in **fig**. **9** ). Because if we start from $\eta > \frac{1}{4}$ and $\varphi = 0$, when $\varphi$ goes to $\frac{\pi}{2}$, the potential deform and the shape like one in $\eta < \frac{-1}{4}$ and $\varphi = 0$, so we can see in **Ref**. [**48**] and the reason of §**5**.**4**.**1**. In this value of $\varphi$, we must have *tanh* and *coth* solution ( i.e. both of them are not $2\pi$-kink) instead of *csch* solution ( i.e. $2\pi$-kink). At $\varphi = \frac{\pi}{2}$, it is the special phase for $\eta > \frac{1}{4}$ because interval of minimum points of the potential change from $2\pi$ to $2\phi_0$ and $2\pi - 2\phi_0$ as we can see **Eq**. (**5**.**2**.**2**.**11**) and §**5**.**2**.**1**. The **Möbius transformation** of sinh also keep $\theta(s = \infty) - \theta(s = -\infty) = 2\pi$ at all $\varphi$, but not for tanh and coth. So At $\varphi = \frac{\pi}{2}$ and $\frac{3\pi}{2}$, it will contradion if we have real solutions of **Möbius transformation** of sinh at $\eta > \frac{1}{4}$. Also *tanh* and *coth* solution can't deform by **Möbius transform** from *csch* solution (i.e. or sinh solution).

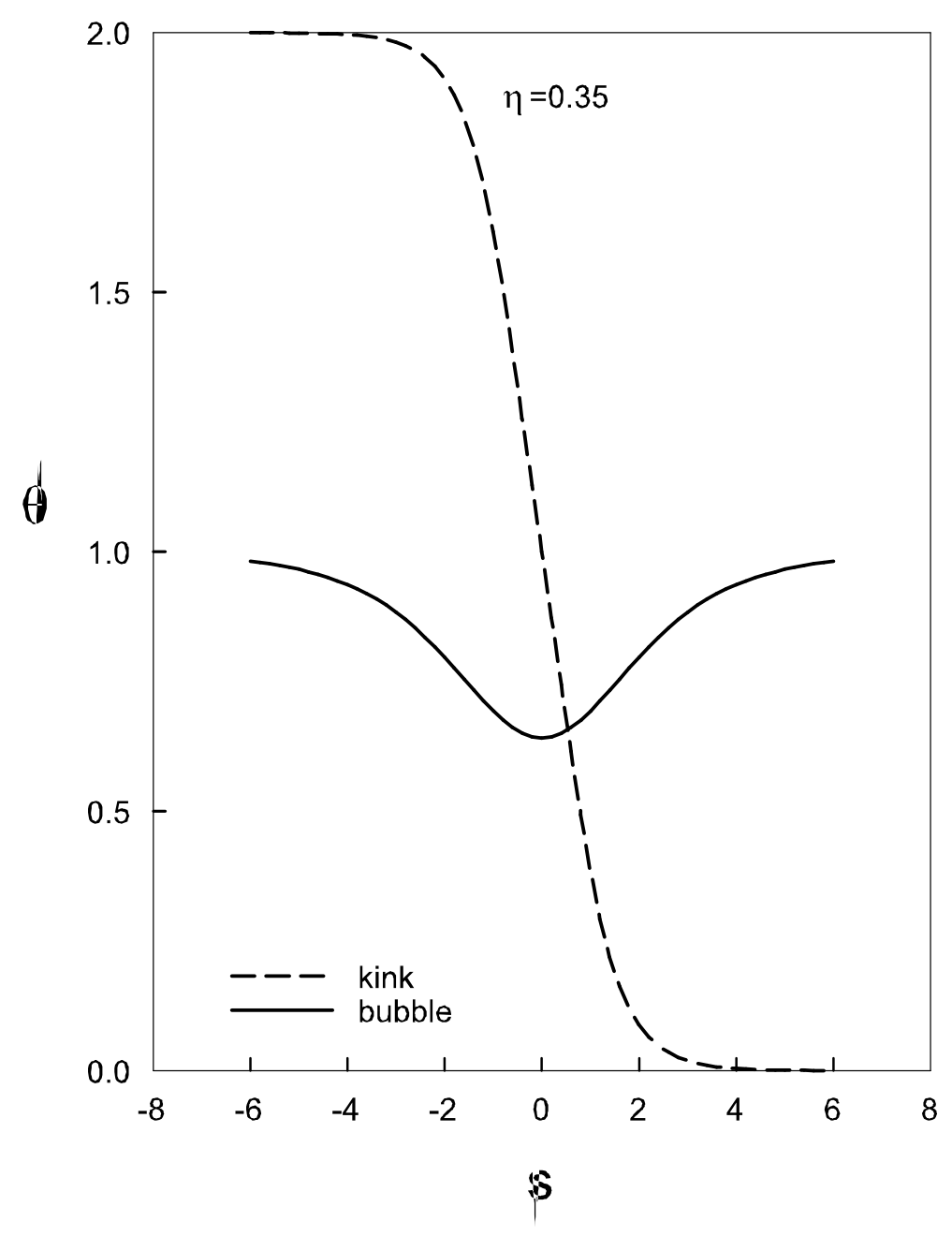

**Fig 14**: . Solutions for $\eta = 0.35$ and $\varphi = 0$. It is the bubble and large kink solution. Notice the unit of $\theta - axis$ is $\pi$.

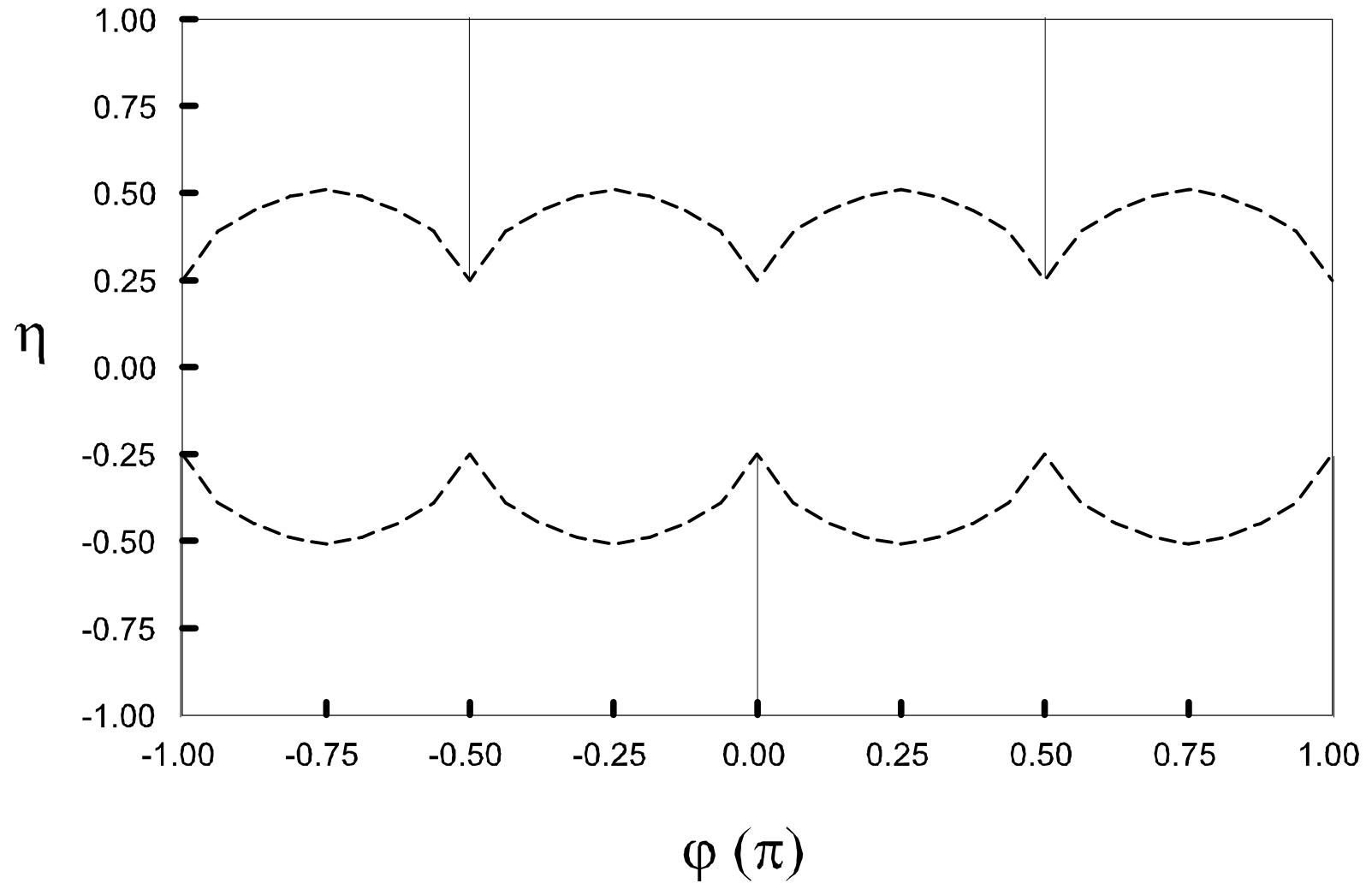

Fig. 15: $\varphi - \eta$ phase diagram

## 5.3.5 Another method to solve asymmetric Double Sine-Gordon Equation:

In this section, I give another method to solve the asymmetric **DSGE**. I want to solve the equation

$$\frac{d^2\theta}{d^2s} = \sin(\theta+\varphi) + 2\eta\sin(2\theta) \qquad 5.3.5.1$$

Assume the solution has the form $\theta = 2\tan^{-1}[h(s)]$, we get

$$2(1+h^2)h'' - 4h(h')^2 = 2h(1+h^2)\cos(\varphi) + (1-h^4)\sin(\varphi) + 8\eta h(1-h^2) \qquad 5.3.5.2$$

Let

$$h = \frac{ay+b}{cy+d} \qquad 5.3.5.3$$

and $y$ satisfy the equation:

$$(y')^2 = ey^4 + fy^2 + g \qquad 5.3.5.4$$

where $e, f, g$ are unknown parameters. It can also be written as

$$y'' = 2ey^3 + fy \qquad 5.3.5.5$$

We will transform the differential equation of $h$ in **Eq**. (**5.3.5.2**) into algebraic equations of $y$ by inserting **eq**. (**5.3.5.3**)~(**5.3.5.5**) into **eq**. (**5.3.5.2**), together with the restriction of the scaling equation:

$$(ad-bc)^2 = 1 \qquad 5.3.5.6$$

Because

$$\frac{d}{ds}\left(\frac{ay+b}{cy+d}\right) = \frac{(ad-bc)}{(cy+d)^2}y' \qquad 5.3.5.7$$

$$\left[\frac{d}{ds}\left(\frac{ay+b}{cy+d}\right)\right]^2 = \frac{ey^4+fy^2+g}{(cy+d)^4} \qquad 5.3.5.8$$

and

$$\frac{d^2}{ds^2}\left(\frac{ay+b}{cy+d}\right) = \frac{(ad-bc)(2dey^3 - fcy^2 + fdy - 2cg)}{(cy+d)^3} \qquad 5.3.5.9$$

we get a polynomial equation of $y$

$$\begin{aligned}
&2(ad-bc)[(a^2+c^2)y^2 + 2(ab+cd)y + (b^2+d^2)](2dey^3 - fcy^2 + fdy - 2cg) \qquad 5.3.5.10\\
&\ -4(ay+b)(ey^4+fy^2+g)\\
=\ &2[ac^2y^3 + (2acd+bc^2)y^2 + (ad^2+2bcd)y + bd^2]\\
&[(a^2+c^2)y^2 + 2(ab+cd)y + (b^2+d^2)]\cos\varphi\\
&\ +(cy+d)[(c^4-a^4)y^4 + 4(c^3d-a^3b)y^3 +\\
&6(c^2d^2-a^2b^2)y^2 + 4(cd^3-ab^3)y + (d^4-b^4)]\sin\varphi\\
&\ +8\eta[ac^3y^3 + (2acd+bc^2)y^2 + (ad^2+2bcd)y + bd^2]\\
&[(c^2-a^2)y^2 + 2(cd-ab)y + (d^2-b^2)]
\end{aligned}$$

Comparing the power series of $y$, we get following equations:

$$4(ad-bc)(a^2+c^2)de-4ae = 2ac^2(a^2+c^2)\cos\varphi + c(c^4-a^4)\sin\varphi + 8\eta ac^2(c^2-a^2) \quad 5.3.5.11a$$

$$\begin{aligned} &2(ad-bc)[-fc(a^2+c^2)+4de(ab+cd)] \ -4be \\ &= 2[2ac^2(ab+cd)+(2acd+bc^2)(a^2+c^2)]\cos\varphi \\ &\quad + [4c(c^3d-a^3b)+d(c^4-a^4)]\sin\varphi \\ &\quad + 8\eta[2ac^2(cd-ab)+(c^2-a^2)(2acd+bc^2)] \end{aligned} \quad 5.3.5.11b$$

$$\begin{aligned} &2(ad-bc)[2de(b^2+d^2)+fd(a^2+c^2)-2fc(ab+cd)] \ -4af \\ &= 2[ac^2(b^2+d^2)+2(ab+cd)(2acd+bc^2)+(a^2+c^2)(ad^2+2bcd)]\cos\varphi \\ &\quad + [6c(c^2d^2-a^2b^2)+4d(c^3d-a^3b)]\sin\varphi \\ &\quad + 8\eta[ac^2(d^2-b^2)+2(cd-ab)(2acd+bc^2)+(c^2-a^2)(ad^2+2bcd)] \end{aligned} \quad 5.3.5.11c$$

$$\begin{aligned} &2(ad-bc)[-2cg(a^2+c^2)+2fd(ab+cd)-fc(b^2+d^2)] \ -4bf \\ &= 2[(2acd+bc^2)(b^2+d^2)+2(ad^2+2bcd)(ab+cd)+bd^2(a^2+c^2)]\cos\varphi \\ &\quad + [4c(cd^3-ab^3)+6d(c^2d^2-a^2b^2)]\sin\varphi \\ &\quad + 8\eta[(2acd+bc^2)(d^2-b^2)+2(ad^2+2bcd)(cd-ab)+bd^2(c^2-a^2)] \end{aligned} \quad 5.3.5.11d$$

$$\begin{aligned} &2(ad-bc)[-4cg(ab+cd)+fd(b^2+d^2)] \ -4ag \\ &= 2[(ad^2+2bcd)(b^2+d^2)+2bd^2(ab+cd)]\cos\varphi \\ &\quad + [c(d^4-b^4)+4d(cd^3-ab^3)]\sin\varphi \\ &\quad + 8\eta[(ad^2+2bcd)(d^2-b^2)+2bd^2(cd-ab)] \end{aligned} \quad 5.3.5.11e$$

$$-4cg(ad-bc)(b^2+d^2) = 2bd^2(b^2+d^2)\cos\varphi + d(d^4-b^4)\sin\varphi + 8\eta bd^2(d^2-b^2) \quad 5.3.5.11f$$

So we have **eqs**. (**5**.**3**.**5**.**6**) and (**5**.**3**.**5**.**11**). Seven equations to solve $a, b, c, d, e, f, g$, seven unknown parameters. The difference is the appearance of $e, f, g$ in **eq**. (**5**.**3**.**5**.**4**). The function $y$ can be some **JEF**, then there are certain relations between $e, f, g$. We rescale $y$ and $s$

$$y(s) = A \cdot z(\beta s) \quad 5.3.5.12$$

so that

$$(y')^2 = A^2\beta^2(z')^2 = eA^4z^4 + fA^2z^2 + g \Rightarrow (z')^2 = \frac{1}{\beta^2}(eA^2z^4 + fz^2 + \frac{1}{A^2}g) \quad 5.3.5.13$$

Having obtained $e, f, g$, we can choose suitable $A$ and $\beta$, so that $z(\beta s)$ is one of standard **JEF**s. Note that different **JEF**s satisfy different differential equations. This implies that only certain **JEF**s can satisfy **eq**. (**5**.**3**.**5**.**4**).

## 5.3.6 Discuss and Conclusion of DSGE in an infinite system

We summarize our results on the $\varphi - \eta$ phase diagram in **Fig**. **15**. The kink solution in **eq**. (**5**.**3**.**3**.**1**) can be found in any place except for those vertical lines. On the vertical lines, the form in **eq**. (**5**.**3**.**4**.**4**) prevails. Note that it corresponds to either $g(s) = \coth(rs)$ or $g(s) = \tanh(rs)$, that respectively, corresponds to large kink or small kink. The bubble solution exist in the region above the upper dashed line or below the lower dashed line.

It had been argued that there is a second order phase transition at $\eta = \eta_c$ for a certain value of $\varphi$. The argument is simple an convincing. In the limit of $\eta \to 0$, the solutions should be those of an ordinary sine-Gordon equation: $\theta = 4\tan^{-1}[\exp(s)] \ = 2\tan^{-1}[\csc h(s)]$. It is compatible with the form in **eq**. (**5**.**3**.**3**.**1**). In the other limit of large $\eta$, one can write **eq**. (**5**.**2**.**1**.**1**) as

$$\frac{1}{4\eta}\frac{\partial^2\theta'}{\partial t^2} - \frac{1}{4\eta}\frac{\partial^2\theta}{\partial x^2} + \frac{1}{2\eta}\sin(\theta'/2+\varphi) + \sin\theta' = 0. \qquad 5.3.6.1$$

where $\theta' = 2\theta$. The third term can be treated as a perturbation. The zeroth order of the solutions should be those of sine-Gordon equation with $\sin(2\theta)$ :

$$\theta = 2\tan^{-1}[\exp(s')] \qquad 5.3.6.2$$

where $s' = 2\sqrt{\eta}\, s$, which implies our $r = 2\sqrt{\eta}$ in **eq**. (**5**.**3**.**3**.**1**). Hence, there must be a phase transition in between. However, the **Möbius transformation** shows differently.

Consider only the leading order terms of **eqs**. (**5**.**3**.**3**.**2**)

$$(a^2+c^2)^2 S + (-a^4+6a^2c^2-c^4)\eta = 0 \qquad 5.3.6.3a$$

$$4(ab+cd)(a^2+c^2)S + [-4a^3b + 6(2a^2cd+2abc^2) - 4c^3d]\eta = 0 \qquad 5.3.6.3b$$

$$[6a^2b^2 + 2(a^2d^2+4abcd+b^2c^2) + 6c^2d^2]S + 6(-a^2b^2+a^2d^2+4abcd+b^2c^2-c^2d^2)\eta = 2r^2 \qquad 5.3.6.3c$$

$$4(ab+cd)(b^2+d^2)S + [-4b^3a + 6(2d^2ab+2cdb^2) - 4d^3c]\eta = 0 \qquad 5.3.6.3d$$

$$(b^2+d^2)^2 S + (-b^4+6b^2d^2-d^4)\eta = 2r^2 \qquad 5.3.6.3e$$

It can be shown easily that none of $a$, $b$, $c$ and $d$ can be 0. Hence we can set $a = pc$ and $b = qd$ where $p$ and $q$ are just two ratio parameters. It can be shown further that $q = -p = \pm 1$ and $|cd| = 1/\sqrt{2}$. Now let $S = -\eta + \delta$ where $\delta = O(\eta^0)$. Substituting into **eq**. (**5**.**3**.**6**.**3e**), one find that $\delta = 2r^2/d^4$. This implies that $b = \pm d = O(\sqrt[4]{\eta})$ and $a = \pm c = O(1/\sqrt[4]{\eta})$. We thus have shown that solutions of the form of **eq**. (**5**.**3**.**3**.**1**) always exist and their parameters $a$, $b$, $c$ and $d$ vary smoothly with $\eta$. There is no phase transition for finite $\eta$ if $\varphi \neq 0$ or $\varphi \neq \pi/2$.

There is another aspect we would like to investigate and that is varying $\varphi$ across $\varphi = \pi/2$ for a fixed $\eta$. We plot in **Fig**. **16** (**a**) **and Fig**. **16** (**b**) the energy as a function of $\varphi$ with $\eta = 0.15$ and $\eta = 0.35$ respectively. One can immediately notice the behavior of energy near $\varphi = \pi/2 + n\pi$. For $\eta = 0.15$, the slope change is large but still smooth. For $\eta = 0.35$, though the energy remains continuous, the slope is not. This indicates that when $|\eta| < 1/4$ there is smooth crossover. But when $|\eta| > 1/4$ there is a second-order phase transition from the energy point of view. The solutions also show different behavior. In Fig. 2, the kinks vary smoothly across the point $\varphi = \pi/2$ at $\eta = 0.15$. For the kinks in **Fig**. **9** (**a**) with $\eta = 0.35$ one finds that near $\varphi \approx \pi/2$, the shapes of the solutions are different. The kinks nearly break into two, but remember their forms are still that of **eq**. (**5**.**3**.**3**.**1**), i. e. , $g(x) = \sinh(rx)$, so the whole range of $\theta$ is still $2\pi$. When $\varphi = \pi/2$ the form of the kinks is that of **eq**. (**5**.**3**.**4**.**4**), i. e., $g(x) = \tanh(rx)$ or $g(x) = \coth(rs)$, the large and small kink solutions. On the other hand, the solutions at $\varphi = \pi/2 \pm \varepsilon$ where $\varepsilon$ is a infinitesimal number, are very similar. Our numerical results also show that the coefficients $a$, $b$, $c$, $d$ and $E$ vary smoothly across the point $\varphi = \pi/2$, not including the point $\varphi = \pi/2$. Thus, the point $\varphi = \pi/2$ is actually a singular point.

In this work, we used the method of " **Möbius transformation**" to solve the **DSGE** with an additional phase $\varphi$. This method transformed the **DSGE** into a set of algebraic equations and solutions can be found for all values of $\varphi$ and $\eta$ which is the ratio between the coefficients of hyperbolic functions. Thus we are able to find the forms of the solution for all the region on the $\varphi - \eta$ plane. We have found there is there is second-order phase transition at certain value of $\eta$ and the singular points at $\varphi = \pi/2 + n\pi$. The resulting forms of our solutions can serve as the basis of various methods, such as form factor perturbation theory, semi-classical method or truncated conformal space approach to study quantum fluctuation. See text for explanation

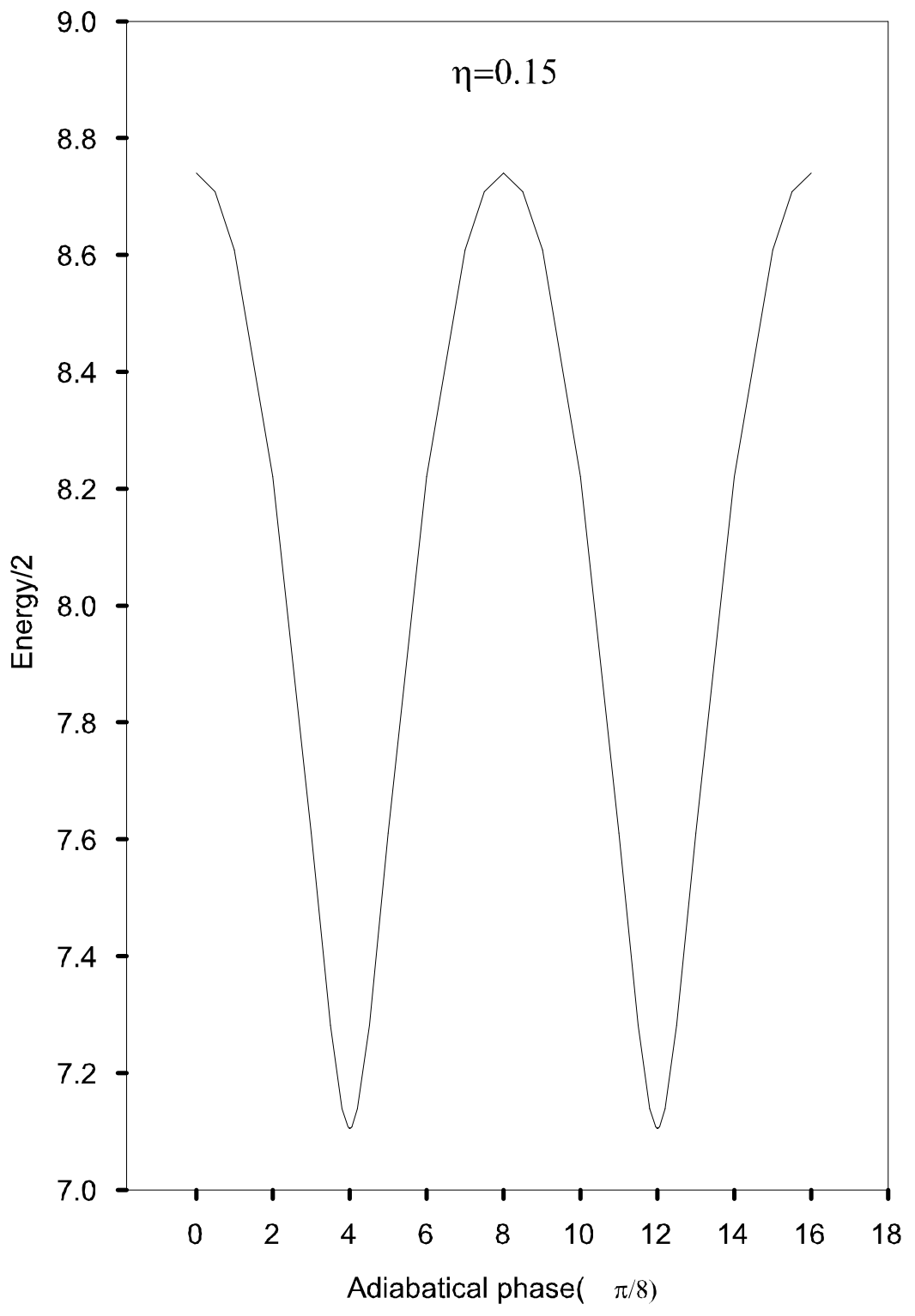

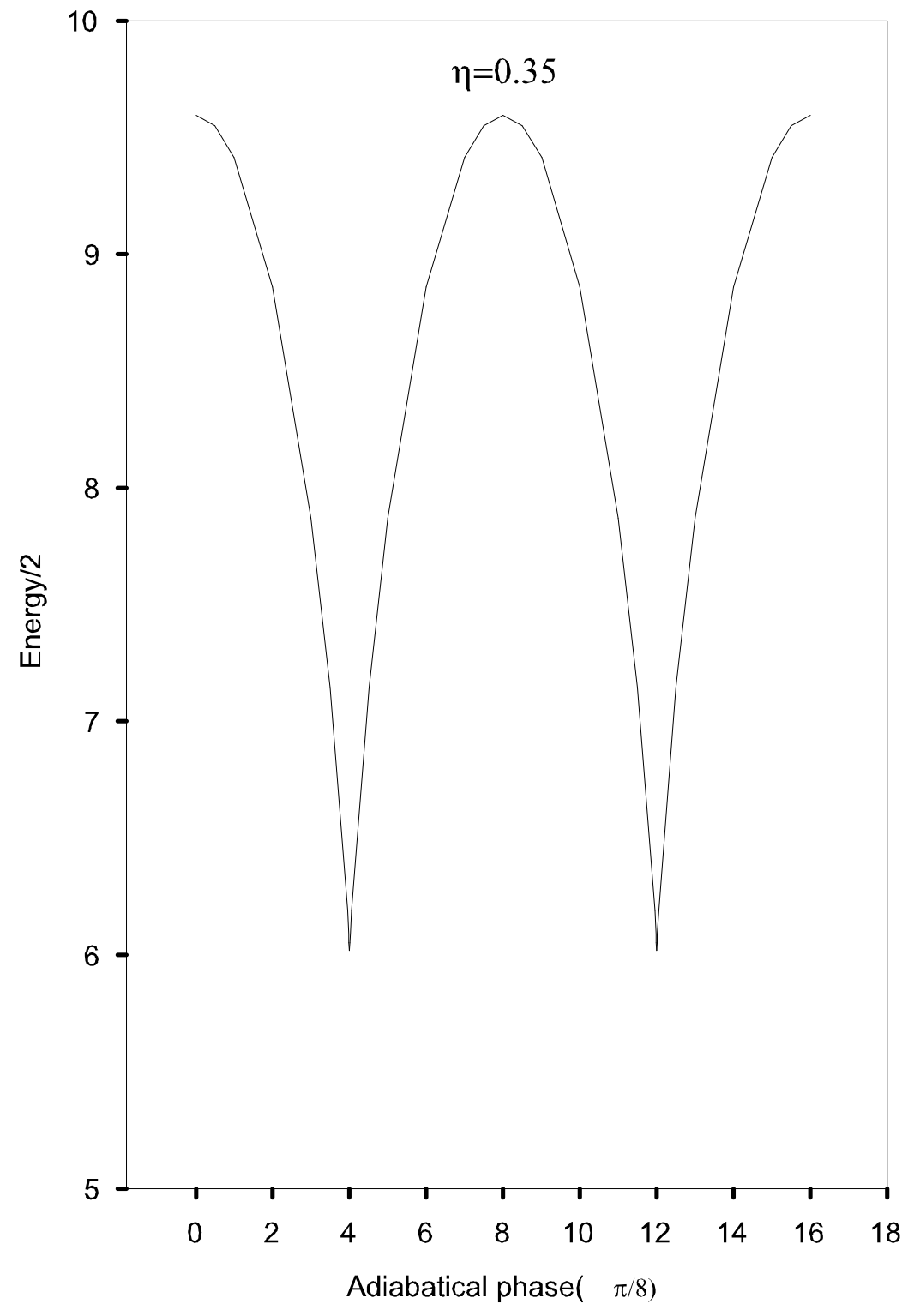

**Fig**. **16**: The energy of the kink versus $\varphi$; **16** (**a**): $\eta = 0.15$; **16** (**b**): $\eta = 0.35$

# 5.4 asymmetric Double Sine-Gordon Equation in finite system:

## 5.4.1 Equations of asymmetric DSGE in a finite system

We have shown that for a finite system, the solution of **SGE** can be of the form $4\tan^{-1}[A_{th}sc(\beta z-\beta z_0)]$. However, for a **DSGE**, this form cannot be the solution. On the other hand, we have seen in §**5.3**, the **Möbius transformation** can be used to solve **DSGE** of an infinite system. Since we know that the limiting forms of JEFs are hyperbolic functions, we are inspired to try to solve **DSGE** of a finite system with this transformation by **JEF**. Suppose we have a finite system where E.O.M. is asymmetric **DSGE**

$$\frac{d^2\theta}{d^2s} = \sin(\theta+\varphi)+2\eta\sin 2\theta \qquad 5.2.1.2$$

with boundary conditions

$$\theta(z=0) = 0 \qquad 5.4.1.1a$$

$$\theta(z=L) = 2\pi \qquad 5.4.1.1b$$

The phase $\varphi$ outside $4\tan^{-1}$ in static soliton of **SGE** we replaced by Möbius parameters: $a, b, c, d$. If we let

$$\theta(z) = 2\tan^{-1}[f(z)] \qquad 5.4.1.2$$

where

$$f(z) = \frac{a\cdot sc[r(z-z_0)]+b}{c\cdot sc[r(z-z_0)]+d} \qquad 5.4.1.3$$

we get

$$(\frac{df}{dz})^2 = \frac{r^2(ad-bc)^2}{(c\cdot sc[r(z-z_0)]+d)^4}\{(1-k_f^2)sc^4[r(z-z_0)]+(2-k_f^2)sc^2[r(z-z_0)]+1\} \qquad 5.4.1.4$$

with the requirement of the scaling constraint

$$(ad-bc)^2 = 1 \qquad 5.4.1.5a$$

Substituting **eq**. (**5.4.1.3**) into **DSGE**, and comparing coefficients, we get the equations which are similar to **eqs**. (**5.3.3.2**)

$$(a^2+c^2)^2S+(a^4-c^4)\cos\varphi+2ac(a^2+c^2)\sin\varphi+(-a^4+6a^2c^2-c^4)\eta = 2r^2(1-k_f^2) \qquad 5.4.1.5b$$

$$\begin{aligned}&4(ab+cd)(a^2+c^2)S+4(a^3b-c^3d)\cos\varphi\\&+2(a^3d+3a^2bc+3ac^2d+bc^3)\sin\varphi\\&+[-4a^3b+6(2a^2cd+2abc^2)-4c^3d]\eta = 0\end{aligned} \qquad 5.4.1.5c$$

$$\begin{aligned}&[6a^2b^2+2(a^2d^2+4abcd+b^2c^2)+6c^2d^2]S+\\&(6a^2b^2-6c^2d^2)\cos\varphi+6(a^2bd+ab^2c+acd^2+bc^2d)\sin\varphi\\&+6(-a^2b^2+a^2d^2+4abcd+b^2c^2-c^2d^2)\eta\\&= 2r^2(2-k_f^2)\end{aligned} \qquad 5.4.1.5d$$

$$4(ab+cd)(b^2+d^2)S+4(b^3a-d^3c)\cos\varphi + 2(b^3c+3b^2ad+3bd^2c+ad^3)\sin\varphi + [-4b^3a+6(2d^2ab+2cdb^2)-4d^3c]\eta = 0 \quad 5.4.1.5e$$

$$(b^2+d^2)^2S+(b^4-d^4)\cos\varphi+2bd(b^2+d^2)\sin\varphi + (-b^4+6b^2d^2-d^4)\eta = 2r^2 \quad 5.4.1.5f$$

We also need to satisfy **eq**. (**5**.**4**.**1**.**1**)

$$rL = 2K(k_f) \quad 5.4.1.5g$$

where $K(k_f)$ is again the elliptic integral.of the first kind. $z_0$ can be calculate directly by **eq**. (**5**.**4**.**1**.**1a**):

$$\theta(z=0) = 0 = 2\tan^{-1}\frac{a\bullet sc[r(-z_0)]\ +b}{c\bullet sc[r(-z_0)]\ +d} \quad 5.4.1.6$$

Now we have $a,b,c,d,r,S,k_f$ seven unknowns, and we also have **eqs**. (**5**.**4**.**1**.**5**) to solve them. In **Table 5**.**1**, **5**.**3**, I list the roots of $L = 1$ and 10. **Table 5**.**2** shows the ” finite-length” properties in constract to the case of infinite length. The main property is that $k_f$ is much less than 1. So in $L = 1$ case, we can compare how different $a,b,c,d,r,S$ from those of $L = \infty$. The $\theta$ curve of $L = 1$ is almost a straight line, while that of L=10 is very similar to the case of $L = \infty$. This can be seen in **Table 5**.**3** in which $k_f$ is very close to **1**. In all, the numbers in **Table 5**.**3** are very close to those in **Table 5**.**1**. This indicates that $L = 10$ is very close to the infinite case. There are interesting things. We have seen in **eq**.(**5**.**2**.**1**.**4**)~(**5**.**2**.**1**.**8**), which are solutions of $L = \infty$, but the solutions is trival at $\eta = \frac{\pm 1}{4}$. Also if $\eta > \frac{1}{4}$, we can solve the solutions by **Möbius transformation** of sinh in all values of $\varphi$ except at $\varphi = \frac{\pi}{2}$. Because at $\varphi = \frac{\pi}{2}$, we have solutions like tanh and coth, which cannot be the **Möbius transformation** of sinh. But in $L = 10$ case, we have data both at $\eta = \frac{\pm 1}{4}$ and $\varphi = \frac{\pi}{2}$ by **Möbius transform** of *sc*. We can even see in **Figs**. **11** (**a**), (**b**) at $\varphi = \frac{\pi}{2}$ when $\eta > \frac{1}{4}$, the $\theta$-plot by my method seems to break into ”large” and ”small” kink, which is described as coth and tanh respectively when $L = \infty$. We can see it is very similar the figures at $\varphi \underset{\neq}{\rightarrow} \frac{\pi}{2}$ when $\eta > \frac{1}{4}$. Those again confirm my method.

**Table 5.2**: For $L = 1, \theta(z=0) = 0, \theta(z=L) = 2\pi$
List datas of $a, b, c, d, r, S, k_f^2, z_0$ for **Eq. (5.4.1.5)**: $f(s) = \frac{a \cdot sc[r(z-z_0)]+b}{c \cdot sc[r(z-z_0)]+d}$ at each $\varphi$

**1.** $\boldsymbol{\varphi = 0}$

| $\eta\backslash$ | $a$ | $b$ | $c$ | $d$ | $r$ | $S$ | $k_f^2$ | $z_0$ |
|---|---|---|---|---|---|---|---|---|
| 0 | 0 | 1 | 1 | 0 | 3.2217 | 19.7582 | 0.0963 | 0.5 |
| 0.15 | 0 | 1.0515 | 0.951 | 0 | 3.5495 | 19.7588 | 0.396 | 0.5 |
| 0.25 | 0 | 1.0703 | 0.9343 | 0 | 3.6684 | 19.7597 | 0.4759 | 0.5 |
| 0.35 | 0 | 1.0859 | 0.9209 | 0 | 3.7671 | 19.761 | 0.5335 | 0.5 |
| 0.45 | 0 | 1.0996 | 0.9094 | 0 | 3.8537 | 19.7626 | 0.5783 | 0.5 |

**2.** $\varphi = \frac{\pi}{8}$

| $\eta\backslash$ | $a$ | $b$ | $c$ | $d$ | $r$ | $S$ | $k_f^2$ | $z_0$ |
|---|---|---|---|---|---|---|---|---|
| 0 | −0.1951 | 0.9808 | 0.9808 | 0.1951 | 3.2216 | 19.7582 | 0.0963 | 0.5641 |
| 0.15 | −0.0112 | 1.0521 | 0.9504 | 0.0027 | 3.5471 | 19.7588 | 0.3942 | 0.5039 |
| 0.25 | −0.0084 | 1.071 | 0.9337 | 0.0004 | 3.6665 | 19.7596 | 0.4747 | 0.503 |
| 0.35 | −0.0072 | 1.0867 | 0.9202 | −0.0017 | 3.7654 | 19.7608 | 0.5326 | 0.5026 |
| 0.45 | −0.0064 | 1.1005 | 0.9087 | −0.0025 | 3.8522 | 19.7624 | 0.5776 | 0.5023 |

**3.** $\varphi = \frac{\pi}{4}$

| $\eta\backslash$ | $a$ | $b$ | $c$ | $d$ | $r$ | $S$ | $k_f^2$ | $z_0$ |
|---|---|---|---|---|---|---|---|---|
| 0 | −0.3827 | 0.9239 | 0.9239 | 0.3827 | 3.2217 | 19.7582 | 0.0963 | 0.6279 |
| 0.15 | −0.0182 | 1.054 | 0.9487 | 0.0019 | 3.5411 | 19.7586 | 0.3898 | 0.5062 |
| 0.25 | −0.014 | 1.0732 | 0.9318 | −0.0027 | 3.6617 | 19.7594 | 0.4717 | 0.5049 |
| 0.35 | −0.0121 | 1.0891 | 0.9183 | −0.0047 | 3.7613 | 19.7605 | 0.5303 | 0.5043 |
| 0.45 | −0.011 | 1.103 | 0.9067 | −0.0058 | 3.8485 | 19.7621 | 0.5758 | 0.504 |

**4.** $\varphi = \frac{3\pi}{8}$

| $\eta\backslash$ | $a$ | $b$ | $c$ | $d$ | $r$ | $S$ | $k_f^2$ | $z_0$ |
|---|---|---|---|---|---|---|---|---|
| 0 | −0.5556 | 0.8315 | 0.8315 | 0.5556 | 3.2217 | 19.7582 | 0.0963 | 0.6913 |
| 0.15 | −0.0181 | 1.0575 | 0.9457 | −0.0041 | 3.5348 | 19.7585 | 0.3852 | 0.5062 |
| 0.25 | −0.0149 | 1.077 | 0.9286 | −0.0075 | 3.6568 | 19.7592 | 0.4687 | 0.5052 |
| 0.35 | −0.0134 | 1.093 | 0.915 | −0.0091 | 3.7571 | 19.7602 | 0.528 | 0.5047 |
| 0.45 | −0.0125 | 1.1071 | 0.9034 | −0.0099 | 3.8448 | 19.7617 | 0.574 | 0.5045 |

**5.** $\varphi = \frac{\pi}{2}$

| $\eta\backslash$ | $a$ | $b$ | $c$ | $d$ | $r$ | $S$ | $k_f^2$ | $z_0$ |
|---|---|---|---|---|---|---|---|---|
| 0 | –0.7071 | 0.7071 | 0.7071 | 0.7071 | 3.2217 | 19.7582 | 0.0963 | 0.754 |
| 0.15 | –0.0118 | 1.0624 | 0.9415 | –0.0134 | 3.5321 | 19.7584 | 0.3832 | 0.5040 |
| 0.25 | –0.0116 | 1.082 | 0.9244 | –0.0135 | 3.6547 | 19.7591 | 0.4674 | 0.504 |
| 0.35 | –0.0113 | 1.0982 | 0.9107 | –0.0137 | 3.7553 | 19.7601 | 0.527 | 0.504 |
| 0.45 | –0.0111 | 1.1124 | 0.8991 | –0.0138 | 3.8432 | 19.7615 | 0.5732 | 0.504 |

**Table 5.3**: For $L = 10, \theta(z = 0) = 0, \theta(z = L) = 2\pi$

List datas of $a, b, c, d, r, S, k_f^2, z_0$ for **Eq. (5.4.1.5)**: $f(s) = \frac{a \cdot sc[r(z-z_0)]+b}{c \cdot sc[r(z-z_0)]+d}$ at each $\varphi$

**1**. $\varphi = 0$

| $\eta\backslash$ | $a$ | $b$ | $c$ | $d$ | $r$ | $S$ | $k_f^2$ | $z_0$ |
|---|---|---|---|---|---|---|---|---|
| 0 | 0 | 1 | 1 | 0 | 1.0004 | 1.0015 | 0.999275 | 5 |
| 0.15 | 0 | 1.1247 | 0.8891 | 0 | 1.265 | 1.1503 | 0.999949 | 5 |
| 0.25 | 0 | 1.1892 | 0.8409 | 0 | 1.4142 | 1.2501 | 0.999988 | 5 |
| 0.35 | 0 | 1.2447 | 0.8034 | 0 | 1.5492 | 1.35 | 0.999997 | 5 |
| 0.45 | 0 | 1.2936 | 0.77301 | 0 | 1.6733 | 1.45 | 0.999999 | 5 |

**2**. $\varphi = \frac{\pi}{8}$

| $\eta\backslash$ | $a$ | $b$ | $c$ | $d$ | $r$ | $S$ | $k_f^2$ | $z_0$ |
|---|---|---|---|---|---|---|---|---|
| 0 | −0.1951 | 0.9808 | 0.9808 | 0.1951 | 1.0004 | 1.0015 | 0.999275 | 7.6873 |
| 0.15 | −0.1112 | 1.1209 | 0.8887 | 0.0334 | 1.2317 | 1.1218 | 0.999928 | 7.5606 |
| 0.25 | −0.0835 | 1.1923 | 0.8389 | −0.0023 | 1.3797 | 1.2119 | 0.999984 | 7.57 |
| 0.35 | −0.0659 | 1.2525 | 0.7994 | −0.0198 | 1.5163 | 1.3054 | 0.999996 | 7.6006 |
| 0.45 | −0.054 | 1.305 | 0.7675 | −0.0287 | 1.6425 | 1.4007 | 0.999999 | 7.6387 |

**3**. $\varphi = \frac{\pi}{4}$

| $\eta\backslash$ | $a$ | $b$ | $c$ | $d$ | $r$ | $S$ | $k_f^2$ | $z_0$ |
|---|---|---|---|---|---|---|---|---|
| 0 | −0.3827 | 0.9239 | 0.9239 | 0.3827 | 1.0004 | 1.0015 | 0.999275 | 8.3865 |
| 0.15 | −0.238 | 1.106 | 0.89 | 0.0656 | 1.1257 | 1.0417 | 0.999793 | 8.0102 |
| 0.25 | −0.1736 | 1.2041 | 0.8337 | −0.0216 | 1.2697 | 1.1012 | 0.999951 | 7.9251 |
| 0.35 | −0.133 | 1.2828 | 0.7858 | −0.0601 | 1.4133 | 1.1747 | 0.999988 | 7.904 |
| 0.45 | −0.1063 | 1.3485 | 0.7476 | −0.0772 | 1.548 | 1.2563 | 0.999997 | 7.9104 |

**4**. $\varphi = \frac{3\pi}{8}$

| $\eta\backslash$ | $a$ | $b$ | $c$ | $d$ | $r$ | $S$ | $k_f^2$ | $z_0$ |
|---|---|---|---|---|---|---|---|---|
| 0 | −0.5556 | 0.8315 | 0.8315 | 0.5556 | 1.0004 | 1.0015 | 0.999275 | 8.8078 |
| 0.15 | −0.4252 | 1.0515 | 0.9072 | 0.1082 | 0.92 | 0.9304 | 0.998379 | 8.2227 |
| 0.25 | −0.2883 | 1.2393 | 0.8302 | −0.1003 | 1.0511 | 0.9325 | 0.999564 | 7.9423 |
| 0.35 | −0.2019 | 1.3767 | 0.7516 | −0.172 | 1.2211 | 0.969 | 0.99992 | 7.8568 |
| 0.45 | −0.1515 | 1.4798 | 0.6953 | −0.1905 | 1.3819 | 1.0268 | 0.999984 | 7.8473 |

**5.** $\varphi = \frac{\pi}{2}$

| $\eta\backslash$ | $a$ | $b$ | $c$ | $d$ | $r$ | $S$ | $k_f^2$ | $z_0$ |
|---|---|---|---|---|---|---|---|---|
| 0 | –0.7071 | 0.7071 | 0.7071 | 0.7071 | 1.0004 | 1.0015 | 0.999275 | 9.119 |
| 0.15 | –0.8916 | 0.5608 | 0.8916 | 0.5608 | 0.6305 | 0.8595 | 0.969782 | 9.0604 |
| 0.25 | –0.3611 | 1.6877 | 0.7634 | –0.7984 | 0.5988 | 0.7833 | 0.958144 | 6.5151 |
| 0.35 | –0.1806 | 2.3593 | 0.4904 | –0.8689 | 0.906 | 0.7452 | 0.998134 | 6.4743 |
| 0.45 | –0.103 | 3.072 | 0.3554 | –0.8906 | 1.14 | 0.7522 | 0.999821 | 6.4466 |

It is especially interesting the $\theta$-plot at $\varphi = \pi/2$ because for an infinite system, (see **Table.5.1** and **Fig. 11 (a)** and **11 (b)**), we have no parameters at $\varphi = \pi/2$ and $\eta > 1/4$. The $\theta$-plot is broken into large and small kinks. Both kinks are **Möbius transformation** of tanh and different from other regions. But at a finite case, we have parameters of **Möbius transformation** of sinh, see **Table 5.3**. I find that the curve does not break.

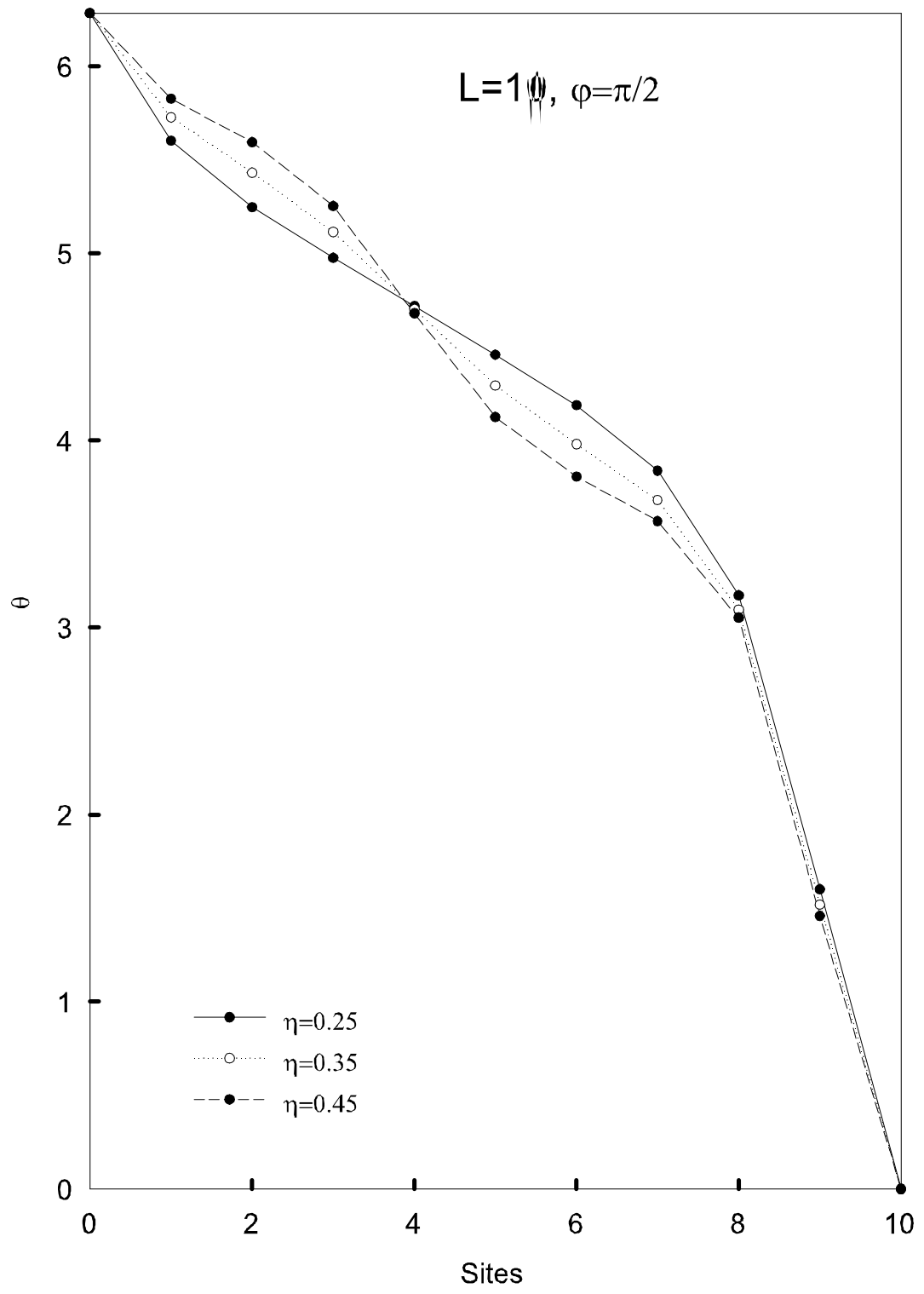

Fig. 17: $\theta$-plot of L=10, $\varphi = \pi/2$ at $\eta = 0.25, 0.35, 0.45$

### 5.4.2 The particular problem: two sets of parameters for the solutions in finite system:

In §**5.3.2**, we have proved there is no real solutions of **Möbius transformation** equations of sinh at $\eta > \frac{1}{4}$ and $\varphi = \frac{\pi}{2}$ for an infinite system. But in this subsection, we get real solutions in the corresponding region for a finite system. Like the situation for an infinite system, we still have the symmetry in **eqs**. (**5.3.3.2**) and (**5.3.4.1**). If we assume that the solution has the properties: $a = -c, b = d$ at $\varphi = \frac{\pi}{2}$ and $\eta > \frac{1}{4}$, then **Eq.** (**5.3.4.2**) should be modified as:

$$a^4(S + 1 - \eta) = \frac{r^2(1 - k_f^2)}{2} \quad 5.4.2.1a$$

$$a^2 b^2 = \frac{r^2(2 - k_f^2)}{4S - 12\eta} \quad 5.4.2.1b$$

$$b^4(S + 1 + \eta) = \frac{r^2}{2} \quad 5.4.2.1c$$

$$4a^2b^2 = 1 \qquad 5.4.2.1d$$

From **eq**. (**5**.**4**.**2**.**1b**,**d**), we obtain:

$$S - 3\eta = r^2(2 - k_f^2) \qquad 5.4.2.2$$

From **eq**. (**5**.**4**.**2**.**1a**,**c**,**d**),we obtain:

$$(S + \eta)^2 - 1 = 4r^4(1 - k_f^2) \qquad 5.4.2.3$$

From **eq**. (**5**.**4**.**2**.**2**) **and** (**5**.**4**.**2**.**3**), we get the solution of $r$ as we have done in **Eq**. (**5**.**3**.**4**.**3**):

$$r^2 = \frac{-4\eta(2 - k_f^2) \pm \sqrt{64\eta^2(1 - k_f^2) + k_f^4}}{k_f^4} \qquad 5.4.2.4$$

But the numerator is negative if $\eta > \frac{1}{4}$, the same situations as that of a infinite system. We can not find solutions at $\varphi = \frac{\pi}{2}$ and $\eta > \frac{1}{4}$ from the equations of *sc* **Möbius transformation**. So my previous assumption is wrong in finite case! In view of **eq**. (**5**.**4**.**1**.**3**), we can construct another solution by making the following substitution

$$a = \sqrt{k_f'}\,\tilde{b};\; b = \frac{-\tilde{a}}{\sqrt{k_f'}} \qquad 5.4.2.5$$

$$c = \sqrt{k_f'}\,\tilde{d};\; d = \frac{-\tilde{c}}{\sqrt{k_f'}}$$

Then

$$\theta(z) = 2\tan^{-1}\{\frac{a \cdot sc[r(z - z_0)] + b}{c \cdot sc[r(z - z_0)] + d}\} \qquad 5.4.2.6$$

$$= 2\tan^{-1}\{\frac{\sqrt{k_f'}\,\tilde{b} \cdot sc[r(z - z_0)] - \frac{\tilde{a}}{\sqrt{k_f'}}}{\sqrt{k_f'}\,\tilde{d} \cdot sc[r(z - z_0)] - \frac{\tilde{c}}{\sqrt{k_f'}}}\} = 2\tan^{-1}\{\frac{\tilde{a} \cdot sc[r(z - z_0) - \mathbf{K}] + \tilde{b}}{\tilde{c} \cdot sc[r(z - z_0) - \mathbf{K}] + \tilde{d}}\}$$

It is clear that $(\tilde{a}\tilde{d} - \tilde{b}\tilde{c})^2 = 1$ and $rL/2 = K(k_f)$. So another solution has the parameter set $\{\tilde{a}, \tilde{b}, \tilde{c}, \tilde{d}, z_0 + \frac{L}{2}\}$ compared to original ones: $\{a, b, c, d, z_0\}$. This only appears in a finite system, because the procedure is not suitable for an infinite system since $K(k_f = 1) = \infty$!

For $\eta > \frac{1}{4}$ and $\varphi \neq \pi/2$, both set of parameters can give us solutions. But at $\varphi = \pi/2$, they cannot give us solution separately. The reason is the same as that for an infinite system. The following two substitutions fail to give compatible form at $\varphi = \pi/2$.

$$\begin{bmatrix} a \\ b \\ c \\ d \end{bmatrix} \; of \;\; \varphi = \begin{bmatrix} -c \\ d \\ -a \\ b \end{bmatrix} \; of \;\; \pi - \varphi \qquad 5.4.2.7$$

$$\begin{bmatrix} \tilde{a} \\ \tilde{b} \\ \tilde{c} \\ \tilde{d} \end{bmatrix} \; of \;\; \varphi = \begin{bmatrix} -\tilde{c} \\ \tilde{d} \\ -\tilde{a} \\ \tilde{b} \end{bmatrix} \; of \;\; \pi - \varphi \qquad 5.4.2.8$$

However, we can connect the first set of parameters to the second set by the following substitution at $\varphi = \frac{\pi}{2}$:

$$\widetilde{a} = -c, \widetilde{b} = d, \widetilde{c} = -a, \widetilde{d} = b \qquad 5.4.2.9$$

So at $\varphi = \frac{\pi}{2}$, we have two solutions

$$\theta(z) = 2\tan^{-1}\{\frac{d\sqrt{k_f'} \cdot sc[r(z-z_0)] + b}{b\sqrt{k_f'} \cdot sc[r(z-z_0)] + d}\} \qquad 5.4.2.10a$$

and

$$\theta(z) = 2\tan^{-1}\{\frac{-b\sqrt{k_f'} \cdot sc[r(z-z_0-\frac{L}{2})] + d}{-d\sqrt{k_f'} \cdot sc[r(z-z_0-\frac{L}{2})] + b}\} \qquad 5.4.2.10b$$

The relation, **eq**. (**5.4.2.9**) and the additional set of parameters of solution solves the problem.

In reality, we can also find solution of *sn* **Möbius transformation** for $\eta > \frac{1}{4}$ and $\varphi = \frac{\pi}{2}$. This correspond to tanh and coth solutions because when $k_f \to 1$, we have known $sn(x, k_f) \to \tanh(x)$. It remains to be checked that whether this is identical to the solution above.

In the final part of §**5.3.4**, I gave the mathematical reasons why we cannot find sinh-type solutions in infinite system at $\varphi = \frac{\pi}{2}$ and $\eta > \frac{1}{4}$, but we can find *sc*-type in all $\varphi$ and $\eta$. Here, I will give physical reason as I have mentioned in §**4.6**. The key point is " boundary conditions ". In infinite case, when $z \to \pm\infty$, $\theta(z = \pm\infty)$ is always located at $V_{\min}$ ( i.e. the lowest potential ). So when $\varphi = \frac{\pi}{2}$ and $\eta > \frac{1}{4}$, the locations of $V_{\min}$ is not within $2\pi$ anymore and it will make contradict to *csch*-type solution. But in finite system, the boundary conditions here is: $\theta(z = 0) = 0, \theta(z = L) = 2\pi$, so by what I have said in §**6.6**, we can have stable solutions if one of locations of $V_{\min}$ is in the system ( i.e. $\in [0, L]$ ). But we still have one question, which concerns the *sn*-type solution. Why cannot we find solutions except at $\varphi = \frac{\pi}{2}$ and $\eta > \frac{1}{4}$ or $\varphi = 0$ and $\eta < \frac{-1}{4}$? The reason is that $|sn[r(z-z_0)| \leq 1$. So that $|2\tan^{-1}[A \cdot sn[r(z-z_0)]]| < \pi$. Hence, this form cannot satisfy the boundary condition: $\theta(z = 0) = 0$ and $\theta(z = L) = 2\pi$. On the other hand, this form can provide a solution in the range: ( $\varphi = \pi/2, \eta > 1/4$ and $\varphi = 0, \eta < -1/4$), like the bubble solution in §**5.2**.

# 5.5 Perturbation method for $\eta \to 0$:

## 5.5.1 Introduction:

Although in one-dimensional field theory, it is not suitable to do perturbation and from renormalization group analysis, the coupling strength of the relevant operator increases to infinity if the system is infinity. The perturbation below is not bad in finite system and $\eta \to 0$. The following perturbation is not the traditional one. I "perturb" the whole state by solving **eq**. (**5.4.1.5**) with $\eta$ being a small parameter, without using propagators in field theory.

## 5.5.2 Perturbation for a finite system:

Here, like in §**3.3.2**, we want to find the degenerate solution with opposite spin. From **Table 5.2**; **5.3**, we observe that at $\eta = 0$, that $\begin{pmatrix} a & b \\ c & d \end{pmatrix} = \begin{pmatrix} -\sin\frac{\varphi}{2} & \cos\frac{\varphi}{2} \\ \cos\frac{\varphi}{2} & \sin\frac{\varphi}{2} \end{pmatrix}$.

So

$$\theta(z) = 2\tan^{-1}\left(\frac{-\sin\frac{\varphi}{2}sc[r(z-z_0)] + \cos\frac{\varphi}{2}}{\cos\frac{\varphi}{2}sc[r(z-z_0)] + \sin\frac{\varphi}{2}}\right) = -\pi + \varphi + 2\tan^{-1}\{sc[r(z-z_0)]\} \qquad 5.5.2.1$$

and another solution is

$$\theta(z) = -\pi + \varphi - 2\tan^{-1}\{sc[r(z-z_0)]\} = -2\tan^{-1}\left(\frac{\sin\frac{\varphi}{2}sc[r(z-z_0)] + \cos\frac{\varphi}{2}}{\cos\frac{\varphi}{2}sc[r(z-z_0)] - \sin\frac{\varphi}{2}}\right) \qquad 5.5.2.2$$

$$= 2\tan^{-1}\left(\frac{-\sin\frac{\varphi}{2}sc[r(z-z_0)] - \cos\frac{\varphi}{2}}{\cos\frac{\varphi}{2}sc[r(z-z_0)] - \sin\frac{\varphi}{2}}\right) = 2\tan^{-1}\left(\frac{\sin\frac{\varphi}{2}sc[r(z-z_0)] + \cos\frac{\varphi}{2}}{-\cos\frac{\varphi}{2}sc[r(z-z_0)] + \sin\frac{\varphi}{2}}\right)$$

As $\eta \to 0$, we expect $\begin{pmatrix} a & b \\ c & d \end{pmatrix} = \begin{pmatrix} -\sin\frac{\varphi}{2} & \cos\frac{\varphi}{2} \\ \cos\frac{\varphi}{2} & \sin\frac{\varphi}{2} \end{pmatrix} + \begin{pmatrix} \Delta_a & \Delta_b \\ \Delta_c & \Delta_d \end{pmatrix}$, where $\Delta_{a,b,c,d}$ are small quantities. I insert these into **eq.(5.5.1.5)** and keep only the first order of $\Delta_{a,b,c,d}$. Solving the four linear equations of $\Delta_a, \Delta_b, \Delta_c, \Delta_d$ coming from **eqs**. (**5.4.1.5b**), (**5.4.1.5c**), (**5.4.1.5e**) and (**5.4.1.5f**), then using the formulas

$$\Delta_a = \frac{\Delta_1}{\Delta}; \Delta_b = \frac{\Delta_2}{\Delta}; \Delta_c = \frac{\Delta_3}{\Delta}; \Delta_d = \frac{\Delta_4}{\Delta} \qquad 5.5.2.3$$

we get

$$
\Delta = \begin{vmatrix}
\begin{array}{c} -4\sin\tfrac{\varphi}{2}S \\ -4\sin^3\tfrac{\varphi}{2}\cos\varphi + \\ 2\sin\varphi(\sin\varphi\sin\tfrac{\varphi}{2} \\ +\cos\tfrac{\varphi}{2}) \\ +(-4\sin^3\tfrac{\varphi}{2} + \\ 12\sin\tfrac{\varphi}{2}\cos^2\tfrac{\varphi}{2})\eta \end{array} & 0 & \begin{array}{c} 4\cos\tfrac{\varphi}{2}S - \\ 4\cos^3\tfrac{\varphi}{2}\cos\varphi \\ -2\sin\varphi(\sin\varphi\cos\tfrac{\varphi}{2} \\ +\sin\tfrac{\varphi}{2}) \\ +(-4\cos^3\tfrac{\varphi}{2} + \\ 12\cos\tfrac{\varphi}{2}\sin^2\tfrac{\varphi}{2})\eta \end{array} & 0 \\
0 & \begin{array}{c} 4\cos\tfrac{\varphi}{2}S + \\ 4\cos^3\tfrac{\varphi}{2}\cos\varphi \\ +2\sin\varphi(\sin\varphi\cos\tfrac{\varphi}{2} \\ +\sin\tfrac{\varphi}{2}) \\ +(-4\cos^3\tfrac{\varphi}{2} + \\ 12\cos\tfrac{\varphi}{2}\sin^2\tfrac{\varphi}{2})\eta \end{array} & 0 & \begin{array}{c} 4\sin\tfrac{\varphi}{2}S - \\ 4\sin^3\tfrac{\varphi}{2}\cos\varphi \\ +2\sin\varphi(\sin\varphi\sin\tfrac{\varphi}{2} \\ +\cos\tfrac{\varphi}{2}) \\ +(-4\sin^3\tfrac{\varphi}{2} + \\ 12\sin\tfrac{\varphi}{2}\cos^2\tfrac{\varphi}{2})\eta \end{array} \\
\begin{array}{c} 4\cos\tfrac{\varphi}{2}S + \\ 3\sin 2\varphi\sin\tfrac{\varphi}{2} \\ +6\sin^3\tfrac{\varphi}{2}\sin\varphi - \\ 3\sin^2\varphi\cos\tfrac{\varphi}{2} \\ +(-18\sin\varphi\cos\tfrac{\varphi}{2} \\ +12\cos^3\tfrac{\varphi}{2})\eta \end{array} & \begin{array}{c} -4\sin\tfrac{\varphi}{2}S - \\ 4\sin^3\tfrac{\varphi}{2}\cos\varphi \\ +3\sin^2\varphi\sin\tfrac{\varphi}{2} \\ +2\cos^3\tfrac{\varphi}{2}\sin\varphi \\ +(4\sin^3\tfrac{\varphi}{2} - \\ 6\sin\varphi\cos\tfrac{\varphi}{2})\eta \end{array} & \begin{array}{c} 4\sin\tfrac{\varphi}{2}S - \\ 3\sin 2\varphi\cos\tfrac{\varphi}{2} \\ +6\cos^3\tfrac{\varphi}{2}\sin\varphi - \\ 3\sin^2\varphi\sin\tfrac{\varphi}{2} \\ +(-18\sin\varphi\cos\tfrac{\varphi}{2} \\ +12\sin^3\tfrac{\varphi}{2})\eta \end{array} & \begin{array}{c} 4\cos\tfrac{\varphi}{2}S - \\ 4\cos^3\tfrac{\varphi}{2}\cos\varphi \\ -3\sin^2\varphi\cos\tfrac{\varphi}{2} \\ -2\sin^3\tfrac{\varphi}{2}\sin\varphi \\ +(-4\cos^3\tfrac{\varphi}{2} + \\ 6\sin\varphi\sin\tfrac{\varphi}{2})\eta \end{array} \\
\begin{array}{c} 4\cos\tfrac{\varphi}{2}S + \\ 4\cos^3\tfrac{\varphi}{2}\cos\varphi \\ +3\sin^2\varphi\cos\tfrac{\varphi}{2} \\ +2\sin^3\tfrac{\varphi}{2}\sin\varphi \\ +(-4\cos^3\tfrac{\varphi}{2} + \\ 6\sin\varphi\sin\tfrac{\varphi}{2})\eta \end{array} & \begin{array}{c} -4\sin\tfrac{\varphi}{2}S - \\ 3\sin 2\varphi\cos\tfrac{\varphi}{2} \\ +6\cos^3\tfrac{\varphi}{2}\sin\varphi - \\ 3\sin^2\varphi\sin\tfrac{\varphi}{2} \\ +(18\sin\varphi\cos\tfrac{\varphi}{2} - \\ 12\sin^3\tfrac{\varphi}{2})\eta \end{array} & \begin{array}{c} 4\sin\tfrac{\varphi}{2}S - \\ 4\sin^3\tfrac{\varphi}{2}\cos\varphi \\ +3\sin^2\varphi\sin\tfrac{\varphi}{2} + \\ 2\cos^3\tfrac{\varphi}{2}\sin\varphi \\ +(-4\sin^3\tfrac{\varphi}{2} + \\ 6\sin\varphi\cos\tfrac{\varphi}{2})\eta \end{array} & \begin{array}{c} 4\cos\tfrac{\varphi}{2}S - \\ 3\sin 2\varphi\sin\tfrac{\varphi}{2} \\ -6\sin^3\tfrac{\varphi}{2}\sin\varphi + \\ 3\sin^2\varphi\cos\tfrac{\varphi}{2} + \\ (-18\sin\varphi\sin\tfrac{\varphi}{2} + \\ 12\cos^3\tfrac{\varphi}{2})\eta \end{array}
\end{vmatrix}
$$

eq. (5.5.2.4)

$$
\Delta_1 = \begin{vmatrix}
\begin{array}{c} 2r^2(1-k_f^2) \\ -S+1+ \\ (\sin^4\frac{\varphi}{2}+\cos^4\frac{\varphi}{2} \\ -\frac{3}{2}\sin^2\varphi)\eta \end{array} & 0 & \begin{array}{c} 4\cos\frac{\varphi}{2}S \\ -4\cos^3\frac{\varphi}{2}\cos\varphi \\ -2\sin\varphi(\sin\varphi\cos\frac{\varphi}{2} \\ +\sin\frac{\varphi}{2}) \\ +(-4\cos^3\frac{\varphi}{2}+ \\ 12\cos\frac{\varphi}{2}\sin^2\frac{\varphi}{2})\eta \end{array} & 0 \\
\begin{array}{c} 2r^2-S-1+ \\ (\sin^4\frac{\varphi}{2}+\cos^4\frac{\varphi}{2} \\ -\frac{3}{2}\sin^2\varphi)\eta \end{array} & \begin{array}{c} 4\cos\frac{\varphi}{2}S+ \\ 4\cos^3\frac{\varphi}{2}\cos\varphi \\ +2\sin\varphi(\sin\varphi\cos\frac{\varphi}{2} \\ +\sin\frac{\varphi}{2}) \\ +(-4\cos^3\frac{\varphi}{2}+ \\ 12\cos\frac{\varphi}{2}\sin^2\frac{\varphi}{2})\eta \end{array} & 0 & \begin{array}{c} 4\sin\frac{\varphi}{2}S- \\ 4\sin^3\frac{\varphi}{2}\cos\varphi \\ +2\sin\varphi(\sin\varphi\sin\frac{\varphi}{2} \\ +\cos\frac{\varphi}{2}) \\ +(-4\sin^3\frac{\varphi}{2}+ \\ 12\sin\frac{\varphi}{2}\cos^2\frac{\varphi}{2})\eta \end{array} \\
4\eta\sin 2\varphi & \begin{array}{c} -4\sin\frac{\varphi}{2}S- \\ 4\sin^3\frac{\varphi}{2}\cos\varphi \\ +3\sin^2\varphi\sin\frac{\varphi}{2}+ \\ 2\cos^3\frac{\varphi}{2}\sin\varphi \\ +(4\sin^3\frac{\varphi}{2}- \\ 6\sin\varphi\cos\frac{\varphi}{2})\eta \end{array} & \begin{array}{c} 4\sin\frac{\varphi}{2}S- \\ 3\sin 2\varphi\cos\frac{\varphi}{2} \\ +6\cos^3\frac{\varphi}{2}\sin\varphi \\ -3\sin^2\varphi\sin\frac{\varphi}{2} \\ +(-18\sin\varphi\cos\frac{\varphi}{2} \\ +12\sin^3\frac{\varphi}{2})\eta \end{array} & \begin{array}{c} 4\cos\frac{\varphi}{2}S \\ -4\cos^3\frac{\varphi}{2}\cos\varphi \\ -3\sin^2\varphi\cos\frac{\varphi}{2} \\ -2\sin^3\frac{\varphi}{2}\sin\varphi \\ +(-4\cos^3\frac{\varphi}{2}+ \\ 6\sin\varphi\sin\frac{\varphi}{2})\eta \end{array} \\
-4\eta\sin 2\varphi & \begin{array}{c} -4\sin\frac{\varphi}{2}S- \\ 3\sin 2\varphi\cos\frac{\varphi}{2} \\ +6\cos^3\frac{\varphi}{2}\sin\varphi- \\ 3\sin^2\varphi\sin\frac{\varphi}{2} \\ +(18\sin\varphi\cos\frac{\varphi}{2} \\ -12\sin^3\frac{\varphi}{2})\eta \end{array} & \begin{array}{c} 4\sin\frac{\varphi}{2}S- \\ 4\sin^3\frac{\varphi}{2}\cos\varphi \\ +3\sin^2\varphi\sin\frac{\varphi}{2} \\ +2\cos^3\frac{\varphi}{2}\sin\varphi \\ +(-4\sin^3\frac{\varphi}{2}+ \\ 6\sin\varphi\cos\frac{\varphi}{2})\eta \end{array} & \begin{array}{c} 4\cos\frac{\varphi}{2}S \\ -3\sin 2\varphi\sin\frac{\varphi}{2} \\ -6\sin^3\frac{\varphi}{2}\sin\varphi \\ +3\sin^2\varphi\cos\frac{\varphi}{2} \\ +(-18\sin\varphi\sin\frac{\varphi}{2} \\ +12\cos^3\frac{\varphi}{2})\eta \end{array}
\end{vmatrix}
$$

eq. (5.5.2.5b)

$$
\Delta_2 = \begin{vmatrix}
-4\sin\frac{\varphi}{2}S - 4\sin^3\frac{\varphi}{2}\cos\varphi + 2\sin\varphi(\sin\varphi\sin\frac{\varphi}{2} + \cos\frac{\varphi}{2}) + (-4\sin^3\frac{\varphi}{2} + 12\sin\frac{\varphi}{2}\cos^2\frac{\varphi}{2})\eta & 2r^2(1-k_f^2) - S + 1 + (\sin^4\frac{\varphi}{2} + \cos^4\frac{\varphi}{2} - \frac{3}{2}\sin^2\varphi)\eta & 4\cos\frac{\varphi}{2}S - 4\cos^3\frac{\varphi}{2}\cos\varphi - 2\sin\varphi(\sin\varphi\cos\frac{\varphi}{2} + \sin\frac{\varphi}{2}) + (-4\cos^3\frac{\varphi}{2} + 12\cos\frac{\varphi}{2}\sin^2\frac{\varphi}{2})\eta & 0 \\
0 & 2r^2 - S - 1 + (\sin^4\frac{\varphi}{2} + \cos^4\frac{\varphi}{2} - \frac{3}{2}\sin^2\varphi)\eta & 0 & 4\sin\frac{\varphi}{2}S - 4\sin^3\frac{\varphi}{2}\cos\varphi + 2\sin\varphi(\sin\varphi\sin\frac{\varphi}{2} + \cos\frac{\varphi}{2}) + (-4\sin^3\frac{\varphi}{2} + 12\sin\frac{\varphi}{2}\cos^2\frac{\varphi}{2})\eta \\
4\cos\frac{\varphi}{2}S + 3\sin 2\varphi\sin\frac{\varphi}{2} + 6\sin^3\frac{\varphi}{2}\sin\varphi - 3\sin^2\varphi\cos\frac{\varphi}{2} + (-18\sin\varphi\cos\frac{\varphi}{2} + 12\cos^3\frac{\varphi}{2})\eta & 4\eta\sin 2\varphi & 4\sin\frac{\varphi}{2}S - 3\sin 2\varphi\cos\frac{\varphi}{2} + 6\cos^3\frac{\varphi}{2}\sin\varphi - 3\sin^2\varphi\sin\frac{\varphi}{2} + (-18\sin\varphi\cos\frac{\varphi}{2} + 12\sin^3\frac{\varphi}{2})\eta & 4\cos\frac{\varphi}{2}S - 4\cos^3\frac{\varphi}{2}\cos\varphi - 3\sin^2\varphi\cos\frac{\varphi}{2} - 2\sin^3\frac{\varphi}{2}\sin\varphi + (-4\cos^3\frac{\varphi}{2} + 6\sin\varphi\sin\frac{\varphi}{2})\eta \\
4\cos\frac{\varphi}{2}S + 4\cos^3\frac{\varphi}{2}\cos\varphi + 3\sin^2\varphi\cos\frac{\varphi}{2} + 2\sin^3\frac{\varphi}{2}\sin\varphi + (-4\cos^3\frac{\varphi}{2} + 6\sin\varphi\sin\frac{\varphi}{2})\eta & -4\eta\sin 2\varphi & 4\sin\frac{\varphi}{2}S - 4\sin^3\frac{\varphi}{2}\cos\varphi + 3\sin^2\varphi\sin\frac{\varphi}{2} + 2\cos^3\frac{\varphi}{2}\sin\varphi + (-4\sin^3\frac{\varphi}{2} + 6\sin\varphi\cos\frac{\varphi}{2})\eta & 4\cos\frac{\varphi}{2}S - 3\sin 2\varphi\sin\frac{\varphi}{2} - 6\sin^3\frac{\varphi}{2}\sin\varphi + 3\sin^2\varphi\cos\frac{\varphi}{2} + (-18\sin\varphi\sin\frac{\varphi}{2} + 12\cos^3\frac{\varphi}{2})\eta
\end{vmatrix}
$$

eq. (5.5.2.5c)

$$
\Delta_3 = \begin{vmatrix}
-4\sin\tfrac{\varphi}{2}S -4\sin^3\tfrac{\varphi}{2}\cos\varphi +2\sin\varphi(\sin\varphi\sin\tfrac{\varphi}{2} +\cos\tfrac{\varphi}{2}) +(-4\sin^3\tfrac{\varphi}{2} +12\sin\tfrac{\varphi}{2}\cos^2\tfrac{\varphi}{2})\eta & 0 & 2r^2(1-k_f^2) -S+1+ (\sin^4\tfrac{\varphi}{2}+\cos^4\tfrac{\varphi}{2} -\tfrac{3}{2}\sin^2\varphi)\eta & 0 \\
0 & 4\cos\tfrac{\varphi}{2}S + 4\cos^3\tfrac{\varphi}{2}\cos\varphi +2\sin\varphi(\sin\varphi\cos\tfrac{\varphi}{2} +\sin\tfrac{\varphi}{2}) +(-4\cos^3\tfrac{\varphi}{2} + 12\cos\tfrac{\varphi}{2}\sin^2\tfrac{\varphi}{2})\eta & 2r^2 - S - 1 + (\sin^4\tfrac{\varphi}{2}+\cos^4\tfrac{\varphi}{2} -\tfrac{3}{2}\sin^2\varphi)\eta & 4\sin\tfrac{\varphi}{2}S - 4\sin^3\tfrac{\varphi}{2}\cos\varphi +2\sin\varphi(\sin\varphi\sin\tfrac{\varphi}{2} +\cos\tfrac{\varphi}{2}) +(-4\sin^3\tfrac{\varphi}{2} + 12\sin\tfrac{\varphi}{2}\cos^2\tfrac{\varphi}{2})\eta \\
4\cos\tfrac{\varphi}{2}S + 3\sin 2\varphi\sin\tfrac{\varphi}{2} +6\sin^3\tfrac{\varphi}{2}\sin\varphi - 3\sin^2\varphi\cos\tfrac{\varphi}{2} +(-18\sin\varphi\cos\tfrac{\varphi}{2} +12\cos^3\tfrac{\varphi}{2})\eta & -4\sin\tfrac{\varphi}{2}S - 4\sin^3\tfrac{\varphi}{2}\cos\varphi +3\sin^2\varphi\sin\tfrac{\varphi}{2} +2\cos^3\tfrac{\varphi}{2}\sin\varphi +(4\sin^3\tfrac{\varphi}{2} - 6\sin\varphi\cos\tfrac{\varphi}{2})\eta & 4\eta\sin 2\varphi & 4\cos\tfrac{\varphi}{2}S - 4\cos^3\tfrac{\varphi}{2}\cos\varphi -3\sin^2\varphi\cos\tfrac{\varphi}{2} -2\sin^3\tfrac{\varphi}{2}\sin\varphi +(-4\cos^3\tfrac{\varphi}{2} + 6\sin\varphi\sin\tfrac{\varphi}{2})\eta \\
4\cos\tfrac{\varphi}{2}S + 4\cos^3\tfrac{\varphi}{2}\cos\varphi +3\sin^2\varphi\cos\tfrac{\varphi}{2} +2\sin^3\tfrac{\varphi}{2}\sin\varphi +(-4\cos^3\tfrac{\varphi}{2} + 6\sin\varphi\sin\tfrac{\varphi}{2})\eta & -4\sin\tfrac{\varphi}{2}S - 3\sin 2\varphi\cos\tfrac{\varphi}{2} +6\cos^3\tfrac{\varphi}{2}\sin\varphi - 3\sin^2\varphi\sin\tfrac{\varphi}{2} +(18\sin\varphi\cos\tfrac{\varphi}{2} - 12\sin^3\tfrac{\varphi}{2})\eta & -4\eta\sin 2\varphi & 4\cos\tfrac{\varphi}{2}S - 3\sin 2\varphi\sin\tfrac{\varphi}{2} -6\sin^3\tfrac{\varphi}{2}\sin\varphi + 3\sin^2\varphi\cos\tfrac{\varphi}{2} +(-18\sin\varphi\sin\tfrac{\varphi}{2} + 12\cos^3\tfrac{\varphi}{2})\eta
\end{vmatrix}
$$

eq.(5.5.2.5e)

$$\Delta_4 = \begin{vmatrix}
-4\sin\frac{\varphi}{2}S - 4\sin^3\frac{\varphi}{2}\cos\varphi + 2\sin\varphi(\sin\varphi\sin\frac{\varphi}{2} + \cos\frac{\varphi}{2}) + (-4\sin^3\frac{\varphi}{2} + 12\sin\frac{\varphi}{2}\cos^2\frac{\varphi}{2})\eta & 0 & 4\cos\frac{\varphi}{2}S - 4\cos^3\frac{\varphi}{2}\cos\varphi - 2\sin\varphi(\sin\varphi\cos\frac{\varphi}{2} + \sin\frac{\varphi}{2}) + (-4\cos^3\frac{\varphi}{2} + 12\cos\frac{\varphi}{2}\sin^2\frac{\varphi}{2})\eta & 2r^2(1-k_f^2) - S + 1 + (\sin^4\frac{\varphi}{2} + \cos^4\frac{\varphi}{2} - \frac{3}{2}\sin^2\varphi)\eta \\
0 & 4\cos\frac{\varphi}{2}S + 4\cos^3\frac{\varphi}{2}\cos\varphi + 2\sin\varphi(\sin\varphi\cos\frac{\varphi}{2} + \sin\frac{\varphi}{2}) + (-4\cos^3\frac{\varphi}{2} + 12\cos\frac{\varphi}{2}\sin^2\frac{\varphi}{2})\eta & 0 & 2r^2 - S - 1 + (\sin^4\frac{\varphi}{2} + \cos^4\frac{\varphi}{2} - \frac{3}{2}\sin^2\varphi)\eta \\
4\cos\frac{\varphi}{2}S + 3\sin 2\varphi\sin\frac{\varphi}{2} + 6\sin^3\frac{\varphi}{2}\sin\varphi - 3\sin^2\varphi\cos\frac{\varphi}{2} + (-18\sin\varphi\cos\frac{\varphi}{2} + 12\cos^3\frac{\varphi}{2})\eta & -4\sin\frac{\varphi}{2}S - 4\sin^3\frac{\varphi}{2}\cos\varphi + 3\sin^2\varphi\sin\frac{\varphi}{2} + 2\cos^3\frac{\varphi}{2}\sin\varphi + (4\sin^3\frac{\varphi}{2} - 6\sin\varphi\cos\frac{\varphi}{2})\eta & 4\sin\frac{\varphi}{2}S - 3\sin 2\varphi\cos\frac{\varphi}{2} + 6\cos^3\frac{\varphi}{2}\sin\varphi - 3\sin^2\varphi\sin\frac{\varphi}{2} + (-18\sin\varphi\cos\frac{\varphi}{2} + 12\sin^3\frac{\varphi}{2})\eta & 4\eta\sin 2\varphi \\
4\cos\frac{\varphi}{2}S + 4\cos^3\frac{\varphi}{2}\cos\varphi + 3\sin^2\varphi\cos\frac{\varphi}{2} + 2\sin^3\frac{\varphi}{2}\sin\varphi + (-4\cos^3\frac{\varphi}{2} + 6\sin\varphi\sin\frac{\varphi}{2})\eta & -4\sin\frac{\varphi}{2}S - 3\sin 2\varphi\cos\frac{\varphi}{2} + 6\cos^3\frac{\varphi}{2}\sin\varphi - 3\sin^2\varphi\sin\frac{\varphi}{2} + (18\sin\varphi\cos\frac{\varphi}{2} - 12\sin^3\frac{\varphi}{2})\eta & 4\sin\frac{\varphi}{2}S - 4\sin^3\frac{\varphi}{2}\cos\varphi + 3\sin^2\varphi\sin\frac{\varphi}{2} + 2\cos^3\frac{\varphi}{2}\sin\varphi + (-4\sin^3\frac{\varphi}{2} + 6\sin\varphi\cos\frac{\varphi}{2})\eta & -4\eta\sin 2\varphi
\end{vmatrix}$$

eq. (5.5.2.5f)

In numerical calculation, if we give actuate value of $S$, $k_f^2$ list in**Table 5.4**, **5.5** and **5.6**, we will get reasonably good approximation compared with $a, b, c, d$ calculate in **eq.**(**5.4.1.5**) if $\eta \to 0$ ( i.e. usually $|\eta| < 0.1$).

Surely we have still two equations which can determine $S$, $k_f^2$, but they are complex in calculate in reality. I only list the equations:

From **eq.**(**5.4.1.5d**), when calculate to order 1 of $\Delta_{a,b,c,d}$, then:

$$[(-2\cos\frac{\varphi}{2}\sin\varphi - 4\sin^3\frac{\varphi}{2})S$$
$$- 3\sin 2\varphi\cos\frac{\varphi}{2} - 3\sin^2\varphi\sin\frac{\varphi}{2} +$$
$$6\sin\varphi\cos^3\frac{\varphi}{2} + (18\cos\frac{\varphi}{2}\sin\varphi - 12\sin^3\frac{\varphi}{2})\eta] \bullet \Delta_a$$
$$+ [(2\sin\frac{\varphi}{2}\sin\varphi + 4\cos^3\frac{\varphi}{2})S$$
$$+ 3\sin 2\varphi\sin\frac{\varphi}{2} - 3\sin^2\varphi\cos\frac{\varphi}{2} +$$
$$6\sin\varphi\sin^3\frac{\varphi}{2} + (-18\sin\frac{\varphi}{2}\sin\varphi + 12\cos^3\frac{\varphi}{2})\eta] \bullet \Delta_b$$
$$+ [(2\sin\frac{\varphi}{2}\sin\varphi + 4\cos^3\frac{\varphi}{2})S$$
$$- 3\sin 2\varphi\sin\frac{\varphi}{2} + 3\sin^2\varphi\cos\frac{\varphi}{2} - 6\sin\varphi\sin^3\frac{\varphi}{2} +$$
$$(-18\sin\frac{\varphi}{2}\sin\varphi + 12\cos^3\frac{\varphi}{2})\eta] \bullet \Delta_c$$
$$+ [(2\cos\frac{\varphi}{2}\sin\varphi + 4\sin^3\frac{\varphi}{2})S$$
$$- 3\sin 2\varphi\cos\frac{\varphi}{2} - 3\sin^2\varphi\sin\frac{\varphi}{2} +$$
$$6\sin\varphi\cos^3\frac{\varphi}{2} + (-18\cos\frac{\varphi}{2}\sin\varphi + 12\sin^3\frac{\varphi}{2})\eta] \bullet \Delta_d$$
$$= 2r^2(2 - k_f^2) - (\sin^2\varphi + 2\sin^4\frac{\varphi}{2} + 2\cos^4\frac{\varphi}{2})S \qquad 5.5.2.5d$$
$$- 6\eta(-\frac{3}{2}\sin^2\varphi + \sin^4\frac{\varphi}{2} + \cos^4\frac{\varphi}{2})$$

The scaling restriction equation, **eq**. (**5.4.1.5a**) becomes:

$$\sin\frac{\varphi}{2}(\Delta_a - \Delta_d) - \cos\frac{\varphi}{2}(\Delta_c + \Delta_d) = 0 \ \ or \ \ 2 \qquad 5.5.2.5a$$

And surely $r$ is related to $k_f^2$ by **eq**. (**5.4.1.5g**) for the boundary condition:

$$rL = 2K(k_f) \qquad 5.4.1.5g$$

## 5.6.3. Three examples of perturbation method:

I consider three cases to illustate the perturbation method:

**Case** *1:* $L = 1, \eta = 0.1$ *for* $\varphi = 0, \pi/8, \pi/4, 3\pi/8, \pi/2$ *for the first example. I use* ***eqs***. *(**5.5.1.5a**, **b**, **c**, **d**, **e**, **f**, **g**) and obtain the parameters:*

***Table 5.4 (a)***: $(L = 1, \eta = 0.1)$

| $\varphi\backslash$ | $a$ | $b$ | $c$ | $d$ | $r$ | $S$ | $k_f^2$ |
|---|---|---|---|---|---|---|---|
| 0 | 0 | 1.04 | 0.9615 | 0 | 3.4762 | 19.7585 | 0.3401 |
| $\pi/8$ | −0.0146 | 1.0405 | 0.961 | 0.0063 | 3.4734 | 19.7585 | 0.3378 |
| $\pi/4$ | −0.0234 | 1.0421 | 0.9595 | 0.0074 | 3.4662 | 19.7584 | 0.332 |
| $3\pi/8$ | −0.0222 | 1.0453 | 0.9567 | 0.0001 | 3.4585 | 19.7583 | 0.3257 |
| $\pi/2$ | −0.012 | 1.0501 | 0.9525 | −0.0132 | 3.4552 | 19.7583 | 0.323 |

***Table 5.4 (b)***: $(L = 1, \eta = 0.1)$, *perturbation parameters from* **eq. (5.6.2.3)** *with* $r, S, k_f^2$ *from* **Table 5.4 (a)**. $\hat{a} = -\sin\frac{\varphi}{2} + \Delta_a; \hat{b} = \cos\frac{\varphi}{2} + \Delta_b; \hat{c} = \cos\frac{\varphi}{2} + \Delta_c; \hat{d} = \sin\frac{\varphi}{2} + \Delta_d$.

| $\varphi\backslash$ | $\Delta_a$ | $\Delta_b$ | $\Delta_c$ | $\Delta_d$ | $\hat{a}$ | $\hat{b}$ | $\hat{c}$ | $\hat{d}$ |
|---|---|---|---|---|---|---|---|---|
| 0 | 0 | 0.0425 | −0.0363 | 0 | 0 | 1.0425 | 0.9637 | 0 |
| $\pi/8$ | 0.2095 | 0.0836 | 0.0055 | −0.2033 | 0.0144 | 1.0644 | 0.9385 | 0.2834 |
| $\pi/4$ | 0.1278 | 0.0844 | 0.0146 | −0.0993 | −0.2549 | 1.0083 | 0.9385 | 0.2834 |
| $3\pi/8$ | 0.0769 | 0.0685 | 0.0086 | −0.0353 | −0.4787 | 0.9 | 0.8401 | 0.5203 |
| $\pi/2$ | 0.0515 | 0.0527 | 0.001 | −0.0015 | −0.6556 | 0.7598 | 0.7081 | 0.7056 |

**Case** *2:* $L = 1, \eta = 0.01$ *for* $\varphi = 0, \pi/8, \pi/4, 3\pi/8, \pi/2$ *for the first example. I use* **eqs. (5.5.1.5a, b, c, d, e, f, g)** *and obtain the parameters:*

***Table 5.5 (a)***: $(L = 1, \eta = 0.01)$

| $\varphi\backslash$ | $a$ | $b$ | $c$ | $d$ | $r$ | $S$ | $k_f^2$ |
|---|---|---|---|---|---|---|---|
| 0 | 0 | 1.0077 | 0.9924 | 0 | 3.2707 | 19.7582 | 0.1502 |
| $\pi/8$ | −0.0773 | 1.0047 | 0.9897 | 0.0723 | 3.2659 | 19.7582 | 0.1451 |
| $\pi/4$ | −0.1457 | 0.9971 | 0.9831 | 0.1353 | 3.2513 | 19.7582 | 0.1294 |
| $3\pi/8$ | −0.1777 | 0.9921 | 0.9792 | 0.1608 | 3.2265 | 19.7582 | 0.1018 |
| $\pi/2$ | −0.0125 | 1.0098 | 0.9905 | −0.0128 | 3.2025 | 19.7582 | 0.0743 |

***Table 5.5 (b)***: $(L = 1, \eta = 0.01)$, *perturbation parameters from* **eq. (5.6.2.3)** *with* $r, S, k_f^2$ *from* **Table 5.5 (a)**. $\hat{a} = -\sin\frac{\varphi}{2} + \Delta_a; \hat{b} = \cos\frac{\varphi}{2} + \Delta_b; \hat{c} = \cos\frac{\varphi}{2} + \Delta_c; \hat{d} = \sin\frac{\varphi}{2} + \Delta_d$

| $\varphi\backslash$ | $\Delta_a$ | $\Delta_b$ | $\Delta_c$ | $\Delta_d$ | $\hat{a}$ | $\hat{b}$ | $\hat{c}$ | $\hat{d}$ |
|---|---|---|---|---|---|---|---|---|
| 0 | 0 | 0.0078 | −0.0076 | 0 | 0 | 1.0078 | 0.9924 | 0 |
| $\pi/8$ | 0.0419 | 0.0153 | 0.0014 | −0.041 | −0.1532 | 0.9961 | 0.9821 | 0.1541 |
| $\pi/4$ | 0.0154 | 0.01 | 0.0013 | −0.0121 | −0.3673 | 0.9339 | 0.9252 | 0.3706 |
| $3\pi/8$ | 0.0025 | 0.0018 | 0.0007 | −0.0015 | −0.553 | 0.8333 | 0.8321 | 0.5541 |
| $\pi/2$ | −0.0042 | −0.0044 | $\approx 0$ | $\approx 0$ | −0.7113 | 0.7027 | 0.7071 | 0.7071 |

**Case** *3:* $L = 10, \eta = 0.01$ *for* $\varphi = 0, \pi/8, \pi/4, 3\pi/8, \pi/2$ *for the first example. I use* **eqs. (5.5.1.5a, b, c, d, e, f, g)** *and obtain the parameters:*

***Table 5.6 (a)***: $(L = 10, \eta = 0.01)$

| $\varphi\backslash$ | $a$ | $b$ | $c$ | $d$ | $r$ | $S$ | $k_f^2$ |
|---|---|---|---|---|---|---|---|
| 0 | 0 | 1.0099 | 0.9902 | 0 | 1.0201 | 1.0113 | 0.999405 |
| $\pi/8$ | −0.187 | 0.9917 | 0.9751 | 0.1765 | 1.0147 | 1.0085 | 0.999372 |
| $\pi/4$ | −0.3732 | 0.9356 | 0.9272 | 0.3551 | 1.0011 | 1.0017 | 0.99928 |
| $3\pi/8$ | −0.5533 | 0.8379 | 0.8413 | 0.5333 | 0.9865 | 0.9946 | 0.999167 |
| $\pi/2$ | −0.7144 | 0.7 | 0.7144 | 0.7 | 0.9802 | 0.9916 | 0.999113 |

*Table 5.6 (b): $(L = 10, \eta = 0.01)$, perturbation parameters from **eq. (5.6.2.3)** with $r, S, k_f^2$ from **Table 5.6 (a)**. $\hat{a} = -\sin\frac{\varphi}{2} + \Delta_a; \hat{b} = \cos\frac{\varphi}{2} + \Delta_b; \hat{c} = \cos\frac{\varphi}{2} + \Delta_c; \hat{d} = \sin\frac{\varphi}{2} + \Delta_d$*

| $\varphi\backslash$ | $\Delta_a$ | $\Delta_b$ | $\Delta_c$ | $\Delta_d$ | $\hat{a}$ | $\hat{b}$ | $\hat{c}$ | $\hat{d}$ |
|---|---|---|---|---|---|---|---|---|
| 0 | 0 | 0.01 | −0.01 | 0 | 0 | 1.01 | 0.99 | 0 |
| $\pi/8$ | 0.0047 | 0.0103 | 0.0133 | −0.0142 | −0.1903 | 0.991 | 0.9941 | 0.1809 |
| $\pi/4$ | −0.0033 | 0.0016 | 0.0368 | −0.0031 | −0.386 | 0.9255 | 0.9607 | 0.3796 |
| $3\pi/8$ | 0.0435 | 0.0399 | −0.0536 | −0.0716 | −0.5121 | 0.8713 | 0.7779 | 0.484 |
| $\pi/2$ | −0.0074 | −0.0148 | 0.0074 | 0.0006 | −0.7145 | 0.6923 | 0.7145 | 0.7078 |

We can observe three things. They depend on $L, \eta, \varphi$ :

**1**. We can compare **Table 5.5** and **5.6**, with fixed $\eta = 0.01$ is fixed, the exact result and the result of perturbation. We find the data at $L = 10$ is much more precise than $L = 1$. The reason is not clear.

**2**. We can compare **Table 5.4** and **5.5** with fixed $L = 1$. Although it is differecult to see which $y$ gives better peturbation results, we can see $\Delta_a, \Delta_b, \Delta_c, \Delta_d$ is smaller in $\eta = 0.01$. That means $\eta = 0.01$ is better because method of perturbation can be handle well. And this also fit our intuition.

**3**. Consider the entries of each table varing with $\varphi$. Those of $\varphi \leq \pi/4$ are better than those of $\varphi \geq \pi/4$. Taking **Table 5.5** and $\varphi = \pi/2$ for example, although $\{\hat{a}, \hat{b}, \hat{c}, \hat{d}\}$ are very bad from peturbation from $\{a, b, c, d\}$, they are close to the exact solution of $\eta = 0$, see **Table 5.3**. Because $\Delta_a, \Delta_b, \Delta_c, \Delta_d$ is also small, the perturbation is good. The reason is in **Fig.16** (**a**), even in $\varphi = \pi/2$, we have no phase transition, but the derivative of energy at $\varphi = \pi/2$ is large even if $\eta$ small. That means at $\varphi = \pi/2$, the system is very senstive with $\eta$. This make it differecult to find exact solutions wnen I solve equations by computer.

## 5.6 Discuss and Conclusion of DSGE in an infinite and a finite system

In this chapter, I find differences between finite and infinite cases. There is no phase transition as $\varphi$ moves around $\pi/2$. I also develop a perturbation method and discuss the range of values of $L, \varphi, \eta$ in which this perturbation work.

# 6. Conclusions:

This Ph. D. thesis mainly contains four parts. The work of the first part is to transform Wazwaz's [**20**] solutions of **SGE** into another form we are familiar with. This paper is accepted by "Applied Mathematics and Computation". In the second paper, I discuss static soliton which satisfy the fixed boundary condition of **SGE** and the spin transport phenomena through bulk state. This mechanism is different from those of [**34**], [**36**]. Because the static soliton has the unique property: $\theta(z+L)-\theta(z) = 2\pi$, we can connect source and drain. We have spin transport phenomena from source through our system to drain. The third paper is **SGE** with twisted boundary condition. An interacting system with usual boundary condition is equivalent to a noninteracting system with twisted boundary condition. And I find forbidden zones which exist edge state and also have the meaning of topology, similar to [**32**], [**33**]. The mechanism in this paper is similar to [**34**], [**36**]. Finally, I have solved asymmetric **DSGE** by numerically method. I use **Möbius transformation** as the mathematical technique. I also disuss why some solution can't exist in some parameter region in an infinite system. But the corresponding solution is exist at a finite system. I also develop and discuss the perturbation theory of asymmetric **DSGE**.

# Appendix A: Basic properties of Jacobi Elliptic Functions

In this appendix we listed some properties of the Jacobian elliptic functions. See **eqs**.(**5**.**2**.**2**;**5**.**2**.**3**) for the definition.

$$sn^2(u) + cn^2(u) = 1, \quad \text{A-1}$$
$$dn^2(u) + k^2 sn^2(u) = 1, \quad \text{A-2}$$
$$sn(u+K) = cn(u)/dn(u) \quad \text{A-3}$$
$$cn(u+K) = -sn(u)/dn(u) \quad \text{A-4}$$

where $k'^2 = 1 - k^2$. The derivatives of Jacobian elliptic functions are

$$\partial_u sn(u) = cn(u)dn(u), \quad \text{A-5}$$
$$\partial_u cn(u) = -sn(u)dn(u), \quad \text{A-6}$$
$$\partial_u dn(u) = -k^2 sn(u)cn(u). \quad \text{A-7}$$

Using above equations we found the differential equations to be satisfied by the Jacobian elliptic functions and listed them in Table 1 where

$$ns(u) = 1/sn(u), \quad \text{A-8}$$
$$nc(u) = 1/cn(u), \quad \text{A-9}$$
$$nd(u) = 1/dn(u), \quad \text{A-10}$$
$$sc(u) = sn(u)/cn(u), \quad \text{A-11}$$
$$sd(u) = sn(u)/dn(u), \quad \text{A-12}$$
$$cd(u) = cn(u)/dn(u), \quad \text{A-13}$$
$$cs(u) = 1/sc(u), \quad \text{A-14}$$
$$ds(u) = 1/sd(u), \quad \text{A-15}$$
$$dc(u) = 1/cd(u). \quad \text{A-16}$$

Having checked those equations in Table I, one can see that there are many combinations of **JEF** that can satisfy **eqs**.(**5**.**2**.**2**) and (**5**.**2**.**3**) where the discriminant is $\mu^2 + 4\kappa\lambda$ for $k_f$ and $(\mu+1)^2 + 4\kappa\lambda$ for $k_g$.

For $L \longrightarrow \infty$, we find that $k \to 1$ and

$$sn(u) \simeq \tanh(u), \quad \text{A-17}$$

$$cn(u) \simeq dn(u) \simeq \sec h(u), \quad \text{A-18}$$

**Table A.1**:

Differential equations satisfied by Jacobian elliptic functions. See Appendix A for the definations of Jacobian elliptic functions.

| **JEF** | its equation | **JEF** | its equation |
|---|---|---|---|
| $y = sn(u)$ | $(\partial_u y)^2 = (1-y^2)(1-k^2y^2)$ | $y = cn(u)$ | $(\partial_u y)^2 = (1-y^2)(1-k^2+k^2y^2)$ |
| $y = dn(u)$ | $(\partial_u y)^2 = (y^2-1)(1-k^2-y^2)$ | $y = ns(u)$ | $(\partial_u y)^2 = (y^2-1)(y^2-k^2)$ |
| $y = nc(u)$ | $(\partial_u y)^2 = (y^2-1)[(1-k^2)y^2+k^2]$ | $y = nd(u)$ | $(\partial_u y)^2 = (1-y^2)[(1-k^2)y^2-1]$ |
| $y = sc(u)$ | $(\partial_u y)^2 = (y^2+1)(1+k'^2y^2)$ | $y = cd(u)$ | $(\partial_u y)^2 = (y^2-1)(k^2y^2-1)$ |
| $y = sd(u)$ | $(\partial_u y)^2 = (1-k'^2y^2)(1+k^2y^2)$ | $y = cs(u)$ | $(\partial_u y)^2 = (1+y^2)(k'^2+y^2)$ |
| $y = dc(u)$ | $(\partial_u y)^2 = (k^2-y^2)(1-y^2)$ | $y = ds(u)$ | $(\partial_u y)^2 = (y^2-k'^2)(y^2+k^2)$ |

# Appendix B: Periodic theorem of static soliton of SGE

In this Appendix, we prove the following lemma.

**Lemma** *If* $\theta_+(z) = \pi/2 - \varphi + 4\tan^{-1}\{A_{th}sc[\beta(L-z_0);k_f]\}$ *and* $A = A_{th}$*, then* $\theta_+(z=L) - \theta_+(z=0) = 2l\pi$*, no matter what value of* $z_0$ *here. Where* $l$ *is a natural number defined by*

$$\beta L = lK(k_f = \sqrt{1-A_{th}^4}) = n\int_0^1 \frac{dt}{\sqrt{(1-t^2)[1-(1-A_{th}^4)t^2]}} \qquad B\text{-}1$$

*with* $\beta = 1/(1-A_{th}^2)$.

Proof:

The boundary condition requires that at $z = 0$

$$\frac{\varphi}{4} = -\tan^{-1}A_{th}sc(-\beta z_0) \qquad \text{B-2}$$

and at $z = L$

$$\frac{n\pi}{2} - \frac{\varphi}{4} = \tan^{-1}A_{th}sc(\beta(L-z_0)) = \tan^{-1}A_{th}sc(lK-\beta z_0) \qquad \text{B-3}$$

Note that the period of $sc(u)$ is $2K$. Hence, when $z$ increases from 0 to $L$, $l/2$ periods will pass. Since the period of $\tan^{-1}$ function is $\pi$, the term $l\pi/2$ has to be added on the left hand side of eq. (B-3). To see it more explicitly, define

$$\alpha = \tan^{-1}(A_{th}sc(\beta(L-z_0))) = \tan^{-1}(A_{th}sc(lK-\beta z_0)), \qquad \text{B-4}$$

and

$$\varsigma = \tan^{-1}(A_{th}sc(-\beta z_0)), \qquad \text{B-5}$$

then we have

$$\tan(\alpha-\varsigma) = \frac{\tan\alpha - \tan\varsigma}{1+\tan\alpha\tan\varsigma} = A_{th}\frac{sc(lK-\beta z_0) - sc(-\beta z_0)}{1+A_{th}^2 sc(lK-\beta z_0)sc(-\beta z_0)}. \qquad \text{B-6}$$

With eqs. (A-3) and (A-4), we have

$$sc(\beta(nK - z_0)) = \frac{sn(nK-\beta z_0)}{cn(nK-\beta z_0)} = \frac{cn(nK-K-\beta z_0))}{-k_f' sn(nK-K-\beta z_0))}. \qquad \text{B-7}$$

where $k_f^2 + k_f'^2 = 1$. Since $k_f^2 = 1 - A_{th}^4$, we find that $k_f' = A_{th}^2$. Eq. (B-6) becomes

$$\tan(\alpha-\varsigma) = -A_{th}\frac{cs(lK-K-\beta z_0)/k_f' - sc(-\beta z_0)}{1 - sc(-\beta z_0)cs(lK-K-\beta z_0)} \qquad \text{B-8}$$

where $cs(u) = 1/sc(u)$. So the denominator of $\tan(\alpha-\varsigma)$ vanishes as $n = 1$ and the numerator is finite. Thus $\tan(\alpha-\varsigma) = \pm\infty$ and $\alpha - \varsigma = \pm\pi/2$. The sign is determined by the fact that as $z$ increases, $\theta_+$ also increases, thus the positive sign should be chosen and $\theta_+(z=L) - \theta_+(z=0) = 2\pi$. If $l = 2$, we can use eqs. (A-3) and (A-4) again or simply use the fact that the period of $sc(u)$ is $2K$ to find out that the numerator vanishes. Thus, $\tan(\alpha-\varsigma) = \pi$ and $\theta_+(z=L) - \theta_+(z=0) = 4\pi$. Hence, we conclude that $\theta_+(z=L) - \theta_+(z=0) = 4(\alpha-\varsigma) = 2l\pi$.

**End of Proof**.

# Appendix C: Introdction to Möbius transformation

By elliptic function theory, If we want to solve the different equation:

$$(\frac{df}{ds})^2 = \varphi(s) = A(f-f_0)(f-f_1)(f-f_2)(f-f_3) \qquad \text{C1}$$

$\varphi(s)$ can be three or four power polynomials.$f_0,\ f_1,\ f_2,\ f_3$ are roots of $\varphi(s)$.
Do the "**Möbius transformation**":

$$f = \frac{a\zeta + b}{c\zeta + d} \qquad \text{C2}$$

So do every roots:

$$f_i = \frac{a\zeta_i + b}{c\zeta_i + d} \qquad \text{C3}$$

If we take special values of $a, b, c, d,$ we can obtain the standard and solved form

$$f = \frac{f_3(f_1 - f_0) * \zeta - f_1(f_3 - f_0)}{(f_1 - f_0)\zeta - (f_3 - f_0)} \qquad \text{C4}$$

Also the same transformation for every roots.
So (C1) becomes:

$$(\frac{d\varsigma}{ds})^2 = B(\varsigma - \beta_0)(\varsigma - \beta_1)(\varsigma - \beta_2)(\varsigma - \beta_3) \qquad \text{C5}$$

We also have following relation and let

$$\lambda = \frac{f_1 - f_0}{f_1 - f_2}\frac{f_3 - f_2}{f_3 - f_0} = \frac{\beta_1 - \beta_0}{\beta_1 - \beta_2}\frac{\beta_3 - \beta_2}{\beta_3 - \beta_0} \qquad \text{C6}$$

So (C5) is

$$(\frac{d\varsigma}{ds})^2 = B\varsigma(\varsigma - 1)(\lambda\varsigma - 1); B = A(f_3 - f_0)(f_2 - f_1) \qquad \text{C7}$$

This is the stand-form,but we can do further transform:

$$\varsigma = \xi^2 \qquad \text{C8}$$

So (C7) becomes the differential equation form of Jacobi elliptic function:

$$(\frac{d\xi}{ds})^2 = \frac{B}{4}(\xi^2 - 1)(\lambda\xi^2 - 1) \qquad \text{C9}$$

Sure, there are also much difficulties to overcome, but we don't want to say much about them because this is subject of elliptic function theory and further beyond the title of this paper, such as how to deal with degenerate roots and how to further transform $\lambda$ so that it is real and $0 < \lambda < 1,$so that (C9) is actually the Jacobi elliptic differential equation.

But instead (C4) and (C8), I can take a little different transformation which is related to the §6 of our paper. Instead map roots into $0, 1, \frac{1}{\lambda}$in (C7), and do quadratic transform in (C8), we can transform strightforword, which maps roots into:$1, -1, \frac{1}{k}, \frac{-1}{k}.$ So

$$\lambda = \frac{f_1 - f_0}{f_1 - f_2}\frac{f_3 - f_2}{f_3 - f_0} = \frac{\frac{1}{k} - 1}{\frac{1}{k} + 1}\frac{\frac{-1}{k} + 1}{\frac{-1}{k} - 1} = (\frac{1-k}{1+k})^2 \quad \text{C10}$$

So (C1) directly become to:

$$(\frac{d\xi}{ds})^2 = B'(\xi^2 - 1)(k^2\xi^2 - 1) \quad \text{C11}$$